\documentclass[
 reprint,
superscriptaddress,
 amsmath,amssymb,
 aps,
prx,
]{revtex4-2}
\usepackage{graphicx}% Include figure files
\usepackage{dcolumn}% Align table columns on the decimal point
\usepackage{bm}% bold math
\usepackage[mathlines]{lineno}% 
\usepackage[utf8]{inputenc}
\usepackage[T1]{fontenc}
\usepackage{mathptmx}
\usepackage{amsmath}
\usepackage{etoolbox}
\usepackage{nicefrac}
\usepackage[normalem]{ulem}
\DeclareMathAlphabet\mathbfcal{OMS}{cmsy}{b}{n}
\DeclareMathAlphabet{\mathcal}{OMS}{cmsy}{m}{n}
\usepackage{siunitx}
\usepackage{xcolor}
\usepackage{esint}
\usepackage{booktabs}
\DeclareSymbolFont{newfont}{OML}{cmm}{m}{it}% Computer Modern math font
\DeclareMathSymbol{\Epsilon}{0}{newfont}{15}% Symbol 15
\DeclareMathSymbol{\Rho}{0}{newfont}{37}% Symbol 37

\renewcommand\epsilon{\Epsilon}
\renewcommand\varrho{\Rho}

\usepackage{xr}
\renewcommand{\vec}[1]{\boldsymbol{#1}}

\newcommand{\hypgeo}[2]{%
  {\vphantom{F}}_{#1}\kern-\scriptspace F_{#2}%
}
\def\onedot{$\mathsurround0pt\ldotp$}
\def\cdddot#1{% three dots 
  \mathbin{\vcenter{\baselineskip.67ex
    \hbox{\onedot}\hbox{\onedot}\hbox{\onedot}%
  }}%
}
\definecolor{changed}{rgb}{0,0,0}
\definecolor{lightGrey}{rgb}{0.6,0.6,0.6}
\definecolor{darkGrey}{rgb}{0.3,0.3,0.3}
\definecolor{codegreen}{rgb}{0,0.6,0}
\definecolor{codegray}{rgb}{0.5,0.5,0.5}
\definecolor{codepurple}{rgb}{0.58,0,0.82}
\definecolor{backcolour}{rgb}{0,0,0}
\definecolor{mBlue}{rgb}{0.12, 0.47, 0.71}
\definecolor{mRed}{rgb}{0.69, 0.13, 0.13}
\makeatletter
\def\@email#1#2{%
 \endgroup
 \patchcmd{\titleblock@produce}
  {\frontmatter@RRAPformat}
  {\frontmatter@RRAPformat{\produce@RRAP{*#1\href{mailto:#2}{#2}}}\frontmatter@RRAPformat}
  {}{}
}%
\makeatother
\begin{document}
\title[]{Beyond the Electric Dipole Approximation: Electric and Magnetic Multipole Contributions Reveal Biaxial Water Structure from SFG Spectra
 at the Air-Water Interface}
% Force line breaks with \\
\author{Louis Lehmann}
\affiliation{Department of Physics, Freie Universität Berlin, Arnimallee 14, 14195 Berlin, Germany}

\author{Maximilian R. Becker}
\affiliation{Department of Physics, Freie Universität Berlin, Arnimallee 14, 14195 Berlin, Germany}

\author{Lucas Tepper}
\affiliation{Department of Physics, Freie Universität Berlin, Arnimallee 14, 14195 Berlin, Germany}

\author{Alexander P. Fellows}
\affiliation{Fritz-Haber-Institut der Max-Planck-Gesellschaft, Faradayweg 4-6, 14195 Berlin, Germany}

\author{Álvaro Díaz Duque}
\affiliation{Fritz-Haber-Institut der Max-Planck-Gesellschaft, Faradayweg 4-6, 14195 Berlin, Germany}

\author{Martin Thämer}
\affiliation{Fritz-Haber-Institut der Max-Planck-Gesellschaft, Faradayweg 4-6, 14195 Berlin, Germany}

\author{Martin Wolf}
\affiliation{Fritz-Haber-Institut der Max-Planck-Gesellschaft, Faradayweg 4-6, 14195 Berlin, Germany}

\author{Roland R. Netz$^*$}
\email{corresponding author: rnetz@physik.fu-berlin.de}
\affiliation{Department of Physics, Freie Universität Berlin, Arnimallee 14, 14195 Berlin, Germany}

\begin{abstract}
The interpretation of sum-frequency-generation (SFG) spectra has been severely limited by the absence of quantitative theoretical predictions of higher-order multipole contributions. 
Magnetic dipole and electric quadrupole contributions are determined by bulk properties but appear in all experimental SFG spectra, obscuring the connection between measured spectra and interfacial structure. 
We present the simulation-based framework to predict the full set of multipole spectral contributions. 
This framework also yields depth-resolved spectra, enabling the precise spatial localization of spectroscopic features.
Applied to the air-water interface, our approach achieves quantitative agreement with experimental spectra for different polarization combinations in both the bending and stretching regions.
Higher-order multipole contributions are crucial for correctly interpreting SFG spectra: in the bending band, the electric dipole and the magnetic dipole contributions have similar intensities, while the electric quadrupole contribution is significantly larger. 
In the OH-stretch region, the electric quadrupole contribution is found to be \textcolor{changed}{in large part} responsible for the characteristic shoulder at $\SI{3600}{cm^{-1}}$.
Crucially, subtracting the quadrupole and magnetic contributions isolates the second-order electric dipole susceptibility, which is a quantitative probe for interfacial molecular orientational anisotropy. This electric-dipole susceptibility reveals a pronounced biaxial ordering of water at the air-water interface.
By resolving a fundamental limitation of the interpretation of SFG spectroscopy, our framework allows for the detailed extraction of interfacial water ordering from SFG spectra.
\end{abstract}
\maketitle
\section{Introduction}
Interfaces are crucial in biological  \cite{heberleProtonMigrationMembrane1994,agmonProtonsHydroxideIons2016,laageWaterDynamicsHydration2017,renBiologicalMaterialInterfaces2019,saakBiologicalLipidHydration2024} and physicochemical  \cite{houGrapheneBasedElectrochemical2011,tahirElectrocatalyticOxygenEvolution2017,fabbriOxygenEvolutionReaction2018,tongElectrolysisLowgradeSaline2020,liUnconventionalStructuralEvolution2025} applications.
In particular, the air-water interface plays an essential role in catalysis  \cite{narayanWaterUniqueReactivity2005,gawandeBenignDesignCatalystfree2013,bjorneholmWaterInterfaces2016,ruiz-lopezMolecularReactionsAqueous2020,kusakaPhotochemicalReactionPhenol2021}, atmospheric  \cite{zhongAtmosphericSpectroscopyPhotochemistry2019,aultAerosolAcidityNovel2020,limmerMolecularInsightsChemical2024} and prebiotic  \cite{dobsonAtmosphericAerosolsPrebiotic2000,griffithSituObservationPeptide2012} chemistry. 
Undoubtedly, understanding the microscopic structure of interfaces is essential. The inhomogeneous interfacial region is only a few angstroms thick \cite{mcbainOpticalSurfaceThickness1939,beagleholeELLIPSOMETRYLIQUIDSURFACES1983,wangDeterminingSubnanometerThickness2019,fellowsHowThickAirWater2024}, or up to a few nanometers if the interface is charged \textcolor{changed}{and the salt concentration is small} \cite{gonellaSecondHarmonicSumFrequency2016,gonellaWaterChargedInterfaces2021a}, and thus much thinner than the bulk region.
This presents significant challenges in spectroscopy, as signals from the interface must be separated from \textcolor{changed}{the dominating} bulk signals. \\ \\ 
An elegant solution is to measure the second-order susceptibility, which is non-zero only when the spatial inversion symmetry is broken, using sum-frequency-generation (SFG) spectroscopy.
Typically, the bulk medium is isotropic and, therefore, the interface's structure accounts for a significant part of the SFG spectrum.
The interpretation of experimental spectra usually assumes that second-order radiation originates from an interfacial electric dipole layer oscillating at the sum of the applied beams' frequencies. 
In this approximation, the SFG signal is determined by a second-order electric dipole susceptibility   \cite{shenFundamentalsSumFrequencySpectroscopy2016a,moritaTheorySumFrequency2018a} and model calculations allow for the determination of the orientation of chemical bonds   \cite{zhuangMappingMolecularOrientation1999,weiMotionalEffectSurface2001,sunOrientationalDistributionFree2018a}, the effective interfacial dielectric constant \cite{yuPolarizationDependentHeterodyneDetectedSumFrequency2022,yuPolarizationDependentSumFrequencyGeneration2022a,chiangDielectricFunctionProfile2022a}, the position of the dipole layer  \cite{yuPolarizationDependentHeterodyneDetectedSumFrequency2022,yuPolarizationDependentSumFrequencyGeneration2022a}, the surface potential  \cite{geigerSecondHarmonicGeneration2009,pezzottiStructuralDefinitionBIL2018,raySecondHarmonicGenerationProvides2023,speelmanQuantifyingSternLayer2025a}, and water structure around macromolecules \cite{chenSpecificIonEffects2007,bruceNonadditiveIonEffects2019}
from experiments. \\ \\
However, this dipole-layer picture neglects higher-order multipole contributions present in the SFG signal. 
Those multipole contributions to SFG spectra are proportional to higher-order response functions that are non-zero even in isotropic bulk media \cite{pershanNonlinearOpticalProperties1963,adlerNonlinearOpticalFrequency1964,guyot-sionnestBulkContributionSurface1988,shiratoriTheoryQuadrupoleContributions2012,hiranoLocalFieldEffects2024} and cannot be easily experimentally separated from the measured signal. 
To determine the structure of an interface from SFG spectra, one needs to identify the electric dipole component, which serves as a fingerprint of the interface structure.
Thus, accurate theoretical predictions of higher-order multipole contributions are essential. \\ \\ 
Off-resonant multipole contributions, measured in second-harmonic generation, were predicted in previous works  \cite{yamaguchiElectricQuadrupoleContribution2011, lebretonMicroscopicViewPolarizationresolved2024}. 
Resonant multipole SFG contributions were estimated using normal-mode calculations  \cite{moriDevelopmentQuadrupoleSusceptibility2020} and found to be significant for the water bending band  \cite{kunduBendVibrationSurface2016}. 
However, normal-mode methods involve rather drastic approximations, including locality assumptions, and cannot predict the spectral line shape. 
Experimentally, the multipolar origin of the water bending band has been investigated with SFG at surfactant monolayers  \cite{sekiUnveilingHeterogeneityInterfacial2019,ahmedResolvingControversyDipole2020,mollDirectEvidenceSurface2021} and combined \textcolor{changed}{SFG and difference-frequency-generation (DFG)} measurements  \cite{fellowsImportanceLayerDependentMolecular2025b},
yielding conflicting results.
Many studies attribute the bending band to the electric dipole component of the SFG signal  \cite{nagataWaterBendingMode2013,niIRSFGVibrational2015,duttaAssignmentVibrationalSpectrum2017,sekiUnveilingHeterogeneityInterfacial2019,sekiDecodingMolecularWater2020,sekiDisentanglingSumFrequencyGeneration2021}, in contrast to our findings. \\ \\
\begin{figure*}
\centering \includegraphics[width=1\textwidth]{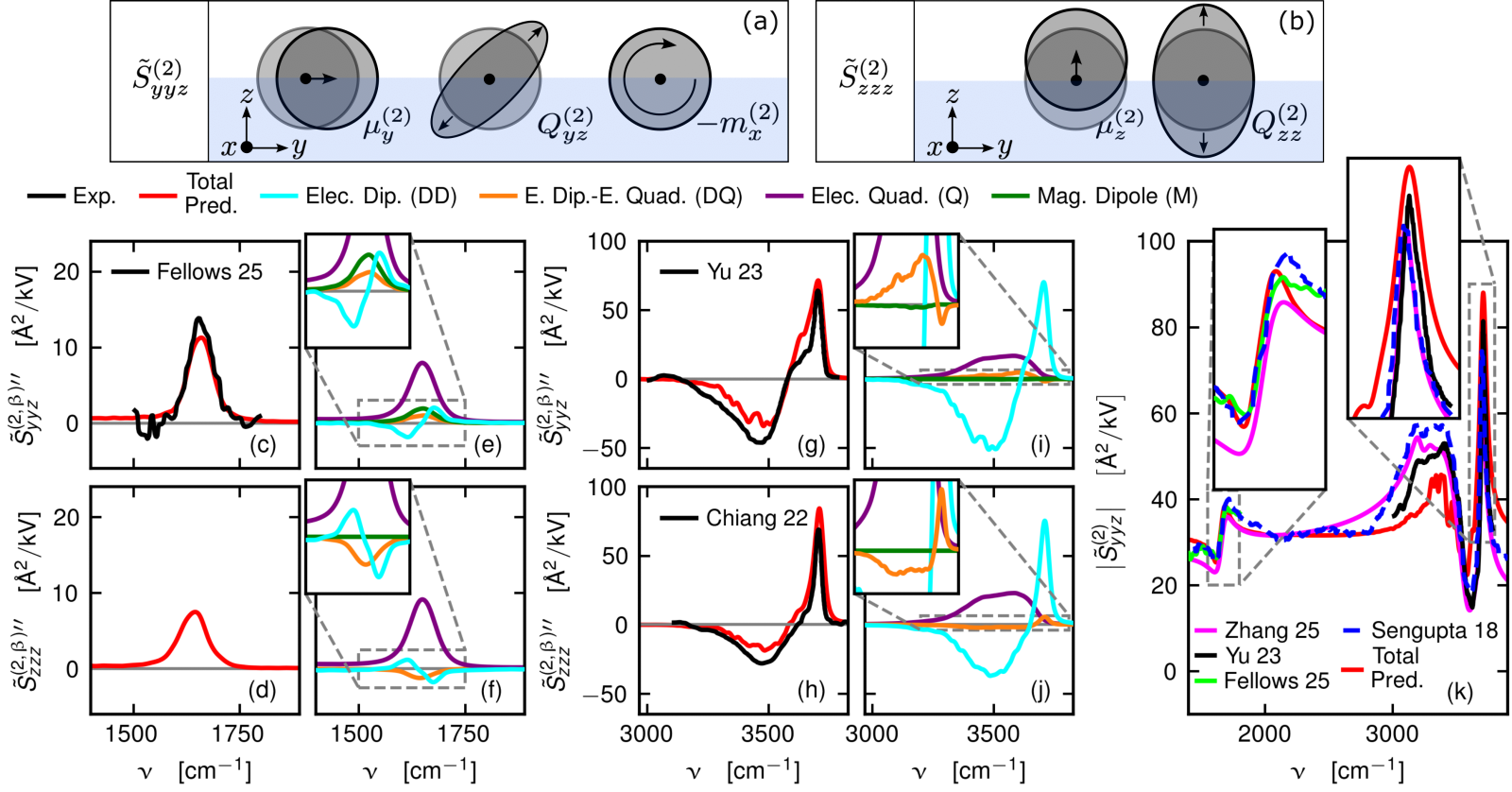}
\caption{SFG spectra and decomposition into multipole components. 
(a) \& (b): Sketch of the different second-order molecular multipole contributions to the total SFG spectra $\tilde{S}^{(2)\prime \prime}_{yyz}$ and $\tilde{S}^{(2)\prime \prime}_{zzz}$, respectively.
Here, $\mu_i^{(2)}$,  $Q_{ij}^{(2)}$ and  $m_i^{(2)}$ denote the induced molecular second-order electric dipole, electric quadrupole, and magnetic dipole moments, respectively.
The blue region represents water and the white region air.
(c): Comparison of predicted and experimental imaginary SFG spectra $\tilde{S}^{(2)\prime \prime}_{yyz}$ in the bending-frequency region. 
The SFG spectrum is decomposed into the pure electric dipole (DD), the electric dipole - electric quadrupole cross (DQ), the electric quadrupole (Q), and the magnetic dipole contribution (M) in (e). (d) \& (f): The same analysis for $\tilde{S}^{(2)\prime \prime}_{zzz}$.
Results for the OH-stretch frequency region are shown in (g)-(j). Experimental data is taken from Fellows \textit{et al.}  \cite{fellowsImportanceLayerDependentMolecular2025b}, Yu \textit{et al.}  \cite{yuFresnelFactorCorrection2023}, and Chiang \textit{et al.}  \cite{chiangDielectricFunctionProfile2022a}. \textcolor{changed}{The predicted absolute spectrum $\big|\tilde{S}^{(2)}_{yyz}\big|$  is compared with various published experimental data \cite{fellowsImportanceLayerDependentMolecular2025b,yuFresnelFactorCorrection2023,senguptaNeatWaterVapor2018a,zhangQuantitativeConsistencyIntensity2025a} in (k).} \textcolor{changed}{We red-shift our predictions for the bending and stretch bands by $\SI{28}{cm^{-1}}$ and $\SI{166}{cm^{-1}}$, respectively, to match the experiments. The boundary between the two shifted regions is set at $\SI{2500}{cm^{-1}}$.
} Grey dashed boxes indicate the regions shown in the insets.}
\label{fig:S2ijk}
\end{figure*}
Our method, based on time-dependent perturbation theory  \cite{kuboFluctuationdissipationTheorem1966,mukamelPrinciplesNonlinearOptical1995}, enables the quantitative prediction of multipolar SFG spectral contributions and constitutes a crucial step toward an accurate determination of the interfacial structure.
We compute all relevant multipole contributions, including the magnetic dipole contributions, which we find to be sizable and of similar intensity as the electric dipole contribution in the bending band. \\ \\
Following the work of Guyot-Sionnest and Shen  \cite{guyot-sionnestGeneralConsiderationsOptical1986}, we formulate the multipolar second-order electric current density as a response to electric displacement (D) fields  perpendicular to the interface and electric (E) fields parallel to the interface, both of which are spatially constant on the relevant length scale. 
In doing so, we eliminate the need to make \textit{ad-hoc} assumptions on the interface structure.
We observe quantitative agreement between simulated and experimental SFG spectra, with significant multipole contributions. \\ \\
Recent experiments investigated the interfacial dielectric constant  \cite{ chiangDielectricFunctionProfile2022a} and the spatial distribution of anisotropically oriented molecules  \cite{ yuPolarizationDependentHeterodyneDetectedSumFrequency2022,fellowsObtainingExtendedInsight2023a, fellowsHowThickAirWater2024}. 
Our work reveals that interfacial water exhibits an $\SI{8}{\angstrom}$ thick biaxial triple-layer structure, as follows from the second-order electric-dipole susceptibility. This susceptibility turns out to be dominated by
the biaxial orientational ordering perpendicular to the water molecular dipole axis in the bending region. \\ \\
\textcolor{changed}{In Section \ref{sec:Quantitative_Comparison}, we show that quantitative agreement between predicted and experimental SFG spectra is obtained once all relevant multipole contributions are considered. Subtracting all higher-order multipole contribution, the electric-dipole contribution is obtained, which in the bending band is demonstrated to reflect interfacial water orientation in Section \ref{sec:Bending_Band_Analysis}. 
In Section \ref{sec:biaxial_orientation}, we show that the interfacial water orientation is significantly biaxial. Sections \ref{sec:Stretch_Band_Analysis} and \ref{sec:Linear_Dielectric_Profile} present the spatially resolved SFG stretch spectrum and the spatially resolved absorption spectrum from the THz to the IR regime, respectively. 
Finally, we discuss the implications of our findings in Section \ref{sec:conclusion} and provide a description of our methods in Section \ref{sec:methods}.}
\section{Results}
\label{sec:Results}
\begin{figure*} \includegraphics[width=1\textwidth]{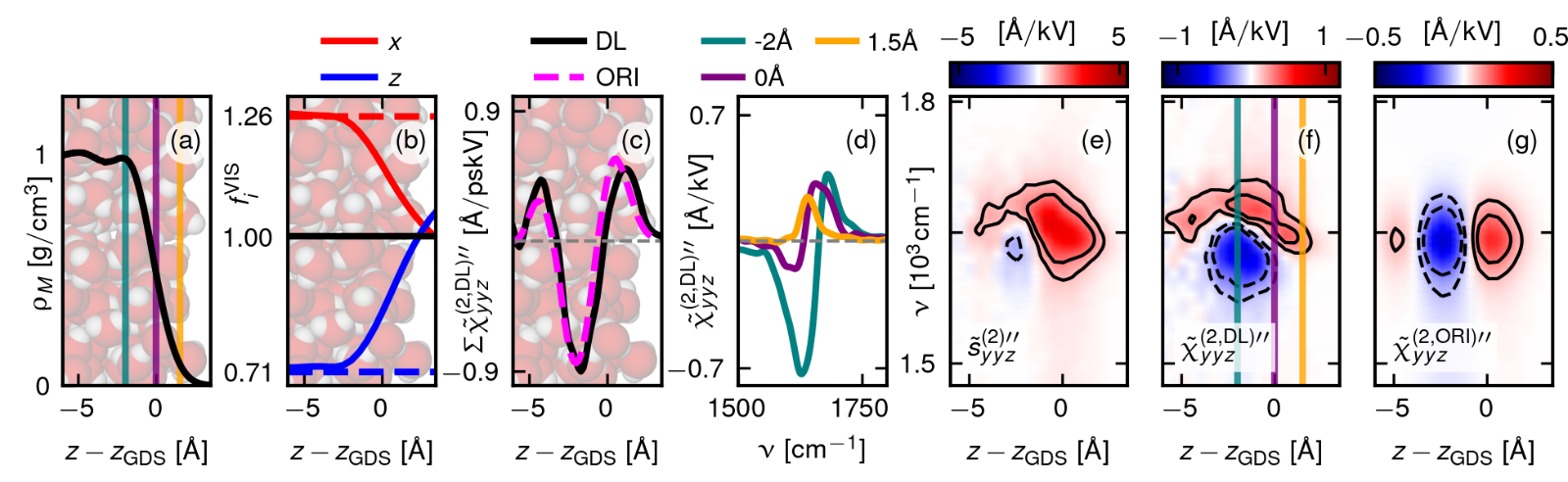}
 \caption{
Depth-dependent analysis of the bending band. The mass density profile is shown in (a), the local field factors in the visual frequency range $f^\mathrm{VIS}_x(z)$ and $f^\mathrm{VIS}_z(z)$ defined in Equation \eqref{eq:def_aver_loc_fac_l} in (b). Lorentz theory predictions in bulk water are denoted by dashed horizontal lines. The black solid line denotes the vacuum value.
 For illustrative purposes, a snapshot from the simulation is shown in the background.
 (c): Comparison of the spatially resolved integrals over the bending band, as defined in Equation \eqref{eq:def_integral_bend}, between $\tilde{\chi}^{(2,\mathrm{DL})\prime \prime}_{yyz}(z)$ and the prediction based solely on molecular orientation, $\tilde{\chi}^{(2,\mathrm{ORI})\prime \prime}_{yyz}(z)$, defined in Equation \eqref{eq:chi2ori}. 
 The electric dipole second-order susceptibility $\tilde{\chi}^{(2,\mathrm{DL})\prime \prime}_{yyz}(z)$, defined in Equation \eqref{eq:chi_2DL}, is shown at selected positions in (d). These positions are marked in (f), where the corresponding two-dimensional profile is presented. 
The second-order response profile of the SFG signal $\tilde{s}^{(2)\prime \prime}_{yyz}(z)$ is presented in (e) and $\tilde{\chi}^{(2,\mathrm{ORI})\prime \prime}_{yyz}(z)$ in (g). The spectra are red-shifted by $\SI{28}{cm^{-1}}$.
}
\label{fig:bending_zres}
\end{figure*}
\subsection{Quantitative Comparison of SFG Spectra with Experiments}
\label{sec:Quantitative_Comparison}
In Figure \ref{fig:S2ijk}, we present SFG spectra $\tilde{S}^{(2)}_{ijk}$ for the polarization combinations $yyz$ and $zzz$, as defined in Equations \eqref{eq:s2_ijk} and \eqref{eq:def_S2ijk}.
Here, the indices $ijk$ denote the polarizations of the second-order electric current density, which produces the observed radiation, the VIS field, and the IR field, respectively. 
The theoretical framework for predicting multipolar SFG spectra is described in detail in Supplemental Information (SI) Sections I–VI. \\ \\
We observe quantitative agreement between our predictions (red lines) and experiments (black lines) in the bending and stretching bands in Figure \ref{fig:S2ijk} (c), (g), (h), (k). 
We analyze the SFG spectrum by dissecting it into its multipole contributions $\tilde{S}^{(2)}_{ijk}=\tilde{S}^{(2,\mathrm{DD})}_{ijk} + \tilde{S}^{(2,\mathrm{DQ})}_{ijk} + \tilde{S}^{(2,\mathrm{Q})}_{ijk} +\tilde{S}^{(2,\mathrm{M})}_{ijk}$. 
This decomposition is described in Section \ref{sec:MM_Contributions}.
Here, $\tilde{S}^{(2,\mathrm{DD})}_{ijk}$ and $\tilde{S}^{(2,\mathrm{DQ})}_{ijk}$ are the pure electric dipole contribution and the electric dipole - electric quadrupole cross contribution, respectively. 
Importantly $\tilde{S}^{(2,\mathrm{DD})}_{ijk}$ describes the interfacial structure, while $\tilde{S}^{(2,\mathrm{DQ})}_{ijk}$ is created by the linear response of electric dipoles to fields from second-order electric quadrupoles; $\tilde{S}^{(2,\mathrm{Q})}_{ijk}$ and $\tilde{S}^{(2,\mathrm{M})}_{ijk}$ are the electric quadrupole and magnetic dipole contributions, which are in contrast
bulk contributions, independent of the interfacial structure.
The sum of these contributions $\tilde{S}^{(2,\mathrm{Q})}_{ijk} +\tilde{S}^{(2,\mathrm{M})}_{ijk}$ is called the interfacial quadrupole bulk contribution (IQB) in the literature \cite{shiratoriTheoryQuadrupoleContributions2012,hiranoLocalFieldEffects2024}. \\ \\ 
To give some intuitive understanding of the physical mechanisms producing the different multipolar SFG contributions we schematically illustrate the second-order molecular multipoles of the valence electron charge cloud in Figure  \ref{fig:S2ijk} (a) and (b). 
Figure \ref{fig:S2ijk} (a) shows the contributions relevant for a y-polarized and (b) for a z-polarized SFG field. 
The second-order molecular electric dipole moment $\mu_i^{(2)}$ is characterized by oscillations of the displacement of the center of the electronic charge distribution and is zero in isotropic media.
In contrast, the second-order electric quadrupole moment, characterized by oscillations of the charge distribution width, and the molecular magnetic dipole moment, characterized by oscillations in its angular momentum, as illustrated in Figure \ref{fig:S2ijk} (a) and (b), are nonzero in bulk.  
The second-order electric quadrupole and magnetic dipole moments consist of electric currents distributions with vanishing mean and contribute to the second-order current density via the gradients of their spatial distributions. \\ \\
The bending band imaginary SFG spectra in Figure \ref{fig:S2ijk} (c)-(f) are dominated by the electric quadrupole contribution (purple), which consists of a positive broad band, 
\textcolor{changed}{in agreement with previous normal mode calculations \cite{kunduBendVibrationSurface2016}.}
The pure electric dipole contribution consists of a rather weak negative-positive double peak (Fig. \ref{fig:S2ijk} (e) \& (f)). 
The observed lineshape of $\tilde{S}^{(2,\mathrm{DD})\prime \prime}_{yyz}$ qualitatively aligns with previous simulation studies  \cite{nagataWaterBendingMode2013,niIRSFGVibrational2015,sekiDecodingMolecularWater2020}, which did not account for higher-order multipole contributions. 
The magnetic dipole contribution appears only in the $yyz$ spectrum and exhibits an intensity similar to the pure electric dipole contribution (Fig. \ref{fig:S2ijk} (e)).  \\ \\ 
The OH-stretch band imaginary SFG spectra are presented in Figure \ref{fig:S2ijk}  (g)-(j).
The pure electric dipole contribution is characterized by a strong negative–positive double peak, in shape similar to the bending band, as can be seen by comparing the cyan lines in Figure \ref{fig:S2ijk} (e) \& (i). 
The difference is that the frequency splitting between the positive and negative components is much larger compared to the spectral linewidth in the OH-stretch region, thus less cancellation takes place and the resulting electric dipole contribution is significantly stronger. 
The positive signal arising from the free OH stretch vibrations at $\SI{3700}{cm^{-1}}$ is largely determined by the pure electric dipole contribution (Fig. \ref{fig:S2ijk} (i) \& (j)), while significant electric quadrupole contributions to the SFG spectra arise from the more slowly oscillating OH bonds between $\SI{3100}{cm^{-1}}$ and $\SI{3650}{cm^{-1}}$. 
In $\tilde{S}^{(2)\prime\prime}_{yyz}$, the electric quadrupole produces most of the positive shoulder at $\SI{3600}{cm^{-1}}$ (Fig. \ref{fig:S2ijk} (i)). \textcolor{changed}{ Comparing $\tilde{S}^{(2,\mathrm{DD})\prime \prime}_{yyz}$ (Fig. \ref{fig:S2ijk} (i)) and $\tilde{S}^{(2,\mathrm{DD})\prime \prime}_{zzz}$ (Fig. \ref{fig:S2ijk} (h)), one sees that $\tilde{S}^{(2,\mathrm{DD})\prime \prime}_{yyz}$ exhibits a small positive shoulder around $\SI{3600}{cm^{-1}}$, whereas $\tilde{S}^{(2,\mathrm{DD})\prime \prime}_{zzz}$ does not. 
This difference is qualitatively consistent with previous predictions, that were based on electric-dipole approximations \cite{mobergTemperatureDependenceAir2018,chiangDielectricFunctionProfile2022a,yuFresnelFactorCorrection2023}.}
In all spectra presented, $\tilde{S}^{(2,\mathrm{DQ})\prime \prime }_{ijk}$ is small, and $\tilde{S}^{(2,\mathrm{DQ})\prime \prime}_{yyz}$ is mainly positive, while $\tilde{S}^{(2,\mathrm{DQ})\prime \prime}_{zzz}$ is mainly negative. \\ \\
\textcolor{changed}{In Figure~\ref{fig:S2ijk} (k), we compare several experimental measurements of the absolute SFG spectrum $\big|\tilde{S}^{(2)}_{yyz}\big|=\sqrt{\left(\tilde{S}^{(2)\prime}_{yyz} \right)^2+\left(\tilde{S}^{(2)\prime \prime}_{yyz}\right)^2}$ \cite{senguptaNeatWaterVapor2018a,yuFresnelFactorCorrection2023,fellowsImportanceLayerDependentMolecular2025b,zhangQuantitativeConsistencyIntensity2025a} with our prediction (red line), showing close agreement within the spread of experimental data across the full frequency range. This comparison also provides an estimate of the experimental uncertainty.}
 \\ \\
\\ \\
Our predicted spectra have been red-shifted to ensure alignment between theory and experiment. 
This is reasonable since it has been demonstrated previously that the MB-Pol water model used by us reproduces the experimental vibrational frequencies when nuclear quantum effects are considered  \cite{meddersInfraredRamanSpectroscopy2015,meddersDissectingMolecularStructure2016,rashmiDissectingMolecularStructure2025}, while for the linear absorption spectra, it has been demonstrated that nuclear quantum effects do not alter the line shape and absolute intensities much  \cite{burnhamVibrationalProtonPotential2008,meddersDissectingMolecularStructure2016}. \textcolor{changed}{Quantum correction factors do in fact not appear in spectra to leading order as discussed in the Methods.} \\ \\
\textcolor{changed}{Our results reveal significant higher-order multipole contributions. Next, we show how the interfacial electric-dipole contribution in the bending band reflects molecular orientation at the air–water interface.}
\subsection{Spatially Resolved Second-Order Response Profile of the Bending Band}
\label{sec:Bending_Band_Analysis}
Figure \ref{fig:bending_zres} presents the depth-dependent second-order response profile of the bending band, relative to the Gibbs dividing surface position $z_\mathrm{GDS}$  \cite{scalfiPropensityHydroxideHydronium2024}. 
Figure \ref{fig:bending_zres} (a) presents the mass density profile with a snapshot of the simulation box in the background. The density transitions from $\SI{1}{g/cm^{3}}$ in bulk to essentially zero over a range of about $\SI{5}{\angstrom}$, centered around $z_\mathrm{GDS}$. \\ \\
\begin{figure*} \includegraphics[width=1\textwidth]{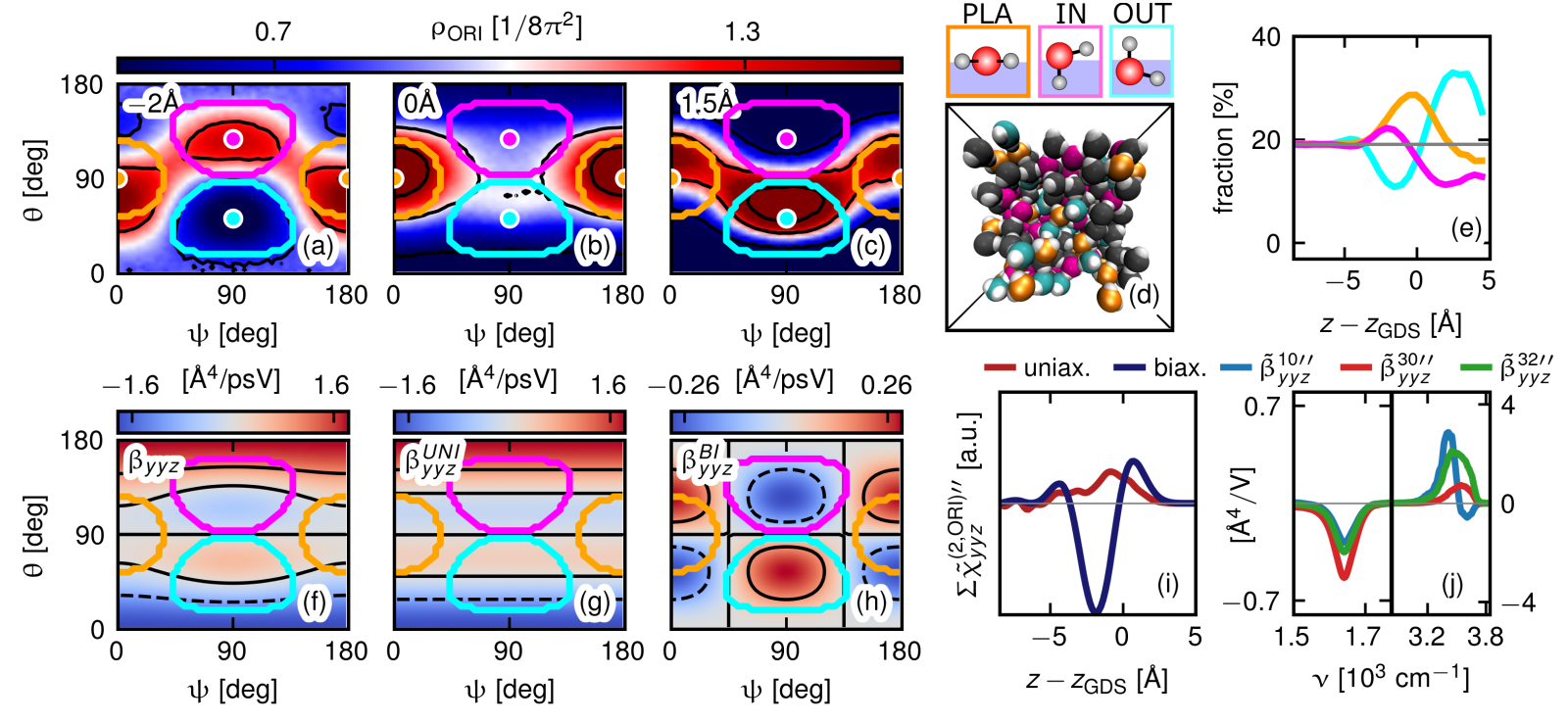}
 \caption{The orientation distribution function $\rho_\mathrm{ORI}(\theta, \psi)$ is shown for the same positions relative to $z_\mathrm{GDS}$ as in Figure \ref{fig:bending_zres} in (a)-(c). 
Three orientational species are distinguished: planar, pointing in, and pointing out, as illustrated at the top of panel (d). These orientations are marked with dots and contours in (a)-(c). 
Additionally, a snapshot of the simulation box viewed from the air is provided in (d), where molecules are color-coded according to their orientational species.  (e) displays the fraction of each orientational species as a function of $z$.
The molecular hyperpolarizability tensor component $\tilde{\beta}_{yyz}(\theta,\psi)$, defined in Equation \eqref{eq:def_beta_ijk}, is shown as a function of the molecular orientation in (f). 
The color indicates the intensity at bending-band frequencies, obtained by integrating the imaginary part $\tilde{\beta}''_{yyz}(\theta,\psi)$ using the same integration boundaries as in Equation~\eqref{eq:def_integral_bend}. This intensity is decomposed into its uniaxial and biaxial components, defined in Equations~\eqref{eq:def_beta_uniax} and~\eqref{eq:def_beta_biax} and shown in (g) and (h), respectively.
The position-resolved profile of the bending band is decomposed into uniaxial and biaxial contributions in (i). The frequency-dependent coefficients $\tilde{\beta}^{lm\prime \prime}_{yyz}$ of $\tilde{\chi}^{(2,\mathrm{ORI})\prime \prime}_{yyz}$ defined in Equation \eqref{eq:def_beta}-\eqref{eq:def_beta_biax} are presented in (j).}
\label{fig:bending_ori}
\end{figure*}
We present the second-order electric dipole susceptibility $\tilde\chi^{(2,\mathrm{DL})}_{ijk}(z)$, which connects the average local electric E-field at the molecular centers to the resulting second-order electric dipole current density in Figure \ref{fig:bending_zres} (d) \& (f). 
The superscript DL stands for dipole and local to distinguish it from regular susceptibilities, which are defined as the response of the macroscopic polarization density to macroscopic electric E-fields  \cite{bornPrinciplesOpticsElectromagnetic1999}. Though not directly experimentally measurable,
$\tilde\chi^{(2,\mathrm{DL})}_{ijk}(z)$  is important since it links the macroscopic second-order electric dipole response to molecular orientation  \cite{armstrongInteractionsLightWaves1962,shenFundamentalsSumFrequencySpectroscopy2016a,moritaTheorySumFrequency2018a}. 
In defining $\tilde\chi^{(2,\mathrm{DL})}_{ijk}(z)$ via Equation \eqref{eq:chi_2DL}, we account for the strength of the local E-field by dividing the second-order response of the electric dipole density $\tilde{s}^{(2,\mathrm{DD})}_{ijk}(z)$, by the local field factors, i.e. the ratios between the amplitudes of the local E-fields $\mathcal{E}^{\mathrm{L},\alpha}_i(z)$ acting on the molecular centers and the amplitudes of the external fields $\mathcal{F}^\alpha_i$, which correspond to D-fields for $i=z$ and to E-fields for $i=x$ or $y$.  Here, $\alpha$ specifies the frequency of the three fields involved, i.e. $\alpha\in\lbrace \mathrm{SFG}, \mathrm{VIS}, \mathrm{IR}\rbrace$. The spatial integral over the second-order response profile $\tilde{s}^{(2,\mathrm{DD})}_{ijk}(z)$ in Equation \eqref{eq:S2ijk_beta} determines the pure electric dipole contribution to the experimentally measurable SFG spectrum $\tilde{S}^{(2,\mathrm{DD})}_{ijk}$; it is related to $\tilde{\chi}^{(2,\mathrm{DL})}_{ijk}(z)$ by
\begin{align} \tilde{s}^{(2,\mathrm{DD})}_{ijk}\left( z  \right)  = f^\mathrm{SFG}_{i}(z ) f^\mathrm{VIS}_{j}(z) f^\mathrm{IR}_{k}(z) \tilde{\chi}^{(2,\mathrm{DL})}_{ijk}\left( z  \right) \, ,
  \label{eq:s2_chi2}
  \end{align} 
where $f^\alpha_i(z)$ are the laterally averaged local field factors, defined by
\begin{align}
     f^\alpha_i(z)= \frac{\mathcal{E}^\mathrm{L,\alpha}_i(z) }{\mathcal{F}^\alpha_i(z)}  \, .
    \label{eq:def_aver_loc_fac_l}
\end{align} 
The local field factors in the optical frequency range $f_i^\mathrm{VIS}(z)$ are presented in Figure \ref{fig:bending_zres} (b). Interestingly, $f_x^\mathrm{VIS}(z)$ exhibits a line shape similar to the mass density profile, while $f_z^\mathrm{VIS}(z)$ is markedly different.
In bulk, we recover values dictated by the Lorentz field approximation  \cite{armstrongInteractionsLightWaves1962,moritaTheorySumFrequency2018a} $f^\mathrm{VIS}_x(-\infty)\approx(2+\tilde \varepsilon^\mathrm{VIS})/3$ and $f^\mathrm{VIS}_z(-\infty) \approx \left(2+\tilde \varepsilon^\mathrm{VIS} \right)/(3\tilde \varepsilon^\mathrm{VIS})$, as indicated by the horizontal dashed lines. Here, $\tilde \varepsilon^\mathrm{VIS}=1.77$ represents the optical dielectric constant of bulk water, extracted from the plateau value of the dielectric profile presented in Figure \ref{fig:eps} (b). 
These findings are qualitatively consistent with the prediction of $f^\mathrm{VIS}_i(z)$ by Shiratori and Morita  \cite{shiratoriMolecularTheoryDielectric2011}. \\ \\
We compare the imaginary second-order response profile $\tilde{s}^{(2)\prime \prime}_{yyz}(z)$ defined in Equation \eqref{eq:s2_ijk} with the second-order electric dipole susceptibility $\tilde{\chi}^{(2,\mathrm{DL})\prime \prime}_{yyz}(z)$ defined in Equation \eqref{eq:chi_2DL}, as well as its prediction based solely on molecular orientation, $\tilde{\chi}^{(2,\mathrm{ORI})\prime \prime}_{yyz}(z)$, in Figure \ref{fig:bending_zres} (e)-(g). 
It is evident that all profiles exhibit a positive contribution for values of $z-z_\mathrm{GDS}\geq 0$ (in red) and a negative contribution (in blue) centered around $z-z_\mathrm{GDS} = \SI{-2}{\angstrom}$. 
We calculate $\tilde{\chi}^{(2,\mathrm{ORI})}_{ijk}(z)$ by summing the molecular hyperpolarizability tensor contributions assigned to each water molecule in its molecular frame, as defined in Equation \eqref{eq:chi2ori}, where 
the molecular hyperpolarizability tensor is extracted from a simulation of bulk water as described in SI Section VII. 
By this, the only interface-specific input to $\tilde{\chi}^{(2,\mathrm{ORI})}_{ijk}(z)$ is the molecular orientation distribution, allowing us to test whether $\tilde{\chi}^{(2,\mathrm{DL})}_{ijk}(z)$ is a marker of the interfacial orientation anisotropy.
To compare the amplitudes of $\tilde{\chi}^{(2,\mathrm{DL})}_{yyz}(z)$ and $\tilde{\chi}^{(2,\mathrm{ORI})}_{yyz}(z)$ we integrate over the bending band according to
\begin{align}
\Sigma \tilde{\chi}^{(2,\mathrm{DL})\prime \prime}_{ijk} (z) = c_0\int\limits_{\nu_1}^{\nu_2} \mathrm{d} \nu \,\tilde{\chi}^{{(2,\mathrm{DL}})\prime \prime}_{ijk}(z,\nu) 
\label{eq:def_integral_bend}
\end{align}
from $\nu_1=\SI{1507}{cm^{-1}}$ to $\nu_2=\SI{1772}{cm^{-1}}$. Here, $c_0$ is the speed of light in vacuum. We find quantitative agreement between $\Sigma \tilde{\chi}^{(2,\mathrm{DL})\prime \prime}_{yyz}(z)$ and $\Sigma \tilde{\chi}^{(2,\mathrm{ORI})\prime \prime}_{yyz}(z)$ in Figure \ref{fig:bending_zres} (c),
demonstrating that $\Sigma \tilde{\chi}^{(2,\mathrm{DL})\prime \prime}_{yyz}(z)$ indeed primarily reflects interfacial orientation within the bending frequency range. 
The characteristic triple-layer structure of $\Sigma \tilde{\chi}^{(2,\mathrm{DL})\prime \prime}_{yyz}(z)$ will be discusses further below.
By comparing $\tilde{\chi}^{(2,\mathrm{DL})\prime \prime}_{yyz}(z)$ (Fig. \ref{fig:bending_zres} (d) \& (f)) with $\tilde{\chi}^{(2,\mathrm{ORI})\prime \prime}_{yyz}(z)$ (Fig. \ref{fig:bending_zres} (g)), it becomes clear that $\tilde{\chi}^{(2,\mathrm{ORI})\prime \prime}_{yyz}(z)$ does not reproduce the $z$-dependent frequency shift, present in $\tilde{\chi}^{(2,\mathrm{DL})\prime \prime}_{yyz}(z)$,
which is thus due to the $z$-dependent anisotropic coordination of interfacial water. 
\subsection{Spectral Signature of Biaxial Interfacial Water Ordering}
\label{sec:biaxial_orientation}
After demonstrating that the second-order electric dipole susceptibility $\tilde{\chi}^{(2,\mathrm{DL})}_{yyz}$ serves as a quantitative probe of the interfacial orientation of molecules, we now analyze the interfacial orientation distribution. \\ \\ The orientation distribution function $\rho_\mathrm{ORI}(\theta,\psi)$, defined in Equation \eqref{eq:def_odf}, is presented at three depth $z-z_\mathrm{GDS}$ in Figure \ref{fig:bending_ori} (a)-(c). 
This function depends on the two Euler angles, $\theta$ and $\psi$.
The angle $\theta$ describes the tilt of the molecular dipole axis relative to the surface normal: for $\theta=0$, the oxygen atom is oriented towards bulk water, and for $\theta=180^\circ$ towards air.
The angle $\psi$ describes the molecular orientation around the dipole axis.
The interfacial water structure is significantly biaxial, as revealed by the pronounced $\psi$-dependence of $\rho_\mathrm{ORI}(\theta,\psi)$. \\ \\
We distinguish three orientational species: the planar-oriented species (orange), where both OH-bond vectors lie in the interfacial plane, the inward-oriented species (magenta), where one OH-bond is orthogonal to the interfacial plane and pointing inward, and the outward-oriented species (cyan), where one OH-bond is orthogonal to the interfacial plane and pointing outward. 
These orientations are sketched on top of Figure \ref{fig:bending_ori} (d) and the corresponding Euler angles are marked with color-coded dots and contours (marking water molecules with angular deviation of $36^\circ$ from the idealized orientations) in the plots of $\rho_{\mathrm{ORI}}(\theta,\psi)$ in Figure \ref{fig:bending_ori} (a)-(c).
A snapshot of the air-water interface viewed from air is presented in Figure \ref{fig:bending_ori} (d); here, the orientational species are color-coded.  \\ \\
As observed in Figure \ref{fig:bending_ori} (a), at $z-z_\mathrm{GDS}=\SI{-2}{\angstrom}$, the inward-oriented species dominates; at $z=z_\mathrm{GDS}$, the planar species prevails (Fig. \ref{fig:bending_ori} (b)), and finally, close to the vapor region at $z-z_\mathrm{GDS}=\SI{1.5}{\angstrom}$, the outward-oriented species becomes dominant  (Fig. \ref{fig:bending_ori} (c)). 
We quantify this observation via the molecular number fraction profiles, which we present in Figure \ref{fig:bending_ori} (e).
Here, we see that water is isotropic $\SI{5}{\angstrom}$ below $z_\mathrm{GDS}$, as the fraction of all orientational species is approximately $\SI{20}{\%}$. 
The inward-oriented fraction peaks at $z-z_\mathrm{GDS}=\SI{-2}{\angstrom}$, at the location of the negative peak of $\tilde{\chi}^{(2,\mathrm{DL})}_{ijk}(z)$ in Figure \ref{fig:bending_zres} (f). The planar-oriented and outward-oriented species peak at $z=z_\mathrm{GDS}$ and near the vapor region at $z-z_\mathrm{GDS}=\SI{2.5}{\angstrom}$, respectively. \\ \\
The uniaxial approximation, which corresponds to averaging the orientation distribution function over the angle $\psi$, neglects the biaxial water ordering. To reveal the spectral effects of water biaxial ordering, we define the orientation and position dependent molecular hyperpolarizability $\tilde{\beta}_{ijk}(z,\theta, \psi)$ by
\begin{multline}
   \varepsilon_0^{-1} \mu_i^{(2)}(t)=e^{-i\omega^\mathrm{SFG}t} \tilde{\beta}_{ijk}(z,\theta, \psi) \mathcal{E}^{\mathrm{L,VIS}}_j(z) \mathcal{E}^{\mathrm{L,IR}}_k(z) \\ + c.c. \, ,
    \label{eq:def_beta_ijk} 
\end{multline}
where $\mu_i^{(2)}(t)$ is the second-order molecular electric dipole moment, $\mathcal{E}^{\mathrm{L,VIS}}_j(z)$ and  $\mathcal{E}^{\mathrm{L,IR}}_k(z)$ are the amplitudes of the local E-fields acting on the molecular center at the position $z$,  and $\theta$ and $\psi$ are the Euler angles, specifying the molecular orientation. Note that $\mu_i^{(2)}(t)$ is a source dipole that induces an additional linear response, as explained in SI Section IV B. 
We extract $\tilde{\beta}_{yyz}(\theta, \psi)$, from simulation trajectories of bulk water, as described in SI Section VII. 
In Figure \ref{fig:bending_ori} (f)  we present the integral of the imaginary part $\tilde{\beta}''_{yyz}(\theta,\psi)$ over the same integration boundaries as in Equation \eqref{eq:def_integral_bend}, to quantify the orientation-dependent hyperpolarizability in the bending band.
We decompose $\tilde{\beta}_{yyz}(\theta, \psi )$ into uniaxial and biaxial components according to
\begin{align}
\tilde{\beta}_{ijk}(\theta, \psi ) &= \tilde{\beta}^{\mathrm{UNI}}_{ijk}(\theta ) + \tilde{\beta}^{\mathrm{BI}}_{ijk}(\theta, \psi ) 
\label{eq:def_beta} \\ 
\tilde{\beta}^\mathrm{UNI}_{ijk}(\theta ) &= q_{10}(\theta) \tilde{\beta}^{10}_{ijk} + q_{30}(\theta) \tilde{\beta}^{30}_{ijk} 
\label{eq:def_beta_uniax} \\
\tilde{\beta}^{\mathrm{BI}}_{ijk}(\theta, \psi ) &= q_{32}(\theta, \psi)  \tilde{\beta}^{32}_{ijk}
\label{eq:def_beta_biax}
\end{align}
shown in Figure \ref{fig:bending_ori} (g)  \&  (h), respectively. 
Here, $q_{10}(\theta)=\cos{\theta}$, $q_{30}(\theta)=\frac{1}{2}\left[ 5 \cos{^3\theta} - 3 \cos{\theta} \right]$ are the first and third Legendre polynomials specifying the dipole distribution and $q_{32}(\theta, \psi)=\cos{\theta} \sin{\theta}^2  \left( \cos{^2\psi}- \sin{^2\psi} \right)$ accounts for the contribution due to the molecular biaxiality. 
Equations \eqref{eq:def_beta}–\eqref{eq:def_beta_biax} express the rotation of the molecular hyperpolarizability into the laboratory frame in an exact way.
The frequency-dependent imaginary parts of the coefficients $\tilde \beta^{10}_{yyz}$, $\tilde \beta^{30}_{yyz}$ and $\tilde \beta^{32}_{yyz}$ are presented in Figure \ref{fig:bending_ori} (j), where we observe that the biaxial contribution $\tilde \beta^{32}_{yyz}$ is significant. \\ \\
We observe that the negative area (blue) at $\theta>90^\circ$ in $\tilde{\beta}_{yyz}(\theta,\psi)$ in Figure \ref{fig:bending_ori} (f) aligns with the inward pointing orientation dominant at $\SI{-2}{\angstrom}$ in Figure \ref{fig:bending_ori} (a) and the positive area (red) at $\theta<90^\circ$ in Figure \ref{fig:bending_ori} (f) aligns with the outwards pointing orientation dominant at $\SI{1.5}{\angstrom}$ in Figure \ref{fig:bending_ori} (c). 
As seen in Figure \ref{fig:bending_ori} (g), the uniaxial contribution changes sign  within the domain of each orientational species, suggesting that uniaxial water ordering by itself does not lead to a SFG signal in the bending region, as will be rigorously demonstrated below. In contrast, the biaxial component projects precisely out the difference between outward- and inward-oriented molecules (Fig. \ref{fig:bending_ori} (h)). 
We note that planar-oriented molecules are not detected by SFG spectroscopy (neither uniaxial nor biaxial), because their mirror plane lies parallel to the interface, making them inversion symmetric.
We dissect the integral over the bending band $\Sigma \tilde{\chi}^{(2,\mathrm{ORI})\prime \prime}_{yyz}(z)$, already presented in Figure \ref{fig:bending_zres} (c), into biaxial and uniaxial contributions in Figure \ref{fig:bending_ori} (i), demonstrating that $\Sigma \tilde{\chi}^{(2,\mathrm{ORI})\prime \prime}_{yyz}(z)$ is dominated by the biaxiality of the orientation distribution. 
\textcolor{changed}{Thus, our analysis of the simulation trajectories shows that interfacial water is significantly biaxial and that the sign of the bending-mode dipole response is an unambiguous indicator of the relative populations of biaxially ordered inward- and outward-oriented water molecules illustrated in Figure~\ref{fig:bending_ori} (d).}
\subsection{Spatially Resolved Second-Order Response Profile of the Stretch Band}
\label{sec:Stretch_Band_Analysis}
 \begin{figure*}
\centering
\includegraphics[width=1\textwidth]{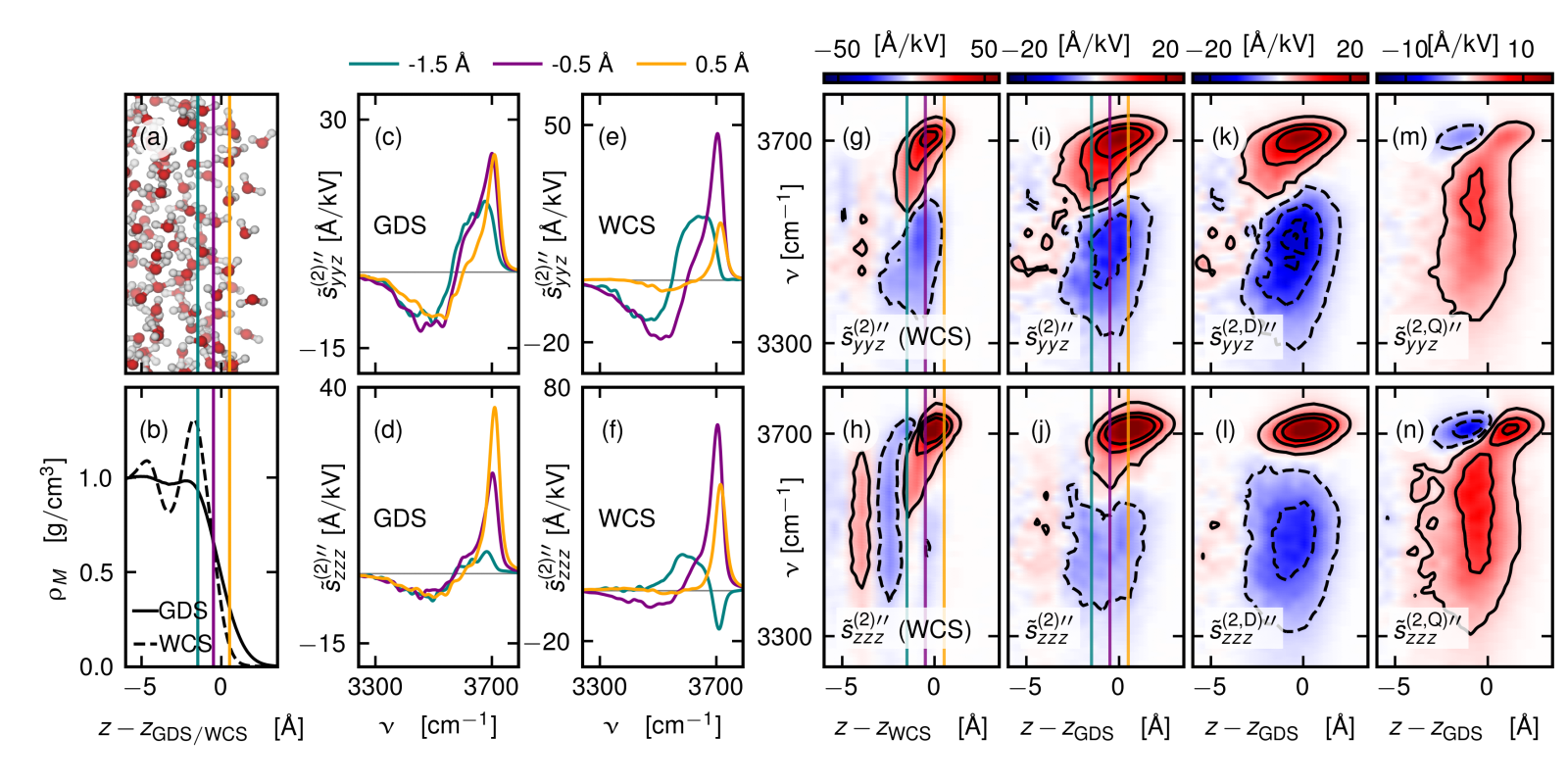}
\caption{
Second-order response profile $\tilde{s}^{(2) \prime \prime}_{ijk}\left( z \right)$ of the OH-stretch band as defined in Equation \eqref{eq:s2_ijk} and the decomposition into its multipole components $\tilde{s}^{(2,\beta)\prime \prime }_{ijk}\left( z \right)$ as defined in Equation \eqref{eq:s2_ijk_beta}. A snapshot of the simulation box is shown in (a).
The mass-density profiles with respect to $z_\mathrm{GDS}$ and $z_\mathrm{WCS}$ are presented in (b).
Slices of $\tilde{s}^{(2) \prime \prime}_{ijk}\left( z \right)$ are shown for selected positions in (c) \& (d) relative to $z_\mathrm{GDS}$ and in (e) \& (f) relative to $z_\mathrm{WCS}$, colored vertical lines in (a), (b), (g)-(j) indicate the positions. The profiles $\tilde{s}^{(2)\prime \prime}_{ijk}\left( z \right)$ are shown relative to $z_\mathrm{WCS}$ in (g) \& (h) and relative to  $z_\mathrm{GDS}$ in (i) \& (j). The profile relative to $z_\mathrm{GDS}$ is dissected into its molecular multipole components $\tilde{s}^{(2,\mathrm{D}) \prime \prime}_{ijk}\left( z \right)$ and $\tilde{s}^{(2,\mathrm{Q}) \prime \prime}_{ijk}\left( z \right)$ in (k) \& (l) and (m) \& (n), respectively. All spectra are red-shifted by $\SI{166}{cm^{-1}}$.
}
\label{fig:s2ijk}
\end{figure*}
We present the imaginary part of the second-order response profile $\tilde{s}^{(2)\prime \prime}_{ijk}(z)$ in the OH-stretch frequency region,  defined in Equation \eqref{eq:s2_ijk}, together with a decomposition into the electric dipole and electric quadrupole contributions according to Equation \eqref{eq:s2_ijk_beta} in Figure \ref{fig:s2ijk}. 
We map the waters center of mass positions in the calculation of $\tilde{s}^{(2)}_{ijk}(z)$ relative to the non-planar Willard-Chandler surface $z_\mathrm{WCS}$   \cite{willardInstantaneousLiquidInterfaces2010} in Figure \ref{fig:s2ijk} (e)-(h) and relative to the planar Gibbs dividing surface $z_\mathrm{GDS}$ in Figure \ref{fig:s2ijk} (c) \& (d) and (i)-(n).
We show a snapshot of the simulation box in Figure \ref{fig:s2ijk} (a).
In Figure \ref{fig:s2ijk} (b),  we present the mass-density profiles relative to $z_\mathrm{GDS}$ and $z_\mathrm{WCS}$. 
In the latter presentation the density profile strongly oscillates, while in the laboratory frame these oscillations are washed out due to the intrinsic roughness of the air-water interface  \cite{sedlmeierNanoroughnessIntrinsicDensity2009b}. \\ \\
Slices of $\tilde{s}^{(2)\prime \prime}_{ijk}(z)$ at selected positions $z-z_{\mathrm{GDS}}$ and $z-z_{\mathrm{WCS}}$ (denoted by the colored lines in (b)) are presented in Figure \ref{fig:s2ijk} (c) \& (d) and (e) \& (f).
Close to the vapor at $z-z_{\mathrm{GDS/WCS}}=\SI{0.5}{\angstrom}$ (orange lines), almost only the free OH stretch peak is visible at $\SI{3700}{cm^{-1}}$.
At $z-z_\mathrm{GDS/WCS}=\SI{-0.5}{\angstrom}$ (purple lines), $\tilde{s}^{(2)\prime \prime}_{ijk}(z)$ resembles the integrated spectra presented in Figure \ref{fig:S2ijk} (g) \& (h).
Finally, the free OH stretch contributions vanish near the bulk at $z-z_\mathrm{GDS/WCS}=\SI{-1.5}{\angstrom}$ (turquoise lines), and we only see the shoulder at $\SI{3600}{cm^{-1}}$ and the negative component at $\SI{3500}{cm^{-1}}$.
As shown in Figure \ref{fig:s2ijk} (h), $\tilde s^{(2)\prime\prime}_{zzz}(z)$ relative to $z_\mathrm{WCS}$ almost perfectly follows the oscillations of the mass density profile in Figure \ref{fig:s2ijk} (b). \\ \\
The imaginary part of the second-order response profile in the laboratory frame $\tilde{s}^{(2)\prime \prime}_{ijk}(z)$ is presented in Figure \ref{fig:s2ijk} (c) \& (d) and (i) \& (j).
Here, both the free OH stretch band at $\SI{3700}{cm^{-1}}$ and the negative contribution at $\SI{3500}{cm^{-1}}$ are located at approximately $z \approx z_\mathrm{GDS}$.
However, the shoulder is created below $z_\mathrm{GDS}$, and the free-OH contributions reaches slightly more into the vapor region.
The profiles of the second-order response of the electric dipole contribution $\tilde{s}^{(2,\mathrm{D})\prime \prime}_{ijk}(z)$ are presented in Figure \ref{fig:s2ijk} (k) \& (l).
These qualitatively agree with previously published depth-resolved SFG spectra \cite{moritaTheoreticalAnalysisSum2000,ishiyamaMolecularDynamicsStudy2006,mobergTemperatureDependenceAir2018,hiranoBoundaryEffectsQuadrupole2022b,fellowsHowThickAirWater2024,fellowsSumFrequencyGenerationSpectroscopy2024,delapuenteNeuralNetworkBasedSumFrequency2024}. \\ \\
The second-order electric dipole contributions to the SFG signal in the stretching band (Fig. \ref{fig:s2ijk} (k) \& (l)) and to the susceptibility in the bending band (Fig. \ref{fig:bending_zres} (f)) undergo significant frequency shifts as a function of position $z$ in the interfacial layer. As shown elsewhere~\cite{brunigTimeDependentFrictionEffects2022}, when going from vapor to liquid, frequency shifts arise from competing effects of non-Markovian friction (which causes blue-shifting) and potential broadening (which causes red-shifting). The former dominates in the bending band, \textcolor{changed}{as seen in Figure~\ref{fig:bending_zres} (f)}, while the latter dominates in the stretch band\textcolor{changed}{, as seen in Figure~\ref{fig:s2ijk} (k) \& (l).}
\\ \\
The electric quadrupole profile $\tilde s^{(2,\mathrm{Q})\prime \prime}_{ijk}(z)$ is presented in Figure \ref{fig:s2ijk} (m) \& (n) and is dominated by a broad positive peak around $z \approx z_\mathrm{GDS}$.
Additionally, we observe a negative contribution closer to the bulk region and a positive contribution closer to the vapor region at the frequency of the free OH vibrations. 
It is important to note that the electric quadrupole profile $\tilde{s}^{(2,\mathrm{Q})\prime \prime}_{ijk}(z)$ strongly depends on position. 
Consequently, the electric quadrupole contribution to the SFG spectra, $\tilde{S}^{(2,\mathrm{Q})}_{ijk}$ as defined in Equation \eqref{eq:S2ijk_beta}, could be used to report on interfacial water structure for a non-vanishing wave vector mismatch $\Delta k_z$, which is an interesting venue for future experimental investigation. \\ \\
\textcolor{changed}{
  While the bending band probes mostly the biaxial interfacial water ordering, as shown in Section~\ref{sec:biaxial_orientation}, the stretch band is instead more sensitive to the local hydrogen-bonding environment, which varies rather abruptly across the interface \cite{tangDefinitionFreeOH2018,chiangDielectricFunctionProfile2022a}. Therefore, the second-order electric-dipole profile of the stretch band is confined to a narrow region around the Gibbs dividing surface, whereas the bending profile extends slightly deeper into the bulk, as seen by comparing Figure~\ref{fig:s2ijk} (k) with Figure~\ref{fig:bending_zres} (f).}
\subsection{Spatially Resolved Linear Dielectric and Absorption Profile}
\label{sec:Linear_Dielectric_Profile}
\begin{figure*}
\centering
\includegraphics[width=1.0\textwidth]{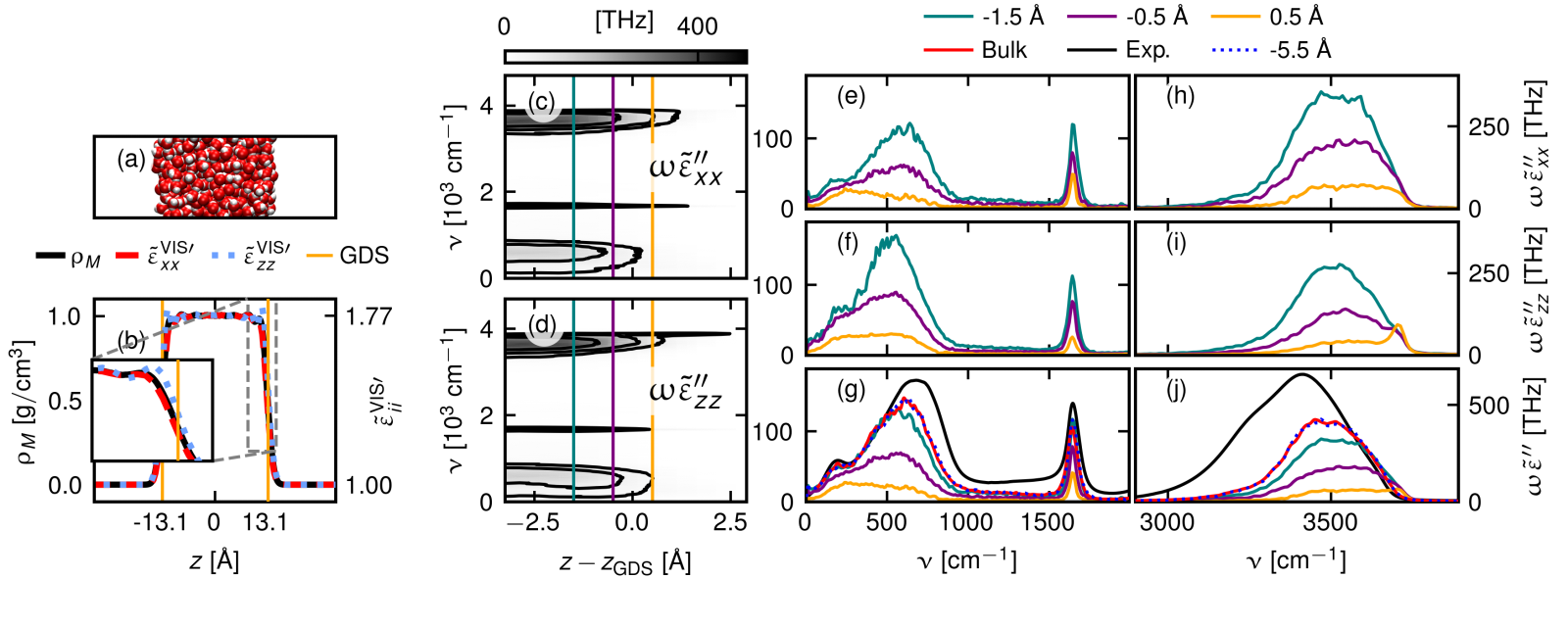}
 \caption{Linear optical dielectric and IR absorption profiles at the air water interface. 
 (a): Snapshot of the simulation.
 (b): The water mass density profile is compared to the real parts of the tensorial optical dielectric profiles.
 (c) \& (d): The linear absorption profiles $\omega \tilde\varepsilon''_{xx}(z)$ and $\omega \tilde\varepsilon''_{zz}(z)$ in the IR and low THz frequency range are shown as a 2D plot.
 (e)-(j): The imaginary part of the linear absorption profile is shown at the same positions relative to the Gibbs dividing surface as in Figure \ref{fig:s2ijk}.
The $xx$-component defined in Equation \eqref{eq:def_eps_xx}, the $zz$-component defined in Equation \eqref{eq:def_eps_zz}, and the isotropic average defined in \eqref{eq:def_eps_iso} are presented in (e), (f), and (g), respectively.
Additionally, the predicted linear absorption spectrum of bulk water and the experimental spectrum  \cite{bertieInfraredIntensitiesLiquids1996} are shown in (g). 
 (h)-(j): Results in the stretch region. The bending band in (e)-(g) is redshifted by $\SI{28}{cm^{-1}}$ and the stretching band in (h)-(j) is red-shifted by $\SI{166}{cm^{-1}}$.} 
\label{fig:eps}
\end{figure*} 
As mentioned before, the $z$-independent external field amplitude $\mathcal{F}^\alpha_z$ corresponds to the D-field, while $\mathcal{F}^\alpha_{x/y}$ corresponds to the E-field.
The relationship between D- and E-fields is determined by the dielectric profile  \cite{sternCalculationDielectricPermittivity2003,bonthuisDielectricProfileInterfacial2011,bonthuisProfileStaticPermittivity2012,bonthuisContinuumHowMolecular2013,gekleNanometerResolvedRadioFrequencyAbsorption2014,locheBreakdownLinearDielectric2018,locheCommentHydrophobicSurface2019a}.  
The real parts of the dielectric profiles $\tilde{\varepsilon}_{xx}^{\mathrm{VIS}\prime}(z)$ (red) and $\tilde{\varepsilon}_{zz}^{\mathrm{VIS}\prime}(z)$ (blue) at optical frequencies are presented in Figure \ref{fig:eps} (b) and have an almost identical shape to the mass density profile (black). 
Thus, in contrast to the static or THz case  \cite{bonthuisDielectricProfileInterfacial2011,beckerInterfacialVsConfinement2024}, the anisotropy of the tensorial interfacial response for optical frequencies is rather small, as demonstrated in the inset.
The bulk plateau value of $1.77$ agrees well with the experimental value of $1.78$ at room temperature under atmospheric pressure  \cite{haynesCRCHandbookChemistry2015a}. \\ \\
% page 818 and 976
We present the imaginary part of the position and frequency-dependent dielectric profiles, $\tilde{\varepsilon}^{\mathrm{IR}\prime \prime}_{xx}(z)$ and $\tilde{\varepsilon}^{\mathrm{IR}\prime \prime}_{zz}(z)$ extracted from molecular dynamics simulations using Equations \eqref{eq:def_eps_zz} and \eqref{eq:def_eps_xx} in Figure \ref{fig:eps} (c) \& (d).
We observe that the dielectric profile $\tilde{\varepsilon}^{\mathrm{IR}\prime \prime}_{ij}(z)$ varies drastically in the interface region and is significantly anisotropic.
We see that the parallel component of the dielectric profile  $\tilde{\varepsilon}^{\mathrm{IR}\prime \prime}_{xx} \left(z  \right)$ (Fig. \ref{fig:eps} (c) \& (e) \& (h)) is characterized by an almost frequency-independent decrease of intensity across the interface, without significant changes in the line shape. \\ \\
The perpendicular profile $\tilde{\varepsilon}^{\mathrm{IR}\prime \prime}_{zz}(z )$ (Fig. \ref{fig:eps} (d) \& (f) \& (i)) is more complex.
The OH-stretch band, which is in bulk centered around $\SI{3500}{cm^{-1}}$, exhibits a blue shift when going from bulk to vapor until only the free OH-stretch contribution at $\SI{3700}{cm^{-1}}$ survives.
The libration band \cite{carlsonExploringAbsorptionSpectrum2020,sheltonCorrelatedLibrationLiquid2024}, centered around $\SI{500}{cm^{-1}}$, is more pronounced in $\tilde{\varepsilon}^{\mathrm{IR}\prime \prime}_{zz}(z)$ than in $\tilde{\varepsilon}^{\mathrm{IR}\prime \prime}_{xx}(z)$, as can be seen in Figures \ref{fig:eps} (e) \& (f). \textcolor{changed}{This can be rationalized by the higher concentration of planar-oriented molecules near the interface (Fig. \ref{fig:bending_ori} (e)), which are more susceptible to $z$-polarized fields.}
\\ \\
We observe that isotropically averaged interfacial and bulk spectra are indistinguishable at $z-z_\mathrm{GDS}=\SI{-5.5}{\angstrom}$ in Figure \ref{fig:eps} (g) \& (j). 
Hence, the second-order response in Figures \ref{fig:bending_zres} \& \ref{fig:s2ijk}, as well as the linear response in Figure \ref{fig:eps}, and the orientational anisotropy in Figure \ref{fig:bending_ori}, can be considered bulk-like at this depth.
The comparison with experiments \cite{bertieInfraredIntensitiesLiquids1996} (black) in Figure \ref{fig:eps} (g) \& (j) shows that the differences between simulation and experiments are rather small around the bending peak and in the OH-stretch region at high frequencies ($>\SI{3500}{cm^{-1}}$) where the SFG stretch signal is most intense, whereas the experimental IR intensity is significantly underestimated at lower frequencies.
\section{Conclusion}
\label{sec:conclusion}
Accounting for higher-order multipole contributions is essential for interpreting SFG spectra and leads to quantitative agreement between simulations and experimental data. 
In the bending band, the electric dipole contribution is almost negligible due to the small frequency shift between mutually canceling positive and negative SFG signals from water at different depths, and has an amplitude comparable to the \textcolor{changed}{usually neglected magnetic dipole contribution}. 
As a result, the total SFG signal is dominated by the electric quadrupole contribution. 
The shoulder appearing at $\SI{3600}{cm^{-1}}$ in the stretch band has been attributed to orientational anisotropy  \cite{jiCharacterizationVibrationalResonances2008a,stiopkinHydrogenBondingWater2011}, or to a combination mode involving intermolecular coupling  \cite{suzukiVibrationalCouplingTopmost2017}, as summarized in the review by Tang \textit{et al.}  \cite{tangMolecularStructureModeling2020}. 
Our results demonstrate that the shoulder primarily originates from electric quadrupole contributions, which do not reflect interfacial structure but rather depend on bulk properties for vanishing wave vector mismatch $\Delta k_z=0$.
Only after subtracting higher-order multipoles can the anisotropic orientation of interfacial molecules be inferred from the SFG signal, by which pronounced biaxial water ordering at the air–water interface is revealed.
This biaxial structure consists of essentially three layers of orientational species, namely an inwards-oriented, a planar-oriented, and an outwards-oriented species. \textcolor{changed}{ This arrangement is surprisingly similar to structural motifs observed at the air-ice interface \cite{odendahlLocalIcelikeStructure2022a}.}
\\ \\
In conclusion, since SFG spectra are significantly influenced by non-interface specific multipole effects, it is crucial to account for and subtract these contributions in order to be able to interpret experimental spectra in terms of interfacial orientations. This can be achieved by subtracting a reference spectrum with the same bulk medium  \cite{fellowsImportanceLayerDependentMolecular2025b}, or by using theoretical predictions, as those provided in this work.
\section{Methods}
\label{sec:methods}
\subsection{Technical Details}
All simulations are done in the NVT ensemble at room temperature $T=\SI{298}{K}$ with the MB-Pol force field  \cite{babinDevelopmentFirstPrinciples2013,babinDevelopmentFirstPrinciples2014,meddersDevelopmentFirstPrinciplesWater2014}. For starting configurations, 94 equally spaced initial configurations are extracted from a $\SI{18.8}{ns}$ long simulation using the SPC/E force field  \cite{berendsenMissingTermEffective1987} and the GROMACS molecular dynamics software \cite{abrahamGROMACSHighPerformance2015}. The SPC/E simulations are done with a time step of $\SI{2}{fs}$ and a velocity rescaling thermostat  \cite{bussiCanonicalSamplingVelocity2007}, 
with a relaxation time of $\SI{1}{ps}$. 
Each of the initial configurations is equilibrated with the MB-Pol force field for $\SI{20}{ps}$, and afterward, production runs with an average runtime of $\SI{0.86}{ns}$ are executed using the LAMMPS software  \cite{thompsonLAMMPSFlexibleSimulation2022}.
Again, a velocity-rescaling thermostat is employed. However, for MB-Pol simulations, the relaxation time is set to $\SI{5}{ps}$, and the time step is $\SI{0.2}{fs}$.
The box size is $\SI{2}{nm}$ in the x and y dimensions and $\SI{6}{nm}$ in the z dimension. This size is chosen to ensure that the box contains more than twice as much air as water, since the water slab spans approximately $\SI{2.6}{nm}$.
The box is filled with $352$ water molecules and the two interfaces lie in the $xy$ plane.  \\ \\
The total dipole moment of the simulation box is computed with the modified TTMF-4 model, which describes the electrostatics of the MB-Pol potential \cite{burnhamVibrationalProtonPotential2008,babinDevelopmentFirstPrinciples2013}. 
The same model is used to predict the linear absorption profile, where we compute polarization density trajectories from monopole and dipole density trajectories.
Molecular polarizabilities are computed with single-molecule quantum chemistry calculations using the software Gaussian 16 \cite{frischGaussian16Rev2016}. 
The theoretical level is CCSD(T)/aug-cc-pVTZ and B3LYP/aug-cc-pVTZ in the parameterization of $\alpha^{n,\mathrm{DD}}_{ij}(t)$ and $\alpha^{n,\mathrm{QD}}_{ijk}(t)$, appearing in Equations \eqref{eq:scf_mu_time} and \eqref{eq:scf_Q_time}, respectively. 
Electrostatic interactions between the induced multipoles are computed as 
described in SI Sections V B and VIII, involving the Ewald summation algorithm included in OpenMM  \cite{eastmanOpenMM8Molecular2024}. \\ \\
The simulation of bulk water is done in a cubic box with a length of $\SI{2.07}{nm}$ consisting of 297 molecules. 
The magnetic dipole contribution is extracted from a different trajectory, because it requires a much smaller write-out frequency. 
The initial configurations for the trajectories used to extract the bulk dielectric constant, molecular hyperpolarizability, and magnetic dipole contribution are taken from a trajectory generated with the SPC/E model after a simulation time of $\SI{20}{ns}$ and $\SI{40}{ns}$, respectively.
The first $\SI{10}{ps}$ of the MB-Pol trajectory is discarded, and a trajectory of length of $\SI{400}{ps}$ is used to calculate the bulk dielectric constant and the molecular hyperpolarizability, and a trajectory of length of $\SI{300}{ps}$ to calculate the second-order magnetic dipole susceptibility.
The simulation parameters of the bulk simulations are the same as for the interface. We use the recommended settings of the MB-Pol potential from the GitHub repository "MBX - Version 0.7" from the Paesani group  \cite{rieraMBXManybodyEnergy2023}. The write-out frequency is $\SI{3.2}{fs}$ for all spectra except $\SI{0.4}{fs}$ for the magnetic dipole contribution.
In all simulations, periodic boundary conditions are imposed. 
However, we correct for the external field modification arising from the periodic replicas as described in SI Section IX. \textcolor{changed}{The smoothing procedure and the error estimation method of the predicted spectra are explained in SI Section X. Details on the comparison between the experimental and simulated absolute spectra presented in Figure \ref{fig:S2ijk}~(k) are given in SI Section~XI.} 
Further references are cited in the SI \cite{bertieInfraredIntensitiesLiquids1989,haleOpticalConstantsWater1973a,ponathNonlinearSurfaceElectromagnetic2012,wolfProgressOpticsVol1977,tenbrinckPolycyclicAromaticHydrocarbons2022,kochChemistsGuideDensity2001,stoneInductionEnergyAssembly1989,hlawatschTimeFrequencyAnalysisConcepts2008,nagataVibrationalSumFrequencyGeneration2010,frenkelUnderstandingMolecularSimulation2002,blackmanMeasurementPowerSpectra1958,zhangDeepPotentialModel2022,smithiiiMathematicsDiscreteFourier2007,mahanManyParticlePhysics1990,caldeiraQuantumTunnellingDissipative1983}.
\label{sec:comp_detail}
\subsection{Prediction of Locally Resolved SFG Spectra}
We consider a planar interface between two media. The system is homogeneous in the $xy$ plane but inhomogeneous along the $z$ dimension. 
The system is illuminated by two monochromatic light sources with frequencies $\omega^\mathrm{VIS}$ and $\omega^\mathrm{IR}$ that induce an electric current density $j_i^{(2)}(z,t)$, oscillating with the sum frequency $\omega^\mathrm{SFG}=\omega^\mathrm{IR} + \omega^\mathrm{VIS}$. 
Here $\omega^\mathrm{IR}$ is in the IR range and invokes equifrequent oscillation of the nuclei, while $\omega^\mathrm{VIS}$ is in the optical frequency range.
The resulting second-order electric current density is determined by the second-order response of a time-dependent perturbation expansion  \cite{mukamelPrinciplesNonlinearOptical1995}
\begin{align}
\varepsilon_0^{-1} j_i^{(2)}(z,t) =  - i \omega^\mathrm{SFG} e^{ -i  \omega^\mathrm{SFG} t } \tilde{s}^{(2)}_{ijk}\left( z  \right) \mathcal{F}^\mathrm{VIS}_j  \mathcal{F}^\mathrm{IR}_k + c.c.  \, ,
\label{eq:s2_ijk}
\end{align}
where $i$,$j$,$k \in \lbrace x,y,z \rbrace$ are Cartesian coordinate indices.
Here $\tilde{s}^{(2)}_{ijk}\left( z  \right)$ is the second-order response profile, the tilde denotes a Fourier transformation, i.e. $\tilde{\phi}(\omega)=\mathrm{FT}[\phi(t)](\omega)$, $\tilde{\phi}'$ and $\tilde{\phi}''$ stand for the real and imaginary parts and $c.c.$ denotes the complex conjugate.
 The external field amplitude $\mathcal{F}^\alpha_i$ corresponds to  \cite{sternCalculationDielectricPermittivity2003,bonthuisDielectricProfileInterfacial2011,bonthuisProfileStaticPermittivity2012,bonthuisContinuumHowMolecular2013,gekleNanometerResolvedRadioFrequencyAbsorption2014}
\begin{align}
\mathcal{F}^\alpha_i = \left( \delta_{ix} + \delta_{iy} \right)  \mathcal{E}^\alpha_i +  \delta_{iz} \varepsilon_0^{-1} \mathcal{D}^\alpha_i \, ,
\label{eq:ext_field_maxwell}
\end{align}
where $\mathcal{E}^\alpha_i$ and $\mathcal{D}^\alpha_i$ are the amplitudes of the E-fields and the D-fields
\begin{align}
    E^\mathrm{\alpha}_i( z ,t)&=\mathcal{E}^\alpha_i (z) e^{-i \omega^\mathrm{\alpha} t } +c.c. 
    \label{eq:def_Ealpha}  \\
        D^\mathrm{\alpha}_i( z ,t)&=\mathcal{D}^\alpha_i (z) e^{-i \omega^\mathrm{\alpha} t } +c.c. \, ,
    \label{eq:def_Dalpha} 
\end{align}
 oscillating with the frequency $\omega^\alpha$, $\alpha \in \lbrace \mathrm{SFG}, \mathrm{VIS}, \mathrm{IR} \rbrace$ and $\delta_{ij}$ is the Kronecker delta. 
Hence, the symbol $\mathcal{F}^\alpha_i$ is a placeholder for the spatially constant amplitude of D-fields or E-fields on the relevant length scale, depending on the polarization. The external fields are given by ${F^\mathrm{\alpha}_i(t)= \mathcal{F}^\mathrm{\alpha}_i e^{ - i \omega^\mathrm{\alpha} t } + c.c.}$. By formulating the theory with respect to these fields, we avoid locality approximations. 
We note that  $\tilde{s}^{(2)}_{ijk}(z)$ is distinct from the second-order susceptibility $\tilde{\chi}^{(2)}_{ijk}(z)$, which defines the response to electric E-fields.
\textcolor{changed}{
  Several studies have defined $\tilde{\chi}^{(2)}_{ijk}(z)$ using constitutive relations that involve an position-independent effective interfacial dielectric constant $\tilde{\varepsilon}^\alpha_\mathrm{eff}$~\cite{chiangDielectricFunctionProfile2022a,yuFresnelFactorCorrection2023,zhangQuantitativeConsistencyIntensity2025a}. Within that framework $\tilde{s}^{(2)}_{ijk}(z)$ and $\tilde{\chi}^{(2)}_{ijk}(z)$ are related by
  \begin{align}
      \tilde{\chi}^{(2)}_{yyz}(z) &= \tilde{\varepsilon}^\mathrm{IR}_\mathrm{eff}\tilde{s}^{(2)}_{yyz}(z) \\
      \tilde{\chi}^{(2)}_{zzz}(z) &= \tilde{\varepsilon}_\mathrm{eff}^{\mathrm{SFG}} \tilde{\varepsilon}_\mathrm{eff}^{\mathrm{VIS}} \tilde{\varepsilon}_\mathrm{eff}^{\mathrm{IR}}  \tilde{s}^{(2)}_{zzz}(z) \, 
      \label{eq:chi2_s2_relation}
  \end{align} 
  with similar relations for other tensor elements.
  The relation in Equation \eqref{eq:s2_chi2} used by us is exact on the linear response level and has the advantage that it fully captures position-dependent dielectric effects.
} \\ \\
In this work, we only consider the signal arising from the interface on the length scale at which the external fields are spatially constant, \textcolor{changed}{therefore we neglect contributions induced by external field gradients.
These are} so-called 
bulk quadrupole contributions
\cite{shiratoriTheoryQuadrupoleContributions2012, moritaTheorySumFrequency2018a, hiranoLocalFieldEffects2024}, which can be experimentally estimated using combined transmission-reflection SFG  \cite{sunSurfaceSumfrequencyVibrational2015}, combined SFG/DFG techniques  \cite{fellowsHowThickAirWater2024,fellowsImportanceLayerDependentMolecular2025b} \textcolor{changed}{or configuration analysis} and are discussed in SI Section IV C. 
 \textcolor{changed}{
 It can be shown that $\tilde{S}^{(2)}_{yyz}$ is largely insensitive to bulk 
quadrupole contributions, provided that the transmitted-beam wavevectors remain approximately parallel \cite{shiratoriTheoryQuadrupoleContributions2012}. 
Therefore, $\tilde{S}^{(2)}_{yyz}$ presented in Figure \ref{fig:S2ijk} (c), (g) and (k) can be assumed to be not significantly perturbed by bulk quadrupole contributions. 
However, this is not the case for $\tilde{S}^{(2)}_{zzz}$ presented in Figure \ref{fig:S2ijk} (d) \& (h). 
We compare our theoretical prediction with a previous experimental configuration analysis
 \cite{ganPolarizationExperimentalConfiguration2006} in SI Section XI and thereby confirm that 
our comparison is not significantly affected by bulk quadrupole contributions.
} \\ \\
The SFG spectrum follows by integration of the second-order response profile over $z$ as
 \begin{align}
     \tilde{S}^{(2)}_{ijk}  &= \int\limits_{-\infty}^\infty \mathrm d z \,  e^{-i \Delta k_z z} \tilde{s}^{(2)}_{ijk}(z )
     \label{eq:def_S2ijk} \, ,
 \end{align}
where $\Delta k_z$ is the wave vector mismatch defined in SI Section II.
We emphasize that Equation \eqref{eq:def_S2ijk} holds whenever the external fields can be assumed to be constant in the region where the dielectric profile is inhomogeneous. This assumption is well satisfied in our case, because the shortest wavelength considered is ${\lambda^\mathrm{SFG} \approx \SI{600}{nm}}$, and the region where the dielectric profile is inhomogeneous is only a few angstroms thick, as demonstrated in this work.
Here we utilize the additional approximation $e^{-i \Delta k_z z} \tilde{s}^{(2)}_{ijk}(z) \approx \tilde{s}^{(2)}_{ijk}(z)$. At the air waiter interface, this is a good approximation, as a typical value for $\Delta k_z$ is $\SI{0.02}{nm^{-1}}$, whereas the thickness of the air-water interface is about $\SI{8}{\angstrom}$  \cite{fellowsHowThickAirWater2024}.
 A derivation of Equations \eqref{eq:s2_ijk}-\eqref{eq:def_S2ijk} is provided in SI Sections I and II.
\subsection{Molecular Multipole Contributions}
\label{sec:MM_Contributions}
The second-order electric current density in Equation \eqref{eq:s2_ijk} can be decomposed into molecular multipole contributions according to
\begin{align}
    j^{(2)}_i(z,t) = j^{(2,\mathrm{D})}_i (z,t) +  j^{(2,\mathrm{Q})}_i(z,t) +  j^{(2,\mathrm{M})}_i(z,t) + ... \, ,
    \label{eq:j2_multi}
\end{align}
where D, Q, and M stand for electric dipole, electric quadrupole, and magnetic dipole, respectively. 
The multipole contributions to the second-order current density are defined as $ { j_i^{(2,\mathrm{D})} (z,t) =   \frac{\partial}{\partial t} \varrho^{(2,\mathrm{D})}_i(z,t) }$, $ { j_i^{(2,\mathrm{Q})} (z,t) =  - \frac{\partial}{\partial t} \frac{\partial}{\partial r_j} \varrho^{(2,\mathrm{Q})}_{ij}(z,t) }$, and ${j_i^{(2,\mathrm{M})} (z,t) =  \epsilon_{ijk}  \frac{\partial}{\partial r_j}   m^{(2)}_{k}(z,t)}$, where $\varrho^{(2,\mathrm{D})}_{i}(z,t)$ and $\varrho^{(2,\mathrm{Q} )}_{ij}(z,t)$ are the electric dipole and electric quadrupole densities, $r_i$ is the $i^\mathrm{th}$ Cartesian coordinate of a vector position, $m^{(2)}_{i}(z,t)$ is the magnetic dipole density and $\epsilon_{ijk}$ is the Levi-Civita symbol  \cite{russakoffDerivationMacroscopicMaxwell1970,jacksonClassicalElectrodynamicsInternational2021}. Higher-order molecular multipole contributions in Equation \eqref{eq:j2_multi} do not contribute to the experimentally detectable spectrum $\tilde{S}^{(2)}_{ijk}$ in Equation \eqref{eq:def_S2ijk} in the limit $z \Delta k_z \rightarrow 0$ and consequently need not be considered  \cite{guyot-sionnestBulkContributionSurface1988}.  \\ \\
Using a timescale-separation approximation, the second-order electric dipole density can be dissected into the contributions due to source electric dipole 
$\varrho^\mathrm{DS}_i(\vec r, t)$ and source quadrupole $\varrho^\mathrm{QS}_{ij}(\vec r, t)$ densities.
Specifically, $\varrho_i^{(2,\mathrm{D})}(z,t)$ can be divided into the pure electric dipole contribution $\varrho^{(2,\mathrm{DD})}_i(z,t)$ induced by $\varrho^\mathrm{DS}_{i}(\vec r, t)$,  and the electric dipole - electric quadrupole cross contribution $\varrho^{(2,\mathrm{DQ})}_{i}(z, t)$, induced by $\varrho^\mathrm{QS}_{ij}(\vec r, t)$.
 As shown in SI Section V B, these contributions can be defined by
 \begin{multline}     
 \varrho^{(2,\mathrm{DD})}_i (z,t ) = \frac{1}{A}\int \mathrm{d} x \int \mathrm{d} y \\ \left[ \varrho^{\mathrm{DS}}_i(\vec r ,t) + \int \mathrm{d} \vec r' \varepsilon_0 \tilde{s}^{\mathrm{NL}}_{ij}(\vec r, \vec r', t) F_j^\mathrm{DS}(\vec r',t) \right]
  \label{eq:def_DD_sNL}
 \end{multline}
 and 
  \begin{align}
  \varrho^{(2,\mathrm{DQ})}_i (z,t ) = \frac{1}{A}\int \mathrm{d} x \int \mathrm{d} y \int \mathrm{d} \vec r' \varepsilon_0 \tilde{s}^{\mathrm{NL}}_{ij}(\vec r, \vec r', t) F_j^\mathrm{QS}(\vec r',t)  \, ,
 \label{eq:def_DQ_sNL}
 \end{align}
 where $F_i^\mathrm{DS}(\vec r,t)$ and $F_i^\mathrm{QS}(\vec r,t)$ are the electrostatic fields created by the densities $\varrho^{\mathrm{DS}}_i(\vec r ,t )$ and $\varrho^{\mathrm{QS}}_{ij}(\vec r ,t )$, $A$ is the interfacial area and $\tilde{s}^{\mathrm{NL}}_{ij}(\vec r, \vec r', t)$ is a nonlocal, linear and instantaneous response function.
We note that the decomposition $\varrho^{(2,\mathrm{D})}_i(z,t)$=$\varrho^{(2,\mathrm{DD})}_i(z,t)+\varrho^{(2,\mathrm{DQ})}_i(z,t)$ does not enter the calculation of $\tilde{S}^{(2)}_{ijk}$ and is only used for interpretation of the SFG spectra.
We define the corresponding electric current densities $j^{(2,\mathrm{DD})}_i(z,t)=\frac{\partial}{\partial t} \varrho_i^{(2,\mathrm{DD})}(z,t)$ and $j^{(2,\mathrm{DQ})}_i(z,t)=\frac{\partial}{\partial t} \varrho_i^{(2,\mathrm{DQ})}(z,t)$. \\ \\
The second-order response profile is decomposed in analogy to Equation \eqref{eq:j2_multi}
into
\begin{align}
    \tilde s^{(2)}_{ijk} (z) &= \tilde s^{(2,\mathrm{D})}_{ijk} (z) + \tilde s^{(2,\mathrm{Q})}_{ijk}(z) + \tilde s^{(2,\mathrm{M})}_{ijk} (z) \\
   \tilde s^{(2,\mathrm{D})}_{ijk} (z) &=  \tilde s^{(2,\mathrm{DD})}_{ijk} (z) + \tilde s^{(2,\mathrm{DQ})}_{ijk} (z) \, ,
\end{align}
where each contribution $\tilde{s}^{(2,\beta)}_{ijk}\left( z \right) $ is defined as 
\begin{align}
\varepsilon_0^{-1} j_i^{(2,\beta)}(z,t) =  - i \omega^\mathrm{SFG} e^{ -i  \omega^\mathrm{SFG} t } \tilde{s}^{(2,\beta)}_{ijk}\left( z  \right) \mathcal{F}^\mathrm{VIS}_j  \mathcal{F}^\mathrm{IR}_k + c.c. \, ,
\label{eq:s2_ijk_beta}
\end{align}
where $\beta \in \lbrace \mathrm{DD},\mathrm{DQ}, \mathrm{D},\mathrm{Q}, \mathrm{M} \rbrace$. Consequently, the spatially integrated multipole contributions to the SFG spectrum are given by
\begin{align}   
\tilde{S}^{(2,\beta)}_{ijk} = \int\limits_{-\infty}^\infty \mathrm{d} z\, e^{-i \Delta k_z z}  \tilde{s}^{(2,\beta)}_{ijk}\left( z  \right) \, .
\label{eq:S2ijk_beta}
\end{align}  
While the total current $j^{(2)}_i(z,t)$ does not depend on the choice of the molecular origin, the individual contributions in Equation \eqref{eq:j2_multi} do  \cite{byrnesAmbiguitiesSurfaceNonlinear2011,hiranoBoundaryEffectsQuadrupole2022b}.
To determine the optimal position of the molecular origin, we consider the SFG signal from an interface with isotropically oriented molecules $\tilde{S}^{(2,\mathrm{ISO})}_{ijk}$, which we call an isotropic interface. 
This isotropic interface is created by cutting bulk water at an arbitrary
$z$ position into two halves. 
Consequently, boundary contributions created by the change in density at the interface are present, but the molecules' orientational distribution is inversion symmetric.
A similar interface was created to predict the Bethe potential  \cite{remsingInfluenceDistantBoundaries2019}.
We test three molecular origins for the calculation of multipoles and find that if we choose the molecular center of mass as the molecular origin, ${ \tilde{S}^{(2,\mathrm{ISO})}_{ijk} \approx \tilde{S}^{(2,\mathrm{Q})}_{ijk}  + \tilde{S}^{(2,\mathrm{M})}_{ijk}}$ does hold in good approximation. 
Thus, choosing the molecular center of mass as the expansion center ensures that the electric dipole contribution is solely due to molecular orientation, as discussed in detail in SI Section VI. 
In contrast, choosing the molecular center different from the center of mass introduces significant boundary contributions in the electric dipole contribution $\tilde{S}^{(2,\mathrm{D})}_{ijk}$. 
Moreover, the center of mass maximizes the decoupling of molecular translations, vibrations, and rotations  \cite{eckartStudiesConcerningRotating1935,herzbergMolecularSpectraMolecular1945,wilsonMolecularVibrationsTheory1980}. \\ \\
The theory of our multipole decomposition is described in SI Sections III and IV. There, we also derive the constitutive relations, including nonlinear multipolar source terms, using the Lorentz field approximation in planar geometry, based on works by Mizrahi and Sipe  \cite{mizrahiLocalfieldCorrectionsSumfrequency1986} and Hirano and Morita  \cite{hiranoLocalFieldEffects2024}.
These equations are solely needed to aid the interpretation of experimental spectra.
\subsection{Fluctuation-Dissipation Relations within the Off-Resonant Approximation}
\label{sec:FDT_s2ijk}
We assume that the VIS field interacts off-resonantly with the system, meaning it does polarize the molecules but does neither excite higher electronic levels nor influence the nuclei trajectories.
This approximation is valid since the VIS field oscillates too rapidly to influence the motion of the nuclei yet slowly enough to act adiabatically on the distribution of the electrons. 
In this limit the SFG signal arises from a first-order perturbation expansion with respect to the external field from the IR light source \cite{moritaRecentProgressTheoretical2008}. \\ \\
The molecular multipoles are determined by the equations
\begin{align}
    \mu^n_i(t)  &= \alpha^{n,\mathrm{DD}}_{ij}(t) f^n_{jk}(t) F^{\mathrm{VIS}}_k (t)
     \label{eq:scf_mu_time} \\
Q^n_{ij}(t)  &= \alpha^{n,\mathrm{QD}}_{ijk}(t)f^n_{kl}(t) F^{\mathrm{VIS}}_l (t)
    \label{eq:scf_Q_time} \, ,
\end{align}
where $\mu^n_i(t)$ and $Q^n_{ij}(t)$ are the induced electric dipole and electric quadrupole moments of the n$^\mathrm{th}$-molecule, and $\alpha^{n,\mathrm{DD}}_{ij}(t)$ and $\alpha^{n,\mathrm{QD}}_{ijk}(t)$ are the  electric dipole and the electric quadrupole polarizabilities of the $n^\mathrm{th}$ molecule. The local field factor $f_{ij}^n(t)$ transforms an external field $F_i(t)$ into the local E-field acting on the n$^\mathrm{th}$ molecule, ${E^n_i(t) = f_{ij}^n(t) F_j(t)}$, and is determined in a self-consistent manner such that the multipoles induced by $E^n_i(t)$ produce the field $E^n_i(t)-F_i(t)$. 
It is demonstrated in SI Section V B that it is incorrect to solve separate self-consistent field equations for the VIS and the SFG field in the time domain, as proposed previously  \cite{moritaRecentProgressTheoretical2008,ishiyamaAnalysisAnisotropicLocal2009,ishiyamaComputationalAnalysisVibrational2017}.
Rather, if a timescale-separation approximation is applied, Equations \eqref{eq:scf_mu_time} and \eqref{eq:scf_Q_time} each split into components at SFG and VIS frequencies, as described in SI Section V B. \\ \\
The polarization contributions of the second-order response profiles, defined in Equation \eqref{eq:s2_ijk_beta}, are determined by a first-order perturbation expansion, for which we introduce the fluctuation-dissipation relations
\begin{widetext}
\begin{align}    s^{(2,\mathrm{D})}_{ijk}(z,t)&=\frac{-\Theta(t)}{A k_B T \varepsilon_0} \frac{\partial}{\partial t}\sum\limits_n^{N_\mathrm{mol}}\left\langle  \alpha^{n,\mathrm{DD}}_{il}(t) f^n_{lj}(t)\delta \left[ z-z^n(t)\right] P_k(0)  \right\rangle 
\label{eq:fdt_s2dijk}  \\
s^{(2,\mathrm{Q})}_{ijk}(z,t)&=\frac{\Theta(t)}{A k_B T \varepsilon_0}  \frac{\partial}{\partial z} \frac{\partial}{\partial t}\sum\limits_n^{N_\mathrm{mol}}\left\langle  \alpha^{n,\mathrm{QD}}_{izl}(t) f^n_{lj}(t) \delta \left[ z-z^n(t)\right] P_k(0)  \right\rangle ,
    \label{eq:fdt_s2qijk}
\end{align}
\end{widetext}
where $\langle ... \rangle$ denotes \textcolor{changed}{classical} ensemble averaging, $T$ is the temperature, $k_B$ is the Boltzmann constant, $\Theta(x)$ is the Heaviside function, $\delta(x)$ is the Dirac delta distribution, and $z^n$ is the z-position of the $n^\mathrm{th}$ molecule. 
These expressions omit the off-resonant hyperpolarizability, which can be included as shown in SI Section V, but does not contribute to the imaginary part of SFG spectra. \\ \\
Equations \eqref{eq:fdt_s2dijk}  and \eqref{eq:fdt_s2qijk} follow from time-dependent perturbation theory using the perturbation Hamiltonian  ${H_{\mathrm{int}}(t)=- P_i F^\mathrm{IR}_i(t)}$, where $P_i$ is the total system's dipole moment. Hirano and Morita  \cite{hiranoBoundaryEffectsQuadrupole2022b} suggest the perturbation expansion using the modified perturbation Hamiltonian ${H_{\mathrm{int}}^{+}(t)=- P^+_i F^\mathrm{IR}_i(t)}$, where $P^+_i(t)$ is the dipole moment of the upper half of the simulation box. It is demonstrated in SI Section V D that this introduces a spurious contribution from the interface between the two halves. \\ \\
\textcolor{changed}{It is shown in SI Section~XII that Equations \eqref{eq:fdt_s2dijk} and \eqref{eq:fdt_s2qijk} provide a leading-order harmonic approximation of the quantum mechanical SFG response profile and that no quantum-correction factor needs to be applied.
The harmonic quantum-correction factor
$Q_\mathrm{HA} = \frac{\hbar \omega \beta}{1 - e^{-\hbar \omega \beta}}$, which appears in the literature \cite{ramirezQuantumCorrectionsClassical2004a,auerVibrationalSumfrequencySpectroscopy2008a,tangMolecularStructureModeling2020},
relates quantum and classical time-correlation functions~\cite{ramirezQuantumCorrectionsClassical2004a}, but does not relate classical and quantum response functions.}
\\ \\ We note that while we treat electric dipole and electric quadrupole contributions in a precise manner, we approximate magnetic dipole contributions by considering only the leading order term of an expansion in terms of electric multipole moments, as outlined in SI Section V C. 
We do not compute the position-resolved profile of magnetic dipole contributions, but only the overall contribution to the SFG signal, which can be extracted from a simulation of a bulk system, according to a fluctuation-dissipation relation presented in SI Section V C. 
Hence, we omit magnetic dipole contributions, whenever we present the response profiles $\tilde{s}^{(2)}_{ijk}(z)$, but consider them when we present the total spectra $\tilde{S}^{(2)}_{ijk}$.
In contrast to the electric dipole and electric quadrupole contributions, the magnetic dipole contribution does depend on ${\omega^\mathrm{VIS}}$, set to $\SI{2730}{THz}$, which is the center frequency of the VIS-field employed in the experimental measurement of the bending band  \cite{fellowsImportanceLayerDependentMolecular2025b} we compare with. 
The VIS frequency in the experimental reference SFG spectra of the stretching band  \cite{chiangDielectricFunctionProfile2022a,yuFresnelFactorCorrection2023} is quite similar.
\subsection{Linear Dielectric and Absorption Profiles}
We define the linear response of the polarization density ${p_i(z,t)=D_i(z,t)-\varepsilon_0 E_i(z,t)}$ to an external field of amplitude $\mathcal{F}^\alpha_i$ as
\begin{align}
    \varepsilon_0^{-1} p^{(1)}_i(z,t) = e^{- i\omega^\alpha t} \tilde{s}^{(1,\mathrm{P}
    )}_{ij}(z) \mathcal{F}_j^\alpha + c.c. \, .
\end{align}
The extraction of $\tilde{s}^{(1,\mathrm{P})}_{ij}(z)$ from molecular dynamics simulation has been described before  \cite{bonthuisDielectricProfileInterfacial2011,bonthuisProfileStaticPermittivity2012,bonthuisContinuumHowMolecular2013,gekleNanometerResolvedRadioFrequencyAbsorption2014,locheBreakdownLinearDielectric2018,locheCommentHydrophobicSurface2019a,beckerInterfacialVsConfinement2024} and is reproduced in SI Sections V B.
At the interface, the dielectric profile is tensorial, and the component perpendicular to the interface is given by
 \begin{align}\tilde \varepsilon_{zz}^{\mathrm{\alpha}}(z ) =  \frac{\varepsilon_0^{-1}\mathcal{D}^\alpha_z}{\mathcal{E}^\alpha_z(z)} =\frac{1}{1 - \tilde{s}^{(1,\mathrm{P})}_{zz}(z )} \,,
\label{eq:def_eps_zz}
\end{align}
while the component parallel to the interface is determined by
\begin{align}
\tilde \varepsilon^\alpha_{xx}(z ) =   \frac{\varepsilon_0^{-1}\mathcal{D}^\alpha_x(z)}{\mathcal{E}^\alpha_x}= 1 +  \tilde s^{(1,\mathrm{P})}_{xx}(z ).
\label{eq:def_eps_xx}
\end{align}
Here, $\mathcal{D}^\alpha_i(z)$ and $\mathcal{E}^\alpha_i(z)$, are defined in Equations \eqref{eq:def_Dalpha} and \eqref{eq:def_Ealpha}, respectively.
Due to the symmetry of our system we have $\tilde \varepsilon^\alpha_{yy}(z )=\tilde \varepsilon^\alpha_{xx}(z )$. In bulk, the dielectric tensor reduces to the isotropic component
\begin{align}    
 \tilde{\varepsilon}^\alpha( z ) = \frac{\tilde \varepsilon^\alpha_{xx}(z ) + \tilde \varepsilon^\alpha_{yy}( z ) + \tilde \varepsilon^\alpha_{zz}(z ) }{3} \, .
 \label{eq:def_eps_iso}
\end{align}
Further information about the dielectric profiles is given in SI Section XIII.
\subsection{Electric Dipole SFG Contribution as a Fingerprint for Interfacial Structure}
\label{sec:tech_frame_elect_dipole}
The response of the second-order electric current density to spatially constant external fields, which are $z$-polarized D-fields and $x$ or $y$-polarized E-fields, is given in Equation \eqref{eq:s2_ijk}.
As we want to relate SFG spectra to the molecular orientation distribution, we are interested in the second-order response of the electric dipole density to the average amplitude of the local E-field $\mathcal{E}^{\mathrm{L},\alpha}_i(z)$ acting on the molecular centers.
We define the second-order electric dipole susceptibility $\tilde{\chi}^{(2,\mathrm{DL})}_{ijk}\left( z  \right)$ by
\begin{multline}    
\varepsilon_0^{-1} j_i^{(2,\mathrm{DD})}(z,t) = \\ - i \omega^\mathrm{SFG} f^\mathrm{SFG}_{i}(z) e^{ -i  \omega^\mathrm{SFG} t } \tilde{\chi}^{(2,\mathrm{DL})}_{ijk}\left( z  \right) \mathcal{E}^\mathrm{L,VIS}_j(z) \mathcal{E}^\mathrm{L,IR}_k(z) + c.c. \, .
\label{eq:chi_2DL}
\end{multline}
The second-order electric dipole susceptibility $\tilde{\chi}^{(2,\mathrm{DL})}_{ijk}\left( z  \right)$ plays a central role for the interpretation of SFG spectra as it allows to interpret the macroscopic electric dipole contribution in terms of molecular orientation \cite{weiMotionalEffectSurface2001,moritaTheorySumFrequency2018a,sunOrientationalDistributionFree2018a,yuPolarizationDependentHeterodyneDetectedSumFrequency2022}.
\textcolor{changed}{$\tilde{\chi}^{(2,\mathrm{DL})}_{ijk}\left( z  \right)$ is related to the electric dipole contribution to the SFG response profile via Equation \eqref{eq:s2_chi2}, which is a frequently employed formalism to relate macroscopic and microscopic nonlinear quantities \cite{armstrongInteractionsLightWaves1962,shiratoriMolecularTheoryDielectric2011,shiratoriTheoryQuadrupoleContributions2012,moritaTheorySumFrequency2018a,hiranoLocalFieldEffects2024}.
}
Note that we do not introduce any locality approximations in the computation of the second-order response $\tilde{s}^{(2)}_{ijk}(z)$, but rather predict $\tilde{\chi}^{(2,\mathrm{DL})}_{ijk}\left( z  \right)$ by dividing the nonlocal second-order response profile $\tilde{s}^{(2,\mathrm{DD})}_{ijk}\left( z  \right)$ by the averaged local field factors $f^\alpha_i(z)$. 
We neglect the frequency dependence of $f^\alpha_i(z)$ by approximating $f^\alpha_i(z) \approx f^\mathrm{VIS}_i(z)$ for $\alpha=\mathrm{IR}$ and $\alpha=\mathrm{SFG}$. \\ \\
 \\ \\
We compare $\tilde{\chi}^{(2,\mathrm{DL})}_{ijk}\left( z  \right)$ with $\tilde{\chi}^{(2,\mathrm{ORI})}_{ijk}\left( z  \right)$,
which is solely based on the anisotropic orientation of the molecules. 
\begin{figure}
\centering
 \includegraphics[width=0.2\textwidth]{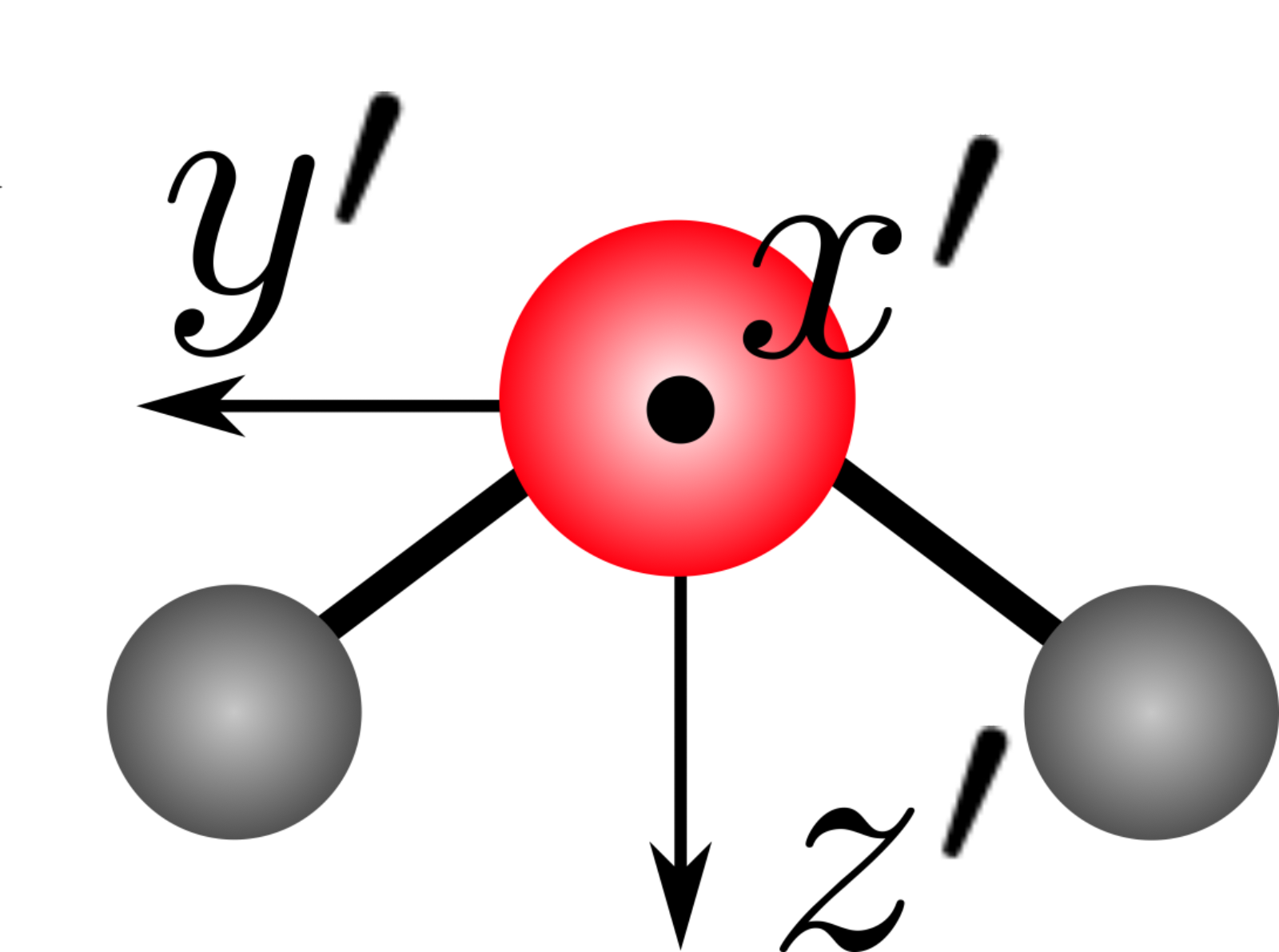}
\caption{Sketch of the molecular Eckart frame for a water molecule.}
\label{fig:mol_frame}
\end{figure}
The Euler angles $\phi$, $\theta$ and $\psi$ specify the molecule's orientation and we employ the $z'y'z'$ convention \cite{grayTheoryMolecularFluids1984}. 
We define the molecular Eckart frame  \cite{eckartStudiesConcerningRotating1935,herzbergMolecularSpectraMolecular1945,adler-goldenFormulasTransformingInternal1985,reyTransformationInternalCoordinates1998} by aligning the molecular $z'$-axis with the permanent dipole vector. The molecular $x'$-axis points out of the molecular plane, while the molecular $y'$-axis is chosen to be perpendicular to both $x'$ and $z'$. The molecular Eckart frame is depicted in Figure \ref{fig:mol_frame}. As shown in SI Section VII, the molecular hyperpolarizability $\tilde{\beta}_{ijk}(\theta, \psi)$ is determined by three orientation-dependent functions $q_{10}(\theta)$, $q_{30}(\theta)$ and $q_{32}(\theta, \psi)$.
The first and third Legendre polynomials  \cite{grayTheoryMolecularFluids1984}
\begin{align}
q_{10}(\theta) &= \cos{\theta} \\
q_{30}(\theta) &= \frac{1}{2} \left[ 5 \cos^3{\theta} - 3 \cos{\theta} \right] \, 
\end{align}
describe the orientation of the molecular dipoles, while $q_{32} (\theta, \psi)$  accounts for the rotation of the molecule around the molecular dipole moment and is only relevant if the molecule is not uniaxial.
Consequently, $q_{32}(\theta, \psi)$ is the biaxiality parameter, given by  
\begin{align}
q_{32} (\theta, \psi) &= \cos(\theta) ( \sin^2{\theta} \cos^2{\psi} - \sin^2{\theta} \sin^2{\psi} ) \, .
\end{align}
We obtain $\tilde{\chi}^{(2,\mathrm{ORI})}_{ijk}(z)$ by averaging all hyperpolarizabilities according to
\begin{align}    
\tilde{\chi}^{(2,\mathrm{ORI})}_{ijk}(z) &= \frac{1}{V}\sum\limits_n^{N_\mathrm{mol}} \tilde{\beta}_{ijk}(\theta^n, \psi^n) \\
&= \rho(z)\left[ q_{10}(z)\tilde{\beta}_{ijk}^{10} +  q_{30}(z) \tilde{\beta}_{ijk}^{30} 
+  q_{32}(z) \tilde{\beta}_{ijk}^{32} \right]\, , \label{eq:chi2ori}
\end{align}
where $V$ is the volume, $\rho(z)$ is the molecular density and $q_{lm}(z)$ is the z-dependent average of $q_{lm}(\theta^n,\psi^n)$ over all molecules. 
The frequency dependent coefficients $\tilde{\beta}_{ijk}^{lm}$ are presented in Figure \ref{fig:bending_ori} (j). 
A detailed description of the orientation analysis can be found in SI Section VII. The full information about the orientation distribution is determined by the orientation distribution function defined by 
\begin{align}
    \rho_\mathrm{ORI}(\theta,\psi) =\frac{1}{2 \pi \sin{\theta} N_\mathrm{mol}}\sum\limits_n^{N_\mathrm{mol}} \left\langle\delta(\theta-\theta^n) \delta(\psi-\psi^n)\right\rangle \, .
    \label{eq:def_odf}
\end{align}
The prefactor ensures that $\rho_\mathrm{ORI}(\theta,\psi)$ is properly normalized as 
$
\int\limits_{0}^{\mathrm{2 \pi}}\mathrm{d} \phi \int\limits_{0}^{\mathrm{\pi}} \sin{\theta} \mathrm{d} \theta \int\limits_{0}^{\mathrm{2 \pi}}\mathrm{d} \psi \, \rho_\mathrm{ORI}(\theta,\psi)=1
$. In an isotropic system $\rho_\mathrm{ORI}(\theta,\psi)$ is constant and is given by $\frac{1}{8 \pi^2}$. 
%SMcitations:
\section{Acknowledgements}
We acknowledge financial support from the Deutsche Forschungsgemeinschaft (DFG) within the grant CRC 1349 and HPC resources on CURTA  \cite{bennettCurtaGeneralpurposeHighPerformance2020}, TRON, and Sheldon from the FU Berlin for data acquisition.
\section{Data Availability}
\textcolor{changed}{A data subset supporting the findings of this study is available on Code Ocean under [URL]. Additional data are available from the corresponding author upon reasonable request.}
\section{Code Availability}
\textcolor{changed}{Code supporting the findings of this study is available on Code Ocean under [URL].}

\bibliography{mybib}
\end{document}

\title[]{Supplemental Information}
% Force line breaks with \\
\author{Louis Lehmann}
\affiliation{Department of Physics, Freie Universität Berlin, Arnimallee 14, 14195 Berlin, Germany.}
\author{Maximilian R. Becker}
\affiliation{Department of Physics, Freie Universität Berlin, Arnimallee 14, 14195 Berlin, Germany.}
\author{Lucas Tepper}
\affiliation{Department of Physics, Freie Universität Berlin, Arnimallee 14, 14195 Berlin, Germany.}
\author{Alexander P. Fellows}
\affiliation{Fritz-Haber-Institut der Max-Planck-Gesellschaft, Faradayweg 4-6, 14195, Berlin, Germany.}
\author{Álvaro Díaz Duque}
\affiliation{Fritz-Haber-Institut der Max-Planck-Gesellschaft, Faradayweg 4-6, 14195, Berlin, Germany.}
\author{Martin Thämer}
\affiliation{Fritz-Haber-Institut der Max-Planck-Gesellschaft, Faradayweg 4-6, 14195, Berlin, Germany.}
\author{Martin Wolf}
\affiliation{Fritz-Haber-Institut der Max-Planck-Gesellschaft, Faradayweg 4-6, 14195, Berlin, Germany.}
\author{Roland R. Netz}%
\affiliation{Department of Physics, Freie Universität Berlin, Arnimallee 14, 14195 Berlin, Germany.}
\email{rnetz@physik.fu-berlin.de}
\date{\today}

\maketitle
\onecolumngrid
\tableofcontents
\newpage
\section{Time-Dependent Nonlinear Perturbation Theory}
\label{sec:TimeDependentPerturbationTheory}
Sum frequency generation (SFG) spectroscopy measures the radiation created by the second-order electric current density induced by the wave mixing of two electric fields. 
In this chapter, we provide the formal link between the Hamiltonian and the resulting second-order electric current density. 
As we will see in Section \ref{sec:lin_spons_born_opp}, when one field interacts off-resonantly with the system, a first-order perturbation expansion is sufficient, which allows application of the fluctuation-dissipation theorem \cite{kuboFluctuationdissipationTheorem1966}.
However, in general, the SFG signal is determined by the second-order time-dependent perturbation expansion of the electric current density \cite{mukamelPrinciplesNonlinearOptical1995} presented here. 
The Hamiltonian 
\begin{align}
H(\vec\Omega, t)  = H_0(\vec \Omega) + H'(\vec\Omega, t) \, ,
\label{eq:def_Ham}
\end{align}
is decomposed into a time-independent part $ H_0(\vec \Omega)$ and a time-dependent perturbation
\begin{equation}
H'(\vec\Omega, t)= -P_i(\vec \Omega) F_i(\vec r, t) - Q_{ij}(\vec \Omega) \frac{\partial}{\partial r_i}  F_j(\vec r, t) + ... \,  \, ,
\label{eq:pert_ham}
\end{equation}
where the external electric field $F_i(\vec{r}, t)$ is time-dependent and varies only weakly in space.
The quantities $P_i(\vec{\Omega})$ and $Q_{ij}(\vec{\Omega})$ represent the dipole and quadrupole moments of the system, respectively, and $N_{\mathrm{part}}$ denotes the number of particles.
The indices $i,j \in \lbrace x,y,z \rbrace$ refer to the Cartesian coordinate axes. 
We do not consider the response to magnetic fields in this work.
The state vector is defined as $\vec{\Omega}=\lbrace \vec{\zeta}^1, ..., \vec{\zeta}^{N_\mathrm{part}}, \vec{\xi}^1 ,...,  \vec{\xi}^{N_\mathrm{part}} \rbrace$ and $\vec{\zeta}^n$ and $ \vec{\xi}^n$ are the positions and momenta of the $n^\mathrm{th}$ particle, respectively. 
Through this work, we implicitly sum over all tensor indices that appear on only one side of an equation and adopt the SI formulation of Maxwell’s equations.
In interfacial systems that are translationally invariant in the $xy$-plane, the $x$- or $y$-polarized external fields correspond to electric (E) fields and $z$-polarized external fields to electric displacement (D) fields, as determined by the relationship 
 \begin{align}
   F_i(\vec r, t) = \left( \delta_{ix} + \delta_{iy} \right) E_i(\vec r, t) + \varepsilon_0^{-1} \delta_{iz} D_z(\vec r, t ) \, .
\label{eq:ext_field_maxwell}
\end{align}
Equation \eqref{eq:ext_field_maxwell} holds for systems which are non-periodic in the $z$-dimension \cite{sternCalculationDielectricPermittivity2003} as described in Section \eqref{app:pbc}.
The external field defined in Equation \eqref{eq:ext_field_maxwell} is constant on length scales relevant for interfacial systems, as follows from Maxwell's equations \cite{bornPrinciplesOpticsElectromagnetic1999,jacksonClassicalElectrodynamicsInternational2021}.
In bulk systems, which are translationally invariant in all three dimensions, the external field can be identified as the E-field
 \begin{align}
    F_i(\vec r, t) = E_i(\vec r, t) \, ,
\label{eq:ext_field_maxwell_bulk}
\end{align}
as follows straightforwardly from Equation \eqref{eq:Fext_PBC} derived by Stern and Feller in 2003 \cite{sternCalculationDielectricPermittivity2003}.
The Liouville operator is defined as 
\begin{align}
\hat{L}( \vec \Omega,t ) \cdot &= - \left\lbrace H(\vec \Omega, t) , \cdot \right\rbrace  \, ,
\end{align}
where 
\begin{align}
\lbrace f(\vec \Omega), g(\vec{\Omega}) \rbrace = \sum\limits_n^{N_\mathrm{part}} \left[ \frac{\partial f(\vec \Omega)}{\partial \zeta_i^n} \frac{\partial g(\vec{\Omega}) }{\partial \xi_i^n}
- 
\frac{\partial f(\vec \Omega )}{\partial \xi_i^n} \frac{\partial g(\vec{\Omega} )}{\partial \zeta_i^n}  \right]
\end{align}
is the Poisson bracket.
The Liouville operator can be separated in the same manner as the Hamiltonian, which is 
\begin{align}
\hat{L}( \vec \Omega , t) &=  \hat{L}_0( \vec \Omega)  + \hat{L}_\mathrm{P}( \vec \Omega , t)  + \hat{L}_\mathrm{Q}( \vec \Omega , t)  + ... \\
\hat{L}_0( \vec \Omega )  &=   - \left\lbrace H_0( \vec \Omega ) , \cdot \right\rbrace  \\
\hat{L}_\mathrm{P}( \vec \Omega , t)  &=  F_i(\vec r, t)  \left\lbrace P_i(\vec \Omega), \cdot \right\rbrace  \\
\hat{L}_\mathrm{Q}( \vec \Omega , t)  &=   \frac{\partial}{\partial r_i}  F_j  (\vec r, t)  \left\lbrace Q_{ij}(\vec \Omega), \cdot \right\rbrace \, ,
\end{align}
where $\hat{L}_0( \vec \Omega )$ does not explicitly depend on time and 
\begin{align}
 \hat{L}'( \vec \Omega ,t) =\hat{L}_\mathrm{P}( \vec \Omega ,t)  + \hat{L}_\mathrm{Q}( \vec \Omega ,t)  + ...    
\end{align}
is the Liouville perturbation operator.
We define the expectation value of a generic observable $ A( \vec \Omega) $, as
\begin{align}
 \langle A(t) \rangle   =  \int\mathrm{d} \vec \Omega \, A(\vec \Omega) \rho(\vec \Omega ,t ) \, ,
  \label{eq:def_observable}
\end{align}
where $\rho(\vec \Omega ,t )$ is the probability distribution.
The time evolution of $\rho(\vec \Omega, t)$ is determined by the Liouville equation
\begin{align}
    \frac{\partial}{\partial t }  \rho(\vec \Omega, t) = - \hat{L}(\vec \Omega, t)\rho(\vec \Omega, t) \,.
\end{align}
The probability distribution can be expanded in the external field $F_i(t)$ analogously to the time-dependent perturbation expansion of the density matrix in quantum mechanics, which can be found in the book by Mukamel \cite{mukamelPrinciplesNonlinearOptical1995}. 
The classical analogy of this derivation is reproduced here.
We introduce the probability distribution in the interaction picture $\rho_\mathrm{I}\left( \vec \Omega ,t,t_0\right)$ via
\begin{align}
    \rho\left( \vec \Omega, t \right)=e^{-(t-t_0) \hat{L}_0(\vec \Omega)} \rho_\mathrm{I}\left( \vec \Omega ,t,t_0\right) \, ,
    \label{eq:def_interaction}
\end{align}
where $e^{-t\hat {L}_0( \vec  \Omega)}$ is the time propagation operator of the unperturbed system and, consequently, the time dependence appearing in $\rho_\mathrm{I}\left( \vec \Omega,t,t_0\right) $ accounts for the perturbation. 
The trajectory of $\rho_\mathrm{I}\left( \vec \Omega, t,t_0 \right)$ is determined by the Liouville equation in the interaction picture
\begin{align}
    \dot\rho_\mathrm{I}\left( \vec \Omega, t,t_0 \right) &= - \hat{L}_\mathrm{I}(\vec \Omega, t,t_0)\rho_\mathrm{I}\left( \vec \Omega, t,t_0 \right) \label{eq:Liouville_interaction_def} \\
    &= - e^{( t-t_0) \hat{L}_0(\vec \Omega)} \hat{L}'(\vec \Omega, t) e^{- (t-t_0) \hat{L}_0(\vec \Omega)} \rho_\mathrm{I}\left( \vec \Omega, t,t_0 \right)  \, ,
    \label{eq:Liouville_interaction}
\end{align}
where $\hat{L}_\mathrm{I}(\vec \Omega, t,t_0)$ is the Liouville operator in the interaction picture.
We introduce the time propagation operator in the interaction picture $\hat{U}_\mathrm{I}(\vec\Omega, t,t_0) $, which is defined by
\begin{align}
\rho_\mathrm{I}(\vec \Omega, t,t_0) =\hat{U}_\mathrm{I}(\vec\Omega, t,t_0) \rho_\mathrm{I}(\vec \Omega, t_0,t_0) \, .
\label{eq:def_propagator}
\end{align}
Inserting Equation \eqref{eq:def_propagator} into Equation \eqref{eq:Liouville_interaction_def}, leads to the differential equation
\begin{align}
    \dot{\hat{U}}_\mathrm{I}(\vec\Omega, t,t_0)=-\hat{L}_\mathrm{I}(\vec \Omega, t,t_0) \hat{U}_\mathrm{I}(\vec\Omega, t,t_0) \, .    \label{eq:time_prop_interaction}
\end{align}
Equation \eqref{eq:time_prop_interaction}  is solved by the Dyson series 
\begin{multline}    
    \hat{U}_\mathrm{I}(\vec\Omega , t,t_0) = 1- \int\limits_{t_0}^t \mathrm{d} \tau_1 \hat{L}_\mathrm{I}(\vec\Omega , \tau_1,t_0)  + \int\limits_{t_0}^t \mathrm{d} \tau_2 \int\limits_{t_0}^{\tau_2} \mathrm{d} \tau_1 \hat{L}_\mathrm{I}(\vec\Omega , \tau_2,t_0) \hat{L}_\mathrm{I}(\vec\Omega , \tau_1,t_0)   \\ - \int\limits_{t_0}^t \mathrm{d} \tau_3 \int\limits_{t_0}^{\tau_3} \mathrm{d} \tau_2 
    \int\limits_{t_0}^{\tau_2} \mathrm{d} \tau_1 \hat{L}_\mathrm{I}(\vec\Omega , \tau_3,t_0) \hat{L}_\mathrm{I}(\vec\Omega , \tau_2,t_0) 
    \hat{L}_\mathrm{I}(\vec\Omega , \tau_1,t_0)   + ... \, ,
    \label{eq:dyson_series_UI}
\end{multline}
where we used the initial condition $\hat{U}_\mathrm{I}(\vec \Omega, t_0,t_0)=1$.
By combining Equations \eqref{eq:def_interaction}, \eqref{eq:Liouville_interaction}, \eqref{eq:def_propagator}, and \eqref{eq:dyson_series_UI}, we obtain the perturbation expansion of the probability distribution
\begin{gather}
    \rho(\vec \Omega, t) = \sum\limits_{n=0}^{\infty} \rho^{(n)}(\vec \Omega, t)\, 
    \label{eq:pert_expansion_rho} \\
    \rho^{(n)}(\vec \Omega, t) = (-1)^n \int\limits_{t_0}^t \mathrm{d} \tau_n \int\limits_{t_0}^{\tau_n} \mathrm{d} \tau_{n-1} ... \int\limits_{t_0}^{\tau_2} \mathrm{d} \tau_{1}  e^{-( t-\tau_n) \hat{L}_0(\vec \Omega) } \hat{L}'(\vec \Omega,\tau_n)  e^{-( \tau_n-\tau_{n-1}) \hat{L}_0(\vec \Omega) } \hat{L}'(\vec \Omega,\tau_{n-1}) ... e^{-( \tau_2-\tau_1) \hat{L}_0(\vec \Omega) }\hat{L}'(\vec \Omega,\tau_{1}) \rho^{(0)}(\vec \Omega) \, ,
\end{gather}
where we assert that the system is in equilibrium at $t=t_0$, which implies $\rho_\mathrm{I}( \vec \Omega, t_0,t_0)=\rho^{(0)}( \vec \Omega)=e^{-\tau\hat{L}_0(\vec \Omega) }\rho^{(0)}( \vec \Omega)$.
Substituting $t_1=\tau_2-\tau_1$, $t_2=\tau_3-\tau_2$ ... and $t_n=t-\tau_n$ leads to
\begin{multline}  
    \rho^{(n)}(\vec \Omega, t) = (-1)^n  \int\limits_{-\infty}^{\infty} \mathrm{d} t_n \int\limits_{-\infty}^{\infty} \mathrm{d} t_{n-1}... \int\limits_{-\infty}^{\infty} \mathrm{d} t_{1}  \Theta(t_n) \Theta(t_{n-1})... \Theta(t_1) e^{-t_n \hat{L}_0(\vec \Omega) } \\\hat{L}'(\vec\Omega,t-t_n)  e^{-t_{n-1} \hat{L}_0(\vec \Omega) } \hat{L}'(\vec \Omega,t-t_n-t_{n-1}) ... e^{-t_1 \hat{L}_0(\vec \Omega) }\hat{L}'(\vec \Omega,t- t_n-t_{n-1}-...-t_1) \rho^{(0)}(\vec \Omega) \, ,
    \label{eq:final_perturb_expandion}
\end{multline}  
where we set $t_0=-\infty$.
We insert the perturbation expansion of $\rho(\vec \Omega,t)$ in Equation \eqref{eq:final_perturb_expandion} into the definition of an expectation value \eqref{eq:def_observable}, which leads to the time-dependent perturbation expansion of $A(\vec \Omega)$
\begin{align}
\langle A(t) \rangle &= \sum\limits_n^\infty A^{(n)}(t) \\
A^{(n)}(t)&=\int \mathrm{d} \vec \Omega \, A(\vec \Omega) \rho^{(n)}(\vec \Omega,t) \, . 
\end{align}
We assume for a moment that there are no external field gradients present, which corresponds to the simplified perturbation Hamiltonian $H'(\vec\Omega,t)=-F_i(t) P_i(\vec\Omega, t)$. 
In this case, we can define a generic response function of $n^\mathrm{th}$-order $\varphi^{(n)}_{i_n...i_1}[A(\cdot),t_n,...,t_1]$ by 
\begin{align}    
    A^{(n)}(t) = \int\limits_{-\infty}^\infty \mathrm{d}t_n\int\limits_{-\infty}^\infty \mathrm{d}t_{n-1} ... \int\limits_{-\infty}^\infty \mathrm{d}t_{1}  
    F_{i_{n}}(t-t_n) 
    F_{i_{n-1}}(t-t_{n}-t_{n-1})...
    F_{i_1}(t-t_{n}-t_{n-1}-...-t_1) \varphi^{(n)}_{i_ni_{n-1}...i_{1}} \left[ A(\cdot),t_n,t_{n-1},...,t_{1}) \right] \, .
    \label{eq:def_response}
\end{align}
Here, the symbol $\cdot$ indicates that $\varphi^{(n)}_{i_n...i_1}[A(\cdot),t_n,...,t_1]$ depends on the function $A(\vec \Omega)$, but not on the state vector $\vec\Omega$ itself.  
By comparison of Equation \eqref{eq:def_response} with Equation \eqref{eq:final_perturb_expandion}, we obtain the general expression for the nonlinear response function
\begin{multline}        \varphi^{(n)}_{i_1,...i_n}[A(\cdot),t_n,t_{n-1},...,t_1] =  (-1)^n\Theta(t_n) \Theta(t_{n-1})...\Theta(t_1) \int\mathrm{d}\vec \Omega \,A(\vec \Omega) e^{-t_n \hat{L}_0(\vec \Omega)} \\ \left \lbrace P_{i_n}(\vec \Omega), e^{-t_{n-1}  \hat{L}_0(\vec \Omega)} \left\lbrace P_{i_{n-1}}(\vec \Omega), ... e^{-t_{1}  \hat{L}_0(\vec \Omega)}  \left\lbrace P_{i_1}(\vec \Omega), \rho^{(0)}(\vec \Omega) \right\rbrace... \right\rbrace \right\rbrace \, .
    \label{eq:generic_response}
\end{multline}
Experimentally, reflected and transmitted electric fields oscillating with the sum frequency can be measured. 
These are determined by the position-dependent second-order electric current density $j^{(2)}_i(\vec r,t)$, as derived in Section \ref{sec:fresnel}.
We expand $j_i(\vec \Omega,t)$ to second order in the external field, yielding
\begin{multline}
j_i^{(2)}(\vec r,t) = \int\mathrm{d}\vec\Omega j_i( \vec\Omega, \vec r) \rho^{(2)}(\vec \Omega, t) 
=\int\limits_{-\infty}^\infty \frac{\mathrm d \omega_2}{2 \pi} \int\limits_{-\infty}^\infty \frac{\mathrm d \omega_1}{2 \pi}  e^{-i(\omega_1 + \omega_2)t} \bigg[  \tilde F_j(\vec r, \omega_2) \tilde F_k(\vec r, \omega_1) \tilde u^{(2,0)}_{ijk}(\vec r,\omega_1 + \omega_2, \omega_1) \\+ 
 \tilde F_j(\vec r, \omega_2) \tilde u^{(2,1)}_{ijkl}(\vec r,\omega_1 + \omega_2, \omega_1) \frac{\partial}{\partial r_k} \tilde F_l(\vec r, \omega_1) + \tilde F_l(\vec r, \omega_1) \tilde u^{(2,2)}_{ijkl}(\vec r,\omega_1 + \omega_2, \omega_1) \frac{\partial}{\partial r_j} \tilde F_k(\vec r, \omega_2)  
\bigg]
\label{eq:m2}
\end{multline}
where $\tilde f(\omega) = \int\limits_{-\infty}^\infty \mathrm d t e^{i \omega t} f(t)$ is an abbreviation for the Fourier transformation and higher orders terms of external field gradients are omitted.
The second-order response functions $u^{(2,0)}_{ijk}(\vec r, t_2, t_1)$, $u^{(2,1)}_{ijkl}(\vec r, t_2, t_1)$ and $u^{(2,2)}_{ijkl}(\vec r, t_2, t_1)$ describe the second-order electric current density due to wave mixing of two external fields, an external field and an external field gradient and an external field gradient and an external field, respectively. 
They are defined as
\begin{align}
u^{(2,0)}_{ijk}(\vec r, t_2, t_1) &= \Theta(t_2) \Theta(t_1)  \int\mathrm{d} \vec \Omega \,  j_i(\vec\Omega,\vec{r})  e^{-t_2 \hat L_0(\vec\Omega)} \left\lbrace P_j(\vec\Omega), e^{-t_1 \hat L_0(\vec\Omega)}   \left\lbrace  P_k(\vec\Omega) ,\rho^{(0)}(\vec\Omega) \right\rbrace\right\rbrace  
\label{eq:uijk_20}\\
u^{(2,1)}_{ijkl}(\vec r, t_2, t_1) &=  \Theta(t_2) \Theta(t_1)   \int\mathrm{d} \vec \Omega \, j_i(\vec\Omega,\vec{r})  e^{-t_2 \hat L_0 (\vec\Omega)}  \left\lbrace P_j(\vec\Omega), e^{-t_1 \hat L_0(\vec\Omega)}  \left\lbrace  Q_{kl}(\vec\Omega) ,\rho^{(0)}(\vec\Omega) \right\rbrace\right\rbrace   
\label{eq:uijk_21}\\
u^{(2,2)}_{ijkl}(\vec r, t_2, t_1) &=   \Theta(t_2) \Theta(t_1)  \int\mathrm{d} \vec \Omega \, j_i(\vec\Omega,\vec{r}) e^{-t_2 \hat L_0(\vec\Omega)} \left\lbrace  Q_{jk}(\vec\Omega), e^{-t_1 \hat L_0 (\vec\Omega)}  \left\lbrace  P_{l}(\vec \Omega), \rho^{(0)}(\vec\Omega) \right\rbrace\right\rbrace    \, ,
\label{eq:uijk_22} 
\end{align}
where $\rho^{(0)}(\vec \Omega)$ is the time-independent equilibrium probability distribution function, namely the Boltzmann distribution. 
We consider monochromatic external fields
\begin{align}
   F^\alpha_i(\vec r, t) = \mathcal{F}^\alpha_i(\vec r) e^{  -i  \omega^\alpha t } +c.c. \, ,
   \label{eq:def_ext_F}
\end{align}
where $\mathcal{F}^\alpha_i(\vec{r})$ denotes the spatially slowly varying amplitude of an external field, $c.c.$ stands for complex conjugate, and $\alpha$ labels the frequency $\omega^\alpha$. This notation is used for all electric fields, with amplitudes represented by calligraphic symbols.
In our case, we have three frequencies $\omega^\mathrm{VIS}$, $\omega^\mathrm{IR}$ and $\omega^\mathrm{SFG}=\omega^\mathrm{VIS} + \omega^\mathrm{IR}$ and thus $\alpha\in\lbrace \mathrm{SFG}, \mathrm{VIS}, \mathrm{IR}\rbrace$.
We can include both pathways to create a second-order current oscillating with the frequency $\omega^\mathrm{SFG}$ by introducing the second-order response profile
\begin{align}
        \tilde{s}^{(2)}_{ijk}\left( \vec r, \omega^\mathrm{VIS}, \omega^\mathrm{IR} \right) = \frac{1}{- i \varepsilon_0 \omega^\mathrm{SFG}} \big[    \tilde{u}^{(2,\mathrm{eff})}_{ijk}(\vec r,\omega^\mathrm{SFG}, \omega^\mathrm{IR}) + \tilde{u}^{(2,\mathrm{eff})}_{ikj}(\vec r,\omega^\mathrm{SFG}, \omega^\mathrm{VIS})  \big] \, .
        \label{eq:def_s2ijk}
\end{align}
Here, the response function $\tilde{u}^{(2,\mathrm{eff})}_{ijk}(\vec r,\omega^\alpha+\omega^\beta, \omega^\mathrm{\alpha})$ includes contributions induced by the gradients of the external field and is determined by
\begin{multline}
    \tilde{u}^{(2,\mathrm{eff})}_{ijk}(\vec r,\omega^\alpha + \omega^\beta, \omega^\alpha)  = \tilde{u}^{(2,0)}_{ijk}(\vec r,\omega^\alpha + \omega^\beta, \omega^\alpha) + \tilde{u}^{(2,1)}_{ijlk}(\vec r,\omega^\alpha + \omega^\beta, \omega^\alpha) \frac{1}{\mathcal{F}^\alpha_k(\vec r)}\frac{\partial}{\partial r_l} \mathcal{F}^\alpha_k(\vec r) \\
    + \tilde{u}^{(2,2)}_{iljk}(\vec r,\omega^\alpha + \omega^\beta, \omega^\alpha) \frac{1}{\mathcal{F}^\beta_j(\vec r)}\frac{\partial}{\partial r_l} \mathcal{F}_j^\beta(\vec r) + ... \, .
\end{multline}
Hence, we can write the second-order electric current density oscillating with $\omega^\mathrm{SFG}$ as
\begin{align}
   \varepsilon_0^{-1} j^{(2)}_i(\vec r,t) = - i  \omega^\mathrm{SFG} e^{-i\omega^\mathrm{SFG} t } \tilde{s}^{(2)}_{ijk} \left( z,\omega^\mathrm{VIS},\omega^\mathrm{IR} \right) \mathcal{F}^\mathrm{VIS}_j(\vec r)  \mathcal{F}^\mathrm{IR}_k(\vec r)  + c.c. \, ,
    \label{eq:s2ijk}
\end{align}
where we assumed that $\tilde{s}^{(2)}_{ijk} \left( z,\omega^\mathrm{VIS},\omega^\mathrm{IR} \right) $ does only depend on $z$, as is the case for interface systems.
In a difference frequency generation (DFG) experiment, one measures the radiation produced by the second-order current of frequency $\omega^\mathrm{DFG}=\omega^\mathrm{VIS} - \omega^\mathrm{IR}$. 
We can retrieve this by replacing $\omega^\mathrm{VIS} \rightarrow - \omega^\mathrm{VIS}$ in Equation \eqref{eq:s2ijk}.
As $\omega^\mathrm{VIS}>\omega^\mathrm{IR}$, the second-order electric current density measured in DFG spectroscopy has the opposite sign compared to the SFG case.
\section{Second-Order Radiation from a Planar Interface}
\label{sec:fresnel}
\begin{figure}
\centering
\includegraphics[width=0.33\textwidth]{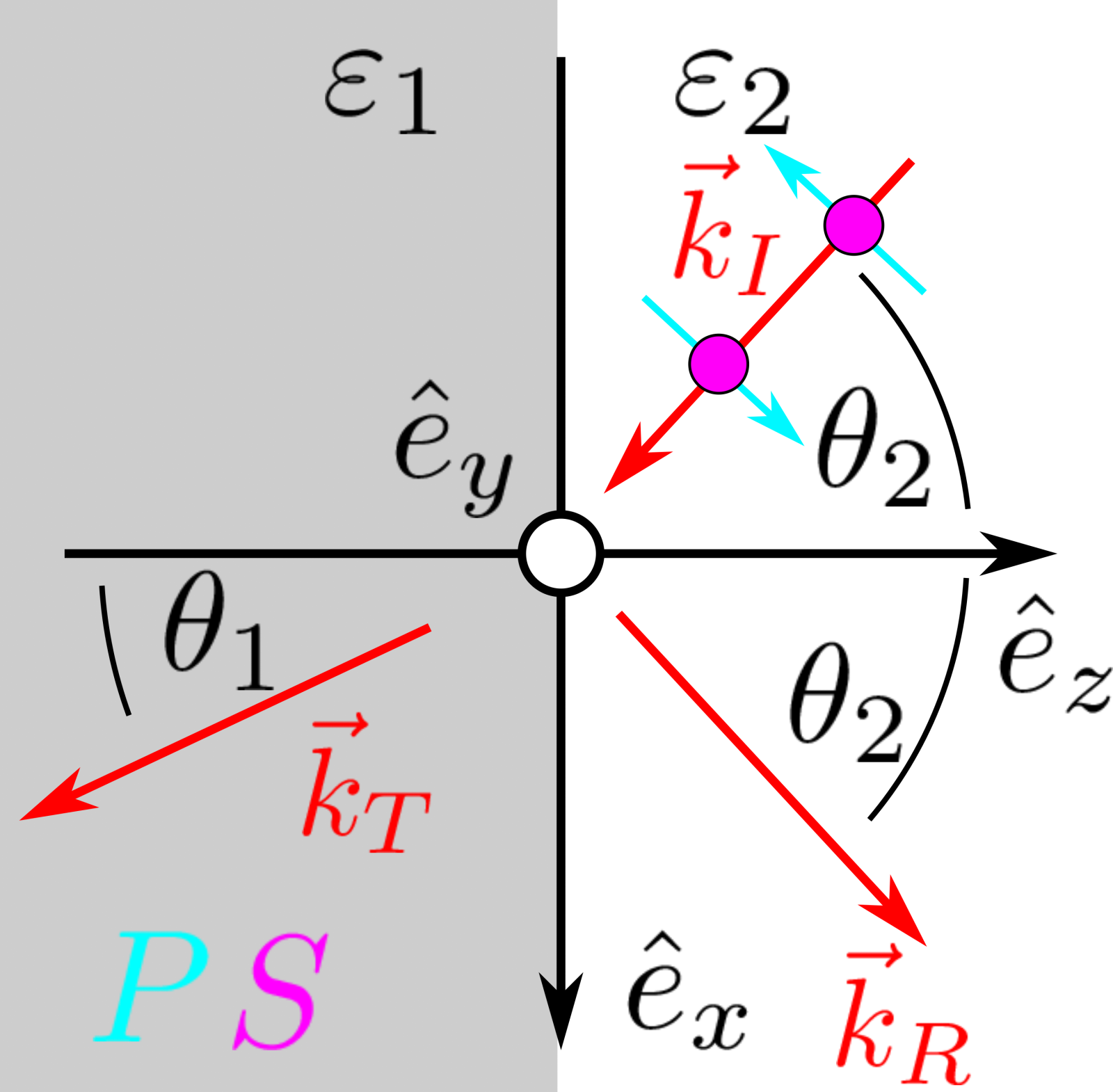}
 \caption{Sketch of the system considered in chapter \ref{sec:fresnel}. The investigated material is on the left side and has the dielectric constant $\tilde\varepsilon_1$. The beams are incident from media $2$, with the incident angle $\theta_2$.   }  
\label{fig:fresnel}
\end{figure}
Here, we derive the radiation outgoing from a second-order electric surface density in planar geometry and relate the intensity of the light sources to the external fields appearing in the perturbation Hamiltonian in Equation \eqref{eq:pert_ham}, without imposing locality approximations. 
We consider an interface between two isotropic bulk media characterized by the spatially constant dielectric constants $\tilde{\varepsilon}^\alpha_1$ and $\tilde{\varepsilon}^\alpha_2$, where the superscript $\alpha$ indicates the frequency of the corresponding external field (SFG, VIS, IR).
In this chapter, we typically assume that the external field amplitudes $\mathcal{F}^\alpha_i(\vec r)$ are constant within the interface region, that is, the area in which the dielectric profile is inhomogeneous. 
As is clear from Equation \eqref{eq:ext_field_maxwell}, this corresponds to the well-known boundary conditions that parallel to the interface the E-field is constant and, in contrast, orthogonal to the interface the D-field is constant, which is regularly applied in optics, for example in the derivation of the well-known Fresnel equations \cite{bornPrinciplesOpticsElectromagnetic1999,jacksonClassicalElectrodynamicsInternational2021}.
The external fields amplitude $\mathcal{F}_i^\alpha(\vec r)$ varies over length scales of the wavelength $\lambda_0^\mathrm{\alpha}=\frac{2 \pi c_0}{\omega^\alpha}$, which is typically considerably larger than the interface region. 
The smallest wavelength in a typical SFG experiment \cite{chiangDielectricFunctionProfile2022a,yuFresnelFactorCorrection2023,fellowsHowThickAirWater2024} is $\lambda_0^\mathrm{SFG}\approx \SI{600}{nm}$, whereas the thickness of the air-water interface is below $\SI{1}{nm}$ \cite{fellowsHowThickAirWater2024}. 
This separation of length scales accounts for both absorption and spatial phase oscillations, which are governed by the real and imaginary parts of the amplitude of the wave vector $k = 2 \pi n^\alpha \lambda_0^{-1}$, respectively, where $n^\alpha$ is the refractive index. 
In water \cite{bertieInfraredIntensitiesLiquids1989}, both the real and imaginary part of the refractive index are smaller than 2.130 at frequencies $\SI{9400}{THz}>\omega>\SI{9}{THz}$ \cite{haleOpticalConstantsWater1973a}.
Figure \ref{fig:fresnel} sketches the system of interest. 
We introduce the electric fields as 
\begin{align}
\vec{E}^{\mathrm{I},\alpha}(\vec r, t) &= \Theta(z) \mathcal{E}^{\mathrm{I},\alpha} e^{-i \left( \omega^\alpha t - \vec{k}^{\mathrm{I},\alpha} \cdot \vec r \right)}
\begin{pmatrix} 
\cos{\phi^\alpha}\cos{\theta^\alpha_2} \\ \sin{\phi^\alpha} \\ \cos{\phi^\alpha}\sin{\theta^\alpha_2} \end{pmatrix} + c.c.
\label{eq:incident} \\
\vec{E}^{\mathrm{T},\alpha} (\vec r, t) &= \Theta(-z) \mathcal{E}^{\mathrm{T},\alpha} e^{-i \left( \omega^\alpha t - \vec{k}^{\mathrm{T},\alpha} \cdot \vec r \right)}
\begin{pmatrix} 
\cos{\phi^\alpha}\cos{\theta^\alpha_1} \\ \sin{\phi^\alpha} \\ \cos{\phi^\alpha}\sin{\theta^\alpha_1} \end{pmatrix} + c.c.
\label{eq:transmitted} \\
\vec{E}^{\mathrm{R},\alpha} (\vec r, t) &= \Theta(z)  \mathcal{E}^{\mathrm{R},\alpha}  e^{-i \left( \omega^\alpha t - \vec{k}^{\mathrm{I},\alpha} \cdot \vec r \right)}
\begin{pmatrix}  
-\cos{\phi^\alpha}\cos{\theta_2^\alpha} \\ \sin{\phi^\alpha} \\ \cos{\phi^\alpha}\sin{\theta_2^\alpha} \end{pmatrix} +c.c. \,,
\label{eq:reflected}
\end{align}
where the angle $\phi$ determines the angle between the $xz$ plane of incidence and the polarization of the electric field. 
Here $\vec{E}^{\mathrm{I},\alpha}(\vec r ,t )$, $\vec{E}^{\mathrm{T},\alpha}(\vec r ,t )$, $\vec{E}^{\mathrm{R},\alpha}(\vec r ,t )$ are the incident, transmitted and reflected electric fields, respectively. The angles are related via Snells law \cite{jacksonClassicalElectrodynamicsInternational2021} as $\tilde n^\alpha_1 \sin{\theta}_1 = \tilde n^\alpha_2 \sin{\theta}_2$, where $ \tilde n^\alpha_a=\sqrt{\tilde{\varepsilon}^\alpha_a}$ is the in general complex refractive index of medium $a\in \lbrace 1 , 2 \rbrace$ at frequency $\omega^\mathrm{\alpha}$.
The corresponding amplitudes are denoted as $\mathcal{E}^{\mathrm{I},\alpha}_\phi$, $\mathcal{E}^{\mathrm{T},\alpha}_\phi$, and $\mathcal{E}^{\mathrm{R},\alpha}_\phi$, representing the incident, transmitted, and reflected components, respectively. The wave vectors defined in Equations \eqref{eq:incident}, \eqref{eq:transmitted} and \eqref{eq:reflected} are
\begin{align}
\vec{k}^{\mathrm{I},\alpha}  =2 \pi \frac{\tilde n_2^\alpha }{\lambda_0^\alpha }
\begin{pmatrix} 
\sin{\theta^\alpha_2} \\ 0 \\ -\cos{\theta_2^\alpha} \end{pmatrix}; \quad 
\vec{k}^{\mathrm{T},\alpha} = 2 \pi \frac{\tilde n_1^\alpha }{\lambda_0^\alpha }
\begin{pmatrix} 
\sin{\theta^\alpha_1} \\ 0 \\ -\cos{\theta^\alpha_1} \end{pmatrix}; \quad
\vec{k}^{\mathrm{R},\alpha}  =2 \pi \frac{\tilde n_2^\alpha }{\lambda_0^\alpha }
\begin{pmatrix} 
\sin{\theta^\alpha_2} \\ 0 \\ \cos{\theta_2^\alpha} \end{pmatrix} \,.
\label{eq:def_k}
\end{align}
It is evident that the aforementioned fields \eqref{eq:incident}-\eqref{eq:reflected} can be written as a linear combination of a beam whose polarization lies in the plane of incidence ($\phi=0$, P-polarized) and one which is perpendicular to the plane of incidence ($\phi=\pi / 2$, S-polarized), we denote the polarization of the beam by the subscript in the amplitudes, e.g. $\mathcal{E}_\mathrm{S}^{\mathrm{I},\alpha}$ and $\mathcal{E}_\mathrm{P}^{\mathrm{I},\alpha}$ are the amplitudes of the S- and P-polarized components of the incident electric field of frequency $\omega^\alpha$. 
Hence, all $y$-polarized quantities are S-polarized, and $x$- or $z$-polarized quantities are P-polarized.
The plane waves in Equations \eqref{eq:incident}-\eqref{eq:reflected} follow Maxwell's equations whenever the dielectric profile $\tilde \varepsilon_{ij}(\omega,z)$ is constant. 
Hence, they hold everywhere except directly at the interface.
We have one incident, transmitted and reflected field, each with the frequency $\omega^\mathrm{VIS}$ and $\omega^\mathrm{IR}$.
The relationships between $\mathcal{E}^{\mathrm{I},\alpha}_\phi$, $\mathcal{E}^{\mathrm{T},\alpha}_\phi$, and $\mathcal{E}^{\mathrm{R},\alpha}_\phi$ are determined by the well-known Fresnel coefficients \cite{jacksonClassicalElectrodynamicsInternational2021}, given below.
We are interested in the SFG signal from the second-order electric current density, determined by the second-order response function defined in Equation \eqref{eq:def_s2ijk}
\begin{align}    
\varepsilon_0^{-1} j^{(2)}_{i}(\vec r,t) =  - i \omega^\mathrm{SFG} e^{- i \left[ \omega^\mathrm{SFG} t - \left( \vec{k}^{\mathrm{T,VIS}} + \vec{k}^{\mathrm{T,IR}} \right) \cdot \vec r \right]} \tilde s^{(2)}_{ijk}(z,\omega^\mathrm{VIS},\omega^\mathrm{IR}) \mathcal{F}^{\mathrm{VIS}}_j(0) \mathcal{F}^{\mathrm{IR}}_k(0) + c.c. \, .
\label{eq:m2_from_ext_2}
\end{align}
Strictly speaking, Equation \eqref{eq:m2_from_ext_2} only holds in medium 1, since we assumed $\mathcal{F}^\alpha_i(\vec r) \propto e^{i \vec{k}^{\mathrm{T},\alpha} \cdot \vec r}$.
However, since $\tilde{s}^{(2)}_{ijk}\left(z, \omega^\mathrm{VIS}, \omega^\mathrm{IR} \right)$, does extend only on an Angstrom scale into medium 2, as shown in the main text, this is an unproblematic assumption.
To relate this second-order current to the setup, we need to relate the amplitudes of the external fields to the amplitudes of the E-fields emitted by the light sources in the experiment.
In the plane of the interface, we have the following relationship between the amplitudes of the electric fields $\mathcal{E}^{\mathrm{I},\alpha}_\phi$, $\mathcal{E}^{\mathrm{R},\alpha}_\phi$ and $\mathcal{E}^{\mathrm{T},\alpha}_\phi$ and the amplitude of the external fields
\begin{align}
\mathcal{F}^\alpha_{x}(0) &= \cos{\theta^\alpha_2} \left( \mathcal{E}_\mathrm{P}^{\mathrm{I},\alpha} - \mathcal{E}_\mathrm{P}^{\mathrm{R},\alpha}  \right) = \cos{\theta^\alpha_1} \mathcal{E}_\mathrm{P}^{\mathrm{T},\alpha}
\label{eq:pre_F_ext_x} \\
\mathcal{F}^\alpha_{y}(0) &=  \mathcal{E}_\mathrm{S}^{\mathrm{I},\alpha}+\mathcal{E}_\mathrm{S}^{\mathrm{R},\alpha} =  \mathcal{E}_\mathrm{S}^{\mathrm{T},\alpha}.
\label{eq:pre_F_ext_y}
\end{align}
As stated in Equation \eqref{eq:ext_field_maxwell}, the z component of the external field corresponds to the electric displacement field. 
Hence, we have
\begin{align}
\mathcal{F}^\alpha_z(0) =  \tilde \varepsilon^\alpha_2 \sin{\theta^\alpha_2} \left( \mathcal{E}_\mathrm{P}^{\mathrm{I},\alpha} + \mathcal{E}_\mathrm{P}^{\mathrm{R},\alpha} \right) =  \tilde \varepsilon_1^\alpha \sin{\theta^\alpha_1} \mathcal{E}_{\mathrm{P}}^\alpha \, .
\label{eq:pre_F_ext_z}
\end{align}
orthogonal to the plane of incidence.
However, the electric field amplitudes are not independent parameters, but are related to each other through the Fresnel coefficients\cite{jacksonClassicalElectrodynamicsInternational2021}
\begin{align}
\mathcal{E}_\mathrm{S}^{\mathrm{R},\alpha}/\mathcal{E}_\mathrm{S}^{\mathrm{I},\alpha} &=\frac{ \tilde n^\alpha_2 \cos{\theta_2^\alpha} - \tilde n^\alpha_1 \cos{\theta^\alpha_1}  }{ \tilde  n^\alpha_1 \cos{\theta^\alpha_1} + \tilde n^\alpha_2 \cos{\theta_2^\alpha}  } \\ 
\mathcal{E}_\mathrm{P}^{\mathrm{R},\alpha}/\mathcal{E}_\mathrm{P}^{\mathrm{I},\alpha} &=\frac{ \tilde n_1^\alpha \cos{\theta_2^\alpha} - \tilde n_2^\alpha \cos{\theta^\alpha_1}  }{ \tilde n_1^\alpha \cos{\theta_2^\alpha} + \tilde n_2^\alpha \cos{\theta^\alpha_1} }
\end{align}
for reflection and 
\begin{align}
\mathcal{E}_\mathrm{S}^{\mathrm{T},\alpha}/\mathcal{E}_\mathrm{S}^{\mathrm{I},\alpha} &=\frac{ 2 \tilde n_2^\alpha\cos{\theta_2^\alpha} }{ \tilde n_1 \cos{\theta^\alpha_1} + \tilde n_2^\alpha\cos{\theta_2^\alpha} }
\label{eq:Fresnel_TS}\\
\mathcal{E}_\mathrm{P}^{\mathrm{T},\alpha}/\mathcal{E}_\mathrm{P}^{\mathrm{I},\alpha} &=\frac{ 2 \tilde n_2^\alpha\cos{\theta_2^\alpha} }{ \tilde n_1 \cos{\theta_2^\alpha} + \tilde n_2^\alpha\cos{\theta^\alpha_1} }
\label{eq:Fresnel_TP}
\end{align}
for transmission.
Hence, we have the following set of equations that connect the experimentally applied incident E-field and the external field appearing in the perturbation Hamiltonian in Equation \eqref{eq:pert_ham} 
\begin{align}
\mathcal{L}^\alpha_{x}=&\mathcal{F}^\alpha_x(0)/\mathcal{E}_\mathrm{P}^{\mathrm{I},\alpha}  =  \frac{  2 \tilde n_2^\alpha\cos{\theta^\alpha_1}\cos{\theta_2^\alpha}  }{  \tilde n_2^\alpha\cos{\theta^\alpha_1} + \tilde n_1 \cos{\theta_2^\alpha}  } 
\label{eq:F_ext_x} \\
\mathcal{L}^\alpha_{y} =& \mathcal{F}_y(0)/\mathcal{E}_\mathrm{S}^{\mathrm{I},\alpha}   =\frac{  2 \tilde n_2^\alpha\cos{\theta_2^\alpha}  }{  \tilde n_1 \cos{\theta^\alpha_1} + \tilde n_2^\alpha\cos{\theta_2^\alpha}  }
\label{eq:F_ext_y} \\
\mathcal{L}^\alpha_{z} =& \mathcal{F}^\alpha_z(0)/\mathcal{E}_\mathrm{P}^{\mathrm{I},\alpha} =\frac{  2 \tilde\varepsilon_1^\alpha \sin{\theta^\alpha_1} \tilde{n}_2^\alpha \cos{\theta_2^\alpha}  }{  \tilde n_2^\alpha \cos{\theta^\alpha_1} + \tilde n_1^\alpha \cos{\theta_2^\alpha}  } \, .
\label{eq:F_ext_z}
\end{align}
The factors $\mathcal{L}^\alpha_i$ are named optical factors in the book by Morita \cite{moritaTheorySumFrequency2018a}.
We can now link the experimental parameters and the second-order electric current density to the electric fields from the light sources. 
In the following, we derive the link between $j^{(2)}_i(\vec r,t)$ and the experimentally measurable second-order radiation.
We write the second-order electric current density in Equation \eqref{eq:m2_from_ext_2}, as
\begin{align}
   j_i^{(2)}(\vec r, t) = \mathcal{J}_i^{(2)}(z) e^{i k^\mathrm{SFG}_x x} e^{- i \omega^\mathrm{SFG} t } + c.c. \, ,
   \label{eq:def_j_prof}
\end{align}
where 
\begin{align}
\mathcal{J}_i^{(2)}(z) = -i \varepsilon_0 \omega^\mathrm{SFG} e^{i \left( k^{\mathrm{T,VIS}}_z + k^{\mathrm{T,IR}}_z \right) z } \tilde s^{(2)}_{ijk}(z,\omega^\mathrm{VIS},\omega^\mathrm{IR}) \mathcal{L}^{\mathrm{VIS}}_j \mathcal{L}^{\mathrm{IR}}_k \mathcal{E}^{\mathrm{I,VIS}} \mathcal{E}^{\mathrm{I,IR}}
\label{eq:J_ampli}
\end{align}
is the vectorial $z$-dependent amplitude of $j_i^{(2)}(\vec r, t)$, valid for S- ($i=x,y$) or P-polarized ($i=z$) incident beams with the incident field amplitudes $ \mathcal{E}^{\mathrm{I,VIS}}$ and $\mathcal{E}^{\mathrm{I,IR}}$.
Typically, in SFG spectroscopy, one measures the amplitude of the second-order reflection $\mathcal{E}_\phi^{\mathrm{R,SFG}}$, where the corresponding electric field amplitude is defined in Equation \eqref{eq:reflected}. We seek a solution for the Green's function $G_{i}(z_0)$ defined by
\begin{align}
\mathcal{E}_\phi^{\mathrm{R,SFG}} = \int\limits_{-\infty}^\infty \mathrm{d} z_0 G_{i}(z_0) \mathcal{J}_i^{(2)}(z_0) \, .
    \label{eq:def_greens_R_SFG}
\end{align}
\begin{figure}
\centering
\includegraphics[width=0.42\textwidth]{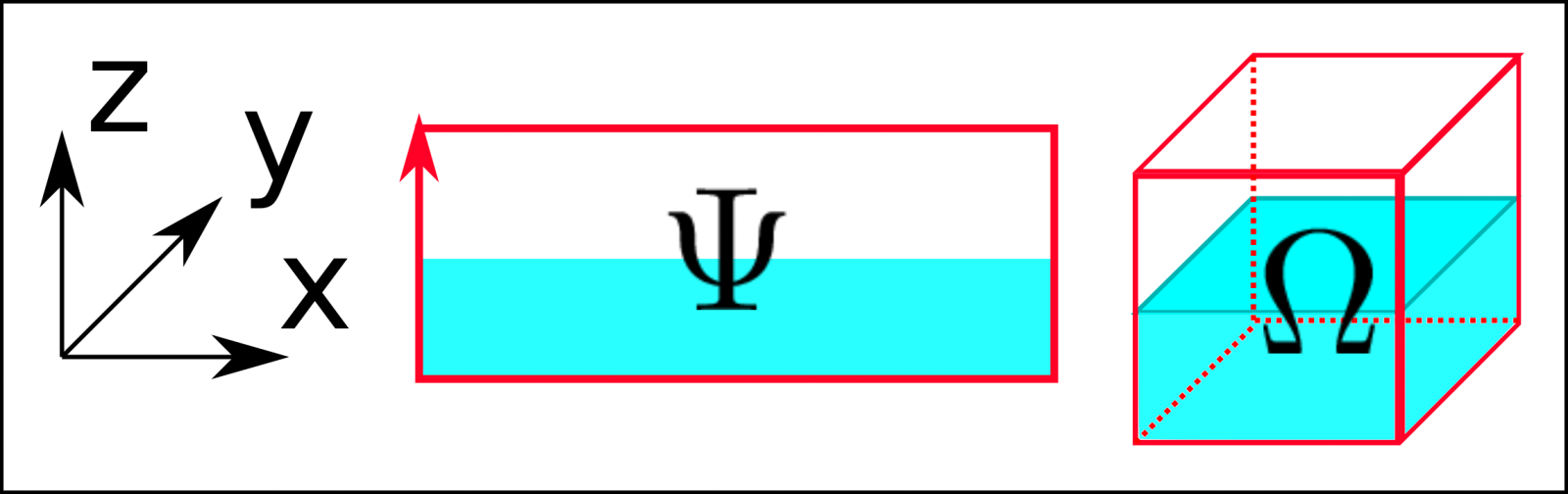}
 \caption{Illustration of the surface $\psi$, with the boundary contour $\partial \psi$ shown in red, alongside the volume $\Omega$, with the boundary surface $\partial \Omega$. The interface lies between the white and the blue regions.}  
\label{fig:maxwells}
\end{figure}
This problem can be solved by applying boundary conditions \cite{ponathNonlinearSurfaceElectromagnetic2012,moritaTheorySumFrequency2018a}.
We work out the Green's function by considering a thin layer $-l_z<z-z_0<l_z$. We split the problem into two parts, a surface current $j^{(2,\mathrm{SURF})}_i(\vec r, t)$ and a bulk current $j^{(2,\mathrm{BULK})}_i(\vec r, t)$, according to
\begin{align}
    j^{(2)}_i(\vec r, t) = j^{(2,\mathrm{SURF})}_i(\vec r, t) + j^{(2,\mathrm{BULK})}_i(\vec r, t) \, .
\end{align}
The surface current is way less than a wavelength away from the interface but is located in a region where the dielectric profile is inhomogeneous. 
The bulk current is located in a region where the dielectric profile is converged to the bulk value but is not necessarily less than a wavelength away from the interface \cite{gonellaWaterChargedInterfaces2021a}.
We begin with the surface current. We set the position of the interface to $z_0=0$. We assume that $l_z<< \lambda_0^\mathrm{SFG}$ and that the dielectric profile is converged to the bulk value at the boundaries, i.e. $\tilde{\varepsilon}_{ij}^\alpha(\pm l_z)=\tilde{\varepsilon}_{2/1}^\alpha$.
For simplicity, we pin the surface current directly at the interface, i.e. 
\begin{align}
    j^{(2,\mathrm{SURF})}_i(\vec r, t) = \delta(z) \mathcal{J}^{(2,\mathrm{SURF})}_i e^{-i \left( \omega^\mathrm{SFG} t - k_x^\mathrm{SFG} x \right)} \, .
    \label{eq:def_surf_current}
\end{align}
However, we will see that the actual position within the interface does not matter. 
First, we work on the boundary condition relevant for the SSP signal, that is the relationship between the $y$ polarized electric field and the $y$ polarized surface current. 
This can be derived from Ampere's law \cite{jacksonClassicalElectrodynamicsInternational2021}
\begin{gather}    
\oint\limits_{\partial \psi} \mathrm d l_i \mathcal B_i(\vec r) = \mu_0 \int\limits_{-l_x}^{l_x} \mathrm d x \int\limits_{-l_z}^{l_z}   \mathrm d z  e^{i k_x^\mathrm{SFG} x}\mathcal{J}^{(2)}_y(z) \\
\implies \Delta \mathcal{B}_x   =  \mu_0 \mathcal{J}^{(2,\mathrm{SURF})}_y,
\label{eq:Delta_Bx}
\end{gather}
where the line integral $\partial \psi$ is around the surface $\psi$, which has the dimensions $2 l_x \times 2 l_z$ and is lying in the plane of the incidence, as depicted in Figure \ref{fig:maxwells}, and $\Delta \mathcal{B}_x=\mathcal{B}_x(+l_z)-\mathcal{B}_x(-l_z)$ is the difference between the amplitudes of the x component of the magnetic field between medium $2$ and medium $1$. 
Here, $\mathcal{J}^{(2,\mathrm{SURF})}_y$ is the integrated amplitude of the second-order current as defined in Equation \eqref{eq:def_surf_current}. 
Because we work in the limit $l_z\rightarrow 0$, the integrals over the $y$ component of the electric field and the $z$ component of the magnetic field vanish. 
Similarly, we have the following result for the permutation of indices
\begin{align}
   \Delta \mathcal{B}_y   =  -\mu_0 \mathcal{J}^{(2,\mathrm{SURF})}_x,
    \label{eq:delta_By}
\end{align}
which is relevant for the SFG signal of a P-polarized current. 
Now, we consider a $z$-polarized second-order electric current density. The magnetic dipole contribution to the z component of current density is zero, i.e. $j^{(2,\mathrm{M})}_z(\vec r,t)=0$.
Hence, the z-polarized second-order current density is entirely determined by the polarization current, and we can use Faraday's law\cite{moritaTheorySumFrequency2018a} to find
\begin{align}
\Delta \mathcal{E}_x&= \frac{\partial}{\partial x} \int\limits_{-l_z}^{l_z} \mathrm d z \mathcal{E}_z (z) =- \frac{\partial}{\partial x} \varepsilon_0^{-1} \int\limits_{-l_z}^{l_z} \mathrm{d} z\, \frac{e^{i k_x^\mathrm{SFG} x}}{-i\omega^\mathrm{SFG}}\mathcal{J}^{(2)}_z(z) \, ,\\
 \Delta \mathcal{E}_x &= \frac{k_x^\mathrm{SFG}}{\varepsilon_0 \omega^\mathrm{SFG}} \mathcal{J}^{(2,\mathrm{SURF})}_z \, ,
\label{eq:Delta_Ex} 
\end{align}
where we use that the $z$ component of the displacement field and the $y$ component of the magnetic field are non-diverging in the layer from $-l_z$ to $l_z$ and $\Delta \mathcal{E}_x$ denotes the difference between the electric field amplitudes between medium $2$ and medium $1$. 
Now, we have three boundary conditions relating the electromagnetic fields above and below an enclosed second-order electric current density.
We know that there is no incident E-field with frequency $\omega^\mathrm{SFG}$, i.e.  $\mathcal{E}^\mathrm{I,SFG}_{\mathrm{S}/\mathrm{P}}=0$.
The amplitudes $\mathcal{E}^\mathrm{R,SFG}_{\mathrm{S}/\mathrm{P}}$ and $\mathcal{E}^\mathrm{T,SFG}_{\mathrm{S}/\mathrm{P}}$ defined in Equations \eqref{eq:reflected} and \eqref{eq:transmitted} are experimentally measurable quantities.
We obtain from  Equations 
\eqref{eq:Delta_Bx} and \eqref{eq:Delta_Ex} and Faraday's law for plane waves the relation
\begin{gather}    
\mathcal{E}^\mathrm{R,SFG}_\mathrm{S} = \mathcal{E}^\mathrm{T,SFG}_\mathrm{S}=\frac{-1}{\varepsilon_0 c_0}\frac{\mathcal{J}^{(2,\mathrm{SURF})}_y}{\tilde n_1^\mathrm{SFG} \cos{\theta^\mathrm{SFG}_1} +  \tilde n_2^\mathrm{SFG} \cos{\theta^\mathrm{SFG}_2}} \, ,
\label{eq:RS_TS_SFG}
\end{gather}
which describes the radiation originating from a second-order S-polarized current. For the P-polarized current we need to consider $\mathcal{J}^{(2,\mathrm{SURF})}_x$ and $\mathcal{J}^{(2,\mathrm{SURF})}_z$.
From the boundary condition in Equation \eqref{eq:delta_By} follows
\begin{gather}
-\frac{1}{c_0 \varepsilon_0} \mathcal{J}^{(2,\mathrm{SURF})}_x =  \tilde n_1^\mathrm{SFG}  \mathcal{E}^\mathrm{T,SFG}_\mathrm{P}  
-  \tilde n_2^\mathrm{SFG}\mathcal{E}^\mathrm{R,SFG}_\mathrm{P} 
\label{eq:delta_ez_insert}
\end{gather}
and we have 
\begin{align}
\frac{ \tilde n_2 \sin{\theta_2}^\mathrm{SFG}}{c_0 \varepsilon_0 } \mathcal{J}^{(2,\mathrm{SURF})}_z &= -\mathcal{E}^{\mathrm{R,SFG}}_\mathrm{P} \cos{\theta_2^\alpha} - \mathcal{E}^{\mathrm{T,SFG}}_\mathrm{P} \cos{\theta^\alpha_1} \, ,
\label{eq:delta_ex_insert}
\end{align}
because of the boundary condition in Equation \eqref{eq:Delta_Ex}.
The solution of Equations \eqref{eq:delta_ez_insert} and \eqref{eq:delta_ex_insert} is given by 
 \begin{align}
\mathcal{E}^{\mathrm{R,SFG}}_\mathrm{P} = \frac{ 1 }{c_0 \varepsilon_0} \frac{\cos{\theta^\mathrm{SFG}_1} \mathcal{J}_x^{(2,\mathrm{SURF})} -\tilde{\varepsilon}_1^\mathrm{SFG}  \sin{\theta^\mathrm{SFG}_1}   \mathcal{J}^{(2,\mathrm{SURF})}_z}{{\tilde n}_1^\mathrm{SFG}\cos{\theta_2^\mathrm{SFG}} + \tilde{n}_2^\mathrm{SFG}\cos{\theta_1^\mathrm{SFG}}} 
\label{eq:RP_SFG}\\
\mathcal{E}^{\mathrm{T,SFG}}_\mathrm{P} = \frac{-1}{c_0 \varepsilon_0} \frac{\cos{\theta^\mathrm{SFG}_2} \mathcal{J}_x^{(2,\mathrm{SURF})} +\tilde{\varepsilon}_2^\mathrm{SFG}  \sin{\theta^\mathrm{SFG}_2}  \mathcal{J}^{(2,\mathrm{SURF})}_z}{\tilde {n}_1^\mathrm{SFG}\cos{\theta_2^\mathrm{SFG}} + \tilde{n}_2^\mathrm{SFG}\cos{\theta_1^\mathrm{SFG}}} \label{eq:TP_SFG} \, .
 \end{align}
 At this point, we have a complete description of the second-order radiation arising from a second-order electric current density located at the interface. 
 The second-order electric current density in the bulk region is not necessarily located directly at the interface. Hence, we must consider that the electric field travels from a second-order current at $z_0$ to the surface. We consider a bulk current localized at $z_0$, i.e.
\begin{align}
    j^{(2,\mathrm{BULK})}_i(\vec r , t) =  \mathcal{J}^{(2,\mathrm{BULK})}_i \delta(z-z_0) e^{-i\left( \omega^\mathrm{SFG} t - k_x^\mathrm{SFG} x \right)} \, .
\end{align}
We know that the electric field directly above and directly below an electric current density in a region with a homogeneous dielectric constant is given by Equations \eqref{eq:RS_TS_SFG}, \eqref{eq:RP_SFG} and \eqref{eq:TP_SFG} when we replace $ n_2^\mathrm{SFG}$, $ \theta_2^\mathrm{SFG}$, with $ n_1^\mathrm{SFG}$, $ \theta_1^\mathrm{SFG}$. 
Furthermore, we know that the radiated E-field in medium 1 arising from the source at $z_0$ is given by
 \begin{align}
     E_i^{\mathrm{I2,SFG}}(\vec r ,t) = \Theta(z - z_0)\Theta(-z) \mathcal{E}_i^{\mathrm{I2,SFG}} e^{- i \left[ \omega^\mathrm{SFG} t - {k}^{\mathrm{T,SFG}}_x x  +  {k}^{\mathrm{T,SFG}}_z (z-z_0) \right]} \, ,
 \end{align}
 where $ \vec{k}^{\mathrm{T,SFG}}$ is defined in Equation \eqref{eq:def_k}. The resulting transmitted E-field in medium 2 with the amplitude $\mathcal{E}^\mathrm{R,SFG}_{\mathrm{S}/\mathrm{P}}$ is then reduced by the transmittance. 
 This attenuation is determined by the corresponding Fresnel relations, i.e., the ratios given in Equations \eqref{eq:Fresnel_TS} or \eqref{eq:Fresnel_TP} when we swap $1,2 \rightarrow 2,1$. 
 Hence, we can write
 \begin{align}
     \mathcal{E}_\mathrm{S}^\mathrm{R,SFG}(z_0)= -   \frac{e^{i {k}^{\mathrm{T,SFG}}_z z_0}}{\varepsilon_0 c_0}  \frac{ \mathcal{J}^{(2,\mathrm{BULK)}}_y }{  \tilde n_1^\mathrm{SFG} \cos{\theta_1^\mathrm{SFG}} +  \tilde n_2^\mathrm{SFG} \cos{\theta_2^\mathrm{SFG}}},
     \label{eq:R_S_SFG_z0}
 \end{align}
 for the radiation created from a S-polarized electric current density $\mathcal{J}^{(2)}_y(z_0)$ located at $z_0$. 
 Similarly, we obtain
  \begin{align}
    \mathcal{E}_\mathrm{P}^\mathrm{R,SFG}(z_0)=  \frac{e^{i {k}^{\mathrm{T,SFG}}_z z_0}}{\varepsilon_0 c_0}  \frac{\cos{\theta_1^\mathrm{SFG}} \mathcal{J}^{(2,\mathrm{BULK)}}_x - \tilde{\varepsilon}_1 \sin{\theta}_1 \mathcal{J}^{(2,\mathrm{BULK)}}_z  }{  \tilde n_1^\mathrm{SFG} \cos{\theta_2^\mathrm{SFG}} +  \tilde n_2^\mathrm{SFG} \cos{\theta_1^\mathrm{SFG}}}  ,
     \label{eq:R_P_SFG_z0}
 \end{align}
 for the intensity of the P-polarized reflection beam. 
 We notice that Equations  \eqref{eq:R_S_SFG_z0} and  \eqref{eq:R_P_SFG_z0} become identical to Equations \eqref{eq:RS_TS_SFG} and \eqref{eq:RP_SFG} if we set $z_0 = 0$. Hence, we can write the second-order radiation originating from a second-order electric current density at an arbitrary position $z_0 \leq 0$ as given in Equation \eqref{eq:def_j_prof} as
 \begin{align}
     \mathcal{E}_\mathrm{S}^\mathrm{R,SFG}(z_0) &= -   \frac{\mathcal{L}_y^\mathrm{SFG} e^{i {k}^{\mathrm{T,SFG}}_z z_0}}{2 \varepsilon_0 c_0 \tilde n_2^\mathrm{SFG} \cos{\theta_2^\mathrm{SFG}} }  \mathcal{J}^{(2)}_y(z_0),
     \label{eq:R_S_SFG_z0_abb} \\
     \begin{split}
          \mathcal{E}_\mathrm{P}^\mathrm{R,SFG}(z_0) &=   \frac{e^{i {k}^{ \mathrm{T,SFG}}_z z_0}}{2 \varepsilon_0 c_0 \tilde n_2^\mathrm{SFG} \cos{\theta_2^\mathrm{SFG}} }  \left[ \mathcal{L}_x^\mathrm{SFG}  \mathcal{J}^{(2)}_x(z_0) - \mathcal{L}_z^\mathrm{SFG} \mathcal{J}^{(2)}_z(z_0) \right] \, ,
     \end{split}
     \label{eq:R_P_SFG_z0_abb}
 \end{align}
 where $\mathcal{L}^\alpha_i$ are the optical factors defined in Equations \eqref{eq:F_ext_x}-\eqref{eq:F_ext_z}.
By combining Equations \eqref{eq:R_S_SFG_z0_abb} and \eqref{eq:R_P_SFG_z0_abb}, we finally obtain an explicit expression for the Green's function defined in Equation \eqref{eq:def_greens_R_SFG}
\begin{align}
      \mathcal{E}^\mathrm{R,SFG} = \frac{ - (1- 2 \delta_{ix} ) \mathcal{L}^\mathrm{SFG}_i }{2 \varepsilon_0 c_0 \tilde n_2^\mathrm{SFG} \cos{\theta_2^\mathrm{SFG}} } \int\limits_{-\infty}^\infty   \mathrm{d} z \, e^{i {k}^\mathrm{T,SFG}_z z} \mathcal{J}_i^{(2)}(z) \, ,
      \label{eq:omit_SP}
\end{align}
where we omit the indices $\phi \in \lbrace \mathrm{S}, \mathrm{P} \rbrace$ because Equation \eqref{eq:omit_SP} holds in both polarizations.
We can insert Equation \eqref{eq:m2_from_ext_2} and the optical factors defined in Equations \eqref{eq:F_ext_x}-\eqref{eq:F_ext_z}, which leads to
\begin{align}    
\mathcal{E}^\mathrm{R,SFG} = \frac{ i \omega^\mathrm{SFG}   (1- 2 \delta_{ix} ) \mathcal{L}^\mathrm{SFG}_i \mathcal{F}^\mathrm{VIS}_j \mathcal{F}^\mathrm{IR}_k   }{2 \varepsilon_0 c_0 \tilde{n}_2^\mathrm{SFG} \cos{\theta_2^\mathrm{SFG}} }  \int\limits_{-\infty}^\infty \mathrm{d} z \, e^{-i \Delta {k}_z z} \tilde{s}_{ijk}^{(2)}\left( z, \omega^\mathrm{VIS} ,\omega^\mathrm{IR} \right) \, ,
\label{eq:R_SFG}
\end{align}
where we introduced the wave vector mismatch 
\begin{align}
\Delta  k_z &= -k_z^\mathrm{T,SFG} -  k_z^\mathrm{T,VIS} -k_z^\mathrm{T,IR} \\ 
&= 2 \pi \left( \frac{\tilde n_1^\mathrm{SFG}}{\lambda_0^\mathrm{SFG}} \cos{\theta_1^\mathrm{SFG}} + \frac{\tilde n_1^\mathrm{VIS}}{\lambda_0^\mathrm{VIS}} \cos{\theta_1^\mathrm{VIS}} + \frac{\tilde n_1^\mathrm{IR}}{\lambda_0^\mathrm{IR}}\cos{\theta_1^\mathrm{IR}}\right)\, .
\label{eq:delta_k}
\end{align} 
In the limit where the VIS laser does not resonate with the system, only $ n^\mathrm{IR}_1$ has a nonzero imaginary part. 
Equation \eqref{eq:R_SFG} is exact and does include polarization, magnetic and multipole contributions. We introduce the SFG signal
 \begin{align}
     \tilde{S}^{(2)}_{ijk}(\omega^\mathrm{VIS},\omega^\mathrm{IR})  &= \int\limits_{-\infty}^\infty \mathrm d z \, e^{-i \Delta k_z z} \tilde{s}^{(2)}_{ijk}(z )
     \label{eq:S2ijk} \, ,
 \end{align}
allowing us to write the second-order reflection in a compact way as
 \begin{align}
\mathcal{E}^\mathrm{R,SFG} &= \gamma_i  \tilde{S}^{(2)}_{ijk}(\omega^\mathrm{VIS}, \omega^\mathrm{IR}) \mathcal{F}_j^\mathrm{VIS} \mathcal{F}_k^\mathrm{IR}  \,  \\
    \mathcal{E}^\mathrm{R,SFG} &= \gamma_i  \tilde{S}^{(2)}_{ijk}(\omega^\mathrm{VIS}, \omega^\mathrm{IR}) \mathcal{L}_j^\mathrm{VIS} \mathcal{E}^\mathrm{I,VIS} \mathcal{L}_k^\mathrm{IR} \mathcal{E}^\mathrm{I,IR}\, ,
    \label{eq:exp_signal}
\end{align}
where the prefactor $\gamma_i$ is given by
\begin{align}
    \gamma_i =  \frac{ i \omega^\mathrm{SFG}   (1- 2 \delta_{ix} )  \mathcal{L}_i^\mathrm{SFG} }{2 \varepsilon_0 c_0 \tilde{n}_2^\mathrm{SFG} \cos{\theta_2^\mathrm{SFG}} } \, ,
\end{align}
and the optical factors $\mathcal{L}_i^\alpha$ are solely determined by \textit{a-priori} known refractive indices of the two bulk media and the experimental setup. 
\section{The Cartesian Multipole Expansion}
Here, we introduce the Cartesian multipole expansion and use it to derive some fundamental relations.
Therefore, we repeat the necessary basic theory of electrostatics, which can be found in textbooks \cite{jacksonClassicalElectrodynamicsInternational2021,bornPrinciplesOpticsElectromagnetic1999,wolfProgressOpticsVol1977}. 
Further, we derive approximate relations for interfacial nonlinear and multipolar constitutive relations, based on the theory by Mizrahi and Sipe in 1986 \cite{mizrahiLocalfieldCorrectionsSumfrequency1986} and Hirano and Morita in 2024  \cite{hiranoLocalFieldEffects2024}.
\subsection{The Multipole Expansion of the Polarization Density}
We consider the spatial averaging operation for an arbitrary function $g(\vec r)$ 
\begin{align}
    g^\mathrm{S} (\vec{r})=\int \mathrm{d} \vec{r}' s(\vec r-\vec{r'}) g(\vec r') \, ,
    \label{def:spat_av}
\end{align}
where $s(\vec r)$ is a normalized smoothing function, for example, a three-dimensional normal distribution. 
The charge density results from the sum over the $N_\mathrm{mol}$ molecular charge densities $\varrho^n(\vec r)$ according to
\begin{align}
\varrho(\vec{r})=\sum_n^{N_{\mathrm{mol}}}\varrho^n(\vec{r}-\vec r^n)  \, .
\label{eq:mol_charge}
\end{align}
Hence, the spatially averaged charge density is determined by
\begin{align}
 \varrho^\mathrm{S}(\vec{r}) &= \sum_n^{N_{\mathrm{mol}}} \int \mathrm{d} \vec{r}' s(\vec{r}-\vec{r}') \varrho^n(\vec{r}'-\vec r^n) \,  \\
    &= \sum_n^{N_{\mathrm{mol}}} \int \mathrm{d} \vec{r}' s(\vec{r}-\vec r^n-\vec{r}') \varrho^n(\vec{r}') \, .
\end{align}
A Taylor expansion of the smoothing function $s(\vec r-\vec r^n - \vec r' )$ around $\vec r-\vec r^n$ leads to
\begin{align}
\varrho^\mathrm{S}(\vec{r}) = \sum\limits_n^{N_\mathrm{mol}}q^n s(\vec r- \vec r^n) -  \frac{\partial} {\partial r_i}    \sum\limits_n^{N_\mathrm{mol}} \mu_i^n     
        s(\vec r- \vec r^n) +   \frac{\partial^2}{\partial r_i \partial r_j}  \sum\limits_n^{N_\mathrm{mol}} Q_{ij}^n s(\vec r- \vec r^n) + ... \, ,   \label{eq:rho_multi_taylor}
\end{align}
where the Cartesian molecular multipoles are defined as
\begin{align}
    q^n &= \iiint \mathrm{d} V \, \varrho^n(\vec r) 
    \label{eq:Q_0n} \\
    \mu_i^n &= \iiint \mathrm{d} V \,r_i \varrho^n(\vec r) 
    \label{eq:Q_In} \\
 Q_{ij}^n &= \frac{1}{2}\iiint \mathrm{d} V \,r_i r_j \varrho^n(\vec r) \, .
  \label{eq:Q_IIn} 
\end{align} 
Except for the first non-zero multipole moment, this expansion does depend on the choice of the molecular origin, i.e., the origin of the molecular charge distributions $\varrho^{n}(\vec r)$. 
We introduce the shorthand notation 
\begin{align}
    \varrho^\mathrm{S}(\vec r) = \varrho^{\mathrm{q}}(\vec r ) - \frac{\partial}{\partial r_i}  \varrho_i^{\mathrm{D}}(\vec r  ) +  \frac{\partial}{\partial r_i} \frac{\partial}{\partial r_j}   \varrho_{ij}^{\mathrm{Q}}(\vec r ) + ... ,
    \label{eq:multipole_expansion} \, ,
\end{align}
where 
\begin{align}
    \varrho^{\mathrm{q}}(\vec r )& =  \sum\limits_n^{N_{\mathrm{mol}}} q^n  s(\vec r - \vec r^n) 
    \label{eq:rho_q} \\
    \varrho^{\mathrm{D}}_i(\vec r)& = \sum\limits_n^{N_{\mathrm{mol}}} \mu^n_i s(\vec r - \vec r^n) 
    \label{eq:rho_D} \\
\varrho^{\mathrm{Q}}_{ij}(\vec r)& =\sum\limits_n^{N_{\mathrm{mol}}} Q^n_{ij} s(\vec r - \vec r^n)
     \label{eq:rho_Q}
\end{align}
are the molecular multipole densities. 
We decompose the E-field $E_i(\vec r)$ into the D-field $D_i(\vec r)$ and the polarization density $p_i(\vec r)$  according to
\begin{align}
   \varepsilon_0 E_i(\vec r) =  D_i (\vec r ) - p_i(\vec r ) \, ,
    \label{eq:def_EField}
\end{align}
where the divergence of the D-field $D_i (\vec r  )$ is determined by the so-called free charge distribution $\varrho_\mathrm{F}(\vec r)$
\begin{align}
    \frac{\partial}{\partial r_i} D_i (\vec r ) = \varrho^\mathrm{F}(\vec r  )\, ,
    \label{eq:def_D}
\end{align}
and the divergence of the polarization density $p_i(\vec r )$ is determined by the remaining bound charge distribution
\begin{align}
  \frac{\partial}{\partial r_i}  p_i(\vec r) &= - \varrho(\vec r) 
  \label{eq:def_P} \\
 p^\mathrm{S}_i(\vec r) &= \varrho_i^\mathrm{D}(\vec r) -\frac{\partial}{\partial r_j} \varrho_{ij}^\mathrm{Q}(\vec r) + ... \, .
    \label{eq:multipol_P}
\end{align}
We note that the D-field $D_i(\vec{r})$ and the polarization density $p_i(\vec{r})$ are not fully determined by Equations~\eqref{eq:def_D} and \eqref{eq:def_P}. 
Here, we assumed that the charge distribution of the polarizable material $\varrho(\vec r)$ contains no molecular monopoles $\varrho^\mathrm{q}(\vec r)=0$ in Equation \eqref{eq:multipol_P}. 
However, under suitable conditions a polarization density created by a monopole density can be defined \cite{bonthuisProfileStaticPermittivity2012}.
 Choosing a three-dimensional delta distribution for $s(\vec r-\vec{r'})=\delta(x-x')\delta(y-y')\delta(z-z')$ leads to $p_i^\mathrm{S}(\vec r)=p_i(\vec r)$. 
This corresponds to assuming that the polarization density $p_i(\vec r)$ is constant on length scales on which $s(\vec r)$ is nonzero.
In this work, we predict the higher-order expectation value of the polarization density, which we assume to be sufficiently smooth on the molecular length scale. 
Even when this assumption is not justified, choosing the delta distribution remains the best option for the spatial averaging function, as it preserves all information about the position. 
Consequently, we replace
\begin{align}
    s(\vec r-\vec{r'})\rightarrow \delta(x-x')\delta(y-y')\delta(z-z') \, .
\end{align} 
Thus, we leave out the superscript $S$ in the following. 
The multipole expansion is a convenient way to relate molecular properties to the fields that enter Maxwell's equations. 
\subsection{Identifying the External Field in Planar Geometry}
We define $\phi^\mathrm{P}(\vec r)$ as the electrostatic potential created by the bound charge density distribution $\varrho(\vec r,t)$ determined by Poisson's equation
\begin{align}
   \nabla^2 \phi^\mathrm{P}(\vec r) = - \varepsilon_0^{-1} \varrho(\vec r ) \, .
   \label{eq:poisson}
\end{align}
Next, we use the Hertz vector method to compute the electrostatic E-field $E_i(\vec r_i)$ imposed by the polarization density $p_i(\vec r)$. 
We define the electric Hertz vector field $\Pi_i(\vec r)$ as \cite{bornPrinciplesOpticsElectromagnetic1999,mizrahiLocalfieldCorrectionsSumfrequency1986} 
\begin{align}
    \phi^\mathrm{P}(\vec r) = -\frac{\partial}{\partial r_i} \Pi_i(\vec r) \, ,
    \label{eq:def_HertzVec}
\end{align}
We insert the polarization density defined in Equation \eqref{eq:def_P} and the Hertz vector field defined in Equation \eqref{eq:def_HertzVec} in the Poisson Equation \eqref{eq:poisson}, which gives us
\begin{align}
 \frac{\partial}{\partial r_i}  \nabla^2    \Pi_i(\vec r) =  - \frac{\partial}{\partial r_i} \varepsilon_0^{-1} p_i(\vec{r  }) \, .
   \label{eq:poisson_Hertz_vec}
\end{align}
Equation \eqref{eq:poisson_Hertz_vec} is satisfied, when
\begin{align}
   \nabla^2   \Pi_i(\vec r) =  -  \varepsilon_0^{-1} p_i(\vec{r  }) \,   \label{eq:poisson_Hertz_vec_2}
\end{align}
holds. Equation \eqref{eq:poisson_Hertz_vec_2} can be solved via a Green's function approach as
\begin{align}
    \Pi_i(\vec r) = \iiint\mathrm{d}\vec r' \frac{p_i(\vec r')}{4 \pi \varepsilon_0  |\vec r- \vec r'|} \, .
\end{align}
Consequently, the total electric field appearing in Equation \eqref{eq:def_EField} is determined by \cite{bornPrinciplesOpticsElectromagnetic1999,mizrahiLocalfieldCorrectionsSumfrequency1986}
\begin{align}
    E_i(\vec r) = \frac{\partial}{\partial r_i} \frac{\partial}{\partial r_j} \iiint\mathrm{d}\vec r' \frac{p_j(\vec r')}{4 \pi \varepsilon_0 |\vec r- \vec r'|} + F_i(\vec r) \, .
    \label{eq:EField}
\end{align}
Equation \eqref{eq:EField} defines the external field $F_i(\vec r)$ as the additional field beside the electrostatic field imposed by the polarization density -$\frac{\partial}{\partial r_i} \phi^\mathrm{P}(\vec r)$
\begin{align}
    E_i(\vec r) =F_i(\vec r)-\frac{\partial}{\partial r_i} \phi^\mathrm{P}(\vec r) \, . 
    \label{eq:def_ext_Field}
\end{align}
If one splits the integral into an infinitesimally small sphere around $\vec r$, which contains the singularity and the remaining volume $\sigma(\vec r)$, one obtains \cite{wolfProgressOpticsVol1977}
\begin{align}
    E_i(\vec r) &=  \iiint\limits_{\sigma(\vec r)}\mathrm{d}\vec r' p_j(\vec r')\frac{\partial}{\partial r_i} \frac{\partial}{\partial r_j}\frac{1}{4 \pi \varepsilon_0 |\vec r- \vec r'|} - \frac{1}{3 \varepsilon_0} p_i(\vec r)+ F_i(\vec r) \\
    &=\iiint\limits_{\sigma(\vec r)}\mathrm{d}\vec r' T^{(2)}_{ij}(\vec r-\vec r')  p_j(\vec r') - \frac{1}{3 \varepsilon_0} p_i(\vec r) + F_i(\vec r) \, ,
    \label{eq:greens_function_E_P}
\end{align}
where we introduced the electrostatic coupling tensor\cite{grayTheoryMolecularFluids1984} 
\begin{align}
\tensor{T}^{(l)}(\vec r)=\nabla^l \frac{1}{4 \pi \varepsilon_0 |\vec r|} \, .
\label{eq:def_T_tensor}
\end{align}
Equation \eqref{eq:greens_function_E_P} is independent of the multipole expansion of $p_i(\vec r)$ defined in Equation \eqref{eq:multipol_P}. 
If the system is only inhomogeneous in one dimension, the polarization density $p_i(\vec r)$ depends only on $z$, and one can solve Equation \eqref{eq:EField} analytically \cite{sternCalculationDielectricPermittivity2003},
which is determined by
\begin{align}
    E_i(z) &= -\frac{\delta_{iz}}{\varepsilon_0} p_i(z)+ F_i(z) \, ,
    \label{eq:greens_function_E_P_only_z}
\end{align}
under the appropriate boundary conditions \cite{sternCalculationDielectricPermittivity2003}. Solving Equation \eqref{eq:greens_function_E_P_only_z} for $F_i(z)$ leads to
\begin{align}
    F_x(z) &= E_x(z) \\
    F_z(z) &= E_z(z) + \frac{1}{\varepsilon_0} p_z(z) \, .
\end{align}
These equations are equivalent to Equation \eqref{eq:ext_field_maxwell}.
\subsection{The Multipole Expansion of the Local Field}
We consider the electric field $E_i^n$ acting on the $n^\mathrm{th}$-molecule imposed by the other molecules in addition to an external field, as follows from Equations \eqref{eq:mol_charge} and \eqref{eq:def_ext_Field} 
\begin{align}
    E_i^{n} &= -\frac{\partial}{\partial r^n_i} \sum\limits_{m \neq n}^{N_{mol}} \phi^m(\vec r^n) + F_i^n \\
    &=-\sum\limits_{m\neq n}^{N_\mathrm{mol}} \iiint \mathrm{d} \vec r' \,T^{(1)}_i(\vec r^n-\vec r^m-\vec r') \varrho^m(\vec r') + F^n_i\, ,
\end{align}
where $F^n_i=F_i(\vec r^n)$ is the external field and $\phi^n(\vec r)$ is the electrostatic potential imposed by the $n^\mathrm{th}$ molecule
\begin{align}
   \nabla^2 \phi^n(\vec r) = -\varepsilon_0^{-1} \varrho^n(\vec r - \vec r^n) \, . 
\end{align}
A Taylor expansion of the electrostatic coupling tensor defined in Equaton \eqref{eq:def_T_tensor} in $\vec r'$  leads to \cite{grayTheoryMolecularFluids1984}
\begin{align}
    E_i^n = \sum\limits_{m\neq n} \left[ -T^{(1)}_i(\vec r^n-\vec r^m) q^m +  T^{(2)}_{ij}(\vec r^n-\vec r^m) \mu_j^m - T^{(3)}_{ijk}(\vec r^n-\vec r^m) Q_{jk}^m + ... \right] + F_i^n\, .
    \label{eq:En}
\end{align}
\subsection{The Multipolar Lorentz-Field In Planar Geometry}
\begin{figure}
     \centering   \includegraphics[width=1.0\linewidth]{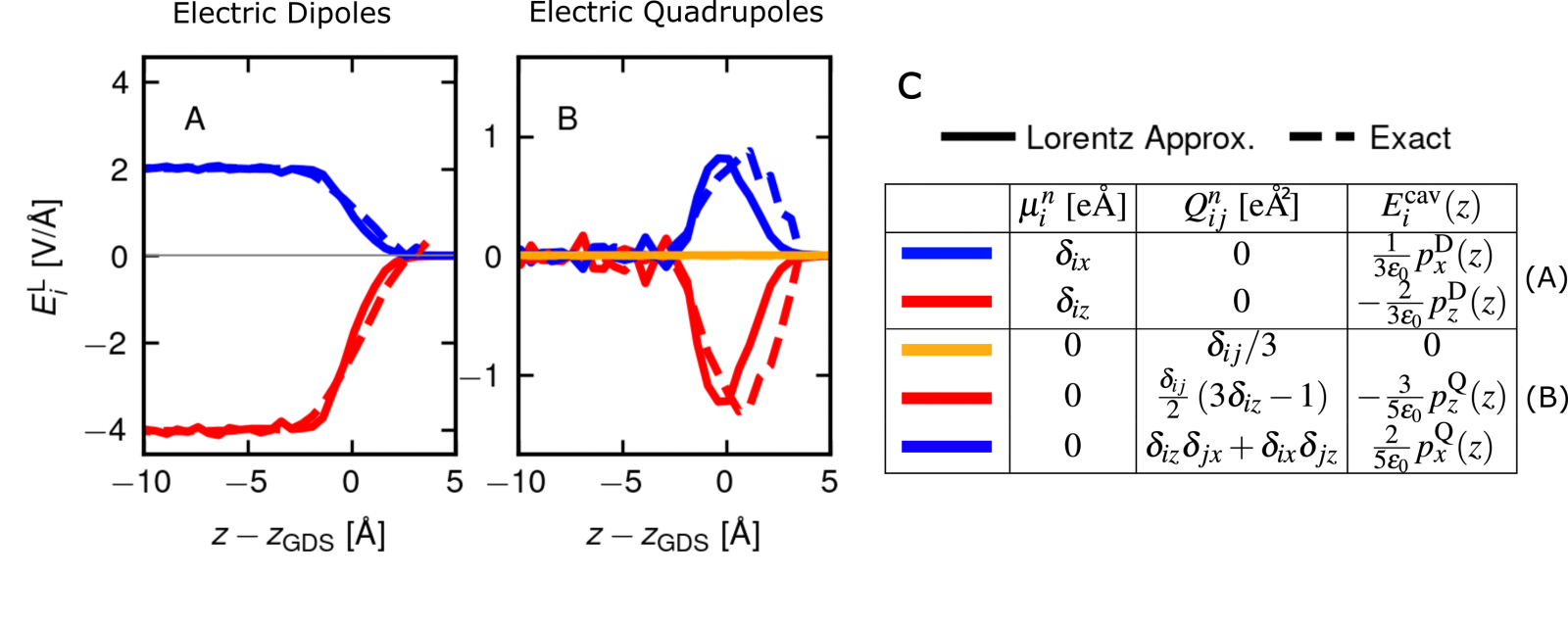}
     \caption{
     The Lorentz-field Approximation \eqref{eq:LorentzField} is tested by  comparing the cavity field $E^\mathrm{cav}_i(z)$ given in Equation \eqref{eq:Ecav} to the average electric field acting on the molecular centers $E^\mathrm{L}_i(z)$, defined in Equation \eqref{eq:aver_field} (dashed lines). 
     For this, identical multipoles, tabulated in C, are placed at the molecular centers.
     The resulting fields $E^\mathrm{cav}_i(z)$ and  $E^\mathrm{L}_i(z)$ are compared in A and in B. 
     The analytically prediction of $E^\mathrm{cav}_i(z)$ based on the density profiles of the multipoles, is presented in Table C.
   The first two rows in C define the dipoles, whose field is shown in A. 
   The last three define the quadrupoles, whose field is shown in B.
     }
\label{fig:validate_Lorentz_approx}
 \end{figure}

Equation \eqref{eq:En} describes the electrostatic field acting on a molecule given a certain molecular configuration. 
In a continuum description, the charge density distribution is described by smooth multipole densities $ \varrho^\mathrm{q}(\vec r)$, $\varrho_{i}^\mathrm{D}(\vec r)$, $\varrho_{ij}^\mathrm{Q}(\vec r)$, ... , defined in Equations \eqref{eq:rho_q}-\eqref{eq:rho_Q}.
In analogy to Equation \eqref{eq:En}, the cavity field can be defined, as the field acting in a cavity carved in a medium characterized by continuous electric multipole densities 
\begin{align}
    E^\mathrm{cav}_i(\vec r) = \iiint\limits_{\sigma(\vec r)} \mathrm{d} \vec r' \left[- T_i^{(1)}(\vec r- \vec r') \varrho^\mathrm{q}(\vec r')+
    T_{ij}^{(2)}(\vec r- \vec r') \varrho^\mathrm{D}_j(\vec r') -  T_{ijk}^{(3)}(\vec r- \vec r') \varrho^\mathrm{Q}_{jk}(\vec r') + ...
    \right] \, + F_i(\vec r) \, .
    \label{eq:Ecav}
\end{align}
Here, we apply the theory by Mizrahi and Sipe \cite{mizrahiLocalfieldCorrectionsSumfrequency1986} to derive the cavity field $E_i^\mathrm{cav}(z)$ for systems that are translationally invariant in the $xy$ plane. 
The goal is to derive an approximate expression for the local electric field acting on a molecular center, based on the molecular position $z$. 
We assume that only the electric dipole and electric quadrupole densities are nonzero. 
We insert the relation
\begin{align}
 -T^{(3)}_{ijk}(\vec r - \vec r') \rho^\mathrm{Q}_{jk}(\vec r')   =  -T^{(2)}_{ij}(\vec r - \vec r') \frac{\partial}{\partial r_k' }\rho^\mathrm{Q}_{jk}(\vec r')  + \frac{\partial}{\partial r_k' }  \left[T^{(2)}_{ij}(\vec r - \vec r') \rho^\mathrm{Q}_{jk}(\vec r') \right]
\end{align}
into Equation \eqref{eq:Ecav}, which leads to
\begin{align}
    E^\mathrm{cav}_i(\vec r) = \iiint\limits_{\sigma(\vec r)} \mathrm{d} \vec r' T^{(2)}_{ij}(\vec r - \vec r') p_j(\vec r') - \iint\limits_{S(\sigma)} \mathrm{d} S_k T_{ij}^{(2)}(\vec r-\vec r') \varrho^{\mathrm{Q}}_{jk}(\vec r') + F_i(\vec r)\, ,
\label{eq:local_field_conti_2} 
\end{align}
where $\mathrm{d}S_k$ is the vector surface element and the area integral is over the sphere's surface $\sigma(\vec r)$. Note that the negative sign is due to the surface normal vector pointing outward from the sphere. 
We assume that the system is only inhomogeneous in the $z$ component and insert the Taylor expansion of the quadrupole density $\varrho^{\mathrm{Q}}_{ij}(z)$ around $z$, which leads to
\begin{align}
    E^\mathrm{cav}_i(z) = \iiint\limits_{\sigma(\vec r)} \mathrm{d} \vec r' T^{(2)}_{ij}(\vec r - \vec r') p_j(z') - \frac{1}{15  \varepsilon_0} \frac{\partial}{\partial z }\varrho^\mathrm{Q}_{{iz}}(z) -\frac{\delta_{iz}}{5 \varepsilon_0} \frac{\partial}{\partial z }\varrho^\mathrm{Q}_{{jj}}(z) + F_i(z)
  \label{eq:local_field_conti_3} 
\end{align}
We add $0=\frac{1}{3 \varepsilon_0} p_i(z) -\frac{1}{3 \varepsilon_0} p_i(z) $ and substitute Equations \eqref{eq:greens_function_E_P} and \eqref{eq:greens_function_E_P_only_z} and the multipole expansion of the polarization density given in Equation \eqref{eq:multipol_P}, which leads to
\begin{align}
    E^{\mathrm{cav}}_i(z) = F_i(z)-\delta_{iz} \varepsilon_0^{-1} p_z(z) +  \frac{1}{3 \varepsilon_0} \varrho^\mathrm{D}_i(z) - \frac{2}{5 \varepsilon_0} \frac{\partial}{\partial z} \varrho^\mathrm{Q}_{iz}(z) - \frac{\delta_{iz}}{5 \varepsilon_0} \frac{\partial}{\partial z} \varrho^\mathrm{Q}_{jj}(z) \, .
    \label{eq:cav_field_final}
\end{align}
We define the average field acting on the molecular centers
 \begin{align}
     E^\mathrm{L}_i (z)=\frac{\sum\limits_n^{N_\mathrm{mol}}\langle \delta(z-z^n) E^n_i \rangle}{\sum\limits_n^{N_\mathrm{mol}}\langle \delta(z-z^n)\rangle} \, ,
     \label{eq:aver_field}
 \end{align}
 where $\langle ... \rangle$ denotes the expectation value, defined in Equation \eqref{eq:def_observable}.
Equation \eqref{eq:cav_field_final} approximates the average local field $E_i^n$ acting on the molecular centers defined in Equation \eqref{eq:aver_field}. 
We refer to the approximation
\begin{align}
 E^\mathrm{L}_i(z)\approx E^\mathrm{cav}_i (z)   \label{eq:LorentzField}
\end{align}
as the Lorentz-field approximation.
This approximation is consistent with the fact that the trace of the molecular electric quadrupole moment does not contribute to the electrostatic field acting on other molecules \cite{grayTheoryMolecularFluids1984}, as can be seen by inserting $\varrho^{\mathrm{Q}}_{ij}(z)=\delta_{ij} \rho(z)$ into Equation \eqref{eq:cav_field_final}.
We check the validity of the Lorentz-field approximation by comparing the average E-field acting on the molecular centers $E^\mathrm{L}_i (z)$ directly computed with Equation \eqref{eq:aver_field} and analytic predictions based on the Lorentz-field approximation \eqref{eq:LorentzField} in Figure \ref{fig:validate_Lorentz_approx}.
We use the air-water interface as the test system. 
Here, we use the center of mass of the water molecules as the molecular centers and set the molecular multipoles to the values tabulated in Figure  \ref{fig:validate_Lorentz_approx} C. 
Afterwards, we compute $E_i^\mathrm{L}(z)$ according to Equation \eqref{eq:aver_field} and 
compare it with the analytically predicted cavity field determined by Equation \eqref{eq:cav_field_final} and tabulated in Figure \ref{fig:validate_Lorentz_approx} C.
The density profile of the molecular centers is proportional to $E^\mathrm{L}_x(z)$ (solid blue line) presented in Figure \ref{fig:validate_Lorentz_approx} A.
As can be seen in Figure \ref{fig:validate_Lorentz_approx} A the Lorentz-field approximation does capture the electrostatic field imposed by distributions of molecular electric dipoles quite well. 
The same holds for a distribution of molecular electric quadrupoles, who create an electric field proportional to their gradients,  as can be seen in  Figure \ref{fig:validate_Lorentz_approx} B.
Clearly, the Lorentz-field approximation in Equation \eqref{eq:LorentzField} provides a reliable estimate for the laterally averaged E-field defined in Equation \eqref{eq:aver_field}. 
The trajectory of the air–water interface system was generated as described in the Methods section of the main text. While the results presented in the main text are averaged over 94 such trajectories, this test uses only one of them, as extensive sampling is not required here. Specifically, we use molecular centers extracted from frames spaced by $\SI{160}{fs}$ over a single $\SI{0.9}{ns}$ trajectory of the air–water interface system. The numerical extraction of $E^\mathrm{L}_i(z)$ is performed using Ewald summation with periodic boundary conditions in all dimensions, as described in Section~\ref{app:polarizability}.
The contribution to $E^\mathrm{L}_i(z)$ from the periodic replicas in the $z$-dimension is removed, as outlined in Section~\ref{app:pbc}.
\subsection{Lorentz-Field Approximation for the Constitutive Relation}
\label{sec:simple_SFG}
Here, we derive constitutive relations that predict the total polarization density of a dipolar dielectric continuum perturbed by the electric dipole source density $\varrho^{\mathrm{DS}}_i(z,t)$, and the electric quadrupole source density, $\varrho^{\mathrm{QS}}_{ij}(z,t)$, using the Lorentz-field approximation defined in Equation~\eqref{eq:LorentzField}. 
We refer to $\varrho^{\mathrm{DS}}_i(z,t)$ and $\varrho^{\mathrm{QS}}_i(z,t)$ as source densities because they act as sources that induce a linear electric dipole density $\varrho_i^\mathrm{DL,\alpha}(z)$. 
We define $\mathcal{P}^\alpha_i(z)$ as the amplitude of $p_i^\alpha(z,t)$
\begin{align}
    p_i^\alpha(z,t) = \mathcal{P}^\alpha_i(z) e^{-i \omega^\alpha t } +c.c. \, .
\end{align}
We are interested in the local electric field oscillating at frequency $\omega^\alpha$ acting on a molecule placed in a spherical cavity within a continuous multipolar charge distribution, within the Lorentz-field approximation in Equation \eqref{eq:LorentzField}, given by
\begin{align}
\mathcal{E}^\mathrm{L,\alpha}_i(z) = \mathcal{F}^{\alpha}_i(z)+ \varepsilon_0^{-1} \left(\frac{1}{3} - \delta_{iz} \right) \left[ \varrho^\mathrm{DS,\alpha}_i(z) + \varrho^\mathrm{DL,\alpha}_i(z) \right] - \varepsilon_0^{-1} \frac{\partial}{\partial z} \left[ \left(\frac{2}{5 } - \delta_{iz} \right)\varrho^\mathrm{QS,\alpha}_{iz}(z) + \frac{\delta_{iz}}{5}\varrho^\mathrm{QS,\alpha}_{jj}(z) \right]  \, ,
\label{eq:local_field_multipole}
\end{align}
where $\varrho_i^\mathrm{DL,\alpha}(z)$ is the linear response to the local E-field determined by
\begin{align}
\varrho_i^\mathrm{DL,\alpha}(z) = \rho(z) \alpha \mathcal{E}_i^\mathrm{L,\alpha}(z) \, ,
    \label{eq:lin2loc}
\end{align}
where $\rho(z)$ is the molecular density and $\alpha$ is 
the molecular electric dipole - electric dipole polarizability, whose average is approximated to be isotropic \cite{bornPrinciplesOpticsElectromagnetic1999}. 
As demonstrated in the main text, the dielectric response of the air-water interface is only weakly anisotropic. 
Although this approximation is not required for the derivation, it simplifies the resulting expressions. 
Given that the goal is to obtain simple relations and that more significant approximations are already introduced through the Lorentz-field approximation, this simplification is justified.
 Now, we eliminate the local field by inserting Equation \eqref{eq:local_field_multipole} into Equation \eqref{eq:lin2loc}, which we solve for the linear response
\begin{multline}    
\varrho_i^\mathrm{DL,\alpha}(z) = \frac{ \varepsilon_0 \rho(z) \alpha }{\varepsilon_0 - \rho(z)\alpha \left[ \frac{1}{3} - \delta_{iz}  \right] }\mathcal{F}^\alpha_i(z) +  \frac{ \rho(z)\alpha \left[ \frac{1}{3} - \delta_{iz}  \right]}{\varepsilon_0 - \rho(z)\alpha \left[ \frac{1}{3} - \delta_{iz}  \right]} \varrho_i^\mathrm{DS,\alpha}(z) \\ -\frac{ \rho(z)\alpha}{\varepsilon_0 - \rho(z)\alpha \left[ \frac{1}{3} - \delta_{iz}  \right]} \frac{\partial}{\partial z}  \left[ \left(\frac{2}{5 } - \delta_{iz} \right)\varrho^\mathrm{QS,\alpha}_{iz}(z) + \frac{\delta_{iz}}{5}\varrho^\mathrm{QS,\alpha}_{jj}(z) \right]     \, .
\label{eq:local_field_multipole_2}
\end{multline}    
We write down the total polarization density as the sum of the polarization of the dielectric medium $\varrho_i^\mathrm{DL,\alpha}(z)$ and the source polarization $\varrho_i^\mathrm{DS,\alpha}(z) -\frac{\partial}{\partial z} \varrho_{iz}^\mathrm{QS,\alpha}(z)$
\begin{align}
    \mathcal{P}_i^{\alpha}(z) &=  \varrho_i^\mathrm{DL,\alpha}(z) + \varrho_i^\mathrm{DS,\alpha}(z) -\frac{\partial}{\partial z} \varrho_{iz}^\mathrm{QS,\alpha}(z) \, .
    \label{eq:total_P_Lorentz}
\end{align}
We insert Equation
\eqref{eq:local_field_multipole_2} into Equation \eqref{eq:total_P_Lorentz}. Subsequently, we insert the Clausius-Mossotti \cite{jacksonClassicalElectrodynamicsInternational2021} relation
\begin{align}
   \frac{\rho(z) \alpha}{3 \varepsilon_0} = \frac{\varepsilon(z)-1}{\varepsilon(z)+2} \, .
 \label{eq:Clausius_Mossoti}
\end{align}
to eliminate $\rho(z)\alpha$, which leads to the constitutive relations
\begin{align}
\mathcal{P}_i^{\alpha}(z) &=  \varrho_i^\mathrm{DL,\alpha}(z) + \varrho_i^\mathrm{DS,\alpha}(z) -\frac{\partial}{\partial z} \varrho_{iz}^\mathrm{QS,\alpha}(z) 
\\
\mathcal{P}_{x/y}^{\alpha}(z)&= \varepsilon_0 \left[ \varepsilon(z) -1 \right] \mathcal{E}^\alpha_{x/y}(z) +\frac{2 + \varepsilon(z)}{3} \varrho_{x/y}^\mathrm{DS,\alpha}(z) - \frac{2 \varepsilon(z)+3}{5} \frac{\partial}{\partial z} \varrho_{x/y}^\mathrm{QS,\alpha}(z)
\label{eq:Lorentz_consti_xy} \\ 
\mathcal{P}_z^{\alpha}(z) &=  \left[ 1 - \varepsilon^{-1}(z) \right] \mathcal{D}^\alpha_z (z) +\frac{1 +2 \varepsilon^{-1}(z)}{3} \varrho_{z}^\mathrm{DS,\alpha}(z) - \frac{2  + 3\varepsilon^{-1}(z)}{5} \frac{\partial}{\partial z}  \varrho_{zz}^\mathrm{QS,\alpha}(z) - \frac{1-\varepsilon^{-1}(z) }{5} \frac{\partial}{\partial z}\varrho_{jj}^\mathrm{QS,\alpha}(z) \, .
\label{eq:Lorentz_consti_z}
\end{align} 
The average local field factor, which relates the external $\mathcal{F}_i^\alpha$ and the local field $\mathcal{E}^{\mathrm{L,\alpha}}_i(z)$ is defined by
\begin{align}
\mathcal{E}^{\mathrm{L,\alpha}}_i(z)= f^\alpha_{i}(z) \mathcal{F}_i^\alpha \, ,
\label{eq:def_aver_loc_field}
\end{align}
It is evident from Equation \eqref{eq:local_field_multipole_2} and \eqref{eq:Clausius_Mossoti}, that in the Lorentz-field approximation $f_{i}(z)$ is determined by
\begin{align}
    f_{x/y}(z) \approx  \frac{2 + \varepsilon(z)}{3}; \quad f_{z}(z) \approx  \frac{1 + 2 \varepsilon^{-1}(z)}{3} \, .
\end{align}
Here, we leave out the superscript $\alpha$ in the local field factor $f_i^\alpha(z)$ as the Clausius-Mossoti Relation \eqref{eq:Clausius_Mossoti} does not predict a frequency dependence of $\varepsilon(z)$.
The local field factor is important in the theory of nonlinear optics, as it relates macroscopic E- and D-fields to the local field acting on the molecular centers \cite{armstrongInteractionsLightWaves1962,shiratoriMolecularTheoryDielectric2011,moritaTheorySumFrequency2018a,hiranoLocalFieldEffects2024}.
\section{Overview of  Multipole Contributions in SFG Spectroscopy}
In this chapter, we describe multipolar contributions to SFG spectra beyond the electric dipole approximation. 
Those contributions are commonly referred to as quadrupole contributions, which serve as an umbrella term for different corrections, of which there are three types. 
First, there are molecular multipole (MM) contributions due to the molecular multipole expansion of the second-order electric source current density $j^{(2)}_i(z,t)$. 
Secondly, there are dielectric multipole (DM) contributions, which account for the inhomogeneity of the local E-field within the interface layer.
MM and DM contributions are located in the interfacial region, within which one can assume that the external fields, namely parallel-polarized E-fields and perpendicular-polarized D-fields, are constant.
Third, there are bulk multipole (BM) contributions induced by the gradients of the external fields far away from the interface.
Molecular and bulk multipole contributions need to be considered for SFG spectra prediction, whereas dielectric multipole contributions are only important for the interpretation of the spectra in terms of the molecular orientation.
\subsection{Molecular Multipole (MM) Contributions}
\label{sec:MM}
\subsubsection{General Expressions for Molecular Multipole Contribution}
Here we provide the theory used to decompose the SFG signal into its multipole contributions. 
We divide the second-order electric current density\cite{jacksonClassicalElectrodynamicsInternational2021} into its multipole contributions
\begin{align}
    j^{(2)}_i(z, t) = j^{(2,\mathrm{D})}_i(z, t) + j^{(2,\mathrm{Q})}_i(z, t), + j^{(2,\mathrm{M})}_i(z, t) + ... \, ,
    \label{eq:j2}
\end{align}
and we do not explicetly list higher order multipoles, as they do not contribute to the spatially integrated SFG spectrum \cite{guyot-sionnestBulkContributionSurface1988}.
Polarization contributions are determined by the time derivative of the second-order polarization density defined in Equation \eqref{eq:def_P}
\begin{align}
    j^{(2,\mathrm{P})}_i(z, t) =  \dot{p}_i^{(2)}(z, t) \, 
    \label{eq:def_pol_flux}
\end{align}
and are composed of the second-order electric dipole current density, which is determined by the time derivative of the second-order electric dipole density $\varrho^{(2,\mathrm{D})}_i$ 
\begin{align}
    j^{(2,\mathrm{D})}_i(z, t) = \dot{\varrho}^{(2,\mathrm{D})}_i(z, t)  \, 
    \label{eq:def_j2D}
\end{align}
and the second-order electric quadrupole current density
\begin{align}
    j^{(2,\mathrm{Q})}_i(z, t) = -\frac{\partial}{\partial z}\dot{\varrho}^{(2,\mathrm{Q})}_{iz}(z, t) \, .
\end{align}
Here, $\dot{\varrho}^{(2,\mathrm{Q})}_{ij}(z, t)$ is the second-order electric quadrupole density.
The second-order magnetic dipole contribution to the electric current density  is determined by the curl of the second-order magnetic dipole density $m_i^{(2)}(z, t)$ 
\begin{align}
    j^{(2,\mathrm{M})}_i(z, t) = \epsilon_{izj} \frac{\partial}{\partial z} m_j^{(2)}(z, t)  \, ,
    \label{eq:def_mag_flux}
\end{align}
where $\epsilon_{ijk}$ is the Levi-Civita symbol. 
We define the decomposition of the response function $\tilde{s}^{(2)}_{ijk}(z,\omega^\mathrm{VIS}, \omega^\mathrm{IR})$ in Equation \eqref{eq:s2ijk} into its multipole contributions by
\begin{align}
\varepsilon_0^{-1} j_i^{(2,\beta)}(z, t) =  - i \omega^\mathrm{SFG} e^{ -i  \omega^\mathrm{SFG} t } \tilde{s}^{(2,\beta)}_{ijk}\left( z, \omega^\mathrm{VIS}, \omega^\mathrm{IR}  \right) \mathcal{F}^\mathrm{VIS}_j  \mathcal{F}^\mathrm{IR}_k + c.c. \, ,
\label{eq:s2ijk_beta}
\end{align}
where $\beta \in \lbrace \mathrm{P}, \mathrm{D}, \mathrm{Q}, \mathrm{M} \rbrace$. 
In a homogeneous bulk medium, the different multipolar densities are related to second-order susceptibilities defined by 
\begin{align}
     \varepsilon_0^{-1} \varrho^{(2,\mathrm{Q})}_{ij}(t) &= e^{-i \omega^\mathrm{SFG} t} \tilde{\chi}^{(2,\mathrm{Q})}_{ijkl} (\omega^\mathrm{VIS},\omega^\mathrm{IR})\mathcal{E}^\mathrm{VIS}_k \mathcal{E}^\mathrm{IR}_l +c.c. 
    \label{eq:chi2Q}\\
    \varepsilon_0^{-1} m^{(2)}_{i}(t) &= e^{-i \omega^\mathrm{SFG} t} \tilde{\chi}^{(2,\mathrm{M})}_{ijk} (\omega^\mathrm{VIS},\omega^\mathrm{IR})\mathcal{E}^\mathrm{VIS}_j \mathcal{E}^\mathrm{IR}_k +c.c. \, .
    \label{eq:chi2M}
\end{align}
These two second-order response functions are nonzero in isotropic media\cite{pershanNonlinearOpticalProperties1963,adlerNonlinearOpticalFrequency1964,guyot-sionnestBulkContributionSurface1988} and represent an intrinsic property of the bulk medium.
We define the different contributions to the SFG signal defined in Equation \eqref{eq:S2ijk} as
\begin{align}
    \tilde{S}^{(2,\beta)}_{ijk} \left( \omega^\mathrm{VIS}, \omega^\mathrm{IR}  \right) = \int\limits_{-\infty}^\infty \mathrm{d} z \,e^{-i\Delta k_z z} \tilde{s}^{(2,\beta)}_{ijk}\left( z,\omega^\mathrm{VIS}, \omega^\mathrm{IR}  \right) \, ,
\label{eq:S2ijk_beta}
\end{align}
Whenever the region where $\tilde{s}^{(2,\beta)}_{ijk}\left( z,\omega^\mathrm{VIS}, \omega^\mathrm{IR}  \right)\neq 0$ is much smaller than all considered  wavelengths $\lambda_0^\alpha$, we can approximate
\begin{align}
z \Delta k_z = 2 \pi  \cos{\theta_1^\mathrm{SFG}} n_1^\mathrm{SFG} z /\lambda_0^\mathrm{SFG} +  2 \pi \cos{\theta_1^\mathrm{IR}} n_1^\mathrm{IR} z/\lambda_0^\mathrm{IR} + 2 \pi \cos{\theta_1^\mathrm{VIS}} n_1^\mathrm{VIS} z/\lambda_0^\mathrm{VIS}\approx 0 \, ,
\end{align}
which we use throughout this work.
In this limit, the MM contributions to the SFG signal are determined by
\begin{align}
\tilde{S}^{(2,\mathrm{Q})}_{ijk} \left(\omega^\mathrm{VIS} , \omega^\mathrm{IR} \right)&= \varepsilon_0^{-1}\tilde{c}^\mathrm{VIS}_j(-\infty) \tilde{c}^\mathrm{IR}_k(-\infty) \tilde{\chi}^{(2,\mathrm{Q})}_{izjk}\left(\omega^\mathrm{VIS}, \omega^\mathrm{IR} \right) 
\label{eq:S2Q_bulk}\\
\tilde{S}^{(2,\mathrm{M})}_{ijk} \left(\omega^\mathrm{VIS}, \omega^\mathrm{IR} \right) &= \varepsilon_0^{-1} \frac{\epsilon_{izl} }{i \omega^\mathrm{SFG}} \tilde c^\mathrm{VIS}_j(-\infty) \tilde{c}^\mathrm{IR}_k(-\infty) \tilde{\chi}^{(2,\mathrm{M})}_{ljk} \left(\omega^\mathrm{VIS}, \omega^\mathrm{IR} \right) \, ,
\label{eq:S2M_bulk}
\end{align}
where 
\begin{align}
\tilde c_i^\alpha(z)  = \frac{\mathcal{E}_i^\alpha(z)}{\mathcal{F}_i^\alpha}   = \delta_{ix} +  \delta_{iy} +   \frac{\delta_{iz}}{\tilde{\varepsilon}_{zz}^{\alpha}(z) }\,.
\label{eq:ext_elec_translation_fac}
\end{align}
is an external field - E-field translation factor.
It is evident from Equations \eqref{eq:S2Q_bulk} and \eqref{eq:S2M_bulk} that MM contributions are independent of the interface.
From Equations \eqref{eq:S2Q_bulk}, \eqref{eq:S2M_bulk} and \eqref{eq:ext_elec_translation_fac} follows that the
MM contributions are determined by
\begin{align}
    \tilde{S}^{(2,\mathrm{Q} )}_{yyz}\left(\omega^\mathrm{VIS}, \omega^\mathrm{IR} \right) &= \left(\frac{1}{\tilde n^\mathrm{IR}_1} \right)^2 \tilde{\chi}^{(2,\mathrm{Q})}_{yzyz}\left(\omega^\mathrm{VIS}, \omega^\mathrm{IR} \right) 
    \label{eq:S2Q_yyz} \\
      \tilde{S}^{(2,\mathrm{Q})}_{zzz}\left(\omega^\mathrm{VIS}, \omega^\mathrm{IR} \right) &= \left(\frac{1}{\tilde n^\mathrm{VIS}_1  \tilde n^\mathrm{IR}_1} \right)^2 \tilde{\chi}^{(2,\mathrm{Q})}_{zzzz}\left(\omega^\mathrm{VIS}, \omega^\mathrm{IR} \right) 
       \label{eq:S2Q_zzz} \\
    \tilde{S}^{(2,\mathrm{M} )}_{yyz}\left(\omega^\mathrm{VIS}, \omega^\mathrm{IR} \right) &= \frac{1}{i \omega^\mathrm{SFG}} \left(\frac{1}{\tilde n^\mathrm{IR}_1} \right)^2 \tilde{\chi}^{(2,\mathrm{M})}_{xyz}\left(\omega^\mathrm{VIS}, \omega^\mathrm{IR} \right) \label{eq:S2M_yyz}\\
      \tilde{S}^{(2,\mathrm{M})}_{zzz}\left(\omega^\mathrm{VIS}, \omega^\mathrm{IR} \right) &= 0 \,.
      \label{eq:S2M_zzz}
\end{align}
Hence, magnetic dipole contributions are not relevant for $  \tilde{S}^{(2)}_{zzz}\left(\omega^\mathrm{VIS}, \omega^\mathrm{IR} \right)$.
For the molecular interpretation of SFG spectra, it is reasonable to distinguish between the electric dipole contribution and MM contributions to the SFG signal, as $\tilde S^{(2,\mathrm{D})}_{ijk}\left(\omega^\mathrm{VIS}, \omega^\mathrm{IR} \right) $ depends on the structure of the interface, whereas $ 
    \tilde S^{(2,\mathrm{Q})}_{ijk}\left(\omega^\mathrm{VIS}, \omega^\mathrm{IR} \right)$ and $ 
    \tilde S^{(2,\mathrm{M})}_{ijk}\left(\omega^\mathrm{VIS}, \omega^\mathrm{IR} \right)$
are, in the considered limit ($z \Delta k_z=0$), entirely determined by the bulk medium. 
However, the division of $\tilde S^{(2)}_{ijk}\left(\omega^\mathrm{VIS}, \omega^\mathrm{IR} \right)$ into $\tilde S^{(2,\mathrm{D})}_{ijk}\left(\omega^\mathrm{VIS}, \omega^\mathrm{IR} \right)$, $\tilde S^{(2,\mathrm{Q})}_{ijk}\left(\omega^\mathrm{VIS}, \omega^\mathrm{IR} \right)$ and $\tilde S^{(2,\mathrm{M})}_{ijk}\left(\omega^\mathrm{VIS}, \omega^\mathrm{IR} \right)$ does depend on the choice of the molecular origin of the molecular multipole expansion, introduced in Equations \eqref{eq:Q_0n}-\eqref{eq:Q_IIn}. 
To draw meaningful conclusions about the structure of the interface of interest, it is usually assumed that $\tilde S^{(2,\mathrm{D})}_{ijk}\left(\omega^\mathrm{VIS}, \omega^\mathrm{IR} \right)$ is induced by the anisotropic orientation distribution of the molecules \cite{zhuangMappingMolecularOrientation1999, sunOrientationalDistributionFree2018a, yuPolarizationDependentSumFrequencyGeneration2022a, moritaTheorySumFrequency2018a}.
In this interpretation, the SFG signal from an interface with isotropically oriented molecules $\tilde S_{ijk}^{(2,\mathrm{ISO})}\left(  \omega^\mathrm{VIS}, \omega^\mathrm{IR}  \right)$ should not have an electric dipole contribution, i.e.
$\tilde S_{ijk}^{(2,\mathrm{ISO})}\left(  \omega^\mathrm{VIS}, \omega^\mathrm{IR}  \right)=\tilde S_{ijk}^{(2,\mathrm{Q})} \left(  \omega^\mathrm{VIS}, \omega^\mathrm{IR}  \right) + \tilde S_{ijk}^{(2,\mathrm{M})} \left(  \omega^\mathrm{VIS}, \omega^\mathrm{IR}  \right)$ should hold.
By imposing an interface in bulk water, we find that this is approximately the case if we choose the molecular center of mass as the molecular origin in Section \ref{sec:origin_sfg}.  
\subsubsection{Additional Decomposition of the Electric Dipole Contribution}
\label{sec:additional_decom_e_dip}
To map SFG spectra onto molecular orientation, it is necessary to introduce a further decomposition of the electric dipole contribution $\tilde{S}^{(2,\mathrm{D})}_{ijk}(\omega^\mathrm{VIS}, \omega^\mathrm{IR})$ defined in Equation \eqref{eq:S2ijk_beta}.
We emphasize that this decomposition is not necessary for predicting SFG signals, but only needed if we want to relate SFG spectra to the interfacial molecular orientation.
We assume that we have two second-order source densities present in our system, of which one is an electric dipole density $\varrho^{ \mathrm{DS}}_i (\vec r,t)$ and the other one an electric quadrupole density $\varrho^{\mathrm{QS}}_{ij} (\vec r,t)$. 
These source densities can be defined by 
\begin{align}   \varrho^\mathrm{DS}_i(\vec r, t) &= \sum\limits_n^{N_\mathrm{mol}} \mu_i^{(2,n)}(t) \delta\left[\vec r-\vec r^n(t)\right] \label{eq:def_rho_DS} \\
    \varrho^\mathrm{QS}_{ij}(\vec r, t) &= \sum\limits_n^{N_\mathrm{mol}} Q_{ij}^{(2,n)}(t) \delta\left[\vec r-\vec r^n(t)\right] \, ,
\label{eq:def_rho_QS}
\end{align}
where $\mu_i^{(2,n)}(t) $ and $Q_{ij}^{(2,n)}(t)$ are second-order molecular multipoles. 
Both densities induce an instantaneous linear response.
Consequently, the second-order electric dipole density $\varrho^{(2,\mathrm{D})}_i(z,t)$ includes the linear response to the electric quadrupole density $\varrho^{ \mathrm{QS} }_{ij} (\vec r,t)$, which cannot be related to molecular orientation.
To account for this, we decompose
\begin{align}   
\varrho^{(2,\mathrm{D})}_i (z,t) = \varrho^{(2,\mathrm{DD})}_i (z,t) + \varrho^{(2,\mathrm{DQ})}_i (z,t) \, ,
    \label{eq:def_j2dd}
\end{align}
where $\varrho^{(2,\mathrm{DD})}_i (z,t)$ is determined by $\varrho^{ \mathrm{DS} }_i (\vec r,t)$ and $\varrho^{(2,\mathrm{DQ})}_i (z,t)$ by  $\varrho^{ \mathrm{QS} }_{ij} (\vec r,t)$.  
The pure electric dipole contribution is determined by
 \begin{align}
 \varrho^{(2,\mathrm{DD})}_i (z,t ) = \frac{1}{L_x L_y}\int \mathrm{d} x \int \mathrm{d} y \left[ \varrho^{\mathrm{DS}}_i(\vec r ,t) + \int \mathrm{d} \vec r' \varepsilon_0\tilde{s}^{\mathrm{NL}}_{ij}(\vec r, \vec r', t) F_j^\mathrm{DS}(\vec r',t) \right] \, .
 \label{eq:def_rho2DD_repeat}
 \end{align}
The electric dipole - electric quadrupole cross contribution is given by
  \begin{align}
 \varrho^{(2,\mathrm{DQ})}_i (z,t ) = \frac{1}{L_x L_y}\int \mathrm{d} x \int \mathrm{d} y \int \mathrm{d} \vec r' \varepsilon_0 \tilde{s}^{\mathrm{NL}}_{ij}(\vec r, \vec r', t) F_j^\mathrm{QS}(\vec r',t)  \, ,
  \label{eq:def_rho2DQ_repeat}
 \end{align}
as derived in Section \eqref{sec:lin_nonloc_resp_func}. 
Here, the area of the interface is denoted by  $L_x L_y$, 
and $F_i^\mathrm{DS}(\vec r, t)$ and $F_i^\mathrm{QS}(\vec r, t)$ are the external fields created by $\varrho^{ \mathrm{DS} }_i (\vec r,t)$ and $\varrho^{ \mathrm{QS} }_i (\vec r,t)$, respectively. 
These fields impose a linear response of the dielectric medium, which is accounted for by the non-local response function $\tilde{s}^{\mathrm{NL}}_{ij}(\vec r, \vec r', t)$.
A simplified relationship can be derived using the constitutive relations \eqref{eq:Lorentz_consti_xy} and \eqref{eq:Lorentz_consti_z} within the Lorentz-field approximation \eqref{eq:LorentzField}, aiding interpretation.
Within this framework, the second-order electric dipole densities determining the electric dipole contribution to the SFG signal are 
\begin{align}
    \varrho_i^{(2,\mathrm{DD})}(z,t) &\approx c_i^\mathrm{VIS} (z) \frac{\varepsilon(z)+2}{3} \frac{1}{L_x L_y}\iint\mathrm{d} x \mathrm{d}y\, \varrho^{\mathrm{DS}}_i(\vec r,t) \\
     \varrho_{x/y}^{(2,\mathrm{DQ})}(z,t) &\approx  \frac{2-2\varepsilon(z)}{5} \frac{\partial}{\partial z} \frac{1}{L_x L_y}\iint\mathrm{d} x \mathrm{d}y\,\varrho^\mathrm{QS}_{x/y}(\vec r,t) \\
     \varrho_z^{(2,\mathrm{DQ})}(z,t) &\approx  \frac{3-3\varepsilon^{-1}(z)}{5} \frac{\partial}{\partial z} \frac{1}{L_x L_y}\iint\mathrm{d} x \mathrm{d}y\, \varrho^\mathrm{QS}_{zz}(\vec r,t) - \frac{1- \varepsilon^{-1}(z) }{5} \frac{\partial}{\partial z} \frac{1}{L_x L_y}\iint\mathrm{d} x \mathrm{d}y\,\varrho_{jj}^\mathrm{QS}(\vec r,t) \, . 
\end{align}
Whether we apply the Lorentz-field approximation or not, we can define the decomposition of the second-order response profile $\tilde{s}^{(2,\mathrm{D})}_{ijk}\left( z, \omega^\mathrm{VIS}, \omega^\mathrm{IR}  \right)$ analogously to Equation \eqref{eq:s2ijk_beta} \begin{align}
\varepsilon_0^{-1} j_i^{(2,\mathrm{DD/DQ})}(z, t) = \varepsilon_0^{-1} \dot{\varrho}^{(2,\mathrm{DD/DQ})}_i(z,t) =  - i \omega^\mathrm{SFG} e^{ -i  \omega^\mathrm{SFG} t } \tilde{s}^{(2,\mathrm{DD/DQ})}_{ijk}\left( z, \omega^\mathrm{VIS}, \omega^\mathrm{IR}  \right) \mathcal{F}^\mathrm{VIS}_j  \mathcal{F}^\mathrm{IR}_k + c.c. \, .
\label{eq:s2ijk_DD_DQ}
\end{align}
and the corresponding contributions to the SFG signal as
\begin{align}
    \tilde{S}^{(2,\mathrm{DD/DQ})}_{ijk} \left( \omega^\mathrm{VIS}, \omega^\mathrm{IR}  \right) = \int\limits_{-\infty}^\infty \mathrm{d} z \,e^{-i\Delta k_z z} \tilde{s}^{(2,\mathrm{DD/DQ})}_{ijk}\left( z,\omega^\mathrm{VIS}, \omega^\mathrm{IR}  \right) \, .
\label{eq:S2ijk_DD_DQ} 
\end{align}
We call $\tilde{S}^{(2,\mathrm{DD})}_{ijk} \left( \omega^\mathrm{VIS}, \omega^\mathrm{IR}  \right)$ the pure electric dipole contribution and $\tilde{S}^{(2,\mathrm{DQ})}_{ijk} \left( \omega^\mathrm{VIS}, \omega^\mathrm{IR}  \right)$ the electric dipole - electric quadrupole cross contribution.
\subsection{Dielectric Multipole (DM) Contributions}
\label{sec:DL}
Dielectric multipole contributions are relevant only when aiming to connect experimental SFG spectra to molecular orientation, they are unnecessary for the prediction of SFG spectra within our framework.
In a fully accurate description, SFG spectra result from the complex many-body dynamics of the system and cannot be explained solely by molecular orientation.
However, under certain approximations, a simplified relationship between molecular orientation and the SFG spectrum can still be established if multipole contributions are subtracted. These approximations are discussed in Section \ref{sec:orientation_analyis}.
In the last chapter, we introduced the second-order response function $\tilde{s}^{(2,\mathrm{DD})}_{ijk}\left(z,\omega^\mathrm{VIS}, \omega^\mathrm{IR} \right)$ as the second-order response of the pure electric dipole density $\varrho^{(2,\mathrm{DD})}_i(z,t)$ to spatially constant external fields $\mathcal{E}_{x/y}^\alpha$ and $\mathcal{D}_{z}^\alpha$.
First, we derive an approximate mapping between molecular hyperpolarizabilities $\tilde\beta^n_{ijk}(\omega^\mathrm{VIS}, \omega^\mathrm{IR})$ and $\tilde{S}^{(2,\mathrm{DD})}_{ijk}(\omega^\mathrm{VIS},\omega^\mathrm{IR})$, introducing the second-order electric dipole susceptibility $\tilde{\chi}^{(2,\mathrm{DL})}_{ijk}(\omega^\mathrm{VIS},\omega^\mathrm{IR})$.
Then we relate $\tilde{\chi}^{(2,\mathrm{DL})}_{ijk}(\omega^\mathrm{VIS},\omega^\mathrm{IR})$ to the SFG spectrum, introducing the DM contributions as correction therms.
We define the molecular hyperpolarizability $\tilde{\beta}_{ijk}^n\left(\omega^\mathrm{VIS}, \omega^\mathrm{IR} \right)$ as 
\begin{align}
   \varepsilon_0^{-1} \mu_i^{(2,n)}(t)=  e^{-i\omega^\mathrm{SFG}t} \tilde{\beta}_{ijk}^n\left(\omega^\mathrm{VIS}, \omega^\mathrm{IR} \right) \mathcal{E}^{\mathrm{L,VIS}}_j(z^n) \mathcal{E}^{\mathrm{L,IR}}_k(z^n) + c.c.\, ,
    \label{eq:def_beta_ijk}
\end{align}
where $z^n$ is the $z$-position of the $n^\mathrm{th}$ molecule. 
Here, $\mu_i^{(2,n)}(t)$ is the second-order molecular electric dipole moment, appearing in Equation \eqref{eq:def_rho_DS},
induced by mixing
the local electric fields $E^{\mathrm{L,VIS}}_i(z,t)=\mathcal{E}^{\mathrm{L,VIS}}_i(z) e^{-i\omega^\mathrm{VIS} t}+c.c.$ and $E^{\mathrm{L,IR}}_i(z,t)=\mathcal{E}^{\mathrm{L,IR}}_i(z) e^{-i\omega^\mathrm{IR} t}+c.c.$. 
The local field factor $f^\alpha_i(z)$ is defined in Equation \eqref{eq:def_aver_loc_field} and
relates the amplitude of the external field  $\mathcal{F}_i^\alpha$ to the
amplitude of the average local field $\mathcal{E}^{\mathrm{L},\alpha}_i(z)$, as  defined in Equation \eqref{eq:aver_field}. 
We relate
\begin{align}
    \varepsilon_0^{-1} \mu_i^{(2,n)}(t) =  e^{-i\omega^\mathrm{SFG}t} \tilde{\beta}_{ijk}^n\left(\omega^\mathrm{VIS}, \omega^\mathrm{IR} \right) f_j^\mathrm{VIS}(z^n) f_k^\mathrm{IR}(z^n) \mathcal{F}^\mathrm{VIS}_j \mathcal{F}^\mathrm{IR}_k+ c.c.\, .\label{eq:def_beta_ijk_LF}
\end{align}
We construct the second-order electric source density introduced in Equation \eqref{eq:def_rho_DS} as
\begin{align}
   \varepsilon_0^{-1} \varrho^\mathrm{DS}_i(z,t)  =  \frac{\mathcal{F}^\mathrm{VIS}_j \mathcal{F}^\mathrm{IR}_k}{L_x L_y}e^{-i \omega^\mathrm{SFG} t} \sum\limits_n^{N_\mathrm{mol}} \tilde{\beta}_{ijk}^n\left(\omega^\mathrm{VIS}, \omega^\mathrm{IR} \right) f_j^\mathrm{VIS}(z^n) f_k^\mathrm{IR}(z^n) \delta(z-z^n) \, .
\end{align}
We can approximately account for the coupling between the second-order source density  $\varrho^\mathrm{DS}_i(z,t)$ and the dielectric medium by multiplication with another local field factor
\begin{align}
\varrho^{(2,\mathrm{DD})}_i(z,t) \approx f_i^\mathrm{SFG}(z)  \varrho^\mathrm{DS}_i(z,t)   \, , 
\end{align}
as shown by Armstrong, Bloembergen, Ducuing and Pershan \cite{armstrongInteractionsLightWaves1962}.
Hence, we have an approximate mapping between $\varrho^{(2,\mathrm{DD})}_i(z,t)$ and the molecular hyperpolarizabilities $\tilde\beta^n_{ijk}(\omega^\mathrm{VIS}, \omega^\mathrm{IR})$. 
We define a second-order electric dipole susceptibility to the local E-field by
\begin{align}
\varepsilon_0^{-1} \varrho_i^{(2,\mathrm{DD})}(z,t) = f_i^\mathrm{SFG}(z) e^{ -i  \omega^\mathrm{SFG} t } \tilde{\chi}^{(2,\mathrm{DL})}_{ijk}\left( z ,\omega^\mathrm{VIS}, \omega^\mathrm{IR} \right) \mathcal{E}^\mathrm{L,VIS}_j(z) \mathcal{E}^\mathrm{L,IR}_k(z) + c.c. \, .
\label{eq:chi_2DL}
\end{align}
We relate $\tilde{\chi}^{(2,\mathrm{DL})}_{ijk}\left( z ,\omega^\mathrm{VIS}, \omega^\mathrm{IR} \right)$ to the second-order external field response function $\tilde{s}^{(2,\mathrm{DD})}_{ijk}\left( z , \omega^\mathrm{VIS}, \omega^\mathrm{IR}  \right)$ by comparing Equations \eqref{eq:chi_2DL} and \eqref{eq:s2ijk_DD_DQ}
\begin{align} \tilde{s}^{(2,\mathrm{DD})}_{ijk}\left( z  ,\omega^\mathrm{VIS}, \omega^\mathrm{IR} \right)  =   f_i^\mathrm{SFG}(z ) f_j^\mathrm{VIS}(z ) f_k^\mathrm{IR}(z ) \tilde{\chi}^{(2,\mathrm{DL})}_{ijk}\left( z ,\omega^\mathrm{VIS}, \omega^\mathrm{IR} \right)  \, .
\label{eq:s2_chi2}
\end{align}
This is the formalism mostly used in SFG theory \cite{shenFundamentalsSumFrequencySpectroscopy2016a,moritaTheorySumFrequency2018a}.
Consequently we have a direct linking between $\tilde{\chi}^{(2,\mathrm{DL})}_{ijk}\left( z,\omega^\mathrm{VIS}, \omega^\mathrm{IR} \right)$ and the second-order response function $\tilde{s}^{(2,\mathrm{DD})}_{ijk}\left( z,\omega^\mathrm{VIS}, \omega^\mathrm{IR} \right)$, without any approximation. 
However, based on Equation \eqref{eq:def_beta_ijk} $\tilde{\chi}^{(2,\mathrm{DL})}_{ijk}\left( z,\omega^\mathrm{VIS}, \omega^\mathrm{IR} \right)$ can be interpreted as the density of molecular hyperpolarizabilities.
\begin{align}
\tilde{\chi}^{(2,\mathrm{DL})}_{ijk}\left( z,\omega^\mathrm{VIS}, \omega^\mathrm{IR} \right) \approx \frac{1}{L_x L_y} \sum\limits_n^{N_\mathrm{mol}} \tilde{\beta}^n_{ijk}\left( \omega^\mathrm{VIS}, \omega^\mathrm{IR} \right) \delta (z-z^n) \, .
\label{eq:chi_2DL_approx_beta}
\end{align}
We perform a Taylor expansion of the local field factors $f_i^\alpha(z)$ around $z_0$
\begin{align}   
   f^\alpha_{i}(z) &= f^\alpha_i(z_0)   + (z-z_0) \frac{\partial}{\partial z_0} f^\alpha_i(z_0) + ... \, .
    \label{eq:taylor_expansion_EField}
\end{align}
Inserting the Taylor Expansion in Equation \eqref{eq:taylor_expansion_EField} into Equation \eqref{eq:s2_chi2} leads to the dielectric multipole expansion of the pure electric dipole contribution to the SFG signal $\tilde{S}^{(2,\mathrm{DD})}_{ijk}\left( \omega^\mathrm{VIS}, \omega^\mathrm{IR} \right)$ defined in Equation \eqref{eq:S2ijk_DD_DQ}
\begin{align}    
\tilde{S}^{(2,\mathrm{DD})}_{ijk}\left( \omega^\mathrm{VIS}, \omega^\mathrm{IR} \right)  =  \tilde{S}^{(2,\mathrm{DL}0)}_{ijk} \left( \omega^\mathrm{VIS}, \omega^\mathrm{IR} \right)   + \tilde{S}^{(2,\mathrm{DL}1)}_{ijk} \left( \omega^\mathrm{VIS}, \omega^\mathrm{IR} \right)   + \tilde{S}^{(2,\mathrm{DL}2)}_{ijk} \left( \omega^\mathrm{VIS}, \omega^\mathrm{IR} \right) + \tilde{S}^{(2,\mathrm{DL}3)}_{ijk} \left( \omega^\mathrm{VIS}, \omega^\mathrm{IR} \right) + ...
    \label{eq:gradient_expansion_S}
\end{align}
in terms of the gradients of the local E-field.
The components up to the first order are given by
\begin{align}
    \tilde{S}^{(2,\mathrm{DL}0)}_{ijk} \left( \omega^\mathrm{VIS}, \omega^\mathrm{IR} \right)&=  f_i^\mathrm{SFG}(z_0) f_j^\mathrm{VIS}(z_0) f_k^\mathrm{IR}(z_0) \tilde{\chi}^{(2,\mathrm{DL0})}_{ijk} \left( \omega^\mathrm{VIS}, \omega^\mathrm{IR} \right) 
    \label{eq:S2ijk_DL0} \\
\tilde{S}^{(2,\mathrm{DL}1)}_{ijk} \left( \omega^\mathrm{VIS}, \omega^\mathrm{IR} \right)&=   f_j^\mathrm{VIS}(z_0) f_k^\mathrm{IR}(z_0) \frac{\mathrm{d}}{\mathrm{d} z_0}f_i^\mathrm{SFG}(z_0)  \tilde{\chi}^{(2,\mathrm{DL1})}_{ijk}  \left( \omega^\mathrm{VIS}, \omega^\mathrm{IR} \right) 
 \label{eq:S2ijk_DL1} \\
\tilde{S}^{(2,\mathrm{DL}2)}_{ijk} \left( \omega^\mathrm{VIS}, \omega^\mathrm{IR} \right)&=  f_i^\mathrm{SFG}(z_0) f_k^\mathrm{IR}(z_0) \frac{\mathrm{d}}{\mathrm{d} z_0}f_j^\mathrm{VIS}(z_0)  \tilde{\chi}^{(2,\mathrm{DL}1)}_{ijk} \left( \omega^\mathrm{VIS}, \omega^\mathrm{IR} \right) 
 \label{eq:S2ijk_DL2} \\
                \tilde{S}^{(2,\mathrm{DL}3)}_{ijk}\left( \omega^\mathrm{VIS}, \omega^\mathrm{IR} \right) &=    f_i^\mathrm{SFG}(z_0) f_j^\mathrm{VIS}(z_0) \frac{\mathrm{d}}{\mathrm{d} z_0}f_k^\mathrm{IR}(z_0)  \tilde{\chi}^{(2,\mathrm{DL}1)}_{ijk} \left( \omega^\mathrm{VIS}, \omega^\mathrm{IR} \right)  \, .
 \label{eq:S2ijk_DL3} 
\end{align}
Here 
\begin{align}   \tilde{\chi}^{(2,\mathrm{DL}n)}_{ijk}\left( \omega^\mathrm{VIS}, \omega^\mathrm{IR} \right) = \frac{1}{n!}\int \limits_{-\infty}^\infty \mathrm{d} z \, (z-z_0)^n \tilde \chi^{(2,\mathrm{DL})}_{ijk}\left( z,\omega^\mathrm{VIS}, \omega^\mathrm{IR} \right) \, \, ,
\end{align}
is the $n^\mathrm{th}$ moment of the dielectric multipole expansion. 
Using the approximation in Equation \eqref{eq:chi_2DL_approx_beta}, the connection between $\tilde{\chi}^{(2,\mathrm{DL0})}_{ijk}\left( \omega^\mathrm{VIS}, \omega^\mathrm{IR}\right)$ and the molecular hyperpolarizabilities reads
\begin{align}
\tilde{\chi}^{(2,\mathrm{DL0})}_{ijk}\left( \omega^\mathrm{VIS}, \omega^\mathrm{IR}\right) \approx \frac{1}{ L_x L_y}\sum\limits_n^{N_{\mathrm{mol}}} \tilde{\beta}^n_{ijk}\left( \omega^\mathrm{VIS}, \omega^\mathrm{IR}\right) \, .
\label{eq:chi2DL0_approx}
\end{align}
Therefore, if all multipole contributions are known, we have a mapping between the SFG spectrum $\tilde{S}^{(2)}_{ijk}(\omega^\mathrm{VIS},\omega^\mathrm{IR})$ and molecular properties, from which we can deduce information about the interfacial structure, e.g., the orientation distribution, as described in Section \ref{sec:orientation_analyis}.
\subsection{Bulk Multipole (BM) Contributions}
\label{sec:BM}
So far, we have ignored that the external fields vary in space.
As given in Equation \eqref{eq:ext_field_maxwell_bulk}, in the homogeneous bulk, the external fields can be identified as E-fields. 
Here, we give a brief overview of the BM contributions arising from medium 1, which is, in our case, bulk water. 
In the infinite and periodic medium 1, the external IR VIS fields appearing in the perturbation Hamiltonian in Equation \eqref{eq:pert_ham} can be identified as the transmitted E-fields, as follows from Equation \eqref{eq:ext_field_maxwell_bulk}.  Consequently, we identify the amplitudes of the external fields as the amplitude of the transmitted E-fields defined in Equation \eqref{eq:transmitted}
\begin{align} \mathcal{E}^{\alpha}_i(\vec r) = \mathcal{E}^{\mathrm{T},\alpha}_ie^{i \vec k^{\mathrm{T},\alpha}\cdot \vec r}  \, .
  \label{eq:ext_bulk_transmitted}
\end{align}
As medium 1 is inversion symmetric, the only nonzero contributions to SFG spectra are due to the gradients of the external fields.
Consequently, BM contributions depend on wavevectors $\vec{k}^{\mathrm{T},\alpha}$, which can be varied experimentally \cite{hiranoLocalFieldEffects2024}. 
This allows for the experimental estimation of BM contributions \cite{sunSurfaceSumfrequencyVibrational2015,fellowsHowThickAirWater2024}. 
In contrast, 
DM and MM contributions must be predicted theoretically.
The gradients of the transmitted IR and VIS fields introduce the following second-order electric current densities in medium 1
\begin{align}
\varepsilon_0^{-1} j_i^{(2,\mathrm{BM0} )}(\vec r, t) &=  i \left( k^\mathrm{T,VIS}_j +  k^\mathrm{T,IR}_j \right) e^{-i  \omega^\mathrm{SFG} t }  
\left[  \epsilon_{ijm}  \tilde{\chi}^{(2,\mathrm{M})}_{mkl}\left(\omega^\mathrm{VIS}, \omega^\mathrm{IR} \right) + i \omega^\mathrm{SFG} \tilde{\chi}^{(2,\mathrm{Q})}_{ijkl}\left(\omega^\mathrm{VIS}, \omega^\mathrm{IR} \right) \right] \mathcal{E}^\mathrm{VIS}_k(\vec r)  \mathcal{E}_l^\mathrm{IR}(\vec r) +c.c. \\
    \varepsilon_0^{-1} j_i^{(2,\mathrm{BM1} )}(\vec r, t) &=  \omega^\mathrm{SFG} k^\mathrm{T,IR}_k e^{-i  \omega^\mathrm{SFG} t } \tilde{\chi}^{(2,\mathrm{BM1})}_{ijkl}\left(\omega^\mathrm{VIS}, \omega^\mathrm{IR} \right) \mathcal{E}^\mathrm{VIS}_j(\vec r)  \mathcal{E}_l^\mathrm{IR}(\vec r) +c.c.\\
    \varepsilon_0^{-1} j_i^{(2,\mathrm{BM2} )}(\vec r, t) &=  \omega^\mathrm{SFG} k^\mathrm{T,VIS}_j e^{-i  \omega^\mathrm{SFG} t }  \tilde{\chi}^{(2,\mathrm{BM2})}_{ijkl}\left(\omega^\mathrm{VIS}, \omega^\mathrm{IR} \right)   \mathcal{E}_k^\mathrm{VIS}(\vec r) \mathcal{E}_l^\mathrm{IR}(\vec r) + c.c. \, ,
\end{align}
where $\tilde{\chi}^{(2,\mathrm{Q})}_{ijkl}\left(\omega^\mathrm{VIS}, \omega^\mathrm{IR} \right)$ and $\tilde{\chi}^{(2,\mathrm{M})}_{ijk}\left(\omega^\mathrm{VIS}, \omega^\mathrm{IR} \right)$ are already defined in Equations \eqref{eq:chi2Q} and \eqref{eq:chi2M}, respectively.
Here, $j^{(2,\mathrm{BM0})}_i(z,t)$ is the contribution due to the inhomogeneity of the electric quadrupole and magnetic dipole density in the bulk region. The other two contributions $j^{(2,\mathrm{BM1})}_i(z,t)$ and $j^{(2,\mathrm{BM2})}_i(z,t)$ are due to the gradients of the IR and VIS field, respectively.
The response functions to external field gradients are introduced in Equations \eqref{eq:uijk_21} and \eqref{eq:uijk_22}. 
By comparing with Equation \eqref{eq:s2ijk}, the susceptibilities $\tilde{\chi}^{(2,\mathrm{BM1})}_{ijkl} \left(\omega^\mathrm{VIS}, \omega^\mathrm{IR} \right)$ and $\tilde{\chi}^{(2,\mathrm{BM2})}_{ijkl} \left(\omega^\mathrm{VIS}, \omega^\mathrm{IR} \right)$ are identified as
\begin{align}
    \tilde{\chi}^{(2,\mathrm{BM1})}_{ijkl} \left(\omega^\mathrm{VIS}, \omega^\mathrm{IR} \right)&= \frac{1}{-i \varepsilon_0 \omega^\mathrm{SFG}} \big[    \tilde{u}^{(2,1)}_{ijkl}( \omega^\mathrm{SFG}, \omega^\mathrm{IR}) + \tilde{u}^{(2,2)}_{iklj}\left( \omega^\mathrm{SFG}, \omega^\mathrm{VIS} \right)  \big] \\
    \tilde{\chi}^{(2,\mathrm{BM2})}_{ijkl} \left(\omega^\mathrm{VIS}, \omega^\mathrm{IR} \right) &= \frac{1}{-i \varepsilon_0 \omega^\mathrm{SFG}} \big[    \tilde{u}^{(2,1)}_{iljk}(  \omega^\mathrm{SFG}, \omega^\mathrm{VIS}) + \tilde{u}^{(2,2)}_{ijkl}(\omega^\mathrm{SFG}, \omega^\mathrm{IR})  \big]  \, .
\end{align}
For further information on BM contributions, we refer to Hirano and Morita's publication \cite{hiranoLocalFieldEffects2024}. 
\section{Linear Response Functions for SFG Spectra Prediction within the Off-Resonant Approximation}
\label{sec:lin_spons_born_opp}
The equations employed for SFG signal prediction in the main text are the ones given in Section \ref{sec:nonlocal_first_and_second_order_pert_exp_pol_dens}.
Here, we derive fluctuation-dissipation relations for the interfacial z-resolved second-order electric current density profiles defined by the equation
 \begin{align}
     \varepsilon_0^{-1} j_i^{(2)}( z , t) = -i \omega^\mathrm{SFG} e^{-i \omega^\mathrm{SFG}t } \tilde{s}^{(2)}_{ijk} \left(z, \omega^\mathrm{VIS}, \omega^\mathrm{IR} \right) \mathcal{F}_j^\mathrm{VIS} \mathcal{F}^\mathrm{IR}_k + c.c. \, ,
     \label{eq:def_s2ijk_F_const}
 \end{align}
 where we assume that the external field amplitudes are constant within the simulation box.
 Corrections to this assumption are the BM contributions introduced in Section \ref{sec:BM}.
 The simulation box can be interpreted as an elementary cell of the macroscopic system.
 A formal expressions for the second-order response function $\tilde{s}^{(2)}_{ijk} \left(z, \omega^\mathrm{VIS}, \omega^\mathrm{IR} \right) $ is given in Equation \eqref{eq:s2ijk}. 
We will see in this chapter that $\tilde{s}^{(2)}_{ijk} \left( z,\omega^\mathrm{VIS}, \omega^\mathrm{IR} \right)$ is a linear response function in the Born-Oppenheimer approximation, which means that we assume that the VIS field does invoke an adiabatic displacement of the electrons charge distribution, but does not excite higher electronic or vibronic energy levels. 
The wavenumber is defined by $\nu^\alpha=\frac{1}{\lambda_0^\alpha}$. The VIS wavenumber range is given by $13,000\,\si{cm^{-1}}< \nu^\mathrm{VIS}< 26,000\,\si{cm^{-1}}$. The fastest nuclei oscillations are around $4,000\,\si{cm^{-1}}$.
In water, the VIS field does not excite higher electronic states as the HOMO-LUMO gap of the water molecule is about $\SI{6.3}{eV}$\cite{tenbrinckPolycyclicAromaticHydrocarbons2022}, which corresponds to a wavenumber of $50,000\,\si{cm^{-1}}$.
If the VIS field oscillates too fast to invoke movements of the nuclei but too slow to excite higher electronic levels, the VIS imposes an instantaneous polarization, but does not alter the trajectory of the nuclei.
Hence, the trajectory of the nuclei is solely determined by the external IR field.
As the response to the VIS field is instantaneous and does not influence the dynamics of the nuclei, we can express the second-order electric current density as a product of a time-dependent effective polarizability profile $a^{(1)}_{ij}\left(z, \omega^\mathrm{VIS},t \right)$ and the external VIS field, which is defined by
\begin{align}
    j^{(2)}_{i}(z,t) = -i \omega^\mathrm{SFG} a^{(1)}_{ij}(z,\omega^\mathrm{VIS},t) e^{-i\omega^\mathrm{VIS} t} \mathcal{F}^\mathrm{VIS}_j + c.c. \, .
    \label{eq:def_a_ij}
\end{align}
Here, the time dependence 
in $a^{(1)}_{ij}(z,\omega^\mathrm{VIS},t)$ is caused by the IR field. 
By comparing Equation \eqref{eq:def_a_ij} and Equation \eqref{eq:def_s2ijk_F_const}, we can establish the relationship  
\begin{align}
\varepsilon_0^{-1} \tilde a^{(1)}_{ij}\left( z,\omega^\mathrm{VIS},\omega^\mathrm{IR} \right) =  \tilde s^{(2)}_{ijk}\left( z,\omega^\mathrm{VIS},\omega^\mathrm{IR} \right)\tilde F_k^{\mathrm{\mathrm{IR}} }(\omega^{\mathrm{IR}}) .
\label{eq:small_s_def}
\end{align}  
between the first-order effective polarizability profile $\tilde a^{(1)}_{ij}\left( z,\omega^\mathrm{VIS},\omega^\mathrm{IR} \right)$ and the second-order response function $\tilde{s}^{(2)}_{ijk}(z,\omega^\mathrm{VIS}, \omega^{\mathrm{\mathrm{IR}}})$.
Consequently, $\tilde{s}^{(2)}_{ijk}\left( z, \omega^\mathrm{VIS}, \omega^\mathrm{IR} \right)$ can be identified as the linear response function of the effective polarizability profile $a_{ij} \left(z,\omega^\mathrm{VIS},\vec\Omega \right)$ to the external IR field.
The first-order time-dependent perturbation expansion of an arbitrary observable in an external IR field can be written as
\cite{kuboFluctuationdissipationTheorem1966}
\begin{align}
\tilde{A}^{(1)} \left(\omega^\mathrm{IR} \right) =\tilde{\varphi}\left[A(\cdot) , P_i( \cdot ), \omega^\mathrm{IR} \right] \tilde{F}_i^\mathrm{IR} \left(\omega^\mathrm{IR} \right) \, ,
\label{eq:expansion}
\end{align}
in the Fourier domain, where $\varphi[A( \cdot ) , P_i( \cdot ), t ]$ is a generic linear response function, depending on the observables $A( \vec{\Omega})$ and the system's electric dipole moment $P_i( \vec{\Omega})$ and the time $t$. 
The dots in the argument indicate that $\varphi[A( \cdot ) , P_i( \cdot ), t ]$ is a functional depending on the functions $A( \vec{\Omega})$ and $P_i( \vec{\Omega})$, but not on the state vector $\vec{\Omega}$.
The fluctuation-dissipation theorem gives the relationship to the equilibrium correlation function, reproduced in Section \ref{app:fd_and_kk}. 
Hence, in the off-resonant limit the second-order response functions $\tilde s^{(2,\beta)}_{ijk}\left( z,\omega^\mathrm{VIS},\omega^\mathrm{IR} \right)$ defined in Equation \eqref{eq:s2ijk_beta} is determined by linear response functions $\varphi[A( \cdot ) , P_i( \cdot ), t ]$ which is derived in this section. 
\subsection{Polarization Contributions}
Polarization contributions are contributions to the SFG signal arising from the polarization current $j^{(2,\mathrm{P})}_i(z,t)=\dot p^{(2)}_i(z,t)$.
In the Born-Oppenheimer approximation, the electronic degrees of freedom depend only parametrically on the phase space vector of the nuclei $\vec{\Omega}$. 
We define the time-dependent effective polarizability $a^{(1,\mathrm{P})}_{ij}(z,t)$ via
\begin{align}
j^{(2,\mathrm{P} )}_i (z, t) = \dot{p}^{(2)}_i(z,t) =  -i \omega^\mathrm{SFG} a^{(1,\mathrm{P})}_{ij}(z,t) F^\mathrm{VIS}_j(t) + c.c. \, .
\end{align}
The difference between the observables $a^\mathrm{P}_{ij}(z,\vec{\Omega})$ and  $a_{ij} \left( z,\omega^\mathrm{VIS},\vec{\Omega} \right)$ defined in Equation \eqref{eq:def_a_ij} is that the former does not include magnetic dipole contributions, while the latter one does.
In the Born-Oppenheimer limit $a^\mathrm{P}_{ij}(z,\vec \Omega)$ is the instantaneous change of the polarization profile defined in Equation \eqref{eq:def_P} to an applied external field, i.e.
\begin{align}
a^\mathrm{P}_{ij}( z, \vec{\Omega} )  =\frac{\partial}{\partial F^\mathrm{VIS}_j} p_i ( z, \vec{\Omega} ) \bigg|_{F^\mathrm{VIS}_j=0} \, ,
\label{eq:def_pol_ex}
\end{align}
for a given set of nuclei coordinates $\vec \Omega$ and does not depend on $\mathrm{\omega}^\mathrm{VIS}$. The first- and second-order response functions of the polarization density defined in Equation \eqref{eq:pol_dens} to external fields are defined by
\begin{align}
\varepsilon_0^{-1} p^{(1)}_i (z, t) & =  e^{-i \omega^\mathrm{\alpha}t } \tilde{s}^{(1,\mathrm{P})}_{ij} \left(z, \omega^\mathrm{\alpha} \right) \mathcal{F}^\mathrm{\alpha}_j + c.c.  
\label{eq:def_s1P_ij} \\
\varepsilon_0^{-1} p^{(2)}_i (z, t) & = e^{-i \omega^\mathrm{SFG}t } \tilde{s}^{(2,\mathrm{P})}_{ijk} \left(z, \omega^\mathrm{IR} \right)  \mathcal{F}_j^\mathrm{VIS} \mathcal{F} ^\mathrm{IR}_k + c.c. \, .  
\label{eq:def_s2P_ijk}
\end{align}
We can relate the first- and second-order response functions $\tilde s^{(1,\mathrm{P})}_{ij}\left( z,\omega^{\mathrm{IR}} \right)$ and $\tilde s^{(2,\mathrm{P})}_{ijk}\left( z,\omega^{\mathrm{IR}} \right)$ to linear response functions by 
\begin{equation}
\tilde s^{(1,\mathrm{P})}_{ij}\left( z,\omega^{\mathrm{IR}} \right)= \varepsilon_0^{-1} \tilde \varphi\left[p_i(z,\cdot) , P_j(\cdot), \omega^\mathrm{IR} \right] + \varepsilon_0^{-1} \left\langle a^{ \mathrm{P}}_{ij}(z,\cdot)  \right\rangle
\label{eq:s1P_ij_fdt}
\end{equation}
and
\begin{equation}
\tilde s^{(2, \mathrm{P})}_{ijk}\left( z,\omega^{\mathrm{IR}} \right)= \varepsilon_0^{-1} \tilde \varphi\left[a^\mathrm{P}_{ij}(z,\cdot) , P_k(\cdot), \omega^\mathrm{IR} \right] +  \varepsilon_0^{-1} \left\langle  b^{\mathrm{P}}_{ijk}(z,\cdot )  \right\rangle \, .
\label{eq:s2P_ijk_fdt}
\end{equation}
The former is the linear response of the polarization density $p_i(z,\vec\Omega)$ and the second the linear response of the effective polarizability profile $a^\mathrm{P}_{ij}(z,\vec\Omega)$ to an external field. 
The instantaneous effective hyperpolarizability profile is defined by
\begin{align}
b^\mathrm{P}_{ijk}\left( z,\vec \Omega  \right)=\frac{\partial^2}{\partial F^\mathrm{VIS}_j \partial F^\mathrm{IR}_k} p_i\left( z, \vec \Omega  \right) \bigg|_{\vec F^\mathrm{VIS}=\vec F^\mathrm{IR}=0} \, .
\label{eq:b_ijk_p}
\end{align}
In this work we investigate the imaginary part of the SFG spectrum, which is independent of $\langle b^{\mathrm{P}}_{ijk}(z) \rangle$. 
In the following, we derive explicit expressions for the polarization and the effective polarizability profile.
\subsubsection{Polarization Profile from a Multipolar Charge Distribution in Planar Geometry}
\label{sec:polarization_dens}
The explicit expressions for the polarization density $p_i^\mathrm{q}(z,\vec \Omega)$ resulting from a monopole density $\varrho^\mathrm{q}(\vec r)$ in planar geometry can be found elsewhere \cite{bonthuisProfileStaticPermittivity2012}.
We use the multipole expansion of the polarization density in Equation \eqref{eq:multipol_P} in planar geometry and extend it with the monopole contribution $p_i^\mathrm{q}(z,\vec \Omega)$
\begin{align}
    p_i(z,\vec{\Omega}) = p_i^\mathrm{q}(z,\vec{\Omega})+p_i^\mathrm{D}(z,\vec{\Omega}) + p_i^\mathrm{Q}(z,\vec{\Omega}) + ... \, .
    \label{eq:pol_dens}
\end{align}
When extracting the linear response of the polarization density to the IR field $\tilde{s}^{(1,\mathrm{P})}_{ijk}(z,\omega^\mathrm{IR})$, we employ the electric monopoles and dipoles of the atoms and pseudo-atoms. 
The electrostatic energy function in the MB-pol model \cite{babinDevelopmentFirstPrinciples2013} is almost identical to the one in the TTM4-F model \cite{burnhamVibrationalProtonPotential2008}. 
Here, point dipoles are located on the hydrogen and oxygen atoms, and point charges on the hydrogen atoms and the so-called M-site.
However, for prediction of the response of the polarization density $p_i(z,\vec\Omega)$ to the VIS field, 
we use the electric multipoles of the entire molecules as described in the following.
The response to the VIS field is determined by an instantaneous and linear response, defined in Equation \eqref{eq:def_pol_ex}.  
In the absence of charge transfer polarizability  $\left( {\frac{ \mathrm d q^n}{\mathrm{d} \mathcal{F}_i}=0} \right) $, we have the corresponding multipole expansion of the effective polarizability profile
\begin{align}
    a^{\mathrm{P}}_{ij}(z,\vec{\Omega}) =  a_{ij}^{\mathrm{D} } (z,\vec{\Omega}) - \frac{\partial}{\partial z} a_{izj}^{\mathrm{Q}} (z,\vec{\Omega}) + ... \,,
    \label{eq:def_axi}
\end{align}
where the effective polarizability profiles $a_{ij}^{\mathrm{D}}(z,\vec{\Omega})$ and $a_{ijk}^{\mathrm{Q}}(z,\vec{\Omega})$ are the instantaneous and linear response of the electric dipole density and the electric quadrupole density to an applied external field. 
The effective polarizability profiles $a_{ij}^{\mathrm{D}}(z,\vec{\Omega})$ and $a_{ijk}^{\mathrm{Q}}(z,\vec{\Omega})$ are determined by the change of the electric dipole and electric quadrupole moments at the $n^\mathrm{th}$ molecular center due to an applied external field
\begin{align}
     \tilde{a}_{ij}^{\mathrm{D} } (z,\vec \Omega)  =\frac{\partial}{\partial F^\mathrm{VIS}_j} \varrho_i^{\mathrm{D}} ( z, \vec{\Omega} ) \bigg|_{F^\mathrm{VIS}_j=0} = \frac{1}{L_x L_y} \sum\limits_n^{N_\mathrm{mol}}  \delta\left[ z-z^n (\vec \Omega )\right] \frac{ \mathrm{d} \mu_i^n(\vec \Omega) }{ \mathrm{d} \mathcal{F}_j }\bigg|_{\mathcal{F}_j=0}
     \label{eq:def_pol_D} \\
    \tilde{a}_{ijk}^{\mathrm{Q}}(z,\vec \Omega) = \frac{\partial}{\partial F^\mathrm{VIS}_k} \varrho_{ij}^{\mathrm{Q}} ( z, \vec{\Omega} ) \bigg|_{\mathcal{F}^\mathrm{VIS}_k=0} =  \frac{1}{L_x L_y} \sum\limits_n^{N_\mathrm{mol}} \delta\left[ z-z^n (\vec \Omega )\right] \frac{ \mathrm{d} Q^{ n }_{ij} (\vec \Omega)}{ \mathrm{d} \mathcal{F}_k }\bigg|_{\mathcal{F}_k=0} \, ,
    \label{eq:def_pol_Q}
\end{align}
assuming that the molecular position $z^n$ depends only on the positions of the nuclei, which is assumed to be independent of $F^\mathrm{VIS}_i(t)$. 
As a change of the dipole moment at the $n$-th site does invoke a change of the electric field acting on all the other molecules, Equations \eqref{eq:def_pol_D} and \eqref{eq:def_pol_Q} need to be solved in a self-consistent manner, as derived in the following.
\subsection{Self-Consistent Field Equations}
\label{sec:SCF_eqs_within_BO}
Here, we introduce the self-consistent field (SCF) equations employed in the following to model the multipoles induced by the VIS field, starting with a molecular Schrödinger equation in the presence of an external electric field.
We consider electronic molecular charge distributions that can be localized on molecules. The wave function of the electrons $\Psi_i^n$ within the $n^\mathrm{th}$ molecule is determined by the solution of the stationary Schrödinger equation.
\begin{align}
    \hat{H}\left[\vec \Omega^n, \vec E^n(\cdot) \right] \Psi_i^n\left[\vec \Omega^n, \vec E^n(\cdot) \right]= H_i\left[\vec \Omega^n, \vec E^n(\cdot) \right] \Psi_i^n\left[\vec \Omega^n, \vec E^n(\cdot)\right] \, ,
\end{align}
which depends in the Born-Oppenheimer approximation only parametrically on the nuclei positions within the molecule $\vec \Omega^n$. The local electric field reads
\begin{align}
    E_i^n(\vec r) = \sum\limits_{m\neq n} \left[ -T^{(1)}_i(\vec r^n+\vec r-\vec r^m) q^m +  T^{(2)}_{ij}(\vec r^n+\vec r-\vec r^m) \mu_j^m - T^{(3)}_{ijk}(\vec r^n+\vec r-\vec r^m) Q_{jk}^m + ... \right] + F_i(\vec r+\vec r^n) \,
    \label{eq:En_r}
\end{align}
and is due to the charge distribution of the neighboring molecules and an external field $F_i(\vec r)$. 
The field $E_i^n$ introduced in Equation \eqref{eq:En} refers to the special case where $E_i^n(\vec{r})$ is evaluated at the molecular center $\vec{r}^n$, i.e., $E_i^n = E_i^n(0)$.
The Hamilton operator is denoted as $\hat{H}\left[\vec \Omega^n, \vec E^n(\cdot) \right]$ and $H_i\left[\vec \Omega^n,  \vec E^n(\cdot)\right]$ is the energy of the $i^\mathrm{th}$ eigenstate.
For water at room temperature, the HOMO-LUMO gab is considerably larger than the thermal energy ($\SI{6.3}{eV}\approx 244 k_B T$) \cite{tenbrinckPolycyclicAromaticHydrocarbons2022}. 
Hence, the charge density of the molecule can, in good approximation, be described as the ground-state charge distribution $\rho^n\left[\vec r,\vec \Omega^n, \vec E^n(\cdot) \right]$, which is fully determined by the ground-state wavefunction $\Psi_0^n\left[\vec \Omega^n, \vec E^n(\cdot) \right]$ \cite{kochChemistsGuideDensity2001}. 
The local electric field is represented by its Taylor expansion
\begin{equation}
    E^n_i(\vec r) = E^n_i(0) + r_j \frac{\partial}{\partial r_j'} E^n_i(\vec r')\bigg|_{\vec r'=0} +... \,
\end{equation}
which leads to the electric multipole expansion of the charge density
\begin{align}
     \rho^n\left[\vec r,\vec \Omega^n, \vec E^n(\cdot)\right] \approx \rho^n\left[\vec r,\vec \Omega^n, 0 \right] + \alpha^{n,\mathrm{\rho D}}_i(\vec r,\vec \Omega^n) E^n_i(0) + \alpha^{n,\mathrm{\rho Q}}_{ij}(\vec r,\vec \Omega^n) \frac{\partial}{\partial r_i'} E^n_j(\vec r')\bigg|_{\vec r'=0} + ... \, . 
     \label{eq:gradient_expansion_charge}
\end{align}
Here $\alpha^{n,\mathrm{\rho D}}_{i} (\vec r,\vec \Omega^n)$ and $\alpha^{n,\mathrm{\rho Q}}_{ij} (\vec r,\vec \Omega^n)$ are the polarizabilities of the molecular electric charge density as a function of the coordinates of the nuclei within the n$^\mathrm{th}$ molecule.
We obtain the molecular polarizabilities by inserting Equation \eqref{eq:gradient_expansion_charge} into the definitions of the molecular multipoles in Equations \eqref{eq:Q_0n}-\eqref{eq:Q_IIn}
\begin{align}
    \mu^n_{i} = \alpha^{n,\mathrm{DD}}_{ij} E^n_j (0)+ \alpha^{n,\mathrm{DQ}}_{ijk} \frac{\partial}{\partial r_j} E^n_k(\vec r)\bigg|_{\vec r=0}+... \,;  \quad  Q^n_{ij} = \alpha^{n,\mathrm{QD}}_{ijk} E^n_{k}(0) + \alpha^{n,\mathrm{QQ}}_{ijkl} \frac{\partial}{\partial r_k} E^n_l(\vec r)\bigg|_{\vec r=0} + ...\, . \,
\label{eq:polarizabilities_qpol}
\end{align}
Here 
\begin{align}
    \alpha^{n,\mathrm{DD}}_{ij}(\vec \Omega^n) = \iiint\mathrm{d} \vec r r_i  \alpha^{n,\mathrm{\rho \mathrm{D}}}_{j}(\vec r,\vec \Omega^n); \quad    \alpha^{n,\mathrm{DQ}}_{ijk}(\vec \Omega^n) = \iiint\mathrm{d} \vec r r_i  \alpha^{n,\mathrm{\rho \mathrm{Q}}}_{jk}(\vec r,\vec \Omega^n)
\end{align}
are the electric dipole - electric dipole and the electric dipole - electric quadrupole polarizabilities, respectively.
The electric quadrupole - electric dipole and the electric quadrupole - electric quadrupole polarizabilities are determined by
\begin{align}
    \alpha^{n,\mathrm{QD}}_{ijk}(\vec \Omega^n) = \frac{1}{2} \iiint\mathrm{d} \vec r r_i r_j  \alpha^{n,\mathrm{\rho \mathrm{D}}}_{k}(\vec r,\vec \Omega^n); \quad    \alpha^{n,\mathrm{QQ}}_{ijkl}(\vec \Omega^n) = \frac{1}{2}\iiint\mathrm{d} \vec r r_i r_j  \alpha^{n,\mathrm{\rho \mathrm{Q}}}_{kl}(\vec r,\vec \Omega^n)\, .
\end{align}
We do not have an electric monopole polarizability as the net charge of our molecules is conserved.
We apply spatially constant external fields, and consequently the electric-field gradients are only induced indirectly and can be assumed to be small, this leads to the employed leading-order approximation of the induced electric dipole and electric-quadrupole at the $n^\mathrm{th}$ molecule
\begin{align}
        \mu^n_{i} \approx \alpha^{n,\mathrm{DD}}_{ij} E^n_j \, ;  \,  \quad  Q^n_{ij} \approx \alpha^{n,\mathrm{QD}}_{ijk} E^n_{k} \, .
    \label{eq:polarizabilities}
\end{align}
The electric field $E^n_i$ that acts on the molecular center $\vec{r}$ is determined by the other molecules alongside a potentially applied external field $\vec{F}(\vec{r}^n)$, as given by Equation \eqref{eq:En}.
Since the electric field produced by the multipoles of the $m^\text{th}$ molecule is influenced by the field produced by the multipoles of the $n^\text{th}$ molecule ($m \ne n$), it must be determined self-consistently using the self-consistent field (SCF) equations
\begin{align}
    \mu^n_i &= \alpha^{n,\mathrm{DD}}_{ij} F^{n}_j +\alpha^{n,\mathrm{DD}}_{ij}  \sum\limits_{m\neq n}^{N_\mathrm{mol}}\left[ -T^{(1)}_j(\vec{r}^{nm})q^m + T^{(2)}_{jk}(\vec r^{nm}) \mu^m_k -  T^{(3)}_{jkl}(\vec r^{nm}) Q^m_{kl} + ... \right] + \mu^{n,\mathrm{S}}_i \label{eq:dip_scf}\\
    Q^n_{ij} &= \alpha^{n,\mathrm{QD}}_{ijk} F^{n}_k+\alpha^{n,\mathrm{QD}}_{ijk}  \sum\limits_{m\neq n}^{N_\mathrm{mol}}\left[  - T^{(1)}_k (\vec{r}^{nm})q^m+ T^{(2)}_{kl}(\vec r^{nm}) \mu^m_l -  T^{(3)}_{klo}(\vec r^{nm}) Q^m_{lo} + ... \, \right]  + Q^{n,\mathrm{S}}_{ij} \, 
    \label{eq:quad_scf}
\end{align}
and so on. Here, $\tensor{T}^{(l)}$ 
is the electrostatic coupling tensor defined in Equation \eqref{eq:def_T_tensor}, $\vec r^{nm}=\vec{r}^n-\vec{r}^m$ is the distance vector between the n$^\mathrm{th}$ and the m$^\mathrm{th}$ molecule and $\mu^{n,\mathrm{S}}_i$ and $Q^{n,\mathrm{S}}_{ij}$ are molecular source electric dipole and electric quadrupole moments (such as permanent multipole moments), respectively. 
These SCF equations are analogous to the minimization of the potential energy of the electronic degrees of freedom \cite{stoneInductionEnergyAssembly1989}.
Hence, the SCF Equation \eqref{eq:dip_scf} must be satisfied at all times, which leads to
\begin{align}
    \mu^n_i(t) &= \alpha^{n,\mathrm{DD}}_{ij}(t) F^{n}_j(t) + \alpha^{n,\mathrm{DD}}_{ij}(t) \sum\limits_{m \neq n}^{N_\mathrm{mol}} \left[ - T^{(1)}_j\left[\vec r^{nm}(t)\right] q^m + T^{(2)}_{jk}\left[\vec r^{nm}(t)\right] \mu^m_k(t) -  T^{(3)}_{jkl}\left[\vec r^{nm}(t)\right] Q^m_{kl}(t) + ... \right] + \mu^{n,\mathrm{S}}_i(t)
    \label{eq:scf_mu_1}\\
     \mu^n_i(t) &= \alpha^{n,\mathrm{DD}}_{ij}(t) E_j^{n}(t) + \mu^{n,\mathrm{S}}_i(t) \, .
      \label{eq:scf_mu_2}
\end{align}
The SCF equation for electric quadrupoles $Q^n_{ij}(t)$ is equivalent. 
\subsubsection{Nonlocal First and Second Order Perturbation Expansion of the Polarization Density}
\label{sec:nonlocal_first_and_second_order_pert_exp_pol_dens}
Here, we derive equations for extracting the second-order polarization contributions. These equations are applied in the spectra prediction presented in the main text. 
As the molecular polarizabilities are only determined by the nuclei positions they cannot oscillate at optical frequencies. 
Consequently, for SFG spectra prediction only the multipoles induced by the VIS field are relevant.
These are determined by the SCF equations
\begin{align}
    \mu^n_i(t) &= \alpha^{n,\mathrm{DD}}_{ij}(t)   F^\mathrm{VIS}_j (t)   + \alpha^{n,\mathrm{DD}}_{ij}(t) \sum\limits_{m\neq n}^{N_\mathrm{mol}} \left[  T^{(2)}_{jk}\left[\vec r^{nm}(t)\right] \mu^m_k(t) -  T^{(3)}_{jkl}\left[\vec r^{nm}(t)\right] Q^m_{kl}(t) + ... \right]
     \label{eq:scf_mu_time}\\
Q^n_{ij}(t) &= \alpha^{n,\mathrm{QD}}_{ijk}(t)  F^\mathrm{VIS}_k (t)  +  \alpha^{n,\mathrm{QD}}_{ijk}(t) \sum\limits_{m\neq n}^{N_\mathrm{mol}} \left[  T^{(2)}_{kl}\left[\vec r^{nm}(t)\right] \mu^m_l(t) -  T^{(3)}_{klo}\left[\vec r^{nm}(t)\right] Q^m_{lo}(t) + ... \right]  \, .
    \label{eq:scf_Q_time}
\end{align}
We can solve Equations \eqref{eq:scf_mu_time} and \eqref{eq:scf_Q_time} in the time domain by applying a constant external field $F^\mathrm{VIS}_i(t)=\mathcal{F}^\mathrm{TEST}_i$ at every time step. The solution can be formally expressed as
\begin{align}
    \mu^n_i(t)  &= \alpha^{n,\mathrm{DD}}_{ij}(t) f^n_{jk}(t) \mathcal{F}^\mathrm{TEST}_k
     \label{eq:scf_mu_time_2}\\
Q^n_{ij}(t) &= \alpha^{n,\mathrm{QD}}_{ijk}(t)f^n_{kl}(t) \mathcal{F}^\mathrm{TEST}_l
    \label{eq:scf_Q_time_2} \, ,
\end{align}
where $f^n_{jk}(t)$ is a local field factor
that relates the external field $\mathcal{F}_i(t)$ with the local E-field defined in Equation \eqref{eq:En}
\begin{align}
    E^n_i(t) = f^n_{ij}(t) F_j(t) \, .
    \label{eq:def_local_field_factor}
\end{align} 
The molecular multipoles are linear in the external field $\mathcal{F}^\mathrm{TEST}_i$. Hence, we can give the explicit expressions
\begin{align}
    a^\mathrm{D}_{ij}\left[ z,\vec \Omega(t) \right] &= \frac{1}{L_x L_y} \sum\limits_n^{N_\mathrm{mol}}\alpha^{n,\mathrm{DD}}_{ik}\left[ \vec \Omega(t) \right] f^n_{kj}\left[ \vec \Omega(t) \right] \delta \left( z-z^n\left[ \vec \Omega(t) \right] \right) 
    \label{eq:adij_mun} \\
    a^\mathrm{Q}_{ijk}\left[ z,\vec \Omega(t) \right]  &= \frac{1}{L_x L_y} \sum\limits_n^{N_\mathrm{mol}} \alpha^{n,\mathrm{QD}}_{ijl}\left[ \vec \Omega(t) \right] f^n_{lk}\left[ \vec \Omega(t) \right] \delta \left( z-z^n\left[ \vec \Omega(t) \right] \right) \, ,
    \label{eq:aqijk_Qn}
\end{align}
 for the effective polarizability profiles $a^\mathrm{D}_{ij}(z,\vec \Omega)$ and $a^\mathrm{Q}_{ijk}(z,\vec \Omega)$, defined in Equations \eqref{eq:def_pol_D} and \eqref{eq:def_pol_Q},
respectively.
We emphasize that they depend on time only via the state vector $\vec \Omega(t)$. 
Hence, the linear response function of the polarization density defined in Equation \eqref{eq:def_s1P_ij} 
is determined by the first-order time-dependent perturbation expansion of the polarization profile in addition to the expectation value of the effective polarizability profile, i.e.
\begin{align}
     \tilde{s}^{(1,\mathrm{P})}_{ij} \left(z,\omega^\alpha \right) = \varepsilon_0^{-1} \tilde{\varphi}\left[ p_i(\cdot), P_j(\cdot), \omega^\alpha \right] + \varepsilon_0^{-1}\langle  a^\mathrm{D}_{ij}\left[ z,\vec \cdot \right] \rangle  -\varepsilon_0^{-1} \frac{\partial}{\partial z} \langle  a^\mathrm{Q}_{izj}\left[ z,\vec \cdot \right] \rangle \, .
     \label{eq:s1P_fdt}
\end{align}
Likewise, the second-order response function of the polarization density, defined in Equation \eqref{eq:def_s2P_ijk}, is determined by the first-order expansion of the polarizability profile and the expectation value of the off-resonant hyperpolarizability profile $b^\mathrm{P}_{ijk}\left[ z,\vec \Omega \right] $
\begin{align}
     \tilde{s}^{(2,\mathrm{P})}_{ijk} \left( z, \omega^\mathrm{IR} \right) = \varepsilon_0^{-1} \tilde{\varphi}\left[ a_{ij}^\mathrm{D}(z,\cdot), P_k(\cdot),\omega^\mathrm{IR} \right] -\varepsilon_0^{-1} \frac{\partial}{\partial z}\tilde{\varphi}\left[ a_{izj}^\mathrm{Q}(z,\cdot), P_k(\cdot) ,\omega^\mathrm{IR}\right] + \varepsilon_0^{-1} \left\langle  b^\mathrm{P}_{ijk}\left[ z,\vec \cdot \right] \right\rangle \, ,
     \label{eq:s2P_fdt}
\end{align}
where we inserted the multipole expansion of the effective polarizability profile in Equation \eqref{eq:def_s2P_ijk}.
According to our formalism defined in Equation \eqref{eq:s2ijk_beta}, we split up the polarization contributions into the second-order electric dipole response 
\begin{align}
     \tilde{s}^{(2,\mathrm{D})}_{ijk} \left( z, \omega^\mathrm{IR} \right) = \varepsilon_0^{-1} \tilde{\varphi}\left[ a_{ij}^\mathrm{D}(z,\cdot), P_k(\cdot),\omega^\mathrm{IR} \right] + \varepsilon_0^{-1} \left\langle  b^\mathrm{D}_{ijk}\left[ z,\vec \cdot \right] \right\rangle \, ,
     \label{eq:s2D_fdt}
\end{align}
and the electric quadrupole response
\begin{align}
     \tilde{s}^{(2,\mathrm{Q})}_{ijk} \left( z, \omega^\mathrm{IR} \right) =  -\varepsilon_0^{-1} \frac{\partial}{\partial z}\tilde{\varphi}\left[ a_{izj}^\mathrm{Q}(z,\cdot), P_k(\cdot) ,\omega^\mathrm{IR}\right]  - \varepsilon_0^{-1} \frac{\partial}{\partial z} \langle  b^\mathrm{Q}_{izjk}\left[ z,\vec \cdot \right] \rangle \, ,
     \label{eq:s2Q_fdt}
\end{align}
where $\langle  b^\mathrm{D}_{ijk}\left[ z,\vec \cdot \right] \rangle$ and $ \langle  b^\mathrm{Q}_{izjk}\left[ z,\vec \cdot \right] \rangle$ are the corresponding off-resonant effective hyperpolarizability profiles.
Nevertheless, for interpretation, we also need to apply the additional decomposition of the electric dipole contribution introduced in Section \ref{sec:additional_decom_e_dip} and factor in the hypothetical response in the scenario that only electric dipole moments are induced by the VIS field, which is equivalent to the SCF equation
\begin{align}
 \mu^n_i(t) &= \alpha^{n,\mathrm{DD}}_{ij}(t)  F^\mathrm{VIS}_j + \alpha^{n,\mathrm{DD}}_{ij}(t) \sum\limits_{m\neq n}^{N_\mathrm{mol}} T^{(2)}_{jk}\left[\vec r^{nm}(t)\right] \mu^m_k(t) \, ,
    \label{eq:SCF_only_dip}
\end{align}
with the solution 
\begin{align}
    \mu^n_i(t) = \alpha^{n,\mathrm{DD}}_{ij}(t) f^{n,\mathrm{D}}_{jk}(t) F^\mathrm{VIS}_k (t) \, .
    \label{eq:def_local_field_D}
\end{align}
Hence, we have the effective pure electric dipole polarizability profile 
\begin{align}
    a^\mathrm{DD}_{ij}\left[ z,\vec \Omega(t) \right] &= \frac{1}{L_x L_y} \sum\limits_n^{N_\mathrm{mol}}\alpha^{n,\mathrm{DD}}_{ik}\left[ \vec \Omega(t) \right] f^{n,\mathrm{D}}_{kj}\left[ \vec \Omega(t) \right] \delta \left( z-z^n\left[ \vec \Omega(t) \right] \right) \, ,
    \label{eq:addij_mun}
\end{align}
which excludes the linear response of the electric dipoles to an electric quadrupole density.
This leads to the pure electric dipole contribution of the second-order response function
\begin{align}   \tilde{s}^{(2,\mathrm{DD})}_{ijk} \left( z, \omega^\mathrm{IR} \right) = \varepsilon_0^{-1} \tilde{\varphi}\left[ a_{ij}^\mathrm{DD}(z,\cdot), P_k(\cdot),\omega^\mathrm{IR} \right] + \varepsilon_0^{-1} \langle  b^\mathrm{DD}_{ijk}\left[ z,\vec \cdot \right] \rangle \, ,  
    \label{eq:s2DD_fdt}
\end{align}
where $b^\mathrm{DD}_{ijk}(z,\vec{\Omega})$ is the equivalent hyperpolarizability profile. From this we can define the electric dipole - electric quadrupole cross contributions
\begin{align}
\tilde{s}^{(2,\mathrm{DQ})}_{ijk}\left(z,\omega^\mathrm{IR} \right) = \tilde{s}^{(2,\mathrm{D})}_{ijk}\left(z,\omega^\mathrm{IR} \right) - \tilde{s}^{(2,\mathrm{DD})}_{ijk}\left(z,\omega^\mathrm{IR} \right) \, ,
\label{eq:s2DQ_def2}
\end{align}
which approximately accounts for the linear response to the second-order electric quadrupole density, as shown numerically in Figure \ref{fig:big_multipole}.
Hence, the corresponding effective polarizability profile reads
\begin{align}
        a^\mathrm{DQ}_{ij}\left[ z,\vec \Omega(t) \right] &= \frac{1}{L_x L_y} \sum\limits_n^{N_\mathrm{mol}}\alpha^{n,\mathrm{DD}}_{ik}\left[ \vec \Omega(t) \right] \left(f^{n}_{kj}\left[ \vec \Omega(t) \right]  - f^{n,\mathrm{D}}_{kj}\left[ \vec \Omega(t) \right] \right) \delta \left( z-z^n\left[ \vec \Omega(t) \right] \right) \, .
        \label{eq:adqij_mun}
\end{align}
The corresponding second-order response profile is determined by the linear response relation
\begin{align}   \tilde{s}^{(2,\mathrm{DQ})}_{ijk} \left( z, \omega^\mathrm{IR} \right) = \varepsilon_0^{-1} \tilde{\varphi}\left[ a_{ij}^\mathrm{DQ}(z,\cdot), P_k(\cdot),\omega^\mathrm{IR} \right] + \varepsilon_0^{-1} \langle  b^\mathrm{DQ}_{ijk}\left[ z,\vec \cdot \right] \rangle \, , 
    \label{eq:s2DQ_fdt}
\end{align}
where $b^\mathrm{DQ}_{ijk}\left[ z,\vec \Omega \right] $ is the associated off-resonant hyperpolarizability profile.
Details on calculating the electric field with periodic boundary conditions and the parametrization of molecular polarizabilities from a set of molecular coordinates are provided in Section \ref{app:polarizability}. 
We predict the linear absorption profile $\tilde{s}^{(1,\mathrm{P})}_{ij}(z,\omega^\mathrm{IR})$ and the total dipole moment $P_i(t)$ from the set of point charges and electric dipole moments
included in the MB-Pol force field \cite{babinDevelopmentFirstPrinciples2013}, based on the TTM4-F model \cite{burnhamVibrationalProtonPotential2008}. 
The multipolar polarizability profiles $a^{\mathrm{DD}}_{ij}(z,t)$ and $a^{\mathrm{QD}}_{ij}(z,t)$ are computed from the trajectories in post-processing as described in this Section.
We arrive at the set of linear response relations, where the dielectric coupling of the molecule with its environment is condensed into the single local field factor $f^n_{ij}(t)$, defined in Equation \eqref{eq:def_local_field_factor}.
We note that, in the theory of nonlinear optics, one frequency-dependent factor appears for each frequency\cite{armstrongInteractionsLightWaves1962}, different from our formulation.
Using a simple model calculation based on the Lorentz-field approximation, we demonstrate that there is no disagreement.
This simple model calculation can be extended to the general case by introducing a time-scale separation.
Within this time-scale separation, we obtain two time-averaged local field factors in the effective polarizability profiles defined in Equations \eqref{eq:adij_mun}, \eqref{eq:aqijk_Qn}, and \eqref{eq:addij_mun}, rather than a single, rapidly-varying one. 
This means that using the time-scale separation transforms the effective polarizability $\alpha^{n,\mathrm{DD}}_{i j'}(t) f^{n,\mathrm{D}}_{j' j}(t)$ into $\bar{f}^n_{i i'}(t) \alpha^{n,\mathrm{DD}}_{i' j'} (t) \bar{f}^{n}_{j' j}(t)$, where $\bar{f}^n_{j' j}(t)$ is the time-averaged local field factor.
Previously, the expression $f^{n,\mathrm{D}}_{i i'}(t) \alpha^{n,\mathrm{DD}}_{i' j'} (t) f^{n,\mathrm{D}}_{j' j}(t)$ was used by others in SFG spectra prediction \cite{moritaRecentProgressTheoretical2008,ishiyamaComputationalAnalysisVibrational2017}, which introduces artifacts, as shown in Figure \ref{fig:big_multipole}.
\subsubsection{Simple Model Calculation}
\label{sec:simple_model_time_aver_loc_field}
In a homogeneous dipolar material, within the Lorentz-field approximation, we obtain the simple SCF equation from Equation \eqref{eq:local_field_multipole}
\begin{align}
    p(t) = \rho \alpha(t) \left[\mathcal{F}^{\mathrm{VIS}} e^{-i\omega^\mathrm{VIS} t} + \frac{1}{3 \varepsilon_0} p(t)    \right] \, ,
    \label{eq:SCF_simple_1}
\end{align}
where $\rho$ is the number density, $\alpha(t)=\alpha_0 + \beta \mathcal{F}^{\mathrm{IR}}e^{-i\omega^\mathrm{IR} t}$ is the polarizability perturbed by the IR laser and $p(t)$ is the resulting polarization
density. Here, we leave out any indices.
We solve Equation \eqref{eq:SCF_simple_1} for $p(t)$ leading to
\begin{align}
    p(t) = \frac{\rho \alpha(t) }{1-\frac{\rho \alpha(t)}{3 \varepsilon_0}} \mathcal{F}^{\mathrm{VIS}} e^{-i\omega^\mathrm{VIS} t} =  \rho \alpha(t) f(t) \mathcal{F}^{\mathrm{VIS}} e^{-i\omega^\mathrm{VIS} t} \, ,
    \label{eq:SCF_simple_2}
\end{align}
where we identified the rapidly-varying local field factor 
\begin{align}
    f(t) = \frac{1 }{1-\frac{\rho \alpha(t)}{3 \varepsilon_0}} \, ,
\end{align}
 by comparison with Equation \eqref{eq:def_local_field_factor} .
Now we perform a Taylor expansion of Equation \eqref{eq:SCF_simple_2} in the external field amplitude $\mathcal{F}^\mathrm{IR}$.
The zeroth-order term is the equifrequent response to the VIS laser
\begin{align}
    p^{(1)}(t) = \frac{\rho \alpha_0 }{1-\frac{\rho \alpha_0}{3 \varepsilon_0} } \mathcal{F}^{\mathrm{VIS}} e^{-i\omega^\mathrm{VIS} t}=\bar{f} \rho \alpha_0 \mathcal{F}^\mathrm{VIS} e^{-i \omega^\mathrm{VIS} t}\, ,
    \label{eq:SCF_simple_3}
\end{align}
where we identified the time-averaged local field factor as 
\begin{align}
    \bar{f} = \frac{1}{1-\frac{\rho \alpha_0}{3 \varepsilon_0}  } = \frac{\varepsilon + 2}{3 } \, .
\end{align}
Here, we used the Clausius-Mossotti Relation \eqref{eq:Clausius_Mossoti} to relate $\bar{f}$ to the dielectric constant.
The first-order term gives rise to the second-order polarization density
\begin{align}
    p^{(2)} (t) = \frac{1}{ \left( 1-\frac{\rho \alpha_0}{3 \varepsilon_0} \right)^2} \rho \beta \mathcal{F}^\mathrm{VIS} \mathcal{F}^\mathrm{IR} e^{-i\omega^\mathrm{SFG} t} = \rho \bar{f} \beta \bar{f} \mathcal{F}^\mathrm{VIS} \mathcal{F}^\mathrm{IR} e^{-i\omega^\mathrm{SFG} t} \, .
    \label{eq:second-order_simple}
\end{align}
which is the conventional formulation of SFG theory\cite{armstrongInteractionsLightWaves1962}.
By comparing Equations \eqref{eq:SCF_simple_2} and \eqref{eq:second-order_simple} we realize that the effective nonlinear polarizability is given by $\alpha(t) f(t)$, whereas the second-order expansion of the effective nonlinear polarizability is given by $\bar{f} \beta  \bar{f}\mathcal{F}^\mathrm{IR} e^{-i\omega^\mathrm{IR} t}$.
Consequently, expansion to the 1$^\mathrm{st}$ order of $\alpha(t) f(t)$ leads to the second-order response of the polarization density.
This is evident in the model calculation, where the linear and nonlinear polarizabilities $\alpha_0$ and $\beta$ are clearly separated. 
In contrast, our equilibrium molecular dynamics simulations provide only trajectories of oscillating molecular polarizabilities, making such a decomposition less straightforward. 
The equivalent calculation for the trajectory obtained from the molecular dynamics simulation is presented in the following section.
\subsubsection{Time-Scale Separation in the Perturbation Expansion of the Approximate Second-Order Polarization}
\label{sec:rigourous_two_field_facs}
The here presented approximation is not applied to the SFG spectra presented in the main text.
Here, we introduce a time-scale separation that leads to an approximate expression to predict the SFG signal from MD simulations, including two time-averaged local field factors. 
This section can be understood as a generalization of the simple model calculation in Section \ref{sec:simple_model_time_aver_loc_field}.
The advantage of the formulation derived here is the possibility of applying approximate expressions for the two time-averaged local field factors, as we do in Section \ref{sec:orientation_analyis}. 
This cannot be done in the exact case, as we cannot give a reasonable estimate for the rapidly-varying local field factor $f^n_{ij}(t)$. 
As usual, we consider the external fields
$F^\alpha_i(t) = \mathcal{F}^\alpha_i e^{-i\omega^\alpha t}$ at frequencies $\omega^\mathrm{IR}$ and $\omega^\mathrm{VIS}$. 
We dissect the time-dependent molecular properties of interest $x^n(t)$ into a slowly moving average $x^{n,0}(t)$ and contributions oscillating at the frequencies of interest $x^{n,\mathrm{IR}}(t)e^{-i \omega^\mathrm{IR} t}$, $x^{n,\mathrm{VIS}}(t)e^{-i \omega^\mathrm{VIS} t}$ and $x^{n,\mathrm{SFG}}(t)e^{-i \omega^\mathrm{SFG} t}$. We define the remainder as
\begin{align}    
    x^{n,\mathrm{R}}(t)=x^n(t) - x^{n,0}(t) -x^{n,\mathrm{IR}}(t)e^{-i \omega^\mathrm{IR}t} - x^{n,\mathrm{VIS}}(t)e^{-i \omega^\mathrm{VIS}t}
    - x^{n,\mathrm{SFG}}(t)e^{-i \omega^\mathrm{SFG}t}\, .
    \label{eq:ansatz_x}
\end{align}
Our observables of interest are the electric dipole, the electric quadrupole and the corresponding polarizabilities, which we decompose by our ansatz in Equation \eqref{eq:ansatz_x} as
\begin{align}
        \mu_i^{n}(t) &=    \mu_i^{n,\mathrm{R}}(t)+ \mu_i^{n,\mathrm{VIS}}(t)e^{-i\omega^\mathrm{VIS}t} + \mu_i^{n,\mathrm{SFG}}(t)e^{-i\omega^\mathrm{SFG}t}  
    \label{eq:mu_n_time-scale} \\
    Q_{ij}^{n}(t) &=Q_{ij}^{n,\mathrm{R}}(t) + Q_{ij}^{n,\mathrm{VIS}}(t)e^{-i\omega^\mathrm{VIS}t} + Q_{ij}^{n,\mathrm{SFG}}(t)e^{-i\omega^\mathrm{SFG}t}  
    \label{eq:Q_n_time-scale} \\
    \alpha^{n,\mathrm{DD}}_{ij}(t)&=\alpha^{n,\mathrm{DD,R}}_{ij}(t) + \alpha^{n,\mathrm{DD,}0}_{ij}(t) + \alpha^{n,\mathrm{DD,IR}}_{ij}(t) e^{-i\omega^\mathrm{IR} t} \label{eq:aDD_n_time-scale} \\
     \alpha^{n,\mathrm{QD}}_{ijk}(t)&=\alpha^{n,\mathrm{QD,R}}_{ijk}(t) + \alpha^{n,\mathrm{QD,0}}_{ijk}(t)+\alpha^{n,\mathrm{QD,IR}}_{ijk}(t) e^{-i\omega^\mathrm{IR} t} \, ,
     \label{eq:aQD_n_time-scale}
\end{align}
where we do not have oscillations of the polarizabilities at the frequencies $\omega^\mathrm{VIS}$ and $\omega^\mathrm{SFG}$ as we neglect electronic hyperpolarizabilities, which produce the off-resonant background in SFG spectroscopy.
We write down the time-dependent SCF Equation \eqref{eq:scf_mu_1} without source terms
\begin{align}
    \mu^n_i(t) &= \alpha^{n,\mathrm{DD}}_{ij}(t)\mathcal{F}^{\mathrm{VIS}}_j e^{-i \omega^\mathrm{VIS} t} +  \alpha^{n,\mathrm{DD}}_{ij}(t) \sum\limits_{m\neq n}^{N_\mathrm{mol}}\left[T^{(2)}_{jk}\left[\vec r^{nm}(t)\right] \mu^m_k(t) -  T^{(3)}_{jkl}\left[\vec r^{nm}(t)\right] Q^m_{kl}(t) + ... \right] \, .
    \label{eq:scf_mu_1_sfg}
\end{align}
Equation \eqref{eq:scf_mu_1_sfg} in the time domain has all frequencies present, in the following we want to extract the SFG component of $\mu^{n}_i(t)$ and  $Q^{n}_{ij}(t)$. 
To invoke the time-scale separation, we introduce a time interval $-\tau<t-\bar{t}<\tau$ in which certain quantities are assumed to be constant; for that, we multiply Equation \eqref{eq:scf_mu_1_sfg}, with a window function $w(t-\bar{t})$ which is normalized according to
\begin{align}
    \int\limits_{-\tau}^\tau\mathrm{d} t \,  w(t) &= 1 
\end{align}
and only nonzero inside the interval $-\tau < t < \tau$. 
This leads to
\begin{align}
    w(t-\bar{t}) \mu^n_i(t) = w(t-\bar{t}) \alpha^{n,\mathrm{DD}}_{ij}(t)\mathcal{F}^\mathrm{VIS}_j e^{-i \omega^\mathrm{VIS} t} +  w(t-\bar{t})\alpha^{n,\mathrm{DD}}_{ij}(t) \sum\limits_{m\neq n}^{N_\mathrm{mol}}\left[T^{(2)}_{jk}\left[\vec r^{nm}(t)\right] \mu^m_k(t) -  T^{(3)}_{jkl}\left[\vec r^{nm}(t)\right] Q^m_{kl}(t) + ... \right] \, .
    \label{eq:scf_mu_2_sfg}
\end{align}
Now, we assert three features on the window function $w(t-\bar{t})$.
First, the time interval $2 \tau$ needs to be short enough so that we can approximate the electrostatic coupling tensor as stationary, i.e.
\begin{equation}
    {w(t-\bar{t}) \mathbf{T}^{(l)}[\vec{r}^{nm}(t)]\approx w(t-\bar{t}) \mathbf{T}^{(l)}[\vec{r}^{nm}(\bar{t})]} \, .
    \label{eq:approx_time_scale_Tlmn}
\end{equation}
Second, we assume that the same holds for the instantaneous amplitudes appearing in Equations \eqref{eq:mu_n_time-scale}-\eqref{eq:aQD_n_time-scale}
\begin{equation}
    {w(t-\bar{t})  x^{n,\alpha}(t)\approx w(t-\bar{t}) x^{n,\alpha}(\bar{t})} \, .
\label{eq:approx_time_scale_xn_alpha}
\end{equation}
Third, we impose that the window function changes only slightly during the time interval $2\pi/\omega^{\mathrm{IR}}$, which implies that 
\begin{align}
    \tilde{w}(\omega) \approx 0 \, \text{, for }|\omega|\geq \omega^\mathrm{IR} .   \label{eq:approx_time_scale_IR}
\end{align}
In the following, we will consider an ideal window function for which the approximations \eqref{eq:approx_time_scale_Tlmn}-\eqref{eq:approx_time_scale_IR} hold exactly.
We consider the following time-scales. The longest considered oscillation period of the external IR field is the bending period, which is approximately $\SI{20}{fs}$. 
However, as evident in Figure \ref{fig:origin_sfg_signal} B, molecular centers do not move significantly at frequencies greater than $\omega =2 \pi \times\SI{6}{THz}$, which corresponds to a period of $\SI{166}{fs}$, from which we conclude that a time-scale separation is only marginally valid and thus needs to be checked numerically.
We define the short-time Fourier transformation (STFT) \cite{hlawatschTimeFrequencyAnalysisConcepts2008} of a generic function $x(t)$ as 
\begin{align}
\tilde{x}(\bar{t},\omega) &= \int\limits_{-\infty}^\infty \mathrm{d} t\, w(t-\bar{t}) e^{i \omega t} x(t) \, .
\label{eq:STFT}
\end{align}
We define the instantaneous amplitude of an oscillating molecular property $x^{n,\alpha}(t)$ as the STFT of $x^{n}(t)$, i.e.
\begin{align}
    x^{n,\alpha}(t)=\tilde{x}^n(t,\omega^\alpha) \, .
\end{align}
Now we show that if the approximations in Equations \eqref{eq:approx_time_scale_xn_alpha} and \eqref{eq:approx_time_scale_IR} are justified, the property $\tilde{x}^{n,\mathrm{R}}_i(\bar{t},\omega^\alpha)=0$ holds, where ${\omega^\alpha \in \lbrace 0, \omega^\mathrm{IR},\omega^\mathrm{VIS}, \omega^\mathrm{SFG} \rbrace}$, which means that $x^{n,\mathrm{R}}(t)$ does not oscillate at these frequencies during the investigated period. For this we compute the STFT of $x^{n}(t)$
\begin{align}
\tilde{x}^{n,\alpha}(\bar{t}) &= \int\limits_{-\infty}^\infty \mathrm{d} t\, w(t-\bar{t}) e^{i \omega^\alpha t} x^n_i(t) \\
&= \int\limits_{-\infty}^\infty \mathrm{d} t\, w(t-\bar{t}) e^{i \omega^\mathrm{\alpha } t} x^{n,\mathrm{R}}(t) + \int\limits_{-\infty}^\infty \mathrm{d} t\, w(t-\bar{t}) x^{n,\alpha}(t) + \sum\limits_{\beta\neq \alpha} \int\limits_{-\infty}^\infty \mathrm{d} t\, w(t-\bar{t}) e^{i \left( \omega^\alpha - \omega^\mathrm{\beta } \right) t}
x^{n,\beta}_i(t) \, .
\label{eq:STFT_xn}
\end{align}
We employ approximation \eqref{eq:approx_time_scale_xn_alpha} and rewrite
\begin{align}
\tilde{x}^{n,\alpha}(\bar{t})=  \tilde{x}^{n,\mathrm{R}}\left( \bar{t},\omega^\mathrm{\alpha} \right) + \tilde{x}^{n,\alpha}(\bar{t}) + \sum\limits_{\beta\neq \alpha} \tilde{x}^{n,\beta}(\bar{t}) \tilde{w}(\omega^\alpha- \omega^\beta) e^{i(\omega^\alpha- \omega^\beta) \bar t }\, .
\label{eq:STFT_mu_SFG_approx}
\end{align}
As the smallest absolute value of the frequency difference $|\omega^\alpha-\omega^\beta|$ is $\omega ^\mathrm{IR}$, we can use the approximation in Equation \eqref{eq:approx_time_scale_IR}, which allows us to write
\begin{align}   
x^{n,\alpha}(\bar{t}) = \tilde{x}^\mathrm{R}     (\bar{t},\omega^\alpha) + x^{n,\alpha}(\bar{t})
\label{eq:approx_STFT_X} \, .
\end{align}
from which follows that 
\begin{align}
\tilde{x}^{n,\mathrm{R}}_i(\bar{t},\omega^\alpha)= 0 \, 
\end{align}
for the ideal window function.
We extract the instantaneous amplitude of the electric dipoles oscillating with frequency $\omega^\mathrm{SFG}$ via STFT of Equation \eqref{eq:scf_mu_1_sfg}, i.e.
\begin{multline}
   \mu^{n,\mathrm{SFG}}_i(\bar{t} ) =  \mathcal{F}^\mathrm{VIS}_j\int \mathrm{d} t \,  w(t-\bar{t})  e^{i \omega^\mathrm{IR} t}  \alpha^{n,\mathrm{DD}}_{ij}(t)  \\
  + \sum\limits_{m\neq n}^{N_\mathrm{mol}} \left(T^{(2)}_{jk} \left[ \vec{r}^{nm}(\bar{t}) \right] \int\mathrm{d} t\,  w(t-\bar{t}) e^{i \omega^\mathrm{SFG} t}   \alpha^{n,\mathrm{DD}}_{ij}(t) \mu^{m}_k(t)  - T^{(3)}_{jkl} \left[ \vec{r}^{nm}(\bar{t}) \right] \int\mathrm{d} t\,  w(t-\bar{t}) e^{i \omega^\mathrm{SFG} t}  \alpha^{n,\mathrm{DD}}_{ij}(t) Q^{m}_{kl}(t) \right) \, .
    \label{eq:SCF_SFG}
\end{multline}
We insert our Ansatz in Equations \eqref{eq:mu_n_time-scale}-\eqref{eq:aQD_n_time-scale} into Equation \eqref{eq:SCF_SFG} leading to
\begin{multline}
\mu^{n,\mathrm{SFG}}_i (\bar{t}) = \tilde{\alpha}^{n,\mathrm{DD,IR}}_{ij} (\bar{t})\left[ \mathcal{F}^\mathrm{VIS}_j
 + \sum\limits_{m \neq n}^{N_\mathrm{mol}} \left( T^{(2)}_{jk} \left[ \vec{r}^{nm}(\bar{t}) \right] \mu^{n,\mathrm{VIS}}_k(\bar{t}) 
- T^{(3)}_{jkl} \left[ \vec{r}^{nm}(\bar{t}) \right] Q^{n,\mathrm{VIS}}_{kl}(\bar{t})  \right)
 \right] 
 \\ +  \tilde{\alpha}^{n,\mathrm{DD,0}}_{ij}(\bar{t}) \left[  \sum\limits_{m\neq n}^{N_\mathrm{mol} } \left(T^{(2)}_{jk} \left[ \vec{r}^{nm}(\bar{t}) \right] \mu^{n,\mathrm{SFG}}_k(\bar{t}) 
- T^{(3)}_{jkl}\left[ \vec{r}^{nm}(\bar{t}) \right]  Q^{n,\mathrm{SFG}}_{kl}(\bar{t}) 
  \right) \right] \\
 + \int \mathrm{d} t \,  w(t-\bar{t}) e^{i \omega^\mathrm{SFG} t} \alpha^{\mathrm{n,DD}}_{ij}(t) \sum\limits_{m \neq n}^{N_\mathrm{mol}} \left(  T^{(2)}_{jk}  \left[ \vec{r}^{nm}(\bar{t}) \right]\mu^{m,\mathrm{R}}_k(t) - T^{(3)}_{jkl} \left[ \vec{r}^{nm}(\bar{t}) \right] Q^{m,\mathrm{R}}_{kl}(t) \right) \, .
 \label{eq:SCF_exact}
\end{multline}
The quadrupole SCF equation is equivalent. Now, we approximate  
\begin{align}
\int \mathrm{d} t \,  w(t-\bar{t}) e^{i \omega^\mathrm{SFG} t} \alpha^{\mathrm{n,DD}}_{ij}(t) \sum\limits_{m \neq n}^{N_\mathrm{mol}} \left(  T^{(2)}_{jk}  \left[ \vec{r}^{nm}(\bar{t}) \right]\mu^{m,\mathrm{R}}_k(t) - T^{(3)}_{jkl} \left[ \vec{r}^{nm}(\bar{t}) \right] Q^{m,\mathrm{R}}_{kl}(t) \right) \approx 0 \, .
    \label{eq:adiabatic_muR}
\end{align}
In addition, we assume that the linear response of the electric quadrupoles is negligible, i.e., $Q^{n,\mathrm{VIS}}_{ij}(\bar{t}) = 0$. 
This is a standard assumption in linear optics \cite{bornPrinciplesOpticsElectromagnetic1999}, where the electric quadrupole contribution vanishes in isotropic media and the electric dipole contribution dominates, giving the electric dipoles a symmetry dominance. 
In contrast, in the second-order case, dipole contributions are zero in isotropic media, making electric quadrupole contributions relevant. 
Hence, we obtain the simplified SCF Equation at the SFG frequency 
\begin{multline}    
\mu^{n,\mathrm{SFG}}_i (\bar{t}) = \tilde{\alpha}^{n,\mathrm{DD,IR}}_{ij}(\bar{t}) \left[ \mathcal{F}^\mathrm{VIS}_j
 + \sum\limits_{m\neq n}^{N_\mathrm{mol}} T^{(2)}_{jk}  \left[ \vec{r}^{nm}(\bar{t}) \right]\mu^{n,\mathrm{VIS}}_k(\bar{t})  
 \right] \\
+  \tilde{\alpha}^{n,\mathrm{DD,0}}_{ij}(\bar{t}) \left[  \sum\limits_{m\neq n}^{N_\mathrm{mol}} \left( T^{(2)}_{jk}\left[ \vec{r}^{nm}(\bar{t}) \right]\mu^{n,\mathrm{SFG}}_k(\bar{t})
- T^{(3)}_{jkl} \left[ \vec{r}^{nm}(\bar{t}) \right] Q^{n,\mathrm{SFG}}_{kl}(\bar{t}) 
  \right) \right] \, .
\label{eq:SCF_mu_after_approx}
\end{multline}
Equivalently, we obtain the SFG component of the electric quadrupoles 
\begin{align}
Q^{n,\mathrm{SFG}}_{ij} (\bar{t}) = \tilde{\alpha}^{n,\mathrm{QD,IR}}_{ijk} (\bar{t}) \left[ \mathcal{F}^\mathrm{VIS}_k
 + \sum\limits_{m\neq n}^{N_\mathrm{mol} }T^{(2)}_{kl}  \left[ \vec{r}^{nm}(\bar{t}) \right]\mu^{n,\mathrm{VIS}}_l(\bar{t})  
 \right] \, ,
  \label{eq:SCF_Q_after_approx}
\end{align}
where we repeat the aforementioned approximations. Both equations depend only on the slow time-scale $\bar{t}$, but not on the regular time-scale $t$. 
Most importantly, the SFG components of the multipoles depend only on the VIS component, and no other frequency is involved. 
Therefore, in the time-scale separation, the time-dependent SCF equation is transformed into two coupled SCF equations, one at frequency $\omega^\mathrm{VIS}$ and one at frequency $\omega^\mathrm{SFG}$, 
which agrees with the typical formulation of nonlinear optics \cite{armstrongInteractionsLightWaves1962,shenFundamentalsSumFrequencySpectroscopy2016a,moritaTheorySumFrequency2018a}.
We compute the STFT of Equation \eqref{eq:scf_mu_1_sfg} at frequency $\omega^\mathrm{VIS}$
\begin{align}    
    \mu_i^{n,\mathrm{VIS}}(\bar{t}) =  \alpha^{n,\mathrm{DD,0}}_{ij} (\bar{t}) \left[ \mathcal{F}^\mathrm{VIS}_j + \sum\limits_{m\neq n}^{N_\mathrm{mol}}T^{(2)}_{jk}  \left[ \vec{r}^{nm}(\bar{t}) \right]  \mu_k^{n,\mathrm{VIS}}(\bar{t}) \right]  \, ,
    \label{eq:scf_VIS}
\end{align}
where we applied the same approximations.
Equation \eqref{eq:scf_VIS} can be solved for every timestep $\bar{t}$, which is given by 
\begin{align}
    \mu_i^{n,\mathrm{VIS}}(\bar{t}) = \alpha^{n,\mathrm{DD},0}_{ij} (\bar{t}) \bar{f}^n_{jk} (\bar{t}) \mathcal{F}^\mathrm{VIS}_k \, ,
    \label{eq:def_time_aver_loc_field}
\end{align}
where 
$\bar{f}^n_{ij} (\bar{t})$ is the time-averaged variant of the the local field factor $f^{n,\mathrm{D}}_{ij} (t)$, defined in Equation \eqref{eq:def_local_field_factor}, but with the rolling mean of the polarizabilities $\alpha^{n,\mathrm{DD},0}_{ij}(\bar{t})=\tilde{\alpha}^{n,\mathrm{DD}}_{ij}(\bar{t},0)$, instead of the fully resolved polarizabilities $\alpha^{n,\mathrm{DD}}_{ij}(t)$. 
Consequently, we can define the source multipoles oscillating at frequency $\omega^\mathrm{SFG}$ as
\begin{align}
\mu^{n,\mathrm{DS,SFG}}_i\left(  \bar{t} \right) &=  \alpha^{n,\mathrm{DD},\mathrm{IR}}_{ij}( \bar{t}) \bar{f}_{jk}^n \left( \bar{t}\right) \mathcal{F}_k^{\mathrm{VIS}} 
 \label{eq:source_dip}\\
     Q^{n,\mathrm{S,SFG}}_{ij}\left(  \bar{t} \right)   &= \alpha^{n,\mathrm{QD},\mathrm{IR}}_{ijk}( \bar{t}) \bar{f}_{kl}^{n}\left( \bar{t}\right) \mathcal{F}_l^{\mathrm{VIS}}  \, .
\end{align}
We introduce the induced molecular electric dipoles from the electric quadrupole sources according to
\begin{align}
\mu^{n,\mathrm{QS,SFG}}_i\left(  \bar{t} \right) = -\alpha^{n,\mathrm{DD},0}_{ij}(\bar{t})  \sum\limits_{m\neq n}^{N_{\mathrm{mol}}} T^{(3)}_{jkl}\left[ \vec r^{nm} \left(\bar{t}  \right) \right] Q^{m,\mathrm{S,SFG}}_{kl}(\bar{t})  \, . 
   \label{eq:source_quad}
 \end{align}
Hence, we can rewrite Equation \eqref{eq:SCF_mu_after_approx} as
\begin{align}
 \mu^{n,\mathrm{SFG}}_{i}\left(  \bar{t} \right) &= \alpha^{ n,\mathrm{DD},0 }_{i j}(\bar{t})  \sum\limits_{m\neq n}^{N_\mathrm{mol}} T_{jk}^{(2)}\left[ \vec r^{nm} \left(\bar{t}  \right) \right]  \mu^{m,\mathrm{SFG}}_{k}\left(  \bar{t}\right)  + \mu^{n,\mathrm{DS,SFG}}_i\left(  \bar{t} \right) + \mu^{n,\mathrm{QS,SFG}}_i\left(  \bar{t} \right)
    \label{eq:scf_dip_sfg_const_F} \, .
\end{align} 
Equation \eqref{eq:scf_dip_sfg_const_F} can be formally solved analogously to the solution of the dipolar SCF Equation given in the appendix of the work by Armstrong, Bloembergen, Ducuing and Pershan \cite{armstrongInteractionsLightWaves1962} and the book from Morita \cite{moritaTheorySumFrequency2018a}.
We
introduce the local field factor $f^{nm}_{ij}$, which relates the local field that acts on the $n^\mathrm{th}$ molecule to the external field that acts on the $m^\mathrm{th}$ molecule 
\begin{align}
E^n_i=\sum\limits_{m}^{N_\mathrm{mol}}f^{nm}_{ij} \mathcal{F}^m_j \, .
    \label{eq:def_fnm}
\end{align}
We can write the formal solution of Equation \eqref{eq:scf_dip_sfg_const_F} as follows
\begin{align}
     \mu^{n,\mathrm{SFG}}_{i}(\bar{t}) &=  \sum\limits_m^{N_\mathrm{mol} }\left[\bar{f}_{ji}^{mn}  (\bar{t})  \mu^{m,\mathrm{DS,SFG}}_{j}(\bar{t}) + \bar{f}_{ji}^{mn}  (\bar{t}) \mu^{m,\mathrm{QS,SFG}}_{j} (\bar{t}) \right] \, ,
    \label{eq:inverted_dip_scf_sfg_fnm}
\end{align}
where we again use the overbar to indicate that we use the time-averaged polarizabilities $\alpha_{ij}^{n,\mathrm{DD},0}(\bar{t})$ in the computation of $\bar{f}_{ij}^{nm}(\bar t)$.
Now, we insert the expressions for the source multipoles from Equations \eqref{eq:source_dip} and \eqref{eq:source_quad} into Equation \eqref{eq:inverted_dip_scf_sfg_fnm}, leading to
\begin{align}
     \mu^{n,\mathrm{SFG}}_{i}(\bar{t}) &=  \sum\limits_m^{N_\mathrm{mol}}  \left( \bar{f}_{ki}^{mn}  (\bar{t})  \alpha^{m,\mathrm{DD},\mathrm{IR}}_{kl}( \bar{t}) \bar{f}_{lj}^m \left( \bar{t}\right) \mathcal{F}_j^{\mathrm{VIS}}  - \bar{f}_{ki}^{mn}  (\bar{t}) \alpha^{m,\mathrm{DD},0}_{ko}(\bar{t})  \sum\limits_{l\neq m}^{N_{\mathrm{mol}}} T^{(3)}_{opq}\left[ \vec r^{ml} \left(\bar{t}  \right) \right] \alpha^{l,\mathrm{QD},\mathrm{IR}}_{pqr}( \bar{t}) \bar{f}_{rj}^l \left( \bar{t}\right) \mathcal{F}_j^{\mathrm{VIS}} \right) \, .
    \label{eq:inverted_dip_scf_sfg_fnm_2}
\end{align}
Consequently, we can approximate the effective polarizability profiles for which we give exact expressions in Equations \eqref{eq:adij_mun}, \eqref{eq:aqijk_Qn},  \eqref{eq:addij_mun} and \eqref{eq:adqij_mun} as
\begin{align} 
    a^\mathrm{DD}_{ij}\left[ z,\vec \Omega \right] &= \frac{1}{L_x L_y} \sum\limits_n^{N_\mathrm{mol}}\alpha^{n,\mathrm{DD}}_{ik}\left[ \vec \Omega \right] f^{n,\mathrm{D}}_{kj}\left[ \vec \Omega \right] \delta \left( z-z^n\left[ \vec \Omega \right] \right)  \\ 
    &\approx \frac{1}{L_x L_y} \sum\limits_{n}^{N_\mathrm{mol}}   \delta \left( z-z^n\left[ \vec \Omega \right] \right)\sum\limits_{m}^{N_\mathrm{mol}}  \bar{f}_{ki}^{mn}  \left[ \vec{\Omega} \right] \alpha^{m,\mathrm{DD}}_{kl}\left[ \vec{\Omega} \right] \bar{f}_{lj}^m \left( \vec \Omega \right)  
    \label{eq:adij_mun_approx_1} \\
a^\mathrm{DQ}_{ij}\left[ z,\vec \Omega \right]  &=   \frac{1}{L_x L_y} \sum\limits_n^{N_\mathrm{mol}}\alpha^{n,\mathrm{DD}}_{ij}\left[ \vec \Omega \right] \left(f^n_{jk}\left[ \vec \Omega \right] - f^{n,\mathrm{D}}_{jk}\left[ \vec \Omega \right]  \right) \delta \left( z-z^n\left[ \vec \Omega \right] \right)  \\ 
    &\approx  -\frac{1}{L_x L_y} \sum\limits_{n}^{N_\mathrm{mol}}  \delta \left( z-z^n\left[ \vec \Omega \right] \right)\sum\limits_{m}^{N_\mathrm{mol}}  \bar{f}_{ki}^{mn}  \left( \vec \Omega \right) \alpha^{m,\mathrm{DD},0}_{ko}\left( \vec \Omega \right)  \sum\limits_{l\neq m}^{N_{\mathrm{mol}}} T^{(3)}_{opq}\left[ \vec r^{ml} \left( \vec \Omega \right) \right] \alpha^{l,\mathrm{QD}}_{pqr}\left( \vec \Omega \right) \bar{f}_{rj}^l \left( \vec \Omega \right) 
\label{eq:adij_mun_approx_2} \\
    a^\mathrm{Q}_{ijk}\left[ z,\vec \Omega \right]  &= \frac{1}{L_x L_y} \sum\limits_n^{N_\mathrm{mol}} \delta \left( z-z^n\left[ \vec \Omega \right] \right)\alpha^{n,\mathrm{QD}}_{ijl}\left[ \vec \Omega \right] f^n_{lk}\left[ \vec \Omega \right]  \\
    &\approx \frac{1}{L_x L_y} \sum\limits_n^{N_\mathrm{mol}} \delta \left( z-z^n\left[ \vec \Omega \right] \right)\alpha^{n,\mathrm{QD}}_{ijl}\left[ \vec \Omega \right] \bar{f}^n_{lk}\left[ \vec \Omega \right] \, .
    \label{eq:aqijk_approx}
\end{align}
Finally, we assume that the distance between an electric dipole source at position $z^n$ and the induced equifrequent electric dipole at position $z^m$ is not too large, which means that we replace $ \delta(z-z^n)\rightarrow\delta(z-z^m)$.
This approximation holds exactly in the Lorentz-field picture, where electric dipole and electric quadrupole densities at a given $z$-position do not act on molecules at other $z$-positions, as can be seen in Equations \eqref{eq:Lorentz_consti_xy}-\eqref{eq:Lorentz_consti_z}. 
Errors introduced by this approximation do not affect the SFG spectrum $\tilde{S}^{(2)}_{ijk}(\omega^\mathrm{VIS},\omega^\mathrm{IR})$ itself, but lead to inaccuracies in the corresponding profile $\tilde{s}^{(2)}_{ijk}(z,\omega^\mathrm{VIS},\omega^\mathrm{IR})$. 
This is because, when integrated, it makes no difference whether we write $\delta(z-z^n)$ or $\delta(z-z^m)$. 
The resulting final expressions are 
\begin{align} 
    a^\mathrm{DD}_{ij}\left[ z,\vec \Omega \right] & \approx \frac{1}{L_x L_y}  \sum\limits_{m}^{N_\mathrm{mol}}  \bar{f}_{ki}^{m} \left[ \vec{\Omega} \right] \alpha^{m,\mathrm{DD}}_{kl}\left[ \vec{\Omega} \right] \bar{f}_{lj}^m \left( \vec \Omega \right)  \delta \left( z-z^m\left[ \vec \Omega \right] \right)
    \label{eq:addij_mun_approx_final} \\
  a^\mathrm{DQ}_{ij}\left[ z,\vec \Omega \right] 
    &\approx -\frac{1}{L_x L_y}  \sum\limits_{m}^{N_\mathrm{mol}}  \delta \left( z-z^m\left[ \vec \Omega \right] \right)   \bar{f}_{ki}^{m}  \left( \vec \Omega \right) \alpha^{m,\mathrm{DD},0}_{ko}\left( \vec \Omega \right)  \sum\limits_{l\neq m}^{N_{\mathrm{mol}}} T^{(3)}_{opq}\left[ \vec r^{ml} \left( \vec \Omega \right) \right] \alpha^{l,\mathrm{QD}}_{pqr}\left( \vec \Omega \right) \bar{f}_{rj}^l \left( \vec \Omega \right) \label{eq:adij_mun_approx_final} \\
    a^\mathrm{Q}_{ijk}\left[ z,\vec \Omega \right]  &\approx \frac{1}{L_x L_y} \sum\limits_n^{N_\mathrm{mol}} \alpha^{n,\mathrm{QD}}_{ijl}\left[ \vec \Omega \right] \bar{f}^n_{lk}\left[ \vec \Omega \right] \delta \left( z-z^n\left[ \vec \Omega \right] \right) \, .
    \label{eq:aqijk_approx_final}
\end{align}
These align well with the formalism commonly employed in nonlinear optics \cite{armstrongInteractionsLightWaves1962,hiranoLocalFieldEffects2024}.
However, several approximations were required to arrive at this description, and as a result, the framework used to predict SFG spectra should be applied with caution.
In the main text, only the results obtained using the more precise Equations
\eqref{eq:adij_mun}, \eqref{eq:aqijk_Qn}, \eqref{eq:addij_mun} and \eqref{eq:adqij_mun} are presented.
However, the advantage of the approximate Equations \eqref{eq:addij_mun_approx_final}-\eqref{eq:aqijk_approx_final} over the more precise form
is that we can introduce approximations for the time-filtered local field factors like the Lorentz-field approximation
\begin{align}
    \bar{f}_{ij}(t) = \delta_{ij} \frac{\varepsilon +2 }{3 - \delta_{iz}(3-3\varepsilon)} \, ,
\end{align}
which saves computation and programming time. On the other hand, the time-dependent local field factors $f^n_{ij}(t)$ need to be extracted from each simulation frame separately, as they include both effects due to the linear response to the second-order source, and the equifrequent modification of the external VIS field. 
We use only the precise equations for the data presented in the main text. 
However, we test the approximate equations in the next section.
We can extract the slower time-scale of the system's response to a second-order source by
removing the IR component of the local field factors via the application of a low-pass filter on the trajectories of the molecular polarizabilities, i.e., 
\begin{align}a_{ij}^{n,\mathrm{DD},0} (t) &=\frac{1}{2 \pi} \int\limits_{-\infty}^\infty \mathrm{d}\omega \, e^{-i \omega t} \Pi\left(\frac{\omega}{2 \omega^\mathrm{CUT}}\right) \tilde{a}_{ij}^{n,\mathrm{DD}} (\omega) \, ,
\label{eq:bar_aij}
\end{align}
where 
\begin{align}
    \Pi(x)=\begin{cases}
    1 & \text{if } |x| < 0.5 \\
    0.5& \text{if } |x| = 0.5 \\
    0& \text{if }  |x| > 0.5 
\end{cases}
\label{eq:rect_func}
\end{align}
is the rectangular function. 
We select $\omega^\mathrm{CUT}/2 \pi=\SI{6.0}{THz}$, ensuring that it remains less than the square root of the smallest frequency of interest. This frequency cutoff is indeed somewhat arbitrary, and other values for $\omega^\mathrm{CUT}$ within the same order of magnitude are conceivable, as long as they are smaller than the square root of the lowest frequency of interest and higher than the dynamics dominating the trajectory of the dielectric response to second-order source multipoles.
Furthermore, other choices of window functions are possible. However, within the time-scale considered, the clean assignment of Fourier components of $\tilde{a}_{ij}^{n,\mathrm{DD}}(\omega)$ below a threshold to the linear response remains the most natural choice.
\subsubsection{Comparison of Equations used for SFG Spectra Prediction}
\begin{figure}
    \centering
\includegraphics[width=1\linewidth]{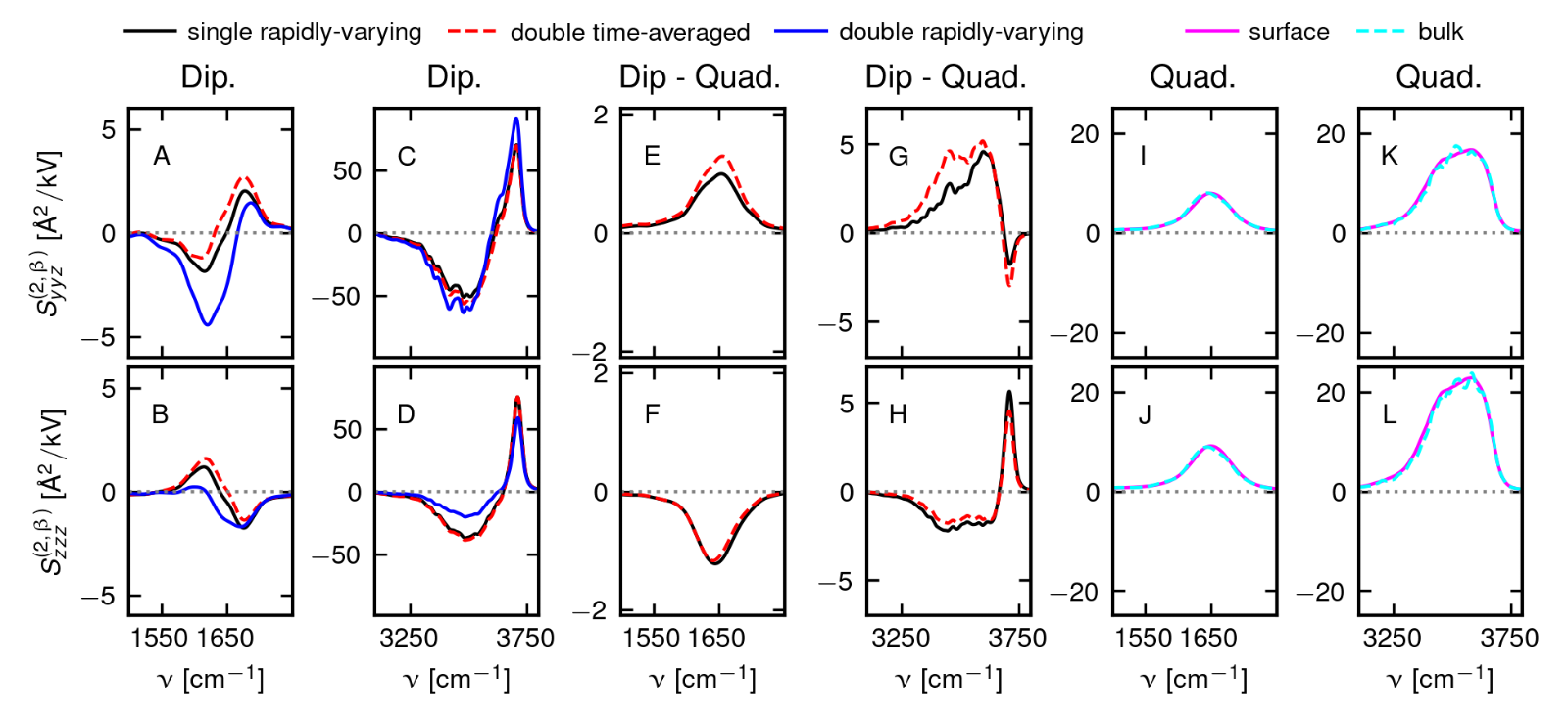}
    \caption{Comparison between different methods to extract multipolar polarization contributions, defined in Equations \eqref{eq:S2ijk_beta} and \eqref{eq:S2ijk_DD_DQ}, from fluctuation-dissipation relations. The pure dipole component $\tilde{S}^{(2,\mathrm{DD})}_{ijk} (\omega^\mathrm{IR})$ is shown in A-D. 
    Here, we compare the three different Equations \eqref{eq:comp_add_exact},\eqref{eq:comp_add_approx} \eqref{eq:comp_add_not_equal} for the effective polarizabilities. 
    These equations are 
    i) the accurate prediction without time-scale separation, involving a single rapidly-varying local field factor, given in Equation \eqref{eq:comp_add_exact} (black), 
    ii) the approximate prediction, involving two time-averaged local field factors given in Equation \eqref{eq:comp_add_approx} (red), 
    iii) The same formula as (ii), but with two rapidly-varying local field factors, as defined in Equation \eqref{eq:comp_add_not_equal} (blue). 
    Clearly, the red and black lines show good agreement, whereas the blue line deviates noticeably. This indicates that the time-scale separation approximation performs well at the air–water interface, while the use of Equation~\eqref{eq:comp_add_not_equal} leads to artifacts.
    In E-H, we compare the electric dipole - electric quadrupole cross contribution $\tilde{S}^{(2,\mathrm{DQ})}_{ijk}(\omega^\mathrm{IR})$ predicted by the fluctuation-dissipation relations  \eqref{eq:S2DQ_fdt} and \eqref{eq:S2DQ_fdt_approx}. 
    There, the line colors have the same meaning as in A-D.  Again, we see that the time-scale separation approximation performs well.
    In I-L we compare the extraction of the electric quadrupole contribution $\tilde{S}^{(2,\mathrm{Q})}_{ijk}(\omega^\mathrm{IR})$ from the air-water interface as defined by the accurate Equation \eqref{eq:S2Q_fdt} and from bulk water as defined in Equation \eqref{eq:S2Q_bulk} using the approximate fluctuation-dissipation relation in Equation \eqref{eq:chi2Q_fdt_corr_func}. Here, we see that the approximations leading to Equation \eqref{eq:chi2Q_fdt_corr_func} are not introducing significant deviation from the exact treatment.}
    \label{fig:big_multipole}
\end{figure}
We introduced the fluctuation-dissipation relations to predict multipolar SFG spectra in Equations \eqref{eq:s2D_fdt}, \eqref{eq:s2Q_fdt} and \eqref{eq:s2DD_fdt}. 
These equations involve the first-order perturbation expansion of the polarizability profiles $a^{\mathrm{D}}_{ij}(z,t)$, $a^{\mathrm{Q}}_{ij}(z,t)$, $a^{\mathrm{DD}}_{ij}(z,t)$ and $a^{\mathrm{DQ}}_{ij}(z,t)$ as defined in Equation \eqref{eq:adij_mun}, \eqref{eq:aqijk_Qn}, \eqref{eq:addij_mun} and \eqref{eq:adqij_mun},  respectively. 
In all relations, a single, rapidly-varying local field factor $f^n_{ij}(t)$, which relates the local field acting on the $n^\mathrm{th}$ molecule to the external VIS field, appears.
It is shown in Section \ref{sec:rigourous_two_field_facs} that whenever a time-scale separation is feasible, one can approximate the effective polarizability profiles with an expression involving two time-averaged local field factors $\bar{f}^n_{ij}(t)$.  A similar expression with two rapidly-varying local field factors was used in the literature for SFG spectra prediction \cite{moritaRecentProgressTheoretical2008,ishiyamaComputationalAnalysisVibrational2017,moritaTheorySumFrequency2018a}.
Here, we compare the three formulations for the example of the pure electric dipole contribution $\tilde{S}^{(2,\mathrm{DD})}_{ijk}(\omega^\mathrm{VIS})$ defined in Equation \eqref{eq:S2ijk_DD_DQ}, obtained by integrating the fluctuation-dissipation relation \eqref{eq:s2DD_fdt} from the middle of the simulated water slab at $z=0$ to infinity
\begin{align}    
    S^{(2,\mathrm{DD})}_{ijk} (t)&= -\frac{\Theta(t)}{L_x L_y k_B T \varepsilon_0} \frac{\partial}{\partial t}\left\langle \sum\limits_n^{N_\mathrm{mol}}\alpha^{n,\mathrm{DD}}_{il}(t)f^{n,\mathrm{D}}_{lj} (t)\Theta \left[ z^n\left( t \right) \right] P_k(0) \right\rangle \label{eq:comp_add_exact}  \\ 
        S^{(2,\mathrm{DD})}_{ijk} (t)&= -\frac{\Theta(t)}{L_x L_y k_B T \varepsilon_0} \frac{\partial}{\partial t}\left\langle \sum\limits_n^{N_\mathrm{mol}}\bar{f}_{li}^{n} (t) \alpha^{n,\mathrm{DD}}_{lm} (t)\bar{f}_{mj}^n (t) \Theta \left[ z^n\left( t \right) \right] P_k(0) \right\rangle 
\label{eq:comp_add_approx} \\
 S^{(2,\mathrm{DD})}_{ijk} (t) &= -\frac{\Theta(t)}{L_x L_y k_B T \varepsilon_0} \frac{\partial}{\partial t}\left\langle \sum\limits_n^{N_\mathrm{mol}}f^{n,\mathrm{D}}_{li} (t) \alpha^{n,\mathrm{DD}}_{lm} (t) f^{n,\mathrm{D}}_{mj} (t) \Theta \left[ z^n\left( t \right) \right] P_k(0) \right\rangle  \, ,
    \label{eq:comp_add_not_equal}
\end{align}
where we ignore off-resonant contributions, meaning that we set $\langle b^\mathrm{DD}_{ijk}(z) \rangle =0$. 
These equations assume the absence of periodic boundary conditions along the
$z$-dimension, corresponding to the presence of an infinite vacuum in that direction. 
However, since the elongation of the simulation box along $z$ is finite in our case, the periodic boundary correction introduced in Section \ref{app:pbc} is applied to each local field factor, as well as to the correlation function itself.
Strictly speaking, none of the three equations is used in their exact form in the existing literature, except for Equation~\eqref{eq:comp_add_exact}, which is employed in our earlier works \cite{fellowsHowThickAirWater2024,fellowsSumFrequencyGenerationSpectroscopy2024,fellowsImportanceLayerDependentMolecular2025b}. 
This is due to differences in the treatment of the boundary (as discussed in Section~\ref{sec:TheTreatmendBoundary}) and to the application of cutoffs to intramolecular correlations \cite{nagataVibrationalSumFrequencyGeneration2010,hiranoBoundaryEffectsQuadrupole2022b}.
However, here we want to compare the effect of using different expressions for the effective polarizabilities $\alpha^{n,\mathrm{DD}}_{ik}(t)f^{n,\mathrm{D}}_{kj} (t)$, $\bar{f}_{ki}^{n} (t) \alpha^{n,\mathrm{DD}}_{kl} (t)\bar{f}_{lj}^n (t)$ and $f^{n,\mathrm{D}}_{ki} (t) \alpha^{n,\mathrm{DD}}_{kl} (t) f^{n,\mathrm{D}}_{lj} (t)$, appearing in Equations \eqref{eq:comp_add_exact}-\eqref{eq:comp_add_not_equal}, and assess the validity of the time-scale separation. 
Hence, we modify only the effective polarizability while keeping the remaining parts unchanged.
Equation \eqref{eq:comp_add_exact} is derived from the first-order expansion of the effective polarizability $\alpha^{n,\mathrm{DD}}_{ik}(t) f^{n,\mathrm{D}}_{kl}(t)$ in Section \ref{sec:nonlocal_first_and_second_order_pert_exp_pol_dens}. 
On the other hand, Equation \eqref{eq:comp_add_approx} follows from a time-scale separation between the fast oscillating molecular polarizability and the slowly varying dielectric response as described in Section \ref{sec:rigourous_two_field_facs}.
Equation \eqref{eq:comp_add_not_equal}, previously used in the literature \cite{moritaRecentProgressTheoretical2008, ishiyamaComputationalAnalysisVibrational2017, moritaTheorySumFrequency2018a}, is compared here with the exact Equation \eqref{eq:comp_add_exact} and the approximate Equation \eqref{eq:comp_add_approx} in Figure \ref{fig:big_multipole} A–D.
We see that the prediction of $\tilde{S}^{(2,\mathrm{DD})}_{ijk}(\omega^\mathrm{IR})$ with the approximative Equation \eqref{eq:comp_add_approx} closely matches the exact result given by Equation \eqref{eq:comp_add_exact}, verifying that the time-scale separation is feasible.
But, Equation \eqref{eq:comp_add_not_equal} predicts significantly different spectra. 
We conclude that the approximate Equation~\eqref{eq:comp_add_approx} can be employed, whereas the application of the theoretically unjustified Equation~\eqref{eq:comp_add_not_equal} should be avoided.
Now we discuss the electric dipole - electric quadrupole cross-contribution $\tilde{S}^{(2,\mathrm{DQ})}_{ijk}(\omega^\mathrm{IR})$, defined in Equation \eqref{eq:S2ijk_DD_DQ}.
Here, the approximate treatment can provide insight into the underlying mechanism.
In the exact formulation presented in Section \ref{sec:nonlocal_first_and_second_order_pert_exp_pol_dens}, this contribution accounts for the electric dipoles induced by neighbouring electric quadrupoles and is determined by the fluctuation-dissipation relation as follows from Equations \eqref{eq:S2ijk_DD_DQ}, \eqref{eq:s2DD_fdt}, \eqref{eq:s2Q_fdt} and \eqref{eq:adqij_mun}
\begin{align}
    S^{(2,\mathrm{DQ})}_{ijk} (t)&= -\frac{\Theta(t)}{L_x L_y k_B T \varepsilon_0} \frac{\partial}{\partial t}\left\langle \sum\limits_n^{N_\mathrm{mol}}\alpha^{n,\mathrm{DD}}_{il}(t) \left[ f^{n}_{lj} (t)  -f^{n,\mathrm{D}}_{lj} (t) \right]\Theta \left[ z^n\left( t \right) \right] P_k(0) \right\rangle
\, .    \label{eq:S2DQ_fdt}
\end{align}
In the approximate treatment introduced in Section \ref{sec:rigourous_two_field_facs}, this contribution is related to the linear response of the electric dipole density to an electric quadrupole source density oscillating at frequency $\omega^\mathrm{SFG}$, with the fluctuation-dissipation relation
\begin{align}
    S^{(2,\mathrm{DQ})}_{ijk} (t)&= \frac{\Theta(t)}{L_x L_y k_B T \varepsilon_0} \frac{\partial}{\partial t}\left\langle 
    \sum\limits_{n}^{N_\mathrm{mol}}  \Theta \left[ z^n(t) \right]  \bar{f}_{li}^{n}  \left(t\right) \alpha^{n,\mathrm{DD},0}_{lo}\left( t \right)  \sum\limits_{m\neq n}^{N_{\mathrm{mol}}} T^{(3)}_{opq}\left[ \vec r^{nm} \left( t \right) \right] \alpha^{m,\mathrm{QD}}_{pqr}\left( t \right) \bar{f}_{rj}^m \left( t \right)  P_k(0) \right\rangle \, ,
    \label{eq:S2DQ_fdt_approx}
\end{align}
as follows from Equations \eqref{eq:S2ijk_DD_DQ}, \eqref{eq:s2DD_fdt},  \eqref{eq:s2Q_fdt} and \eqref{eq:adij_mun_approx_final}. Again, we neglected the off-resonant part, as it is of no importance in this work.
We observe that the approximations made in Section \ref{sec:rigourous_two_field_facs} are quite good, as indicated by the small differences between the electric quadrupole - electric dipole cross contribution $\tilde{S}^{(2,\mathrm{DQ})}_{ijk}(\omega^\mathrm{IR})$ predicted with Equations \eqref{eq:S2DQ_fdt} and \eqref{eq:S2DQ_fdt_approx}, as visible in Figure \ref{fig:big_multipole} E-H.
The last non-zero polarization contribution is the electric quadrupole contribution $\tilde{S}^{(2,\mathrm{Q})}_{ijk}(\omega^\mathrm{IR})$, defined in Equation \eqref{eq:S2ijk_beta}, given by the integral from $z=z_0$ to infinity of $\tilde{s}^{(2,\mathrm{Q})}_{ijk}(z,\omega^\mathrm{IR})$, which is determined by the fluctuation-dissipation relation in Equation \eqref{eq:s2Q_fdt}
\begin{align}
    S^{(2,\mathrm{Q})}_{ijk} (t)&= -\frac{\Theta(t)}{L_x L_y k_B T \varepsilon_0} \frac{\partial}{\partial t}\left\langle \sum\limits_n^{N_\mathrm{mol}}\alpha^{n,\mathrm{QD}}_{izl}(t)  f^{n}_{lj} (t) \delta \left[ z_0 - z^n\left( t \right) \right] P_k(0) \right\rangle \, .
    \label{eq:S2Q_fdt}
\end{align}
Here $z_0$ is a $z$-position in the bulk region and thus the electric quadrupole contribution $\tilde{S}^{(2,\mathrm{Q})}_{ijk} (\omega^\mathrm{IR})$ does not depend on it.
We average
$\tilde{S}^{(2,\mathrm{Q})}_{ijk} (\omega^\mathrm{IR})$ over the region $-\SI{2.5}{\angstrom}< z_0 < \SI{2.5}{\angstrom
}$.
However, $\tilde{S}^{(2,\mathrm{Q})}_{ijk}(\omega^\mathrm{IR})$ can also be predicted from a simulation of a bulk system using Equation \eqref{eq:S2Q_bulk}.
Here, we introduce a linear-response equation for the second-order electric quadrupole susceptibility extracted in bulk. 
As the system is homogeneous, we can employ the Lorentz-field approximation \eqref{eq:LorentzField}, i.e.  $\bar{f}_{ij}^{n}(t)=\delta_{ij}\frac{\tilde{\varepsilon}+2}{3}$, which leads to the following formula for the extraction of the second-order quadrupole susceptibility from bulk media
\begin{align}
     \tilde{\chi}^{(2, \mathrm{Q})}_{ijkl}(\omega^\mathrm{IR}) =\frac{\tilde{\varepsilon}^\mathrm{VIS}+2}{3 \varepsilon_0 V} \sum\limits_n^{N_{mol}}   \tilde \varphi\left[  \alpha^{ n,\mathrm{QD} }_{ijk}(\cdot ), P_l( \cdot ), \omega^\mathrm{IR} \right] +\varepsilon_0^{-1}
     \left\langle b_{ijkl}^{\mathrm{Q}}(\cdot) \right\rangle 
     \label{eq:chi2q_fdt}
\end{align}
where $V$ is the volume.
Setting the off-resonant contribution $\varepsilon_0^{-1}
     \left\langle b_{ijkl}^{\mathrm{Q}}(\cdot) \right\rangle$ to zero, Equation \eqref{eq:chi2q_fdt} becomes equivalent to the expression
\begin{align}
    \chi^{(2,\mathrm{Q})}_{ijkl} (t)&= -\frac{\Theta(t)}{V k_B T \varepsilon_0} \frac{\tilde{\varepsilon}^\mathrm{VIS}+2}{3} \frac{\partial}{\partial t} \left\langle \sum\limits_n^{N_\mathrm{mol}}\alpha^{n,\mathrm{QD}}_{ijk}(t) P_l(0) \right\rangle \, .
    \label{eq:chi2Q_fdt_corr_func}
\end{align}
In Figure \ref{fig:big_multipole} I–L, we compare $\tilde{S}^{(2,\mathrm{Q})}_{ijk}(\omega^\mathrm{IR})$ predicted from simulations of a water slab and bulk water using Equations \eqref{eq:S2Q_fdt} and \eqref{eq:chi2Q_fdt_corr_func}, respectively. 
Note that for the prediction from the bulk system we conditionally divide $\chi^{(2,\mathrm{Q})}_{ijkl}(\omega^\mathrm{IR})$ by the dielectric constant in order to receive $\tilde{S}^{(2,\mathrm{Q})}_{ijk}(\omega^\mathrm{IR})$, as dictated by Equation \eqref{eq:S2Q_bulk}.
Both predictions overlap almost perfectly, stressing that $\tilde{S}^{(2,\mathrm{Q})}_{ijk}(\omega^\mathrm{IR})$ is independent of the structure of the interface, that the timescale separation works well, and that the Lorentz field approximation can be applied in bulk.
\subsubsection{Linear Nonlocal Response Function}
\label{sec:lin_nonloc_resp_func}
Here, we rewrite the solution of the SCF Equation \eqref{eq:inverted_dip_scf_sfg_fnm} as a non-local instantaneous equifrequent response function to external fields $\tilde{s}^{ \mathrm{NL} }_{ij}(\vec r, \vec r',t)$. 
We do this to provide a translation of the SCF equation into the language of optics, utilizing electric fields that appear in constitutive relations. 
Consequently, we need to apply the time-scale separation introduced in Section \ref{sec:rigourous_two_field_facs}, as the concept of an equifrequent response to a nonlinear source exists only on this level of approximation.
We assume that the charge-density distribution is represented by point multipoles defined in Equations \eqref{eq:rho_q}-\eqref{eq:rho_Q}. 
Consequently, the external fields acting on the $m^\mathrm{th}$-molecule, imposed by the electric dipole and quadrupole densities $\varrho^{\mathrm{DS}}_i(\vec r ,t )$ $\varrho^{\mathrm{QS}}_{ij}(\vec r ,t )$ oscillating at frequency $\omega^\mathrm{SFG}$, which are defined in Equations \eqref{eq:def_rho_DS} and \eqref{eq:def_rho_QS}, are given by
\begin{align}
 F_i^\mathrm{DS}(\vec r, t) &= \int_{\sigma(\vec r)} \mathrm{d} \vec r' T^{(2)}_{ij} (\vec r - \vec r') \varrho^{\mathrm{DS}}_j(\vec r',t) \approx F_i^\mathrm{DS,SFG}(\vec r, t) e^{-i\omega^\mathrm{SFG} t} \\
 F_i^\mathrm{DS,SFG}(\vec r, t) &= \int_{\sigma(\vec r)} \mathrm{d} \vec r' T^{(2)}_{ij} (\vec r - \vec r') \sum\limits_n^{N_\mathrm{mol}} \mu^{n,\mathrm{DS,SFG}}_j(t) \delta[\vec r'- \vec r^n(t)]
 \label{eq;def_FDS} \\
 F_i^\mathrm{QS}(\vec r, t) &= -\int_{\sigma(\vec r)} \mathrm{d} \vec r' T^{(3)}_{ijk} (\vec r - \vec r') \varrho^{\mathrm{QS}}_{jk}(\vec r',t) \approx F_i^\mathrm{QS,SFG}(\vec r, t) e^{-i\omega^\mathrm{SFG} t} \\
  F_i^\mathrm{QS,SFG}(\vec r, t) &= - \int_{\sigma(\vec r)} \mathrm{d} \vec r' T^{(3)}_{ijk} (\vec r - \vec r') \sum\limits_n^{N_\mathrm{mol}} Q^{n,\mathrm{S,SFG}}_{jk}(t) \delta[\vec r'-\vec r^n(t)] \, .
 \label{eq:def_FQS}
\end{align}
We integrate over the entire volume except for a small sphere centered at 
$\vec r$ to exclude any contribution from a multipole at $\vec r$ acting on itself.
Here, $F_i^\mathrm{DS,SFG}(\vec r, t)$ and $F_i^\mathrm{QS,SFG}(\vec r, t)$ are the slowly varying amplitudes of the external fields imposed by the electric source dipoles and quadrupoles, respectively. 
We use the symbol $F$ as opposed to $E$, for $F_i^\mathrm{DS,SFG}(\vec r, t)$ and $F_i^\mathrm{QS,SFG}(\vec r, t)$, since these fields are external to the linearly induced electric dipoles, which will be introduced in following. 
We leave out the slow time-scale $\bar{t}$ introduced in Section \ref{sec:rigourous_two_field_facs} and simply assume that all amplitudes are sufficiently slowly varying, as we already convinced ourselfs that the time-scale separation works well in Figure \ref{fig:big_multipole}.
We want to dissect the electric dipoles oscillating at frequency $\omega^\mathrm{SFG}$ into the contribution due to the instantaneous linear response to the electric quadrupole density $\varrho^{ (2,\mathrm{DQ} )}_{i}(z ,t )$ and the remaining pure electric dipole contributions $\varrho^{ (2,\mathrm{DD} )}_{i}(z ,t )$. 
Therefore, we take the SCF Equation \eqref{eq:scf_dip_sfg_const_F} and
subtract the source electric dipoles $\mu_i^{n,\mathrm{DS,SFG}}(t)$, leading to the SCF equation for the linear induced dipoles $ \mu^{n,\mathrm{L},\mathrm{SFG}}_{i}\left(  t \right)=  \mu^{n,\mathrm{SFG}}_{i}\left(  t \right) -  \mu^{n,\mathrm{DS},\mathrm{SFG}}_{i}\left(  t \right)$
\begin{align}
\mu^{n,\mathrm{L},\mathrm{SFG}}_{i}\left(  t \right) &= \alpha^{ n,\mathrm{DD},0 }_{i j}(t)  \sum\limits_{m\neq n}^{N_\mathrm{mol}} T_{jk}^{(2)}\left[ \vec r^{nm} \left(t  \right) \right]  \mu^{m,\mathrm{L,SFG}}_{k}\left(  t\right) +  \alpha^{ n,\mathrm{DD},0 }_{i j}(t)  \sum\limits_{m\neq n}^{N_\mathrm{mol}} \left( T_{jk}^{(2)}\left[ \vec r^{nm} \left(t  \right) \right]  \mu^{m,\mathrm{DS},\mathrm{SFG}}_{k}\left(  t\right) -  T_{jkl}^{(3)}\left[ \vec r^{nm} \left(t  \right) \right]  Q^{m,\mathrm{S},\mathrm{SFG}}_{kl}\left(  t\right)  \right)  \, .
\label{eq:SCF_eq_ext_source}
\end{align}
We replace the right-hand side by the external field due to the multipolar sources $F_i^\mathrm{DS,SFG}\left[\vec r^n(t), t\right]+F_i^\mathrm{QS,SFG}\left[\vec r^n(t), t\right]$, leading to 
\begin{align}
\mu^{n,\mathrm{L},\mathrm{SFG}}_{i}\left(  t \right) &= \alpha^{ n,\mathrm{DD},0 }_{i j}(t)  \sum\limits_{m\neq n}^{N_\mathrm{mol}} T_{jk}^{(2)}\left[ \vec r^{nm} \left(t  \right) \right]  \mu^{m,\mathrm{L,SFG}}_{k}\left(  t\right) +  \alpha^{ n,\mathrm{DD},0 }_{i j}(t) F_j^\mathrm{DS,SFG}\left[\vec r^n(t), t\right] +  \alpha^{ n,\mathrm{DD},0 }_{i j}(t) F_j^\mathrm{QS,SFG}\left[\vec r^n(t), t\right] \, .
\end{align}
This equation is formally solved by introduction of the local field factors $\bar{f}^{nm}_{ij} (t)$ defined in Equation \eqref{eq:def_fnm}
\begin{align}
\mu_i^{n,\mathrm{L,SFG}}(t) = \alpha^{ n,\mathrm{DD},0 }_{i j}(t) \sum\limits_{m}^{N_\mathrm{mol}}\bar{f}^{nm}_{jk} (\bar t) \left( F_k^\mathrm{DS,SFG}\left[\vec r^m(t), t \right]+ F_k^\mathrm{QS,SFG}\left[\vec r^m(t), t\right] \right) \, .
\end{align}
We rewrite this expression as an integral over space
\begin{align*}
\mu_i^{n,\mathrm{L,SFG}}(t)  = \int\mathrm{d} \vec r'  \alpha^{ n,\mathrm{DD},0 }_{i k}(t) \left[ F_j^\mathrm{DS,SFG}(\vec r',t) +  F_j^\mathrm{QS,SFG}(\vec r',t) \right] \sum\limits_{m}^{N_\mathrm{mol}} \bar{f}^{nm}_{kj}(t) \delta[\vec r'- \vec r^m(t) ]   \, .
\end{align*}
Now we compute the linear dipole density oscillating at frequency $\omega^\mathrm{SFG}$ determined by
\begin{align*}
\varrho^\mathrm{LD}_i(\vec r, t)  &= \int\mathrm{d} \vec r' \left[ F_j^\mathrm{DS,SFG}(\vec r',t) +  F_j^\mathrm{QS,SFG}(\vec r',t) \right]  \sum\limits_{n}^{N_{\mathrm{mol}}} \sum\limits_{m}^{N_{\mathrm{mol}}} \alpha^{ n,\mathrm{DD},0 }_{i k}(t) \bar{f}^{nm}_{kj}(t)  \delta[ \vec r-r^n(t) ]  \delta[\vec r'- \vec r^m(t) ]   \, \\
&= \varepsilon_0 \int\mathrm{d} \vec r' \tilde{s}^{\mathrm{NL}}_{ij}(\vec r, \vec r', t)  \left[ F_j^\mathrm{DS,SFG}(\vec r',t) +  F^\mathrm{QS,SFG}_j(\vec r',t) \right]  \, ,
\end{align*}
where
\begin{align}
\tilde{s}^{\mathrm{NL}}_{ij}(\vec r, \vec r', t)  =    \varepsilon_0^{-1} \sum\limits_{n}^{N_{\mathrm{mol}}} \sum\limits_{m}^{N_{\mathrm{mol}}} \alpha^{ n,\mathrm{DD},0 }_{i k}(t) \bar{f}^{nm}_{kj}(t)  \delta[ \vec r-\vec r^n(t) ]  \delta[\vec r'- \vec r^m(t) ]  
\end{align}
is the instantaneous, linear and nonlocal dipolar response. 
Finally, we decompose the second-order electric dipole density $\varrho^{(2,\mathrm{DD})}_i (z,t ) $ into the pure dipole contribution 
 \begin{align}
 \varrho^{(2,\mathrm{DD})}_i (z,t ) = \frac{1}{L_x L_y}\int \mathrm{d} x \int \mathrm{d} y \left[ \varrho^{\mathrm{DS}}_i(\vec r ,t) + \int \mathrm{d} \vec r'\varepsilon_0 \tilde{s}^{\mathrm{NL}}_{ij}(\vec r, \vec r', t) F_j^\mathrm{DS}(\vec r',t) \right]
 \label{eq:def_rho2DD}
 \end{align}
 and similarly obtain the second-order electric dipole contributions induced by the electric quadrupole source density 
  \begin{align}
 \varrho^{(2,\mathrm{DQ})}_i (z,t ) = \frac{1}{L_x L_y}\int \mathrm{d} x \int \mathrm{d} y \int \mathrm{d} \vec r' \varepsilon_0 \tilde{s}^{\mathrm{NL}}_{ij}(\vec r, \vec r', t) F_j^\mathrm{QS}(\vec r',t)  \, .
  \label{eq:def_rho2DQ}
 \end{align}
These equations are not used in the extraction of the corresponding response functions, which are determined by the second-order pure electric dipole response $\tilde{s}^{(2,\mathrm{DD})}_{ijk}(z,\omega^\mathrm{IR})$ defined in Equation \eqref{eq:s2DD_fdt} and its difference from the full second-order electric dipole response $\tilde{s}^{(2,\mathrm{DQ})}_{ijk}(z,\omega^\mathrm{IR})=\tilde{s}^{(2,\mathrm{D})}_{ijk}(z,\omega^\mathrm{IR})-\tilde{s}^{(2,\mathrm{DD})}_{ijk}(z,\omega^\mathrm{IR})$ defined in Equation \eqref{eq:s2DQ_fdt}. However, these equations help us to understand the mechanism behind the contribution $\tilde{S}^{(2,\mathrm{DQ})}_{ijk}(\omega^\mathrm{IR})$, as
Equation \eqref{eq:def_rho2DD} excludes the second-order electric dipoles induced by the second-order electric quadrupoles while Equation \eqref{eq:def_rho2DQ} takes this contribution into account.
\subsection{Magnetic Dipole Contribution}
\subsubsection{Linear Response Equations for Interfacial Magnetic Dipole Contributions}
\begin{figure} \includegraphics[width=1\textwidth]{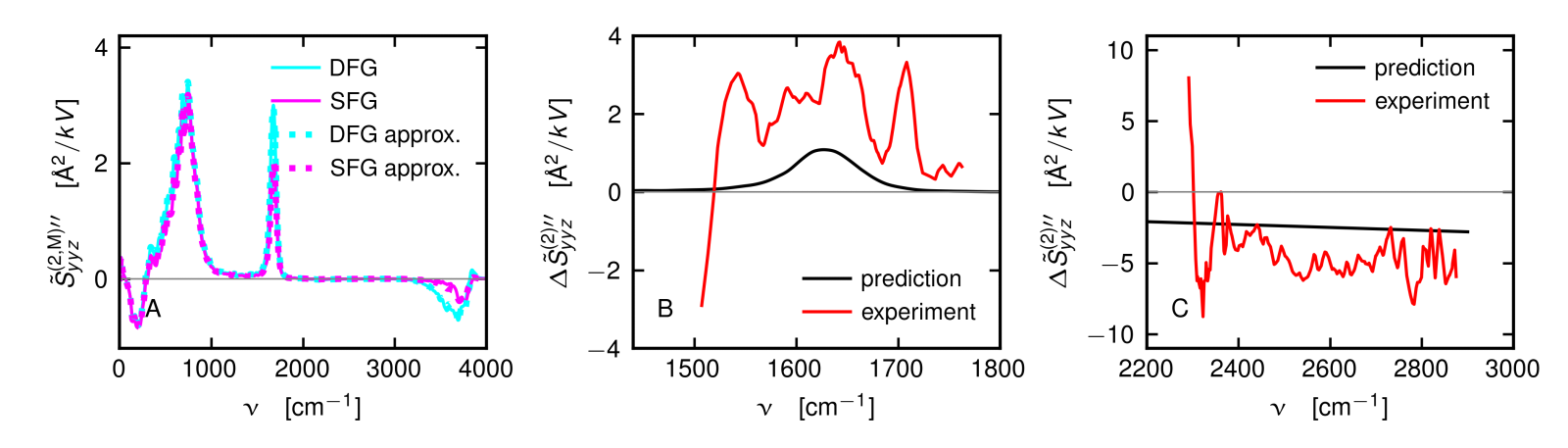}
 \caption{
 Predicted magnetic dipole contribution to the SFG and DFG signal, according to Equation \eqref{eq:S2M_yyz}, within the electric dipole approximation introduced in Equation \eqref{eq:mag_dip_leading_order}. 
 In (A), we present $\tilde{S}^{(2,\mathrm{M})}_{yyz}\left( \omega^\mathrm{VIS}, \omega^\mathrm{IR} \right)$ and $\tilde{S}^{(2,\mathrm{M})}_{yyz}\left( -\omega^\mathrm{VIS}, \omega^\mathrm{IR} \right)$ that are measurable in DFG and SFG experiments, respectively. 
 We compare two methods for computing $\tilde{\chi}^{(2,\mathrm{M})}_{ijk}(\omega^\mathrm{VIS}, \omega^\mathrm{IR})$ as given in Equations~\eqref{eq:chi2m_exact} (solid lines) and~\eqref{eq:chi2m_approx} (broken lines). In the approximate expression, labeled "DFG/SFG approx." (Equation~\eqref{eq:chi2m_approx}), we assume that oscillations of the molecular polarizability tensor do not contribute to the magnetic dipole contribution.
 In B  \&  C, we compare the predicted difference between DFG and SFG spectra defined in Equation \eqref{eq:delta_S_dfg_sfg_mag}. 
 We compare the experimental DFG-SFG difference spectra of the bending band in (B) with the theoretical prediction using Equation \eqref{eq:chi2m_exact}. 
 In (C), we compare the off-resonant response from H$_\mathrm{2}$O at the eigenfrequency of the D$_\mathrm{2}$O stretch vibrations and the prediction by Equation \eqref{eq:delta_S_dfg_sfg_mag_off_res} 
 The experimental data is published \cite{fellowsHowThickAirWater2024,fellowsImportanceLayerDependentMolecular2025b}. The theoretically predicted spectrum in (B) is red-shifted by $\SI{28}{cm^{-1}}$.}
\label{fig:mag}
\end{figure}
We defined the contribution of the magnetic dipole density to the second-order electric current density in Equation \eqref{eq:def_mag_flux}.
Here, we derive a relation of the second-order magnetic dipole density profile $m_i^{(2)}(z,t)$ to external electric fields
\begin{align}
   m_i^{(2)}(z,t) = e^{-i \omega^\mathrm{SFG} t} \tilde{g}^{(2,\mathrm{M})}_{ijk}\left( z, \omega^\mathrm{VIS},\omega^\mathrm{IR} \right) \mathcal{F}^\mathrm{VIS}_j \mathcal{F}^\mathrm{IR}_k + c.c.,
   \label{eq:def_g2m_ijk}
\end{align}
where  $\tilde{g}^{(2,\mathrm{M})}_{ijk}\left( z, \omega^\mathrm{VIS},\omega^\mathrm{IR} \right)$ is the second-order response function of the magnetic dipole density to spatially constant external electric fields. 
The molecular observable of interest is the effective molecular magnetic dipole moment
\begin{align}
    m^n_i(t) = \frac{\epsilon_{ijk}}{2} \iiint \mathrm{d} V  r_j j^{n}_k(\vec r, t) + \epsilon_{ijk} \mu^{n}_j(t) \dot{R}^{n}_k \, ,
    \label{eq:def_mag_dio}
\end{align}
 where $j^n_i(\vec r, t)$ is the electric current density of the charge density of the $n^\mathrm{th}$ molecule relative to the molecular origin $R_i^n(t)$ and $\mu^{n}_i(t)$ is the molecular dipole moment \cite{russakoffDerivationMacroscopicMaxwell1970,jacksonClassicalElectrodynamicsInternational2021} .
 The first term of Equation \eqref{eq:def_mag_dio} is the definition of the molecular magnetic moment in the stationary frame, and the second term needs to be considered whenever the movements of the molecular origin at the frequency of interest cannot be neglected \cite{jacksonClassicalElectrodynamicsInternational2021}.
As the nuclei do not oscillate with the frequency $\omega^\mathrm{VIS}$ and the core electrons can be assumed to be non-polarizable as they are tightly bound to the nuclei, only the valence electrons can contribute meaningfully to oscillations of the electric current density of frequency $\omega^\mathrm{SFG}$.
We assume that the electric current density of the valence electrons within the $n^\mathrm{th}$ molecule can be written as
\begin{align}
j^n_i (\vec r, t)= \varrho^n(\vec r, t) v^n_i(\vec r, t)    \label{eq:electron_current},
\end{align}
where $\varrho^n(\vec r, t)$ and $v^n_i(\vec r, t)$ are the valence electrons charge density and drift velocity relative to the molecular origin. 
Without further assumptions, the velocity field $v^n_i(\vec r, t)$ cannot be calculated within the Born-Oppenheimer approximation. 
We perform a Taylor expansion of the velocity field in $\vec r$, i.e.
\begin{align}
    v^n_i(\vec r , t) &=   v^{0,n}_i( t) + r_a  v^{a,n}_i( t) + ... \, , 
    \label{eq:taylor_drift_velocity} \\  
    v^{0,n}_i( t) &=  v^n_i(0,t) \, , \\
       v^{a,n}_i( t) &= 
       \frac{\partial}{\partial r_a} v^{n}_i(\vec r,  t)\big |_{\vec r=0} \, .
\end{align}
We insert the expansion introduced in Equation \eqref{eq:taylor_drift_velocity} into the definition of the molecular magnetic dipole moment in Equation \eqref{eq:def_mag_dio}. Subsequently, we relate the magnetic dipole to the electric multipole series defined in Equation \eqref{eq:Q_0n}-\eqref{eq:Q_IIn} as
\begin{align}
    m^n_i(t) = \frac{\epsilon_{ijk}}{2} \left[ v^{0,n}_k( t)  \mu^n_j (t) + 2 v^{a,n}_k( t) Q^n_{aj}( t) + ... \, \right] +  \epsilon_{ijk} \mu^{n}_j(t) \dot{R}^n_k \, ,
    \label{eq:mag_expansion_v}
\end{align} 
where $\mu^n_j$ and $Q^n_{aj}$ are the electric dipole and electric quadrupole moments of the valence electrons within the $n^\mathrm{th}$ molecule, respectively.
Note that here only the charge density of the valence electrons is considered, which is not charge neutral. Hence, $\mu^n_i(t)$ depends on the choice of the molecular center.
This series converges when the drift velocity does not vary too much in space and when the electric multipole expansion converges. 
We compute the effective molecular magnetic moment for the leading-order term
\begin{align}
   m^n_i(t) \approx \frac{\epsilon_{ijk}}{2}  \mu^n_j ( t) v^{0,n}_k( t) +  \epsilon_{ijk} \mu^{n}_j(t) \dot{R}^n_k\, .
    \label{eq:mag_dip_leading_order}
\end{align}
We relate the velocity $\vec v^0(t)$ to the time derivative of the dipole moment, that is,
\begin{align}
    v^{0,n}_i(t) = \frac{1}{N  q^e } \dot{ \mu}^{n}_i(t) \, ,
    \label{eq:v0_is_mu}
\end{align}
where the number of valence electrons per molecule is denoted as $N^\mathrm{e}$ (for water $N^\mathrm{e}=8$)  and ${q^\mathrm{e}\approx -\SI{1.6E-19}{C}}$ is the electron charge. 
Combining Equations \eqref{eq:mag_dip_leading_order} and \eqref{eq:v0_is_mu} leads to the expression
\begin{align}
    m_i^n(t) = \frac{\epsilon_{ijk}}{2 N^\mathrm{e} q^\mathrm{e}}  \mu^n_j(t) \dot{\mu}^n_k(t)  + \epsilon_{ijk}  \mu_j^n(t) \dot{R}^n_k (t)
    \label{eq:mag_dip_elec}
\end{align}
for the magnetic moment $m_i^n$ of the $n$-th molecule. 
The approximation in Equation \eqref{eq:mag_dip_elec} only considers the electric currents characterized by charge displacements, i.e. the transport of the charge density distribution due to applied external fields.
In the main text, it is demonstrated that the contribution in Equation \eqref{eq:mag_dip_elec} contributes significantly to the SFG signal in the bending region. 
We will demonstrate here that it predicts the difference between experimental SFG and DFG spectra of water rather well. 
Most importantly, it ensures the independence of the SFG signal from the choice of the molecular center, as demonstrated in Section \ref{sec:origin_sfg}. 
We showed in Section \ref{sec:TimeDependentPerturbationTheory} that we can retrieve the SFG component of the second-order multipoles by considering the created complex second-order current when we apply two external fields $\mathcal{F}_j^\mathrm{VIS} e^{-i \omega^\mathrm{VIS} t}$ and $\mathcal{F}_k^\mathrm{IR} e^{-i \omega^\mathrm{IR} t}$. We consider the complex valence electron current in the presence of the two just mentioned complex external fields, namely
\begin{align}
    \mu^n_i(t) = \mu^{0,n}_i(t) + a^n_{ij}(t)   \left( e^{-i \omega^\mathrm{VIS} t} \mathcal{F}^\mathrm{VIS}_j  +  e^{-i \omega^\mathrm{IR} t} \mathcal{F}^\mathrm{IR}_j\right) \, ,
    \label{eq:ele_n}
\end{align}
where $a^n_{ij} \approx \mathrm{d} \mu^n_{i} / \mathrm{d} \mathcal{F}^\mathrm{VIS}_j$ is the effective polarizability of the $n^\mathrm{th}$ molecule and does not depend on the choice of the molecular origin.  We insert Equation \eqref{eq:ele_n} into Equation \eqref{eq:mag_dip_elec} and add the complex conjugate afterwards, leading to
\begin{multline}
     m^n_i(t) = \frac{\epsilon_{ijk}}{2 N^\mathrm{e} q^\mathrm{e}} \bigg[  \mu^{0,n}_j(t) \dot{a}^n_{kl}(t) - \dot{\mu}^{0,n}_j(t) a^n_{kl}(t) 
     - i \omega^\mathrm{VIS}  \mu^{0,n}_j(t) a^n_{kl}(t) + 2 N^\mathrm{e} q^\mathrm{e} \alpha^n_{jl}(t) \dot{R}_k^n(t) 
     \\  + \left[a^n_{jm}(t) \dot{a}^n_{kl}(t) - \dot{a}^n_{jm}(t) a^n_{kl}(t) 
     -i \omega^\mathrm{DFG} a^n_{jm}(t) a^n_{kl}(t) \right] e^{-i \omega^\mathrm{IR} t} \mathcal{F}^\mathrm{IR}_m \bigg] e^{-i\omega^\mathrm{VIS} t} \mathcal{F}^\mathrm{VIS}_l + c.c. + ... \, ,
     \label{eq:mag_dip_elec_2}
\end{multline}
where we do not write out contributions not oscillating with the sum frequency $\omega^\mathrm{SFG}$ and use the shorthand notation ${\omega^\mathrm{DFG}=\omega^\mathrm{VIS}-\omega^\mathrm{IR}}$. Furthermore, we assert that the molecular center position $R^n_i(t)$ is only a function of the nuclei coordinates and consequently does not oscillate at optical frequencies.
We consider the magnetic dipole moment of the $n^\mathrm{th}$ molecule $\check{m}^n_i(t)$ with a different molecular origin $\check R_i^n(t) \neq R_i^n(t)$.
The difference between $\check{m}^n_i(t)$ and $m^n_i(t)$ is determined by
\begin{align}
    \Delta \check{m}^n_i(t) &= \check{m}^n_i(t) - m^n_i(t) \\
    &= -\frac{\epsilon_{ijk}}{2} \frac{\partial}{\partial t} \left[ \check{R}^n_j(t)  - R^n_j(t) \right] a^n_{kl}(t)  \mathcal{F}_l e^{-i\omega^\mathrm{VIS} t} + c.c. + ... \, ,
    \label{eq:delta_m_orign}
\end{align}
where we write all terms oscillating with frequency $\omega^\mathrm{SFG}$.
This term guarantees the origin independence of the SFG signal, as long as quadrupole contributions are properly accounted for.
We introduce the effective molecular magnetic dipole polarizabilities
\begin{align}    
h^n_{il}\left(\omega^\mathrm{VIS}, \vec \Omega\right) &= \frac{\epsilon_{ijk}}{2 N^\mathrm{e} q^\mathrm{e}} \bigg[   \mu^{0,n}_j(\vec \Omega) \dot{a}^n_{kl}(\vec \Omega) - \dot{\mu}^{0,n}_j(\vec \Omega) a^n_{kl}(\vec \Omega)   - i \omega^\mathrm{VIS} \mu^{0,n}_j(\vec \Omega) a^n_{kl}(\vec \Omega) + 2 N^\mathrm{e} q^\mathrm{e} \alpha^n_{jl}(t) \dot{R}_k^n(t) \bigg] \\
    l^n_{ilm}\left( \omega^\mathrm{VIS},\omega^\mathrm{IR},\vec \Omega \right)  &=  -\frac{\epsilon_{ijk}i  \omega^\mathrm{DFG} }{2 N^\mathrm{e} q^\mathrm{e}}    a^n_{jm}(\vec \Omega)  a^n_{kl}(\vec \Omega)    \, ,
\end{align}
where $h^n_{il}(\vec \Omega)$ is the external IR field-driven effective molecular magnetic dipole polarizability The off-resonant instantaneous magnetic dipole hyperpolarizability is $l^n_{ilm}(\vec \Omega)$. 
We discard the contribution to $l^n_{ilm}(\vec \Omega)$ from  $a^n_{jm}(\vec \Omega) \dot{a}^n_{kl}(\vec \Omega) - \dot{a}^n_{jm}(\vec \Omega) a^n_{kl}(\vec \Omega)$, because its expectation value is zero due to the time reversibility property in equilibrium.
We introduce the effective magnetic dipole polarizability profile
\begin{align}
   h_{ij}\left(z,\omega^\mathrm{VIS}, \vec \Omega\right) = \frac{1}{L_x L_y} \sum\limits_n^{N_\mathrm{mol}} \delta\left[ z^n(\vec \Omega) - z \right] h^n_{ij}\left(\omega^\mathrm{VIS}, \vec \Omega\right)\, 
\end{align}
and the effective magnetic dipole hyperpolarizability profile 
\begin{align}
   l_{ijk}(z,\omega^\mathrm{VIS},\omega^\mathrm{IR},\vec \Omega) =  \frac{1}{L_x L_y} \sum\limits_n^{N_\mathrm{mol}}  \delta\left[ z^n(\vec \Omega) - z \right] l^n_{ijk}\left(\omega^\mathrm{VIS}, \omega^\mathrm{IR}, \vec \Omega\right) \, .
\end{align}
We expand $h_{ij}(z,\omega^\mathrm{VIS})$ to first order in the IR field, which leads to the definition of the second-order magnetic dipole response profile
\begin{align}
    \tilde{g}_{ijk}^{(2,\mathrm{M})}(z,\omega^\mathrm{VIS}, \omega^\mathrm{IR} ) = \tilde{\varphi}\left[ h_{ij}\left(z,\omega^\mathrm{VIS}, \cdot \right) ,P_k(\cdot ) ,  \omega^\mathrm{IR} \right]+ \langle l_{ijk}\left(z,\omega^\mathrm{VIS},\omega^\mathrm{IR} \right) \rangle \, .
\end{align}
Consequently, the magnetic dipole contribution to the second-order response profile $\tilde{s}^{(2)}_{ijk} \left( z ,\omega^\mathrm{VIS},\omega^\mathrm{IR} \right)$ is equal to
\begin{align}
\tilde{s}^{(2,\mathrm{M})}_{ijk} \left( z ,\omega^\mathrm{VIS},\omega^\mathrm{IR} \right) = \varepsilon_0^{-1} \frac{\epsilon_{izl}}{ - i \omega^\mathrm{SFG} }  \frac{\partial}{\partial z} \tilde{g}_{ljk}^{(2,\mathrm{M})}(z,\omega^\mathrm{VIS}, \omega^\mathrm{IR} )  \, .
    \label{eq:s2m_ijk_fdt}
\end{align}
In fact, within our length scale separation, the interface layer is substantially smaller than the wavelength, and the magnetic dipole contribution is entirely determined by the second-order magnetic dipole susceptibility in the isotropic bulk medium as given in Equation \eqref{eq:S2M_bulk}. 

\subsubsection{Magnetic Dipole Moment in an Isotropic Bulk Medium}
\begin{figure} \includegraphics[width=0.3\textwidth]{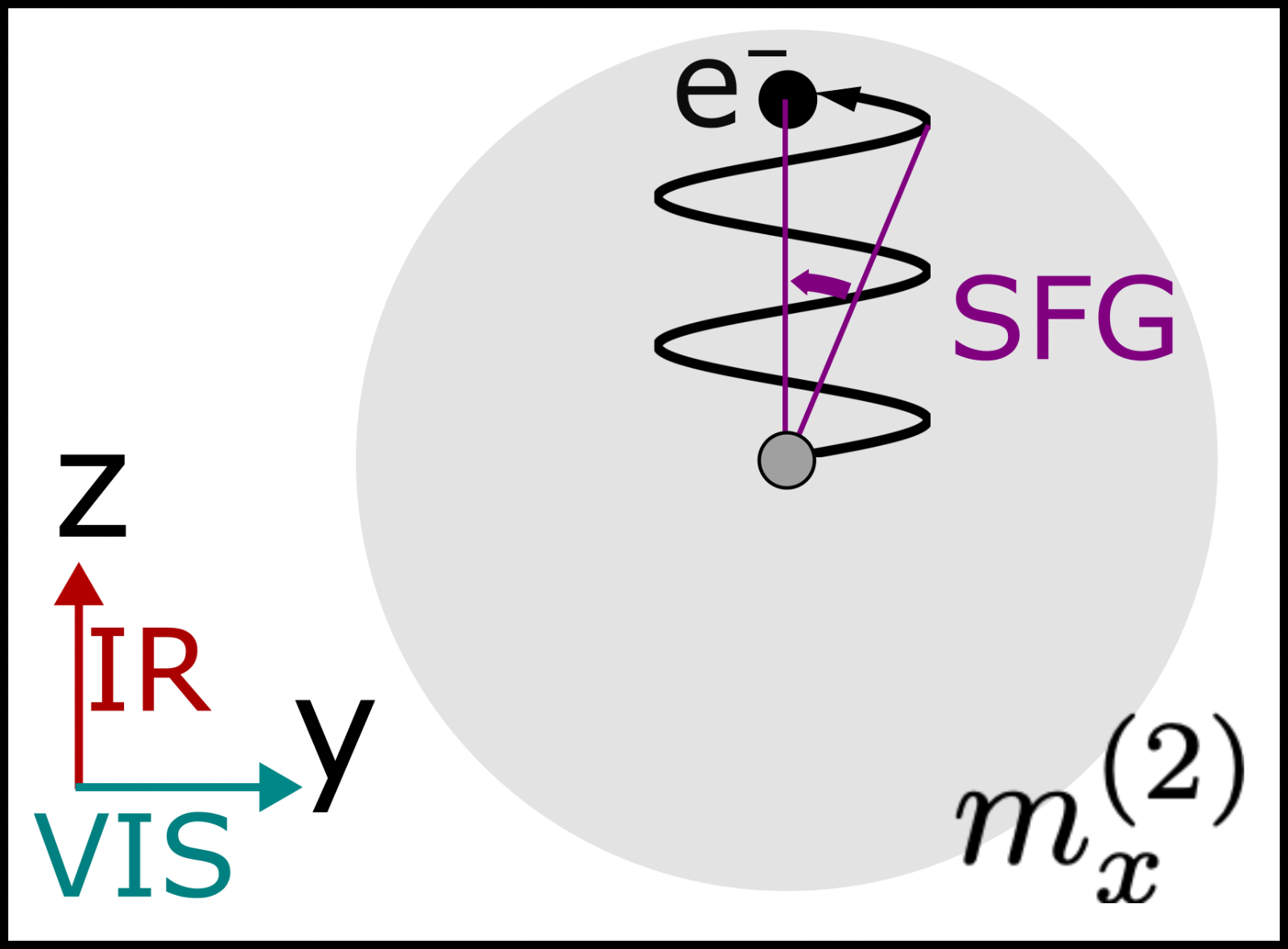}
 \caption{Mechanism of the nonlinear magnetic dipole contribution.
Valence electrons are periodically displaced along the $z$-axis by the IR field (red arrow) and simultaneously along the $y$-axis by the VIS field (teal arrow). The resulting trajectory of the electron is sketched by the black line.
This motion generates with frequency  $\omega^\mathrm{SFG}$ oscillating angular momentum (illustrated by the purple arrow).}
\label{fig:mag_mechanism}
\end{figure}

In a bulk medium, we are interested in the total magnetic dipole moment, i.e. the sum of all molecular magnetic dipole moments $m_i^n(t)$
\begin{align}
    M_i(t)=\sum\limits_n^{N_\mathrm{mol}} m_i^n(t) \,.
\end{align}
Here, we derive an equation predicting the second-order magnetic dipole susceptibility $\tilde{\chi}^{(2,\mathrm{M})}_{ijk} \left( \omega^\mathrm{VIS}, \omega^\mathrm{IR} \right)$, defined in Equation \eqref{eq:chi2M}.
In an isotropic medium  $\tilde{\chi}^{(2,\mathrm{M})}_{ijk} \left( \omega^\mathrm{VIS}, \omega^\mathrm{IR} \right)=\epsilon_{ijk} \tilde{\chi}^{(2,\mathrm{M})}_{xyz} \left( \omega^\mathrm{VIS}, \omega^\mathrm{IR} \right)$ holds \cite{pershanNonlinearOpticalProperties1963}. Hence, it is sufficient to consider only $M_x(t)$, induced by $F_i^\mathrm{VIS}(t)=\delta_{iy} \mathcal{F}^\mathrm{VIS}_y e^{-i\omega^\mathrm{VIS} t }$ and $F_i^\mathrm{IR}(t)=\delta_{iz} \mathcal{F}^\mathrm{IR}_z e^{-i\omega^\mathrm{IR} t }$, which is given by
\begin{align}    
    M_x(t)=\sum\limits_n^{N_{mol}} m^n_x(t)=H_{xy}\left( \omega^\mathrm{VIS}, t \right) e^{-i \omega^\mathrm{VIS} t} \mathcal{F}_y^\mathrm{VIS} + L_{xyz} \left( \omega^\mathrm{VIS}, \omega^\mathrm{IR} \right)e^{-i \omega^\mathrm{SFG} t} \mathcal{F}_y^\mathrm{VIS} \mathcal{F}_z^\mathrm{IR}  +c.c. \, ,
\end{align}
where $H_{xy}\left(\omega^\mathrm{VIS},t \right)$ is the effective magnetic dipole polarizability driven by the IR field and $L_{xyz}\left(\omega^\mathrm{VIS},\omega^\mathrm{IR} \right)$ is the effective magnetic dipole hyperpolarizability. These can be defined as
\begin{align}    
\begin{split}
    H_{xy}\left(\omega^\mathrm{VIS}, \vec \Omega\right) =& \frac{1}{2 N^\mathrm{e} q^\mathrm{e}} \sum\limits_n^{N_\mathrm{mol}} \bigg[  \mu^{0,n}_y(\vec \Omega) \dot{a}^n_{zy}(\vec \Omega) - \mu^{0,n}_z(\vec \Omega) \dot{a}^n_{yy}(\vec \Omega) - \dot{\mu}^{0,n}_y(\vec \Omega) a^n_{zy}(\vec \Omega) + 2 N^\mathrm{e} q^\mathrm{e} \left[ a^n_{yy}(t) \dot{R}_z^n(t)  -  a^n_{zy}(t) \dot{R}_y^n(t) \right] \, ,  \\ & + \dot{\mu}^{0,n}_z(\vec \Omega) a^n_{yy}(\vec \Omega) - i \omega^\mathrm{VIS} \left[ \mu^{0,n}_y(\vec \Omega) a^n_{zy}(\vec \Omega)   - \mu^{0,n}_z(\vec \Omega) a^n_{yy}(\vec \Omega) \right] \bigg]  
 \end{split} \\
L_{xyz}\left(\omega^\mathrm{VIS},\omega^\mathrm{IR}, \vec \Omega\right) =&  \frac{i \omega^\mathrm{DFG} }{2 N^\mathrm{e} q^\mathrm{e}} \sum\limits_n^{N_\mathrm{mol}} \left[ a^n_{yy}(\vec\Omega)  a^n_{zz}(\vec\Omega) - a^n_{yz}(\vec\Omega)  a^n_{zy}(\vec\Omega)  \right]  \, .
     \label{eq:H_xy_exact}
\end{align}
These expressions can be simplified by neglecting molecular polarizability fluctuations, that is, by setting $a^n_{ij}(t) \approx \delta_{ij} a_{\mathrm{iso}}$, where $a_{\mathrm{iso}}$ is the time average of the isotropic component of the effective polarizability tensor, which can be related to the isotropic component of the molecular electric dipole - electric dipole polarizability $\alpha_{\mathrm{iso}}$, via
\begin{align}
    a^n_{\mathrm{iso}} = \frac{2 + \tilde\varepsilon^\mathrm{VIS}}{3}\alpha^n_{\mathrm{iso}} \, ,
\end{align}
as follows from the Lorentz-field approximation in Equation \eqref{eq:LorentzField}.
Furthermore, it can be assumed that the motion of the molecular centers does not contribute significantly.
Unlike the approximation in Equation \eqref{eq:mag_dip_leading_order}, these assumptions are not necessary, but they greatly simplify the expression and thus help clarify its physical origin.
We check the validity of these approximations later on. They lead to the simplified expressions 
\begin{align}
    H_{xy}(\omega^\mathrm{VIS},\vec \Omega) &\approx  \frac{i \omega^\mathrm{DFG} }{2 N^\mathrm{e} q^\mathrm{e}}  \sum\limits_n^{N_\mathrm{mol}} a^n_\mathrm{iso} \mu^{0,n}_z(\vec \Omega)  \, , \\     L_{xyz}\left(\omega^\mathrm{VIS},\omega^\mathrm{IR}, \vec \Omega\right) &\approx \frac{i \omega^\mathrm{DFG}  }{2 N^\mathrm{e} q^\mathrm{e}} \sum\limits_n^{N_\mathrm{mol}}a^n_{\mathrm{iso}} a^n_{\mathrm{iso}} \, .
\end{align}
Consequently, the approximative second-order magnetic dipole susceptibility reads
\begin{align}
\tilde{\chi}^{(2,\mathrm{M})}_{xyz}(\omega^\mathrm{VIS}, \omega^\mathrm{IR}) \approx  \frac{i \omega^\mathrm{DFG} } {2 V \varepsilon_0 N^\mathrm{e} q^\mathrm{e}} \sum\limits_n^{N_\mathrm{mol}} a^n_\mathrm{iso}\left(  \tilde{\varphi}\left[  \mu_z^{0,n}(\cdot), P_z(\cdot ) , \omega^\mathrm{IR}  \right] + a^n_\mathrm{iso}   \right) \, ,
\label{eq:chi2m_approx}
\end{align}
 where $V$ is the volume of the system. 
 Hence, the second-order magnetic dipole susceptibility is proportional to the response of the $z$-component of the polarization density of the valence electrons to the applied IR laser.
This polarization is displaced in the $y$-direction by the VIS laser, inducing an electric current with angular momentum oscillating at frequency $\omega^\mathrm{SFG}$.
As a result, a second-order magnetic dipole moment is generated at the molecular centers, as sketched in Figure \ref{fig:mag_mechanism}.
 If we do not neglect the fluctuations of the effective polarizability tensor and the motion of the molecular center, we obtain
\begin{align}
\tilde{\chi}^{(2,\mathrm{M})}_{xyz}(\omega^\mathrm{VIS}, \omega^\mathrm{IR}) = \frac{1 } { V \varepsilon_0  } \left(  \tilde{\varphi}\left[ H_{xy}(\omega^\mathrm{VIS}, \cdot ), P_z(\cdot ) , \omega^\mathrm{IR}  \right] + \left\langle L_{xyz}\left( \omega^\mathrm{VIS},\omega^\mathrm{IR} \right)\right\rangle \right)  \, ,
    \label{eq:chi2m_exact}
\end{align}
Later, in Section \ref{sec:origin_sfg}, where we investigate the origin independence of the multipolar SFG spectrum we compute the second-order magnetic dipole susceptibility for a generic molecular center determined by the nuclei positions $\check{\vec R^n}$ as
\begin{align} \tilde{\check{\chi}}^{(2,\mathrm{M})}_{ijk}\left(\omega^\mathrm{VIS} , \omega^\mathrm{IR}\right) =   \tilde{\chi}^{(2,\mathrm{M})}_{ijk}\left(\omega^\mathrm{VIS} , \omega^\mathrm{IR}\right) + i \omega^\mathrm{SFG} \sum\limits_n^{N_\mathrm{mol}} \tilde{\varphi}\left[ \frac{\epsilon_{ilm}}{2} \left[ \check{R}^n_l(\cdot)  - R^n_l(\cdot) \right] a^n_{mj}(\cdot), P_k(\cdot) , \omega^\mathrm{IR}\right]\, ,
\label{eq:chi2m_origin}
\end{align}
where we used Equation \eqref{eq:delta_m_orign}, $\vec R^n$ denotes molecular center of mass and $\tilde{\chi}^{(2,\mathrm{M})}_{ijk}\left(\omega^\mathrm{VIS} , \omega^\mathrm{IR}\right)$ is the second-order magnetic dipole susceptibility with the molecular center of mass as the molecular center.
The SFG signal is related to $\chi^{(2,\mathrm{M})}_{xyz}(\omega^\mathrm{SFG}, \omega^\mathrm{VIS})$ by Equation \eqref{eq:S2M_yyz}, i.e.
\begin{align}
    \tilde{S}^{(2,\mathrm{M})}_{yyz} \left( \omega^\mathrm{VIS}, \omega^\mathrm{IR} \right) =& \frac{1}{  \left( \tilde n_1^\mathrm{IR} \right)^2 i \omega^\mathrm{SFG} } \tilde{\chi}^{(2,\mathrm{M})}_{xyz}(\omega^\mathrm{VIS}, \omega^\mathrm{IR}) \, .
     \label{eq:S_yyz_m} 
\end{align}
In the approximation used to derive Equation \eqref{eq:chi2m_approx}, the magnetic dipole contribution in SFG is proportional to $ \omega^\mathrm{DFG} / \omega^\mathrm{SFG}$ and, in general, depends on $\omega^\mathrm{VIS}$. As the DFG signal can be obtained by replacing $\omega^\mathrm{VIS} \rightarrow -\omega^\mathrm{VIS}$ the magnetic contribution in DFG using the approximation in Equation \eqref{eq:chi2m_approx} is proportional to $ \omega^\mathrm{SFG} / \omega^\mathrm{DFG}$. If we do not apply the approximation in Equation \eqref{eq:chi2m_approx}, but use Equation \eqref{eq:chi2m_exact} instead, we still have a dependence on $\omega^\mathrm{VIS}$, but the functional dependence is more complicated. In both cases, we can test the accuracy of Equations \eqref{eq:chi2m_approx} and \eqref{eq:chi2m_exact} by comparing experimental DFG and SFG difference measurements, theoretically determined by
\begin{align}
    \Delta \tilde{S}^{(2)}_{ijk}(\omega^\mathrm{VIS},\omega^\mathrm{IR}) = \tilde{S}^{(2)}_{ijk}(-\omega^\mathrm{VIS},\omega^\mathrm{IR}) - \tilde{S}^{(2)}_{ijk}(\omega^\mathrm{VIS},\omega^\mathrm{IR}) \, .
     \label{eq:delta_S_dfg_sfg_mag}
\end{align}
At frequencies where $\omega^\mathrm{IR}$ does not resonate with the system nuclei, we obtain the simple relation using Equation \eqref{eq:chi2m_approx}
\begin{align}
    \Delta \tilde{S}^{(2)}_{yyz} (\omega^\mathrm{VIS},\omega^\mathrm{IR}) \approx  \frac{N^\mathrm{mol}}{V} \frac{a_\mathrm{iso}^2}{2 \left(  \tilde n_1^\mathrm{IR} \right)^2  \varepsilon_0 N^\mathrm{e} q^\mathrm{e} } \left( \frac{\omega^\mathrm{SFG}}{\omega^\mathrm{DFG}  } - \frac{\omega^\mathrm{DFG}}{\omega^\mathrm{SFG}} \right) \, .
    \label{eq:delta_S_dfg_sfg_mag_off_res}
\end{align}
We present the predicted magnetic dipole contributions to the SFG and DFG signal in Figure \ref{fig:mag}. 
There we apply the Lorentz-field approximation in Equation \eqref{eq:LorentzField} to relate the effective molecular polarizability $a^n_{ij}(t)$ to the electric dipole - electric dipole polarizability tensor $\alpha^{n,\mathrm{DD}}_{ij}(t)$, defined in Equation \eqref{eq:polarizabilities}, i.e. $a^n_{ij}(t)=\alpha^{n,\mathrm{DD}}_{ij}(t)\frac{\tilde{\varepsilon}^\mathrm{VIS}+2}{3}$.
In Figure \ref{fig:mag} A we present $\tilde{S}^{(2,\mathrm{M})}_{yyz} (\omega^\mathrm{VIS},\omega^\mathrm{IR})$ and $\tilde{S}^{(2,\mathrm{M})}_{yyz} (-\omega^\mathrm{VIS},\omega^\mathrm{IR})$, where we compare the two ways of calculating $\tilde{\chi}^{(2,\mathrm{M})}_{xyz}(\omega^\mathrm{VIS}, \omega^\mathrm{IR})$, using the approximate Equation \eqref{eq:chi2m_approx} and Equation \eqref{eq:chi2m_exact}. 
As it is visible, both agree quite well.
We test the electric dipole approximation defined in Equation \eqref{eq:mag_dip_leading_order} by comparing with experimentally measured difference spectra in the bending frequency region \cite{fellowsImportanceLayerDependentMolecular2025b} in Figure \ref{fig:mag} B and experimentally measured difference spectra in the off-resonant frequency region \cite{fellowsHowThickAirWater2024} in Figure \ref{fig:mag} C.
As is visible, the electric dipole approximation rather accurately predicts the DFG/SFG difference spectra.
We conclude that our leading-order treatment captures the essence of magnetic dipole contributions in SFG spectroscopy. 
\subsection{The Treatment of the Boundary}
\label{sec:TheTreatmendBoundary}
\begin{figure*}
\centering
 \includegraphics[width=1\textwidth]{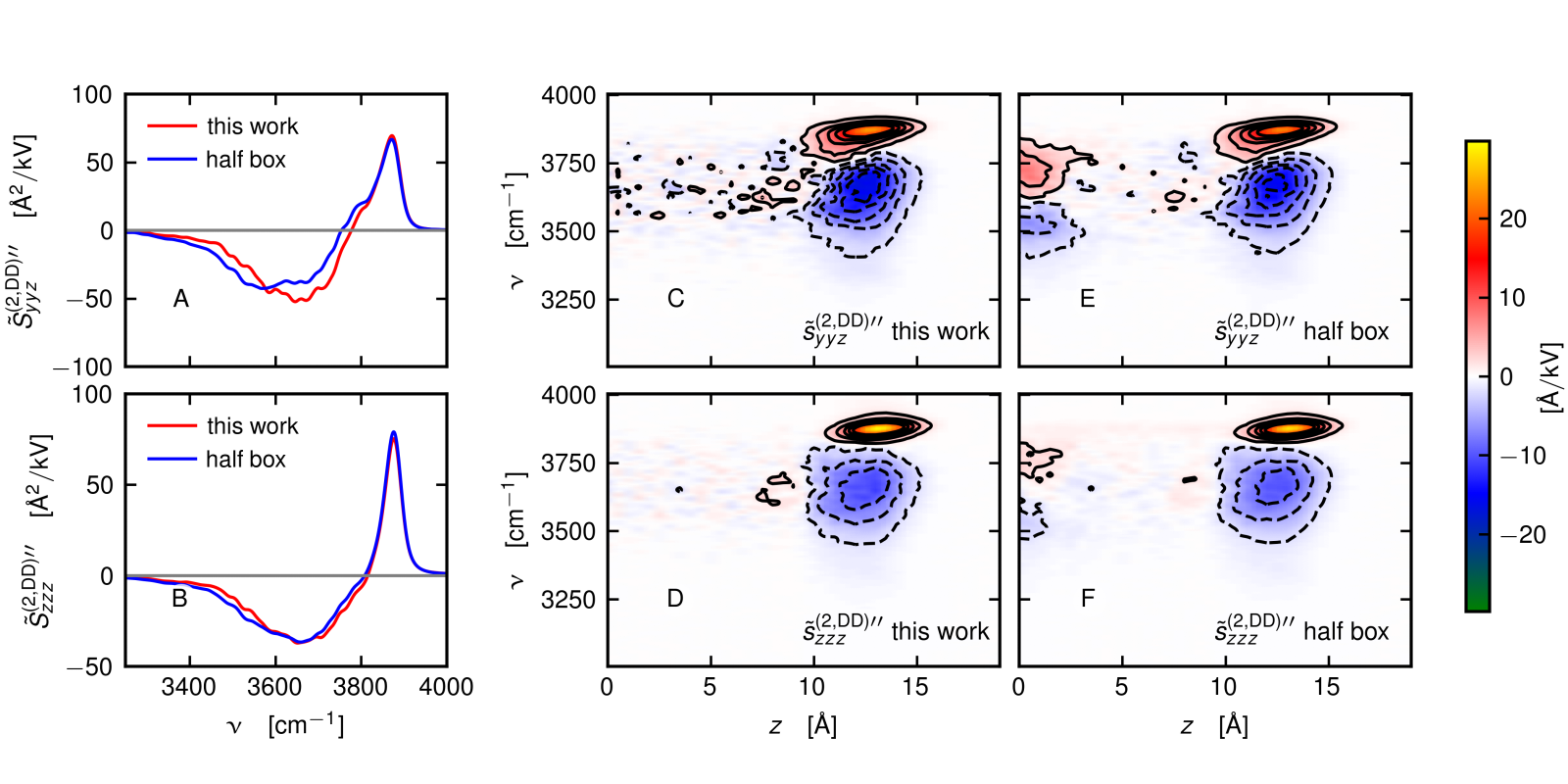}
\caption{
Comparison of the different ways to resolve the problem of the presence of two canceling interface contributions in a molecular dynamics simulation. We compare the approach proposed by Hirano and Morita\cite{hiranoBoundaryEffectsQuadrupole2022b}, where one correlates the polarizability profile with the electric dipole moment of the upper half of the box in Equation \eqref{eq:def_s2ijk+}, with our approach in Equation \eqref{eq:md_2_D}, where we correlate the polarizability profile with the full dipole moment as dictated by the perturbation Hamiltonian in Equation \eqref{eq:pert_ham}. We compare the predicted electric dipole components of the SFG signals  using both approaches in A  \&   B. We present the second-order response function profiles in C-F. We use our boundary treatment defined in Equation \eqref{eq:md_2_D} in C  \&   D and the one proposed by Hirano and Morita defined in Equation \eqref{eq:def_s2ijk+} in E  \&   F. The latter produces a spurious contribution at the boundary at $z=0$.
}
\label{fig:boundary}
\end{figure*}
We introduced the fluctuation-dissipation relations, which relate the second-order response of the electric dipole density, the electric quadrupole density and the magnetic dipole density to equilibrium correlation functions between effective polarizabilities and the total dipole moment of the simulation box $P_i(\vec \Omega)$ in Equations \eqref{eq:s2D_fdt}, \eqref{eq:s2Q_fdt} and \eqref{eq:chi2m_exact}, respectively.
Here, we explain why we compute correlation functions between the effective polarizabilities of interest and the electric dipole moment of the entire system instead of the electric dipole moment of only one half of the system, as suggested by Hirano and Morita \cite{hiranoBoundaryEffectsQuadrupole2022b}.
We consider the perturbation Hamiltonian defined in Equation \eqref{eq:pert_ham} in the absence of external field gradients
\begin{align}
    H'(\vec \Omega, t)=-F^\mathrm{IR}_i(t) P_i( \vec{\Omega} ) \, .
    \label{eq:pert_ham_classic_no_grad}
\end{align}
Thus, an arbitrary first-order observable is determined by
\begin{align}
    \tilde{O}^{(1)}(\omega^\mathrm{IR}) = \tilde{\varphi} \left[ O(\cdot) , P_i(\cdot), \omega^\mathrm{IR} \right] \tilde{F}^\mathrm{IR}_i \, .
\end{align}
We summarize the fluctuation-dissipation theorem \cite{kuboFluctuationdissipationTheorem1966}, relating $\varphi \left[ O(\cdot) , P_i(\cdot), t \right] $ and the equilibrium correlation functions $ C_{O P_i} (t) $ in Section \ref{app:fd_and_kk}. The fluctuation-dissipation theorem states
\begin{align}
\varphi[O(\cdot), P_i(\cdot), t]  = -\beta \Theta(t) \dot{ C}_{O P_i} (t) \, .
\end{align}
Hence, all introduced linear response functions are determined by equilibrium correlation functions involving the electric dipole moment of the total system. We introduce the second-order response function $\tilde s^{(2,\mathrm{MD})}_{ijk}(z, \omega^\mathrm{IR} )$ specifying the hypothetical second-order electric current density in the simulation box in the presence of two external fields as
\begin{align}
   \varepsilon_0^{-1} j^{(2, \mathrm{MD})}_{i}(z,t) = - i \omega^\mathrm{SFG} e^{-i \omega^\mathrm{VIS} t } \tilde{s}^{(2,\mathrm{MD} )}_{ijk} \left( z, \omega^\mathrm{IR} \right) \mathcal{F}^\mathrm{VIS}_j \mathcal{F}^\mathrm{IR}_k + c.c. \, .
\end{align}
We assume that our system has a single interface in this work. 
However, in our molecular dynamics simulations, two identical interfaces create inverted SFG signals as they are mirrored. Hence, the signal from the full system in the molecular dynamics simulation is zero, i.e.
\begin{align}
    \int\limits_{-\infty}^\infty \mathrm d z \, \tilde s^{(2,\mathrm{MD})}_{ijk}(z, \omega^\mathrm{IR} ) = 0 \, .
\end{align}
The problem of cancelling contributions is easily avoided by looking at the signals from the two surfaces separately. 
For large enough systems, the two interfaces are independent of each other, and the predicted signal from each interface corresponds to the signal we would expect from a system with a single interface.
Hence, the full second-order response is given by
\begin{align}
  \tilde{s}^{(2)}_{ijk}(z, \omega^\mathrm{IR} )  = \frac{\Theta(z)}{2} \big[\tilde{s}^{(2,\mathrm{MD})}_{ijk}(z, \omega^\mathrm{IR} ) 
  - \tilde{s}^{(2,\mathrm{MD})}_{ijk}(-z, \omega^\mathrm{IR} ) \big]   \, .
  \label{eq:md_2_D}
\end{align}
In contrast, Hirano and Morita \cite{hiranoBoundaryEffectsQuadrupole2022b} suggest correlating all the relevant observables with the electric dipole moment of the upper half of the system $P^+_i(t)$ instead. 
The respective pure electric dipole polarization profile is then 
\begin{equation}
\tilde s^{(2, \mathrm{DD}+)}_{ijk}(z,\omega^{\mathrm{IR}})= \varepsilon_0^{-1} \tilde{\varphi} \left[ a^{\mathrm{DD}}_{ij}(z,\cdot), P^+_k(\cdot) , \omega_{\mathrm{IR}} \right] +  \varepsilon_0^{-1} \left\langle b^{\mathrm{DD}}_{ijk}(z)  \right\rangle \, .
\label{eq:def_s2ijk+}
\end{equation}
The pure electric dipole contribution to the SFG spectrum is then determined by
\begin{align} \tilde{S}^{(2,\mathrm{DD}+)}_{ijk}\left(\omega^\mathrm{IR}\right) = \int\limits_0^\infty \mathrm{d} z \, \tilde{s}^{(2,\mathrm{DD}+)}_{ijk}\left(z,\omega^\mathrm{IR}\right) \, 
\end{align}
and is averaged over both interfaces.
Hence, $\tilde{S}^{(2,\mathrm{DD}+)}_{ijk}\left(\omega^\mathrm{IR}\right)$ involves only correlations between polarizabilities and dipole moments of molecules at $z$-positions in the upper half of the simulation box, while $\tilde{S}^{(2,\mathrm{DD})}_{ijk}\left(\omega^\mathrm{IR}\right)$ does involve correlations between the polarizabilities of molecules in the upper half of the simulation box and the dipole moments of all molecules.
In Figure \ref{fig:boundary} we compare the profile $s^{(2, \mathrm{DD}+)}_{ijk}(z,\omega^{\mathrm{IR}})$, with $s^{(2, \mathrm{DD})}_{ijk}(z,\omega^{\mathrm{IR}})$ calculated using Equations \eqref{eq:def_s2ijk+} and \eqref{eq:md_2_D}, respectively. 
We see in Figure \ref{fig:boundary} E  \&   F that $s^{(2, \mathrm{DD}+)}_{ijk}(z,\omega^{\mathrm{IR}})$ has an artificial contribution located at the boundary at $z=0$.
This artifact is absent in $s^{(2, \mathrm{DD})}_{ijk}(z,\omega^{\mathrm{IR}})$ computed according to Equation \eqref{eq:md_2_D} presented in Figure \ref{fig:boundary} C  \&   D. 
It can be understood by considering isotropically coordinated molecules sitting at $z=0^+$:
Suppose that we correlate the effective polarizabilities of these molecules with the full electric dipole moment of the system. 
In that case, we obtain no second-order response as these molecules are equally coordinated by other molecules sitting above and below $z=0$, and therefore we do not have a contribution to the SFG signal from below $\SI{8}{\angstrom}$.
However, suppose we correlate the effective polarizabilities of these molecules only with the dipole moments of the molecules above $z=0$. 
In that case, we introduce an artificial asymmetric coordination, which gives rise to the artifact presented in Figure \ref{fig:boundary} E  \&   F. 
We present the influence of this artifact on the integrated pure dipole contribution to the SFG signal defined in Equation \eqref{eq:S2ijk_beta} in Figure \ref{fig:boundary} A  \&   B. 
\section{Expansion Point Dependence of MM Contributions}
\label{sec:expandion_point_dependence}
\subsection{A Simple Model System}
\begin{figure}
    \centering
    \includegraphics[scale=1]{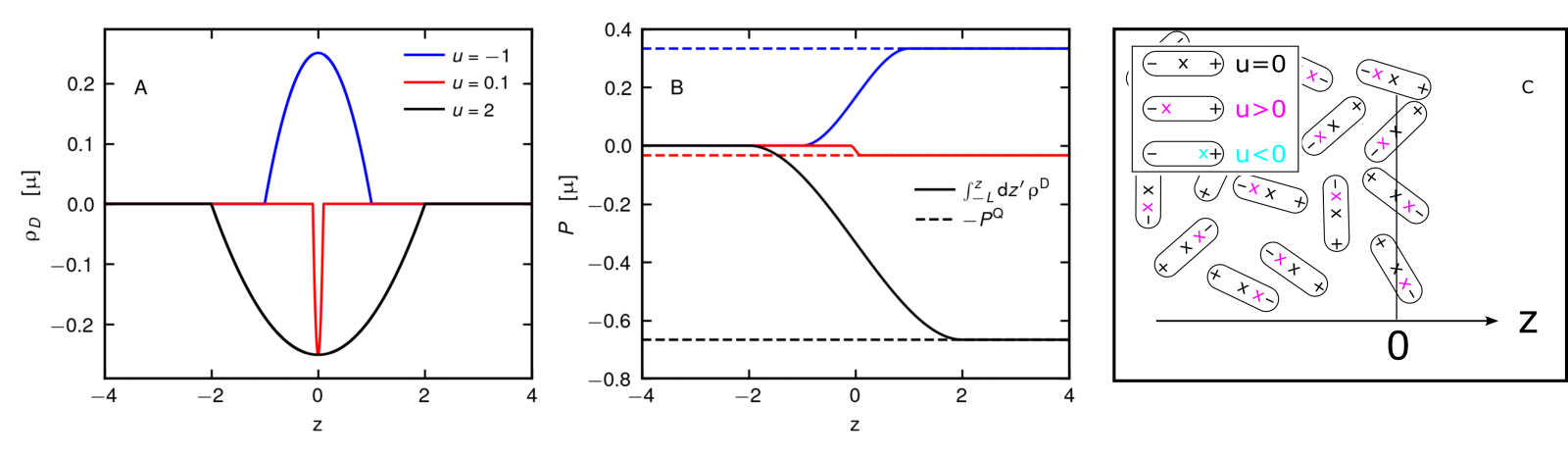}
    \caption{ The electric dipole density profile of our simple model according to Equation \eqref{eq:rhoD_model} is shown in A and its running integral in B. 
    Both are presented for various options of values of the offset parameter $u$, introduced in Equation \eqref{eq:rho_mol_model}. 
    The quadrupole contribution is given in Equation \eqref{eq:P2Q_model}, and its negative value is depicted in B. 
    C: Sketch of the the position of the molecular center for the cases $u=0$, $u>0$ and $u<0$ and the mechanism behind the $u$-dependent dipolar ordering. If $u>0$, the concentration of antiparallel oriented electric dipole moments is enhanced at the interface.}
    \label{fig:model_system}
\end{figure}
Here we demonstrate using a simple model calculation that the choice of the molecular expansion center influences the partitioning of the integrated polarization density defined in Equation \eqref{eq:def_P} into the electric dipole and molecular multipole contributions but not their sum.
We consider the charge density of an electric dipole created by two charges that are displaced along the $z$-coordinate in the molecular frame
\begin{align}    
    \varrho^\mathrm{mol}(\vec r,u) = q \delta(x) \delta(y) \delta[ z - a - u] - q \delta(x) \delta(y)\delta[z + a - u],
    \label{eq:rho_mol_model}
\end{align}
where $2a$ is the distance between the charges $q$ and $-q$ and $u$ is a displacement from the molecular center. 
The dependence of the molecular frame on $u$ is sketched in the inset in Figure \ref{fig:model_system} C.
Hence, the electric dipole and electric quadrupole moments in the molecular frame according to their definitions in Equations \eqref{eq:Q_In} and \eqref{eq:Q_IIn} are
\begin{align}
    \mu_i = \delta_{iz} \mu; \quad Q_{ij}(u) = \delta_{ij} \delta_{iz} u \mu,
\end{align}
where we define $\mu=2 a q$. 
We assume that the molecular centers ($u=0$) of the electric dipoles are confined to $z<0$. 
In the following, we do a test calculation to confirm that the integral over the polarization density
\begin{align}
    P_z= \int \limits_{-L}^\infty \mathrm{d}z \, p_z( z  ) =  \int \limits_{-L}^\infty \mathrm{d}z   \varrho^{\mathrm{D}}_z( z, u ) + \varrho_{zz}^\mathrm{Q}(-L,u) 
    \label{eq:def_int_p_2_model}
\end{align}
is independent of $u$, where $-L<0$ is an arbitrary point in the bulk region. Here, 
$\varrho^{\mathrm{D}}_z( z, u )$ and $\varrho_{zz}^\mathrm{Q}(z,u)$ are the electric dipole density and the electric quadrupole density, defined in Equations \eqref{eq:rho_D} and \eqref{eq:rho_Q}, respectively.
For simplicity, we assume that the molecular centers of the electric dipoles are homogeneously distributed in the region $z<0$ and that they are isotropically orientated around their molecular center, which implies that the integral in Equation \eqref{eq:def_int_p_2_model} should evaluate to zero.  Hence, we have the distribution function
\begin{align}
    \varrho^\mathrm{ISO}(z,\theta) = \frac{1}{4 \pi} \Theta(-z) \, ,
    \label{eq:def_rho_iso}
\end{align}
where $\theta$ is the angle between the molecular and the laboratory $z$-axis. 
We construct the function
\begin{align}
    G_z(z,u | z_0, \theta) = \mu \cos{\theta} \delta(z + u \cos{\theta} - z_0  ) \, ,
\end{align}
which is the electric dipole density according to Equation \eqref{eq:rho_D} of a single molecule with molecular center at $z_0$, offset by $u$ times the projection of the electric dipole axis on the $z$-axis of the laboratory frame $\cos{\theta}$. 
Because the electric dipole density is determined by the sum of the molecular dipole densities, we can construct the electric dipole density defined in Equation \eqref{eq:def_int_p_2_model}
\begin{align}   \varrho^\mathrm{D}_z(z,u)  = 2 \pi \int\limits_{-\infty}^\infty \mathrm{d} z_0 \int\limits_0^\pi \mathrm{d} \theta \sin{\theta}  \varrho^\mathrm{ISO}(z_0,\theta)  G_z(z ,u | z_0, \theta) \, .
\end{align}
After integration over $z_0$ and substituting $x=\cos{\theta}$, we arrive at
\begin{align}
\varrho^\mathrm{D}_z(z,u) = \frac{\mu}{2} \int \limits_{-1}^1 \mathrm{d} x \, x \Theta( - ux - z) \, .
    \label{eq:rhoD_model}
\end{align}
This integral can be solved using the relationship
\begin{align}
    \int\limits_a^b \mathrm{d} x f(x) \Theta(x) =
     \Theta(b) \Theta(-a) \int\limits_0^b \mathrm{d} x f(x)  +   \Theta(a) \int\limits_a^b \mathrm{d} x f(x) \, , 
\end{align}
valid for all $a,b \in \mathbb{R}$ and $a\leq b$.
Hence, we arrive at
\begin{align}
    \varrho^\mathrm{D}_z(z,u) =   \frac{\mu \mathrm{sign}(u)}{4} \Pi\left(\frac{z}{2 u} \right) \left(  \frac{z^2}{u^2}-1  \right) \, .
\end{align}
We obtain the following result for the electric dipole and electric quadrupole contributions to the integral in Equation \eqref{eq:def_int_p_2_model}
\begin{align}
    \int \limits_{-L}^\infty \mathrm{d}z   \varrho^{\mathrm{D}}_z( z, u ) &= - \frac{u \mu }{3}  
    \label{eq:P2D_model}  \\
    \varrho_{zz}^\mathrm{Q}(-L,u)  &= \frac{Q_{ii}(u)}{3} =  \frac{u \mu}{3}  \, .
     \label{eq:P2Q_model} 
\end{align}
Here, $\frac{u \mu}{3} $ in Equation \eqref{eq:P2Q_model} represents the quadrupole density in the isotropic region, since the number fraction in the bulk is set to unity in Equation \eqref{eq:def_rho_iso} and the isotropic component of the quadrupole tensor is one-third times the trace \cite{grayTheoryMolecularFluids1984}.
The analytically calculated electric dipole profiles for $u \in \lbrace -1,0.1,2 \rbrace$ and their running integrals are presented in Figure \ref{fig:model_system} A and B, respectively. In Figure \ref{fig:model_system} A, we see that positive displacements $u$ lead to negative interfacial dipole contributions, and negative displacements $u$ lead to positive interfacial dipole contributions. 
This can be understood by considering electric dipoles with their molecular center located at a given $z$, as sketched in Figure~\ref{fig:model_system} C.  
If a dipole is aligned along the laboratory z-axis ($\vec{e}_z$), its center is shifted by a distance $-u$.  
Therefore, when $u > 0$, dipoles oriented parallel to $\vec{e}_z$ are pushed to the left side, while those oriented antiparallel are pushed to the right side. 
As there are no molecular centers at $z>0$, this leads to an increased concentration of antiparallel-oriented dipoles located at the interface.
The opposite occurs when $u < 0$.
This model calculation demonstrates that the integral over the polarization density $P_i$ is independent of the expansion point; however, the assignment into electric dipole and electric quadrupole contributions is not. 
A displacement from the molecular center induces an artificial electric dipole polarization, which is corrected by the electric quadrupole contribution. 
Hence, if we neglect the contributions of electric quadrupoles, $P_i$ depends on $u$. 
In the context of the SFG spectrum, it is assumed in Section~\ref{sec:orientation_analyis} that the electric dipole contribution can be related to the molecular orientation.  
To make this connection meaningful, the molecular origin must be defined such that an interface composed of isotropically oriented molecules produces no net electric dipole contribution. 
In this model calculation, this would correspond to setting $u=0$.
\subsection{On the Origin Dependence of the Multipole Decomposition}
\label{sec:origin_sfg}
\begin{figure}
    \centering    \includegraphics[scale=1]{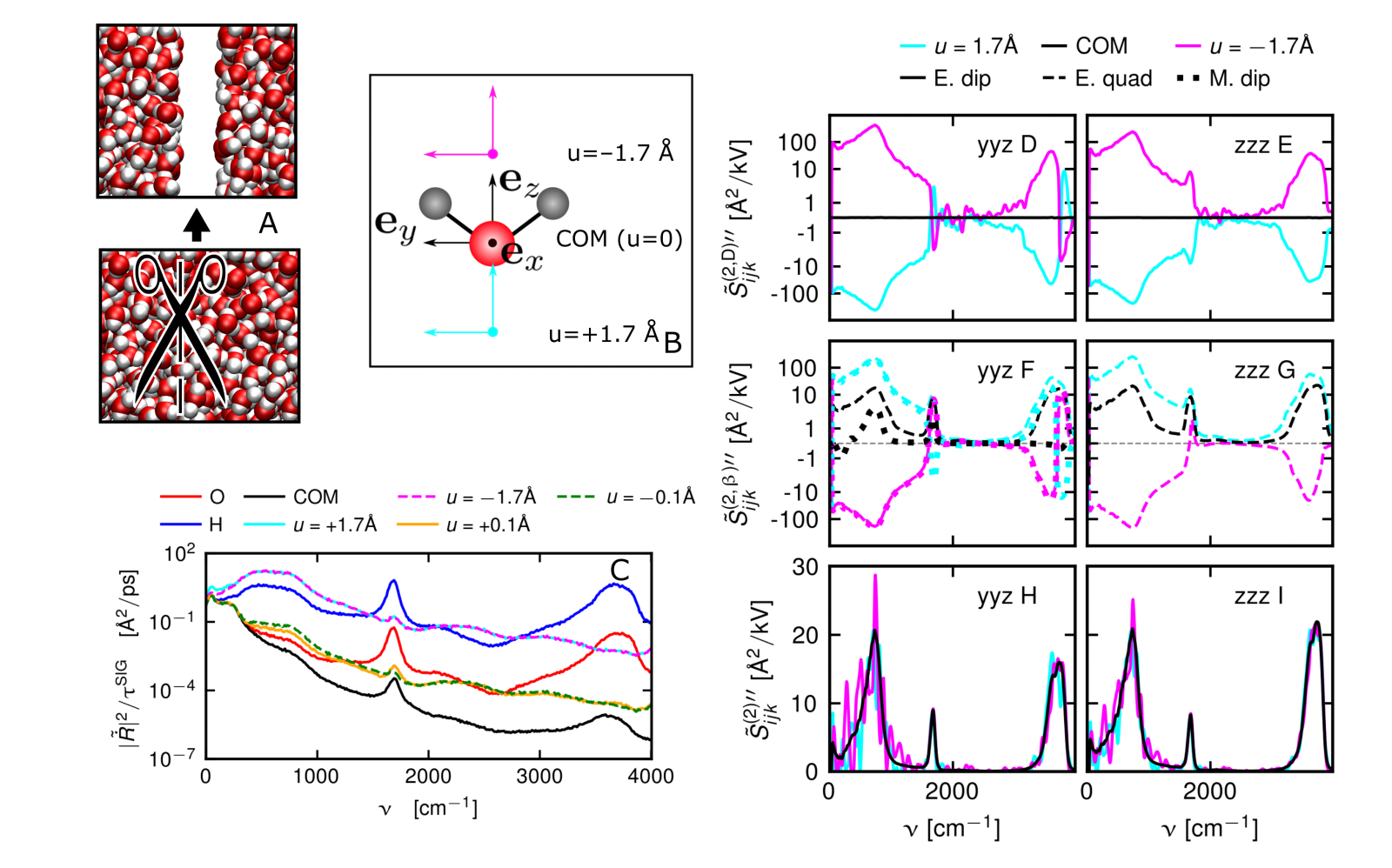}
    \caption{ 
    The isotropic interface, obtained by cutting bulk water according to the molecular center positions at a planar interface, is sketched in A.
    The investigated molecular frames aligned relative to the Eckart reference frame are presented in B and defined in Equation \eqref{eq:check_R_pm}.  
    In C, we present the power spectrum of the hydrogen atom, the oxygen atom, the center of mass and different deviations from the center of mass $u$ in bulk water.
   In D-I, we present the multipolar SFG spectrum from the isotropic interface using different molecular origins. 
   In D \& E, we present the electric dipole contribution, and in F \& G, the electric quadrupole and magnetic dipole contributions. 
   The total signal calculated with respect to different molecular origins is presented in Figure H  \&   I, demonstrating that it does not depend on the molecular origin. }    \label{fig:origin_sfg_signal}
\end{figure}
Here, we demonstrate numerically that the SFG signal is independent of the molecular origin if all multipole contributions are considered. 
We show numerically that if the molecular origin chosen for the multipole expansion coincides with the center of mass, there is no electric dipole contribution to the SFG signal from an artificially created interface with isotropic molecular orientation, 
which we call an isotropic interface. 
The multipole expansion of electrostatic interactions between non-overlapping charge distributions is independent of the choice of origin in the molecular frame, provided that all orders are included, as can be shown analytically \cite{grayTheoryMolecularFluids1984}.
The isotropic interface is created by cutting bulk water at an arbitrary
$z$-position into two halves as sketched in Figure \ref{fig:origin_sfg_signal} A. 
We assign the molecules to the left or right half space, as sketched in Figure \ref{fig:origin_sfg_signal} A, based on their center of mass position. 
Hence, we get an isotropic air-water interface extracted from a simulation of an isotropic bulk system. 
In this test, we apply the Lorentz-field Approximation \eqref{eq:LorentzField} for the time-averaged local field factors, meaning we approximate in Equations \eqref{eq:adij_mun} and \eqref{eq:aqijk_Qn}
\begin{align}
  \alpha^{n,\mathrm{DD}}_{ik}(t) f_{kj}^n(t) &\approx   \alpha^{n,\mathrm{DD}}_{ij}(t) \frac{2 + \tilde \varepsilon^\mathrm{VIS}}{3}  \, , \\ 
 \alpha^{n,\mathrm{QD}}_{ijl}(t) f_{lk}^n(t) &\approx   \alpha^{n,\mathrm{QD}}_{ijk}(t) \frac{2 + \tilde \varepsilon^\mathrm{VIS}}{3} \, .
\end{align}
Here, $c^\mathrm{VIS}_i$ is the external field - electric field translation factor defined in Equation \eqref{eq:ext_elec_translation_fac} and we leave out the $z$-dependence, as the bulk system is homogeneous.
We note that it would be wrong to account for the additional contribution due to the linear response to the second-order source  polarization density discussed in Section \ref{sec:lin_nonloc_resp_func}, as the second-order polarization density does not exist in the isotropic bulk system in the first place.
For the extraction of the second-order response profile $\tilde{s}^{(2,\mathrm{D})}_{ijk}(z,\omega^\mathrm{IR})$ we employ Equation \eqref{eq:s2D_fdt}. 
The multipole contributions $\tilde{S}^{(2,\mathrm{Q})}_{ijk}$ and $\tilde{S}^{(2,\mathrm{M})}_{ijk}$ are computed with Equations \eqref{eq:chi2m_origin}, \eqref{eq:chi2Q_fdt_corr_func}, \eqref{eq:S2Q_bulk} and \eqref{eq:S2M_bulk}. 
In all contributions, we account for the different identification of the external field in interface and bulk systems, according to Equations \eqref{eq:ext_field_maxwell} and \eqref{eq:ext_field_maxwell_bulk}, by transforming
$\tilde S^{(2,\beta)}_{ijk}(\omega^\mathrm{VIS}, \omega^\mathrm{IR}) \rightarrow \tilde c_j^\mathrm{VIS} \tilde c_k^\mathrm{IR} \tilde S^{(2,\beta)}_{ijk}(\omega^\mathrm{VIS}, \omega^\mathrm{IR})$.
We decompose the SFG signal from the isotropic interface $\tilde{S}^{(2,\mathrm{ISO})}_{ijk} \left( \omega^\mathrm{VIS}, \omega^\mathrm{IR} \right)$ into the multipole contributions 
\begin{align}   \tilde{S}^{(2,\mathrm{ISO})}_{ijk} \left( \omega^\mathrm{VIS}, \omega^\mathrm{IR} \right) = \tilde{S}^{(2,\mathrm{D})}_{ijk} \left(  \omega^\mathrm{IR} \right) + \tilde{S}^{(2,\mathrm{Q})}_{ijk} \left(  \omega^\mathrm{IR} \right) + \tilde{S}^{(2,\mathrm{M})}_{ijk} \left( \omega^\mathrm{VIS}, \omega^\mathrm{IR} \right) \, .
    \label{eq:s2_iso}
\end{align}
We compare origins displaced by a distance $u$ from the center of mass along the bisector axis $\hat{e}^n_{z'}$
in the Eckart reference frame shown in Figure \ref{fig:mol_frame}. 
Hence, we have the relation between the center of mass $\vec{R}^{n}(t)$ and the displaced center $\vec{\check{R}}^{n}(t)$ 
\begin{align}
    \vec{\check{R}}^{n}(t) = \vec{R}^{n}(t) \mp u \vec{e}_z(t) \, , 
    \label{eq:check_R_pm}
\end{align}
in the laboratory frame.
Figure \ref{fig:origin_sfg_signal} B depicts three different expansion points within the molecular frame. 
We compare the power spectra ${\tilde{C}_{VV}(\omega)=\frac{1}{\tau^\mathrm{max}} | \tilde{ \dot{R} }_x(\omega) |^2}$ of the nuclei, the center of mass, and the molecular origins defined in Equation \eqref{eq:check_R_pm} in Figure \ref{fig:origin_sfg_signal} C.
Here, we observe that the oscillations of the center of mass are negligible compared to the oscillations of the nuclei, as long as $\nu^\mathrm{IR} \geq \SI{250}{cm^{-1}}$. However, this observation no longer holds if the center of mass is shifted along the Eckart bisector axis, as can be seen in Figure \ref{fig:origin_sfg_signal} C.
We verify the independence of the expansion point of the SFG signal from an isotropic reference interface defined in Equation \eqref{eq:s2_iso} in Figure \ref{fig:origin_sfg_signal} D-I. 
Here we test three molecular centers $u=0$ and $u=\pm\SI{1.7}{\angstrom}$. 
The value of $u=-\SI{1.7}{\angstrom}$ is chosen because it approximately minimizes the electric quadrupole and magnetic dipole contributions in the OH-stretch region, as shown in Fig. \ref{fig:origin_sfg_signal} F, which would be an alternative criterion of choosing the molecular center. 
The electric dipole contributions are shown in Figure \ref{fig:origin_sfg_signal} D  \&   E.
We observe that if we choose the center of mass as the expansion point, the electric dipole contribution to the SFG spectrum from an isotropic reference interface is zero, as indicated by the flat black line in Figure \ref{fig:origin_sfg_signal} D \& E. However, we obtain massive deviations of the electric dipole contribution from zero if $u$ in Equation \eqref{eq:check_R_pm} is not equal to zero.
Hence, minimizing multipole contributions is not a useful criterion to define the molecular origin.
The electric quadrupole and magnetic dipole contributions are presented in Figure \ref{fig:origin_sfg_signal} F  \&   G. 
Here, we observe significant nonzero magnetic dipole and electric quadrupole contributions for all choices of the molecular centers. 
We verify numerically that the SFG spectrum is independent of the molecular origin, as can be seen by the overlapping black, magenta, and cyan lines in Figure \ref{fig:origin_sfg_signal} H \& I. 
If we choose the center of mass as the expansion point, the relationship
\begin{align}  \tilde{S}^{(2,\mathrm{ISO})}_{ijk} \left(  \omega^\mathrm{IR} \right) \approx \tilde{S}^{(2,\mathrm{Q})}_{ijk} \left( \omega^\mathrm{VIS}, \omega^\mathrm{IR} \right) +  \tilde{S}^{(2,\mathrm{M})}_{ijk} \left( \omega^\mathrm{VIS}, \omega^\mathrm{IR} \right)  \, ,
    \label{eq:S_iso_eq_S_MM}
\end{align} 
holds in a very fine approximation, which follows from $\tilde{S}^{(2,\mathrm{D})}_{ijk}(\omega^\mathrm{IR})\approx 0 $, as can be seen in Figure \ref{fig:origin_sfg_signal} D  \&   E. 
The electric dipole contributions are commonly interpreted to be induced by the anisotropic structure of the interface \cite{guyot-sionnestBulkContributionSurface1988,zhuangMappingMolecularOrientation1999,shenFundamentalsSumFrequencySpectroscopy2016a,sunOrientationalDistributionFree2018a,moritaTheorySumFrequency2018a,yuPolarizationDependentHeterodyneDetectedSumFrequency2022,yuPolarizationDependentSumFrequencyGeneration2022a}.
This is only valid if an isotropic interface does not create an electric dipole contribution. 
On the basis of this criterion we chose the center of mass as the molecular origin of the multipole expansion. 
The multipole contributions presented in Figure \ref{fig:origin_sfg_signal} H \& I are a universal property of bulk water (like the dielectric constant) and are independent of the interface.
Hence, these contributions are the same as for air-water interface presented in the main text in Figure 1.
When the center of mass is chosen as the molecular center, multipole contributions can be viewed as the SFG signal from the isotropic interface.
Unlike the rest of this work, where the electric dipole polarizabilities are parameterized using CCSD(T)/aug-cc-pVTZ single-molecule calculations, we use the B3LYP/aug-cc-pVTZ level here to match the parameterization of the electric quadrupole polarizabilities. 
As shown in Section \eqref{app:polarizability}, the difference between the two parametrizations is minor.
Additional data, then described in the Methods section, is used to extract the electric dipole and quadrupole contributions because of the low signal-to-noise ratio. The results are averaged over nine trajectories, each about $\SI{300}{ps}$ long, using the MB-Pol force field. 
The initial configurations were taken from an SPC/E simulation, spaced by $\SI{2}{ns}$.
We cut the system at $10$ $z$-positions for our artificial interface in the computation of $\tilde{S}^{(2,\mathrm{D})}_{ijk}\left( \omega^\mathrm{IR} \right)$.
\section{Non-Uniaxial Orientation Analysis}
\label{sec:orientation_analyis}
Here, we derive the equations for the prediction of the molecular hyperpolarizability $\tilde{\beta}_{ijk}(\theta, \psi)$ and $\tilde{\chi}^{(2,\mathrm{ORI})}_{ijk}(z, \omega^\mathrm{IR})$, appearing in the main text.
Due to the disappearance of the electric dipole density $j^{(2,\mathrm{D})}_i(z,t)$, defined in Equation \eqref{eq:def_j2D}, in a system with isotropically oriented molecules, the electric dipole susceptibility tensor $\chi^{(2,\mathrm{DL})}_{ijk}(\omega^\mathrm{VIS},\omega^\mathrm{IR})$, defined in Equation \eqref{eq:chi_2DL}, incorporates the fingerprint of interfacial orientation. 
As mentioned previously, the SFG spectrum is determined by complex many-body dynamics and cannot be related solely to molecular orientation in an exact manner.
Here, we introduce the necessary approximations to relate SFG spectra to molecular orientation. 
We consider the fluctuation-dissipation relation of the second-order pure electric dipole  response profile $\tilde{s}^{(2,\mathrm{DD})}_{ijk}(z,\omega^\mathrm{IR})$ in Equation \eqref{eq:s2DD_fdt}. 
We apply the time-scale separation given in Equation \eqref{eq:addij_mun_approx_final} and write 
\begin{align} s^{(2,\mathrm{DD})}_{ijk}(z,t) = \frac{-\Theta(t)}{\varepsilon_0 k_B T L_x L_y} \frac{\partial}{\partial t}\sum\limits_n^{N_\mathrm{mol}} \left\langle \delta[z-z^n(t)] \bar f^n_{i' i}(t) \alpha^{n,\mathrm{DD}}_{i' j'} (t) \bar f^n_{j' j}(t) P_k(0)\right\rangle \, .
  \label{eq:s2DD_ori_start}
\end{align}
This equation assumes an infinite system where there is no field due to the periodic images. For finite system sizes one needs to correct for the unphysical field due to the periodic images as described in Section \ref{app:pbc}. Furthermore, the result in Equation \eqref{eq:s2DD_ori_start} does not depend on $\omega^\mathrm{VIS}$, which is a consequence of the off-resonant approximation, as discussed in Section \ref{sec:lin_spons_born_opp}.
Here, $\bar f^n_{i' i}(t)$ are time-averaged local field factors defined in Equation \eqref{eq:def_time_aver_loc_field} and $\alpha^{n,\mathrm{DD}}_{ij}(t)$ is the trajectory of the molecular electric dipole - electric dipole polarizability defined in Equation \eqref{eq:polarizabilities}.
We rotate the effective molecular polarizability tensor $\bar f^n_{i' i}(t) \alpha^{n,\mathrm{DD}}_{i' j'} (t) \bar f^n_{j' j}(t)$ and the dipole moment of the system $P_i(t)$ into and out of the molecular frame 
\begin{align}
  s^{(2,\mathrm{DD})}_{ijk}(z,t) = \frac{-\Theta(t)}{\varepsilon_0 k_B T L_x L_y} \frac{\partial}{\partial t}\sum\limits_n^{N_\mathrm{mol}} \left\langle \delta[z-z^n(t)]  D^n_{ia}(t) D^n_{i'a}(t) D^n_{jb}(t) D^n_{j'b}(t) \bar f^n_{i'' i'}(t) \alpha^{n,\mathrm{DD}}_{i'' j''} (t) \bar f^n_{j'' j'}(t)  D^n_{kc}(0) D^n_{k'c}(0) P_{k'}(0)\right\rangle \, ,
\end{align}
using direction cosine tensor components $D^n_{ij}(t)$ \cite{grayTheoryMolecularFluids1984}. These are defined by 
the dot product between the Cartesian laboratory basis vector $\hat{e}_i$ and the basis vector $\hat{e}^n_i(t)$ of the $n^\mathrm{th}$ molecular Eckart frame \cite{eckartStudiesConcerningRotating1935}
\begin{align}
    D^n_{ij}(t) = \hat{e}_i \cdot \hat{e}^n_j(t) \, .
    \label{eq:direction_cosine}
\end{align}
Now we impose three approximations that allow us to map $\tilde{s}^{(2,\mathrm{DD})}_{ijk}(z,t)$ to molecular orientation. 
First, we assume that the local field factor is only a function of the molecular $z$-position and consequently determined by the averaged local field factor $f_i^\mathrm{\alpha}(z)$ defined in Equation \eqref{eq:def_aver_loc_field}, i.e. $\bar{f}^n_{ij}(t) \approx \delta_{ij} f^\mathrm{SFG/VIS}_i\left[z^n(t)\right]$.
This leads to 
\begin{align}
  s^{(2,\mathrm{DD})}_{ijk}(z,t) \approx f^\mathrm{SFG}_{i}(z)  f^\mathrm{VIS}_{j}(z) \frac{-\Theta(t)}{\varepsilon_0 k_B T L_x L_y} \frac{\partial}{\partial t}\sum\limits_n^{N_\mathrm{mol}} \left\langle \delta[z-z^n(t)]  D^n_{ia}(t) D^n_{jb}(t) D^n_{kc}(0) D^n_{i' a}(t) D^n_{j'b}(t)\alpha^{n,\mathrm{DD}}_{i' j'} (t)   D^n_{k'c}(0) P_{k'}(0)\right\rangle \, .
  \label{eq:f_z_approx_s2dd}
\end{align}
Note that we can replace $f^\alpha_{i'}(z)$ by $f^\alpha_{i}(z)$, because $D^n_{ia}(t) D^n_{i'a}(t)=\delta_{ii'}$.
We introduce the correlation function between the polarizability and the electric dipole moment corrected for the local field $P_{i}(t) / f^\mathrm{IR}_i([z^n(t)])$ in the molecular frame
\begin{align}
    C^n_{abc}(t) = \left\langle D^n_{i a}(t) D^n_{jb}(t)\alpha^{n,\mathrm{DD}}_{i j} (t) D^n_{kc}(0) \frac{P_{k}(0)}{f^\mathrm{IR}_k([z^n(0)])}  \right\rangle \, .
\end{align}
Second, we assume that $C^n_{abc}(t)$ is not correlated with the position and orientation of the molecule, leading to
\begin{align}
  s^{(2,\mathrm{DD})}_{ijk}(z,t) \approx f^\mathrm{SFG}_{i}(z)  f^\mathrm{VIS}_{j}(z) \frac{-\Theta(t)}{\varepsilon_0 k_B T L_x L_y} \frac{\partial}{\partial t} \left( \sum\limits_n^{N_\mathrm{mol}} \left\langle \delta[z-z^n(t)]  D^n_{ia}(t) D^n_{jb}(t) D^n_{kc}(0) f^\mathrm{IR}_{k}[z^n(0)] \right\rangle C^n_{abc}(t) \right) \, .
  \label{eq:uncorr_approx_s2dd}
\end{align}
Third, we apply the so-called slow-motion limit \cite{weiMotionalEffectSurface2001}, which means that we assume that the left-hand correlation function in Equation \eqref{eq:uncorr_approx_s2dd} varies more slowly than the right-hand one. This leads to the final result
\begin{align}
  s^{(2,\mathrm{DD})}_{ijk}(z,t) \approx \frac{f^\mathrm{SFG}_{i}(z) f^\mathrm{VIS}_{j}(z) f^\mathrm{IR}_{k}(z)}{L_x L_y} \sum\limits_n^{N_\mathrm{mol}} \left\langle \delta(z-z^n)  D^n_{ia} D^n_{jb} D^n_{kc} \right\rangle \frac{-\Theta(t)}{\varepsilon_0 k_B T} \frac{\partial}{\partial t} C^n_{abc}(t)  \, .
\label{eq:uncorr_approx_s2dd_final}
\end{align}
By comparing Equations \eqref{eq:s2_chi2},  \eqref{eq:chi_2DL_approx_beta} and \eqref{eq:uncorr_approx_s2dd_final} we find 
\begin{align}
    \tilde\beta^n_{ijk}(\omega^\mathrm{IR}) =\left\langle \delta(z-z^n)  D^n_{ia} D^n_{jb} D^n_{kc} \right\rangle \frac{-\Theta(t)}{\varepsilon_0 k_B T} \frac{\partial}{\partial t} C^n_{abc}(t)\, 
\end{align}
for the molecular hyperpolarizability tensor, defined in Equation \eqref{eq:def_beta_ijk}.
Consequently, we have an approximate relationship between the SFG signal and the molecular orientation.
From this we can introduce the fluctuation-dissipation relation of the hyperpolarizability tensor in the molecular frame
\begin{align}
    \bar\beta_{abc}(t) = \frac{-\Theta(t)}{\varepsilon_0 k_B T N_\mathrm{mol}   }  \sum\limits_n^{N_\mathrm{mol}} \frac{\partial}{\partial t} C^n_{abc}(t) \, .
    \label{eq:bar_beta}
\end{align}
Here, we assume that all molecules are identical, allowing averaging. 
We extract $\tilde{\bar{\beta}}_{abc}(\omega^\mathrm{IR})$ from a simulation of bulk water, where $f_i^\mathrm{IR}(z)$ is neither a function of the position nor anisotropic. 
We approximate $f_i^\mathrm{IR}$ by the Lorentz-field approximation using $\varepsilon^\mathrm{IR}\approx 1.77$, which leads to $f^\mathrm{IR}_i(z)\approx  1.26\, \delta_{ij}$.
Combining the approximations in Equations \eqref{eq:chi_2DL_approx_beta} and \eqref{eq:uncorr_approx_s2dd_final}, leads to the prediction of $\chi^{(2,\mathrm{DL})}_{ijk}(z,\omega^\mathrm{VIS}, \omega^\mathrm{IR})$ defined in Equation \eqref{eq:chi_2DL}, 
solely based on molecular orientation
\begin{align}
\tilde{\chi}^{(2,\mathrm{ORI})}_{ijk}(z, \omega^\mathrm{IR}) = \frac{\tilde{\bar{\beta}}_{abc}(\omega^\mathrm{IR})}{L_x L_y} \sum\limits_n^{N_\mathrm{mol}} \langle  D^n_{ia} D^n_{jb} D^n_{kc} \delta(z-z^n) \rangle  \, .
\label{eq:def_chi2_ori}
\end{align}
One can approximate
\begin{align}
\tilde{\chi}^{(2,\mathrm{ORI})}_{ijk}(z,\omega^\mathrm{IR}) \approx \tilde{\chi}^{(2,\mathrm{DL})}_{ijk}(z, \omega^\mathrm{IR})   \, ,
\label{eq:chi2_ori}
\end{align}
whenever the second-order electric dipole susceptibility $\tilde{\chi}^{(2,\mathrm{DL})}_{ijk}(\omega^\mathrm{VIS}, \omega^\mathrm{IR})$ introduced in Equation \eqref{eq:chi_2DL} is dominated by orientational anisotropy.
Hence, molecular orientation can be related to the SFG spectrum.
The approximation in Equation \eqref{eq:chi2_ori} is not necessarily good, because molecular hyperpolarizability depends not only on the molecular orientation but also on the surrounding environment, which is anisotropic and differs from the bulk at the interface.
By design, $\tilde{\chi}^{(2,\mathrm{ORI})}_{ijk}(z,\omega^\mathrm{IR})$ predicts the SFG response solely from molecular orientation and known bulk properties.
Within this framework, the SFG spectrum can be related to the expectation values of the elements of the rotation matrix $ \langle  D^n_{ia} D^n_{jb} D^n_{kc} \delta(z-z^n) \rangle$, which is a 6$^\mathrm{th}$-rank tensor with 729 elements. We can reduce the number of independent elements by symmetry considerations, but we remain at the direct product of the three non-zero unique tensor components of $\tilde{\chi}^{(2,\mathrm{ORI})}_{ijk}(\omega^\mathrm{IR})$ and the four nonzero unique components of $\tilde{\bar\beta}_{abc}(\omega^\mathrm{IR})$.
\begin{figure}
\centering
\includegraphics[width=0.2\textwidth]{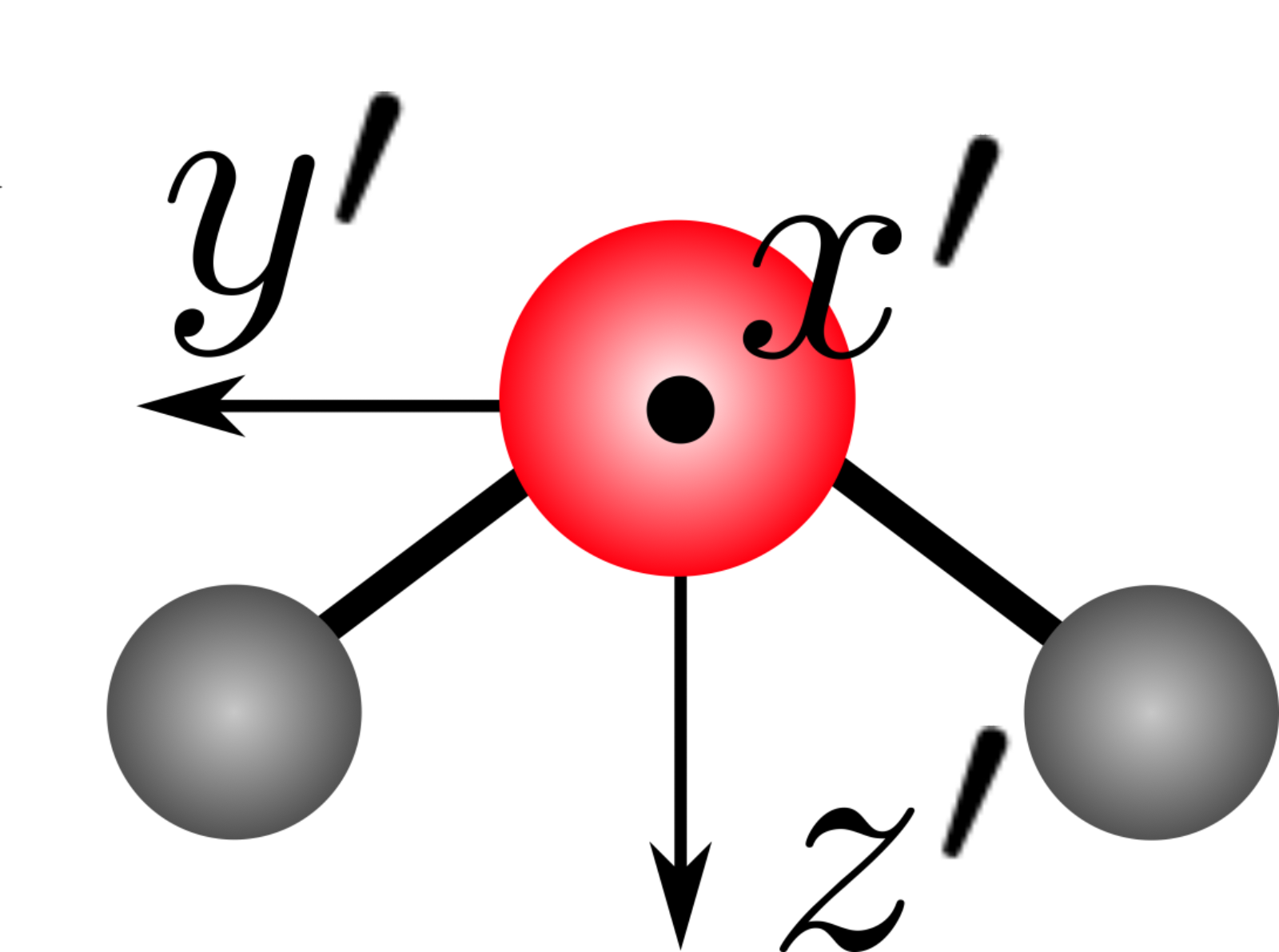}
\caption{Sketch of the body-fixed coordinate frame. The axes with labels $x'y'z'$ are the out of plane axis $\hat{e}_x'$, the normalized permanent electric dipole vector $\hat{e}_z'$ and $\hat{e}_y'=\hat{e}_z'\times \hat{e}_x'$ 
.}
\label{fig:mol_frame}
\end{figure}
One can drastically reduce the number of measured order parameters by assuming that the molecules of interest have uniaxial symmetry, which means that the molecules are rotational symmetric around a symmetry axis 
\cite{zhuangMappingMolecularOrientation1999,weiMotionalEffectSurface2001,yuPolarizationDependentSumFrequencyGeneration2022a,yuPolarizationDependentHeterodyneDetectedSumFrequency2022,moritaTheorySumFrequency2018a}, which is not the case for bulk water.
 Whenever the uniaxial approximation is not well-justified, it is adversible to transform $\tilde{\bar\beta}_{abc}(\omega^\mathrm{VIS},\omega^\mathrm{IR})$ into its irreducible representation, as derived here.
 First, we transform the Cartesian tensor $\tilde{\bar\beta}_{abc}(\omega^\mathrm{VIS},\omega^\mathrm{IR})$ into a reducible spherical tensor.
The irreducible representation of a spherical tensor can be determined using the procedure outlined in the book by Gray and Gubbins \cite{grayTheoryMolecularFluids1984}, which we describe in the following section. %p 493
We consider the Cartesian tensor of third rank $t_{ijk}$, which we transform into the reducible spherical tensor $t^{111}_{n_1 n_2 n_3}$, where ${n_1,n_2,n_3 \in \lbrace -1,0,1 \rbrace}$ according to
\begin{align}
    t^{111}_{n_1 n_2 n_3} = U_{n_1 i} U_{n_2 j} U_{n_3 k}  t_{ijk} \, ,
    \label{eq:Cart2Spher}
\end{align}
and
\begin{align}
   U= \begin{pmatrix}
\frac{1}{\sqrt{2}} & -\frac{i}{\sqrt{2}} & 0 \\
0 & 0 & 1 \\
-\frac{1}{\sqrt{2}} & -\frac{i}{\sqrt{2}} & 0
\end{pmatrix}.
\end{align}
The irreducible representation $t^{\gamma,l}_{n}$ of the tensor $t^{111}_{n_1 n_2 n_3}$ can be formed according to
\begin{align}
    t^{\gamma,l}_{n} = C^{11\gamma}_{n_1 n_2 \kappa} C^{\gamma 1 l}_{\kappa n_3 n } \, t^{111}_{n_1 n_2 n_3} \, ,
    \label{eq:reduction}
\end{align}
where $C^{l_1 l_2 l_3}_{n_1 n_2 n_3}$ is a Clebsch-Gordan coefficient\cite{grayTheoryMolecularFluids1984}.
Using Equations \eqref{eq:Cart2Spher}- \eqref{eq:reduction}, we can relate the irreducible representation of the uniaxial third-rank tensor $u^{\gamma,l}_n$ to its Cartesian representation $u_{ijk}$ via 
\begin{align}
    u^{0,1}_0 &= -\frac{1}{\sqrt{3}} ( 2 u_{yyz} +u_{zzz} )\\
    u^{2,1}_0 &= \frac{2}{\sqrt{15}} \left( u_{yyz} -3u_{yzy} - u_{zzz} \right) \\
    u^{2,3}_0 &=\sqrt{\frac{2}{5}} \left( -u_{yyz} -2 u_{yzy} + u_{zzz} \right) \, .
\end{align}
Here, $u^{0,1}_0$ and $u^{2,1}_0$ transform like vectors (such as dipole moments), while $u^{2,3}_0$ transforms like a third-rank spherical tensor (such as an octupole moment).
 The SFG signal of an isotropic medium is zero because no component transforms like a scalar (such as a monopole moment).
Hence, we can express a Cartesian third rank tensor as a sum of its irreducible representation $u^{\gamma,1}_n$ by using
\begin{align}
    u_{yyz} &= -\frac{1}{\sqrt{3}} u^{0,1}_0 + \frac{1}{\sqrt{15}} u^{2,1}_0 -\frac{1}{\sqrt{10}} u^{2,3}_0 \, 
    \label{eq:uniax_yyz_as_iredu}\\
    u_{yzy} &= -\sqrt\frac{3}{20} u^{2,1}_0 - \frac{1}{\sqrt{10}} u^{2,3}_0\,  \label{eq:uniax_yzy_as_iredu}\\
    u_{zzz} &= -\frac{1}{\sqrt{3}} u^{0,1}_0 -\frac{2}{\sqrt{15}} u^{2,1}_0 +\sqrt{\frac{2}{5}} u^{2,3}_0 \, .
    \label{eq:uniax_zzz_as_iredu}
\end{align}
Our system from which we want to predict the SFG signal has uniaxial symmetry, but the water molecule has three distinguishable axes, depicted in figure \ref{fig:mol_frame}.
The nonzero components of the irreducible representation of the spherical molecular hyperpolarizability tensor are
\begin{align}
    \tilde{\bar{\beta}}^{0,1}_0  &= -\frac{1}{\sqrt{3}} \left(  \tilde{\bar{\beta}}_{x'x'z'} + \tilde{\bar{\beta}}_{y'y'z'} +  \tilde{\bar{\beta}}_{z'z'z'}\right)\label{eq:iredu_bet_010_a_as_cart}  \\
   \tilde{\bar{\beta}}^{2,1}_0&= \frac{1}{\sqrt{15}} \left(  \tilde{\bar{\beta}}_{x'x'z'} + \tilde{\bar{\beta}}_{y'y'z'}  - 3 \tilde{\bar{\beta}}_{y'z'y'} -2 \tilde{\bar{\beta}}_{z'z'z'} \right) \label{eq:iredu_bet_210_a_as_cart}  \\
   \tilde{\bar{\beta}}^{2,3}_0 &= -\frac{1}{\sqrt{10}} \left(  \tilde{\bar{\beta}}_{x'x'z'}+ \tilde{\bar{\beta}}_{y'y'z'}  + 2 \tilde{\bar{\beta}}_{y'z'y'} -2 \tilde{\bar{\beta}}_{z'z'z'} \right) \label{eq:iredu_bet_230_a_as_cart}  \\
   \tilde{\bar{\beta}}^{2,3}_{\pm2} &= \frac{1}{\sqrt{12}}\left(  \tilde{\bar{\beta}}_{x'x'z'} - \tilde{\bar{\beta}}_{y'y'z'}  - 2 \tilde{\bar{\beta}}_{y'z'y'}  \right) \, ,
   \label{eq:iredu_bet_232_a_as_cart}  \\
\end{align}
where we defined the $yz$-plane as the molecular plane.
A spherical tensor in the molecular frame $\bar{t}^{\gamma,l}_m$ is rotated into the laboratory frame according to
\begin{align}
    t^{\gamma,l}_n = \mathcal{D}^{l*}_{nm}(\phi,\theta,\psi) \,  \bar{t}^{\gamma,l}_m \, ,
    \label{eq:rotate_wigner_D}
\end{align}
where $ \mathcal{D}^{l*}_{nm}(\phi,\theta,\psi)$ is the complex-conjugated Wigner rotation matrix \cite{grayTheoryMolecularFluids1984} and $\phi,\theta,\psi$, are the three Euler angles that specify the orientation of the molecular Eckart reference frame\cite{eckartStudiesConcerningRotating1935,herzbergMolecularSpectraMolecular1945,adler-goldenFormulasTransformingInternal1985,reyTransformationInternalCoordinates1998} presented in Figure \ref{fig:mol_frame}. 
Here, we employ the $z'y'z'$ convention, where the molecule undergoes a sequence of three intrinsic rotations. 
First, we rotate the molecule around its $z'$ axis by an angle $\phi$. 
Given the system's uniaxial nature, the orientation distribution around $\phi$ is isotropic. 
Next, we rotate the molecule around the newly rotated $y'$ axis by an angle $\theta$, introducing a tilt between the molecular dipole vector and the interface normal vector. 
Finally, we perform a third rotation around the tilted molecular dipole vector by an angle $\psi$.
From Equations \eqref{eq:uniax_yyz_as_iredu}-\eqref{eq:rotate_wigner_D} follows
the relationship between $\chi^{(2,\mathrm{ORI})}_{ijk}(z,\omega^\mathrm{VIS},\omega^\mathrm{IR})$ and $\tilde{\bar{\beta}}^{\gamma,l}_{m}$
\begin{align}
\begin{split}
     \tilde\chi^{(2,\mathrm{ORI})}_{yyz}(z,\omega^\mathrm{IR}) &=  \frac{1}{L_x L_y }\sum\limits_{n}^{N_\mathrm{mol}}\bigg[ \left\langle \mathcal{D}^{1*}_{00}(\phi^n,\theta^n,\psi^n) \delta(z-z^n) \right\rangle \left( -\frac{1}{\sqrt{3}} \tilde{\bar{\beta}}^{0,1}_0(\omega^\mathrm{IR}) + \frac{1}{\sqrt{15}}  \tilde{\bar{\beta}}^{2,1}_0(\omega^\mathrm{IR}) \right) \\ &-\frac{1}{\sqrt{10}} \left\langle \mathcal{D}^{3*}_{00}(\phi^n,\theta^n,\psi^n) \delta(z-z^n) \right\rangle \tilde{\bar{\beta}}^{2,3}_0(\omega^\mathrm{IR}) \\
   & -\frac{1}{\sqrt{10}} \left\langle \left( \mathcal{D}^{3*}_{02}(\phi^n,\theta^n,\psi^n) + \mathcal{D}^{3*}_{0-2}(\phi^n,\theta^n,\psi^n)  \right) \delta(z-z^n)\right\rangle  \tilde{\bar{\beta}}^{2,3}_2(\omega^\mathrm{IR}) \bigg] \, ,
    \label{eq:chi2_ori_yyz} \end{split} \\
    \begin{split}
     \tilde \chi^{(2,\mathrm{ORI})}_{zzz}(z,\omega^\mathrm{IR}) &=  \frac{1}{L_x L_y }\sum\limits_{n}^{N_\mathrm{mol}}\bigg[\left\langle \mathcal{D}^{1*}_{00}(\phi^n,\theta^n,\psi^n) \delta(z-z^n) \right\rangle \left( -\frac{1}{\sqrt{3}} \tilde{\bar{\beta}}^{0,1}_0(\omega^\mathrm{IR}) - \frac{2}{\sqrt{15}}  \tilde{\bar{\beta}}^{2,1}_0(\omega^\mathrm{IR}) \right) \\ &+ \sqrt{\frac{2}{5}} \left\langle \mathcal{D}^{3*}_{00}(\phi^n,\theta^n,\psi^n) \delta(z-z^n) \right\rangle \tilde{\bar{\beta}}^{2,3}_0(\omega^\mathrm{IR}) \\
    &+ \sqrt{\frac{2}{5}}  \left\langle \left( \mathcal{D}^{3*}_{02}(\phi^n,\theta^n,\psi^n) + \mathcal{D}^{3*}_{0-2}(\phi^n,\theta^n,\psi^n)  \right) \delta(z-z^n)\right\rangle  \tilde{\bar{\beta}}^{2,3}_2(\omega^\mathrm{IR}) \bigg]
     \label{eq:chi2_ori_zzz} \end{split} \\
\begin{split}
    \tilde \chi^{(2,\mathrm{ORI})}_{yzy}(z, \omega^\mathrm{IR}) &= \frac{1}{L_x L_y }\sum\limits_{n}^{N_\mathrm{mol}}\bigg[-\sqrt\frac{3}{20} \langle \mathcal{D}^{1*}_{00} (\phi^n,\theta^n,\psi^n)\delta(z-z^n) \rangle \tilde{\bar{\beta}}^{2,1}_0(\omega^\mathrm{VIS},\omega^\mathrm{IR}) \\ &- \frac{1}{\sqrt{10}} \langle \mathcal{D}^{3*}_{00} (\phi^n,\theta^n,\psi^n) \delta(z-z^n) \rangle \tilde{\bar{\beta}}^{2,3}_0(\omega^\mathrm{IR}) \\ &- \frac{1}{\sqrt{10}}
    \langle \left( \mathcal{D}^{3*}_{02}(\phi^n,\theta^n,\psi^n) + \mathcal{D}^{3*}_{0-2}(\phi^n,\theta^n,\psi^n)  \right) \delta(z-z^n)\rangle  \tilde{\bar{\beta}}^{2,3}_2(\omega^\mathrm{IR})  \bigg] \, ,
     \label{eq:chi2_ori_yzy} 
\end{split}
\end{align}
where we used the system's uniaxial symmetry and the symmetry of the water molecule. The advantage of Equations \eqref{eq:chi2_ori_yyz}-\eqref{eq:chi2_ori_yzy} over Equation \eqref{eq:def_chi2_ori} is that we expressed the orientational SFG tensor $ \tilde\chi^{(2,\mathrm{ORI})}_{ijk}(z,\omega^\mathrm{VIS},\omega^\mathrm{IR})$ in terms of three order parameter profiles $\langle \mathcal{D}^{1*}_{00} \delta(z-z^n) \rangle$, $\langle \mathcal{D}^{3*}_{00} \delta(z-z^n) \rangle$ and $\langle \left( \mathcal{D}^{3*}_{02}(\phi^n,\theta^n,\psi^n) + \mathcal{D}^{3*}_{0-2}(\phi^n,\theta^n,\psi^n)  \right) \delta(z-z^n) \rangle$. 
The second advantage is that we have a built-in symmetry decomposition, i.e. $\langle \left( \mathcal{D}^{3*}_{02}(\phi^n,\theta^n,\psi^n) + \mathcal{D}^{3*}_{0-2}(\phi^n,\theta^n,\psi^n)  \right) \delta(z-z^n)\rangle$ factors in the deviation from uniaxial ordering of the molecules. We make Equations \eqref{eq:chi2_ori_yyz}-\eqref{eq:chi2_ori_zzz} more transparent by defining
\begin{align}
  \rho(z) q_{10}(z)&=\frac{1}{L_x L_y}\sum\limits_{n}^{N_\mathrm{mol}} \left\langle \mathcal{D}^{1*}_{00}(\phi^n,\theta^n,\psi^n) \delta(z-z^n) \right\rangle =\frac{1}{L_x L_y} \sum\limits_{n}^{N_\mathrm{mol}} \left\langle \left(\hat{e}^n_{z'} \cdot \hat{e}_{z}\right) \delta(z-z^n)  \right\rangle \\
 \rho(z) q_{30}(z)&= \frac{1}{L_x L_y}\sum\limits_{n}^{N_\mathrm{mol}} \left\langle \mathcal{D}^{3*}_{00} (\phi^n,\theta^n,\psi^n) \delta(z-z^n) \right\rangle =\frac{1}{L_x L_y}\sum\limits_{n}^{N_\mathrm{mol}}\bigg[\left\langle \frac{1}{2} \left[ 5 \left(\hat{e}^n_{z'} \cdot \hat{e}_{z}\right)^3 - 3 \left(\hat{e}^n_{z'} \cdot \hat{e}_{z}\right) \right]\delta(z-z^n) \right\rangle \\
 \begin{split}
 \rho(z) q_{32}(z)&= \frac{1}{L_x L_y}\sqrt{\frac{4}{30}}\sum\limits_{n}^{N_\mathrm{mol}} \left\langle \left[ \mathcal{D}^{3*}_{02} (\phi^n,\theta^n,\psi^n)+ \mathcal{D}^{3*}_{0-2} (\phi^n,\theta^n,\psi^n) \right] \delta(z-z^n) \right\rangle \\
 &=\frac{1}{L_x L_y} \sum\limits_{n}^{N_\mathrm{mol}} \left\langle \left(\hat{e}^n_{z'} \cdot \hat{e}_{z}\right) \left[ \left( \hat{e}^n_{x'} \cdot \hat{e}_{z} \right)^2 -  \left( \hat{e}^n_{y'} \cdot \hat{e}_{z} \right)^2 \right] \right\rangle \, .
  \end{split}
\end{align}
Hence, $q_{10}(z)$ and $q_{30}(z)$ are the first and third moments of an expansion of the orientation of the dipole moment profile and $q_{32}(z)$ describes the average rotation around the dipole axis $\hat{e}^n_{z^{\prime}}$ for a given dipole orientation $\hat{e}_z \cdot \hat{e}^n_{z'}$. 
Furthermore, we introduce the rescaled irreducible representation
\begin{align}
   \tilde \beta^{0,1}_0 &= \frac{1}{3} \left(  \tilde{\bar{\beta}}_{x'x'z'} + \tilde{\bar{\beta}}_{y'y'z'} +  \tilde{\bar{\beta}}_{z'z'z'}\right) \label{eq:beta010}\\
   \tilde \beta^{2,1}_0&= \frac{1}{15} \left(  \tilde{\bar{\beta}}_{x'x'z'} + \tilde{\bar{\beta}}_{y'y'z'}  - 3 \tilde{\bar{\beta}}_{y'z'y'} -2 \tilde{\bar{\beta}}_{z'z'z'} \right) \label{eq:beta210}\\
  \tilde  \beta^{2,3}_0  &= -\frac{1}{10}  \left(  \tilde{\bar{\beta}}_{x'x'z'}+ \tilde{\bar{\beta}}_{y'y'z'}   + 2 \tilde{\bar{\beta}}_{y'z'y'} -2 \tilde{\bar{\beta}}_{z'z'z'} \right)  \label{eq:beta230}\\
  \tilde \beta^{2,3}_{\pm 2}  &= \frac{1}{4} \left(  \tilde{\bar{\beta}}_{x'x'z'} - \tilde{\bar{\beta}}_{y'y'z'}  - 2 \tilde{\bar{\beta}}_{y'z'y'}  \right) \,.
   \label{eq:beta232}
\end{align}
All of this leads to the rather compact expression
\begin{align}
    \chi^{(2,\mathrm{ORI})}_{yyz}(z,\omega^\mathrm{IR}) / \rho(z) &= q_{10}(z) \left[ \tilde\beta^{0,1}_0(\omega^\mathrm{IR}) + \tilde\beta^{2,1}_0(\omega^\mathrm{IR}) \right] -q_{30}(z)\tilde\beta^{2,3}_0(\omega^\mathrm{IR}) -q_{32}(z) \tilde\beta^{2,3}_2( \omega^\mathrm{IR}) 
    \label{eq:chi2_ori_yyz_friendly}  \\
    \chi^{(2,\mathrm{ORI})}_{zzz}(z,\omega^\mathrm{IR}) / \rho(z) &= q_{10}(z) \left[  \tilde\beta^{0,1}_0(\omega^\mathrm{IR}) - 2 \tilde\beta^{2,1}_0(\omega^\mathrm{IR}) \right] +2 q_{30}(z)\tilde\beta^{2,3}_0(\omega^\mathrm{IR}) + 2q_{32}(z)\tilde \beta^{2,3}_2( \omega^\mathrm{IR})\label{eq:chi2_ori_zzz_friendly}  \\
    \chi^{(2,\mathrm{ORI})}_{yzy}(z,\omega^\mathrm{IR}) / \rho(z) &= -\frac{3}{2} q_{10}(z) \tilde\beta^{2,1}_0(\omega^\mathrm{IR})- q_{30}(z)\tilde\beta^{2,3}_0(\omega^\mathrm{IR}) - q_{32}(z)\tilde \beta^{2,3}_2( \omega^\mathrm{IR}) \, ,
    \label{eq:chi2_ori_yzy_friendly} 
\end{align}
which we simplify even further to
\begin{align}
    \chi^{(2,\mathrm{ORI})}_{ijk}(z,\omega^\mathrm{IR}) &= \rho(z) \left[ q_{10}(z) \tilde{\beta}^{10}_{ijk}\left( \omega^\mathrm{IR} \right) + q_{30}(z) \tilde{\beta}^{30}_{ijk}\left(   \omega^\mathrm{IR} \right) + q_{32}(z) \tilde{\beta}^{32}_{ijk}\left( \omega^\mathrm{IR} \right) \right] \, ,
 \label{eq:chi2_even_more_friendly}
\end{align}
where the molecular hyperpolarizabilities can be determined by comparing Equations \eqref{eq:chi2_ori_yyz_friendly}-\eqref{eq:chi2_ori_yzy_friendly} with Equation \eqref{eq:chi2_even_more_friendly}.
In an experiment one measures the SFG spectrum $S^{(2)}_{ijk}(\omega^\mathrm{VIS},\omega^\mathrm{IR})$. If we know the multipolar contributions to the SFG signal, we can
extract $ \tilde{\chi}^{(2,\mathrm{DL0})}_{ijk}(\omega^\mathrm{IR})$ according to the approximation in Equation \eqref{eq:chi2DL0_approx}. If it is additionally dominated by orientation, we can approximate
\begin{align} \tilde{\chi}^{(2,\mathrm{DL0})}_{ijk}(\omega^\mathrm{IR}) &\approx Q_{10} \tilde{\beta}^{10}_{ijk}\left( \omega^\mathrm{IR} \right) + Q_{30} \tilde{\beta}^{30}_{ijk}\left(   \omega^\mathrm{IR} \right) + Q_{32} \tilde{\beta}^{32}_{ijk}\left( \omega^\mathrm{IR} \right) \, ,
    \label{eq:S2DL_ori_ijk} 
\end{align}
where the order parameters
\begin{align}
    Q_{10} &= \int\limits_{-\infty}^\infty \mathrm{d} z \, \rho(z) q_{10}(z) =  \frac{1}{L_x L_y }\sum\limits_{n}^{N_\mathrm{mol}}\left\langle \left( \hat{e}^n_{z'} \cdot \hat{e}_{z} \right)\right\rangle 
    \label{eq:Q10} \\
    Q_{30} &= \int_{-\infty}^\infty \mathrm{d} z \,  \rho(z) q_{30}(z) = \frac{1}{L_x L_y }\sum\limits_{n}^{N_\mathrm{mol}} \left\langle \frac{1}{2} \left[ 5 \left(\hat{e}^n_{z'} \cdot \hat{e}_{z}\right)^3 - 3  \left( \hat{e}^n_{z'} \cdot \hat{e}_{z} \right) \right] \right\rangle \label{eq:Q30} \\ 
    Q_{32} &= \int_{-\infty}^\infty \mathrm{d} z \, \rho(z) q_{32}(z) = \frac{1}{L_x L_y } \sum\limits_{n}^{N_\mathrm{mol}} \left\langle \left( \hat{e}^n_{z'} \cdot \hat{e}_{z} \right) \left[ \left( \hat{e}^n_{x'} \cdot \hat{e}_{z} \right)^2 -  \left( \hat{e}^n_{y'} \cdot \hat{e}_{z} \right)^2 \right] \right\rangle \label{eq:Q32} \, 
\end{align}
are quantifying the orientational anisotropy of the investigated interface. Within this framework, there are three order parameters: $Q_{10}$, $Q_{30}$, and $Q_{32}$, as well as three combinations of polarization: $\tilde{\chi}^{(2,\mathrm{DL0})}_{yyz}(\omega^\mathrm{IR})$, $\tilde{\chi}^{(2,\mathrm{DL0})}_{yzy}(\omega^\mathrm{IR})$, and $\tilde{\chi}^{(2,\mathrm{DL0})}_{zzz}(\omega^\mathrm{IR})$. Hence, interfacial orientation can be determined from SFG spectroscopy.
\begin{figure}
    \centering
    \includegraphics[scale=1]{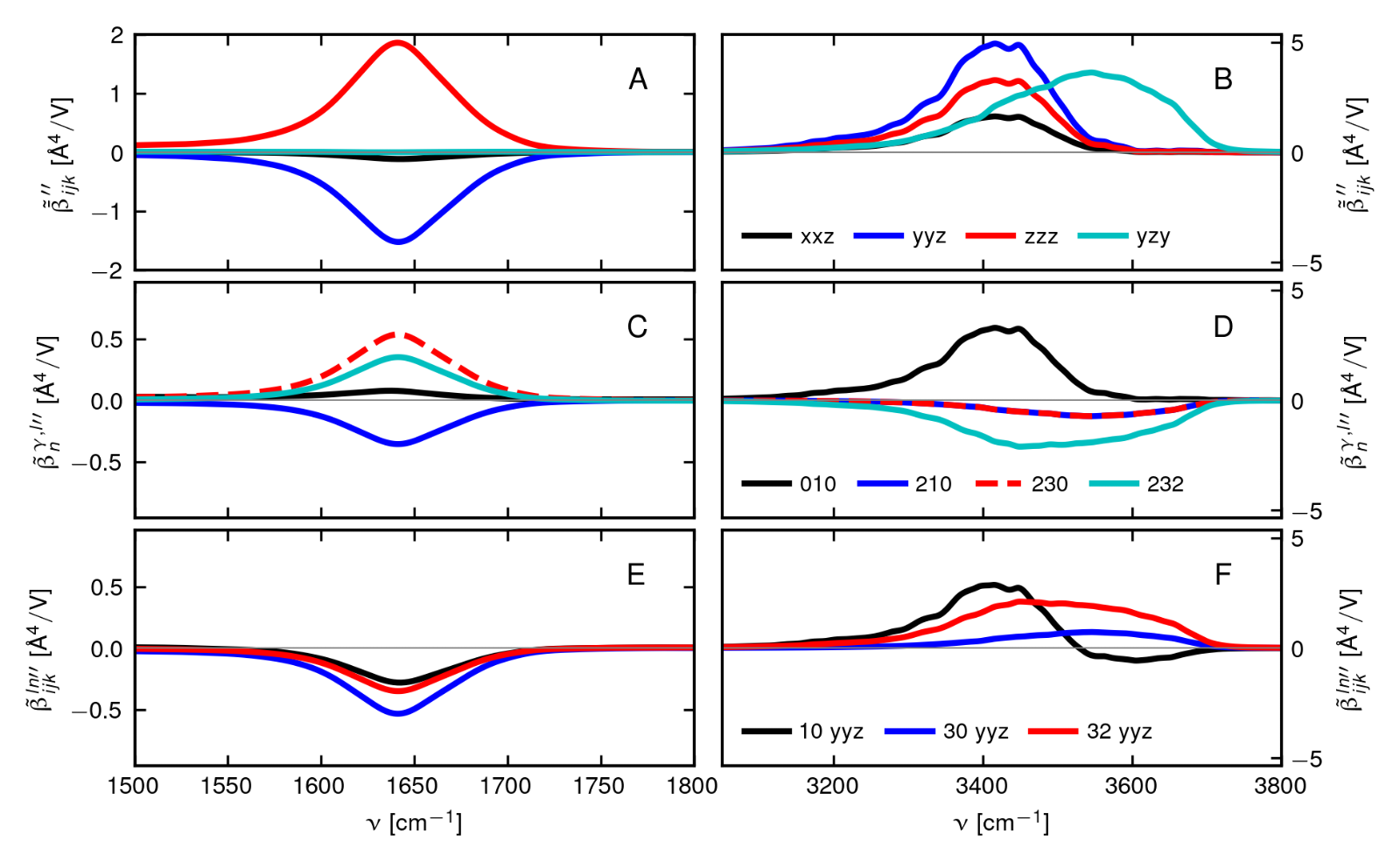}
    \caption{
    Different representation of the water molecule's nonzero molecular hyperpolarizability tensor in the molecular frame, defined in Equation \eqref{eq:def_beta_ijk}, extracted from a simulation of bulk water. 
    The different Cartesian tensor components are shown in A and B. The irreducible representation according to Equations \eqref{eq:beta010}-\eqref{eq:beta230} is shown in C and D. 
    The coefficients $\tilde{\beta}^{ln}_{yyz}$, defined in Equation \eqref{eq:S2DL_ori_ijk}, that relate the SFG spectrum to the interfacial order parameters $Q_{ln}$ defined in Equations \eqref{eq:Q10}-\eqref{eq:Q32} 
    are presented in E and F. 
    As it is visible, all components of the molecular hyperpolarizability tensor are relevant for determining orientational structure, independent of the representation. } 
    \label{fig:different_betas}
\end{figure}
The different molecular hyperpolarizabilities needed to determine the interfacial structure from the SFG spectra are presented in Figure \ref{fig:different_betas}. 
As already mentioned, we extract $\tilde{\bar{\beta}}_{abc}(\omega^\mathrm{VIS},\omega^\mathrm{IR})$ according to equation \eqref{eq:bar_beta} from a simulation of bulk water. Details about the simulation parameters are given in the Methods section of the main text. 
The Cartesian tensor elements are shown in A  \&   B. The irreducible representation, defined in Equations \eqref{eq:beta010}-\eqref{eq:beta230} are shown in C  \&   D. 
The coefficients $\tilde{\beta}^{lm}_{yyz}(\omega^\mathrm{VIS}, \omega^\mathrm{IR})$ that relate the interfacial order parameters $Q_{lm}$ to SFG spectra, as defined in Equation \eqref{eq:S2DL_ori_ijk}, are shown in E  \&   F.
There, $\beta^{10}_{ijk}(\omega^\mathrm{VIS}, \omega^\mathrm{IR})$, $\beta^{30}_{ijk}(\omega^\mathrm{VIS}, \omega^\mathrm{IR})$ and $\beta^{32}_{ijk}(\omega^\mathrm{VIS}, \omega^\mathrm{IR})$ are of the same order of magnitude.
Hence, all three order parameters $Q_{10}$, $Q_{30}$, and $Q_{32}$ need to be taken into account for SFG spectrum prediction.
\section{Computation of Effective Polarizabilities in Periodic Boundary Conditions}
\label{app:polarizability}
\begin{figure}
\centering
\includegraphics[width=1\textwidth]{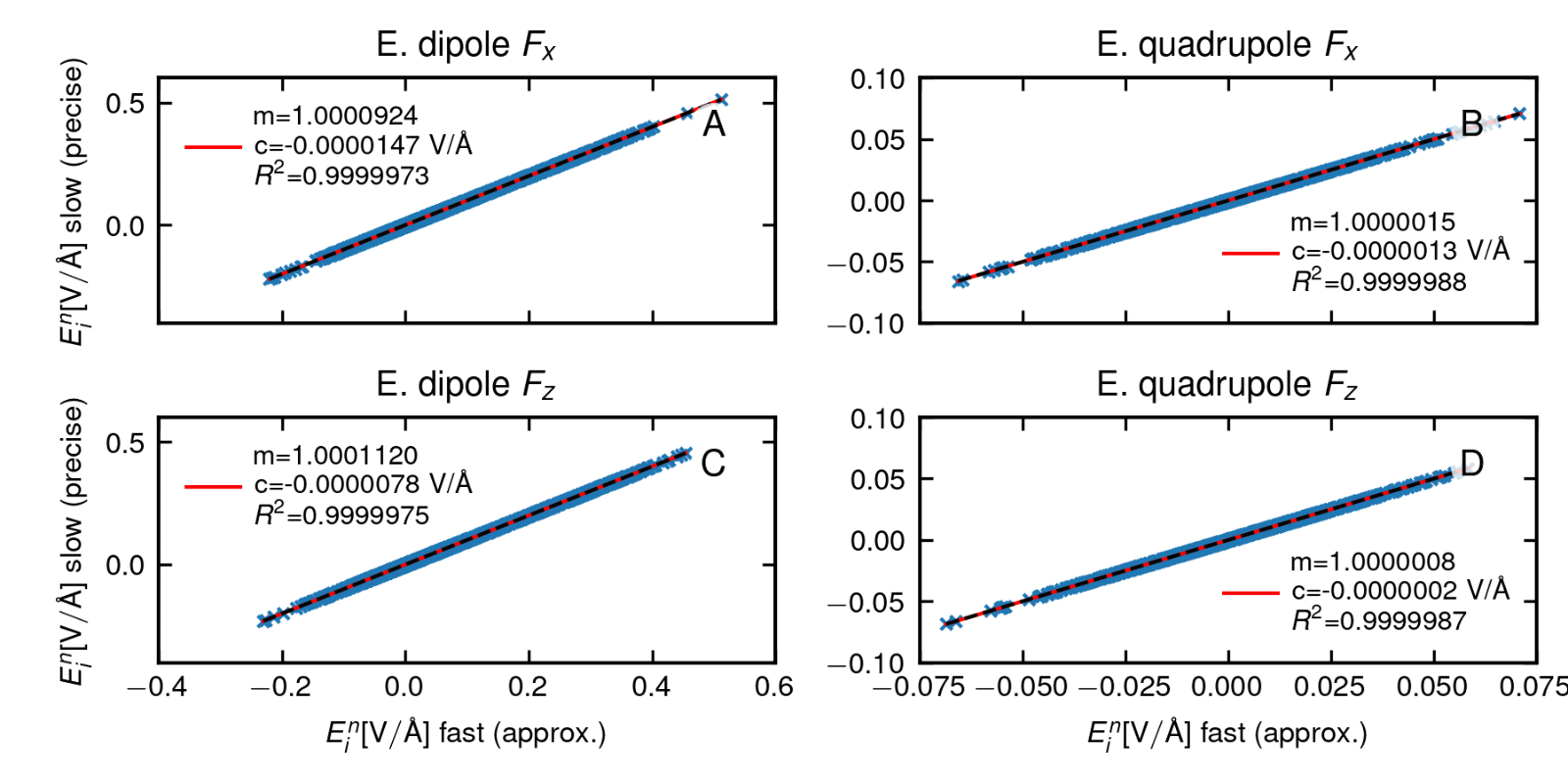}
 \caption{Benchmarking of the monopole representation of the multipolar charge distribution using Equations \eqref{eq:mono_rep_of_multi_mu} and \eqref{eq:mono_rep_of_multi_Q} and the values $d=\SI{0.05}{\angstrom}$ and $d=\SI{0.075}{\angstrom}$ for the electric dipoles and electric quadrupoles respectively. The blue dots show the data points, the black line is the function $x=x$, and the red line is the fit of the data points. The benchmark values obtained by linear regression are presented in the legend.}  
\label{fig:bench_multi_as_mono}
\end{figure}

Here, we describe how we compute the electrostatic field imposed by electric dipole and electric quadrupole moments under periodic boundary conditions and how we parameterize the molecular polarizabilities from single molecule quantum chemistry calculations.
The electrostatic field due to the set of molecular electric dipole moments can be computed via an Ewald summation \cite{frenkelUnderstandingMolecularSimulation2002}
\begin{align}
   \vec E^{n} &= -\vec{\nabla}_{ \mu^n}  H(  \vec \mu^1,   \vec \mu^2 , ...) \\
   &=  \vec E^{n,\mathrm{real}} + \vec E^{n,\mathrm{rec}} +  \vec E^{n,\mathrm{self}} 
   \label{eq:Ewald_dip} \\ 
    \vec E^{n,\mathrm{real}}  &= \sum\limits_{m=1}^N\frac{1}{4 \pi \varepsilon_0}\bigg[ \vec r^{nm} ( \vec \mu^m \cdot \vec r^{nm} ) f_2( r^{nm}, \sigma )  - \vec{\mu}^m f_1( r^{nm} , \sigma )  \bigg]\\
  \vec E^{n,\mathrm{rec}}  &= -\sum\limits_{\vec k\neq 0} \sum\limits_{m=1}^{N} \frac{ \vec k e^{-k^2 / 4 \sigma}}{V \varepsilon_0 k^2} \left(  \vec  \mu^m \cdot \vec k \right)\cos{ \left( \vec k  \cdot \vec r^{nm} \right) } \\
   \vec E^{n,\mathrm{self}} &= \frac{(\sigma / \pi)^{3/2}}{3 \varepsilon_0} \vec{\mu}^n \, ,
\end{align}
where  $\vec{\nabla}_{\mu^n} = \left( \frac{\partial}{\partial \mu^n_{x}} , \frac{\partial}{\partial \mu^n_{y}}, \frac{\partial}{\partial \mu_{iz}} \right)^T$ and $H(  \vec \mu_1,  \vec \mu_2 , ...)$ is the dipole-dipole interaction energy in a periodic system\cite{frenkelUnderstandingMolecularSimulation2002}. 
The quantity $\sigma$ is the so-called Ewald parameter related to the variance of the screening charge added in the Ewald summation. 
The functions $f_1\left(r^{ij},\sigma \right)$ and $f_2\left(r^{ij},\sigma\right)$ serve as shorthand notations for
\begin{align}
    f_1(r^{ij}, \sigma ) &= \frac{\text{erfc}\left( \sqrt{\sigma}  r^{ij} \right)} {\left(r^{ij}\right)^3} + 2 \sqrt{\frac{\sigma}{\pi}} \frac{e^{- \sigma \left( r^{ij} \right)^2}}{\left(r^{ij}\right)^2} \\
     f_2(r^{ij}, \sigma ) &= 3\frac{\text{erfc}\left( \sqrt{\sigma}  r^{ij} \right)} {\left(r^{ij}\right)^5} + 2 \sqrt{\frac{\sigma}{\pi}} \left( 2 \sigma + \frac{3}{ \left(r_{ij}\right)^2} \right) \frac{e^{- \sigma \left(r_{ij}\right)^2}}{\left(r_{ij}\right)^2} .
\end{align}
This is equivalent to the periodic summation of the dipole-dipole tensor $T^{(2)}_{ijk} (\vec r^{nm})$ \cite{frenkelUnderstandingMolecularSimulation2002}. 
The only other multipole interaction we need to implement is the electrostatic field imposed by the molecular quadrupoles described by the dipole-quadrupole tensor $T^{(3)}_{ijk} (\vec r^{nm})$.
We compute $T^{(3)}_{ijk} (\vec r^{nm})$ in real space by direct summation over periodic replicas using a cutoff $r_C=\SI{60}{\angstrom}$.
This summation is very slow and cannot be applied to the entire trajectory in a reasonable time.
To make this summation numerically feasable, we introduce an approximative but faster method and represent the molecular multipoles by monopoles, where we use two-point charges for the electric dipoles and transform
\begin{align}
    \vec \mu^n \delta( \vec r - \vec r^n) \rightarrow\lim\limits_{d\rightarrow 0} \frac{\mu^n}{2 d} \left[ \delta ( \vec r - \vec r^n - d \vec \mu^n/\mu^n ) - \delta ( \vec r - \vec r^n + d \vec \mu^n/\mu^n )\right] \, .
    \label{eq:mono_rep_of_multi_mu}
\end{align}
For the electric quadrupoles, we transform, using seven point charges, 
\begin{align}
\tensor{Q}^{n} \delta(\vec r- \vec r^n) \rightarrow \lim\limits_{d\rightarrow 0}\frac{Q^{n,\mathrm{eig}}_{kk}}{d^2} \left[ \delta(\vec r -\vec r^n - d \vec q^k) +  \delta(\vec r -\vec r^n + d \vec q^k ) - 2 \delta(\vec r - \vec r^n)\right ] \, ,
\label{eq:mono_rep_of_multi_Q}
\end{align}
where $Q^{n,\mathrm{eig}}_{kk}$ is the diagonalized quadrupole tensor in the eigenframe and $\vec q^k_{i}$ are the corresponding normalized eigenvectors. 
This is achieved by placing two particles of the same charge along each of the three eigenvectors of the quadrupole tensor and a seventh particle at the molecular center. The central particle carries a charge equal in magnitude and opposite in sign to the sum of the six other particle charges, thereby canceling the net molecular monopole. 
Once we transform the dipole and quadrupole densities into a monopole density, we can use the Ewald summation implementation in OpenMM \cite{eastmanOpenMM8Molecular2024}. 
The more precise method, referred to as the slow method, computes the E-field from electric dipoles using the self-written Ewald summation defined in Equation~\eqref{eq:Ewald_dip}, and includes the E-field from the electric quadrupoles, predicted by the tensor $T^{(3)}_{ijk}(\vec{r}^{nm})$, with a cutoff of $\SI{60}{\angstrom}$. 
The approximate approach, referred to as the fast method, computes the electrostatic field from a monopole density constructed via Equations ~\eqref{eq:mono_rep_of_multi_mu} and~\eqref{eq:mono_rep_of_multi_Q} using the implementation of the Ewald summation in OpenMM. 
We optimize the values for $d$ so that the difference between the local electric fields acting on the molecular centers predicted by the slow and fast method is minimal. 
We repeat the same with the electric quadrupoles.
The optimal parameters are $d=\SI{0.05}{\angstrom}$ for the electric dipole density and $d=\SI{0.075}{\angstrom}$ for the electric quadrupole density. 
In this test, the accuracy is satisfactory regardless of whether the regular Ewald summation algorithm or the particle mesh Ewald algorithm (PME) \cite{frenkelUnderstandingMolecularSimulation2002} is used.
However, since computing the electric field is not the performance bottleneck in our algorithm, we opt for the regular Ewald summation, as it has fewer control parameters. 
We compare the predictions from both methods of the local E-field acting on the molecular centers defined in Equation \eqref{eq:En} in Figure \ref{fig:bench_multi_as_mono}.
The respective multipoles are taken from the solution of the SCF equations
\eqref{eq:scf_mu_time_2} and \eqref{eq:scf_Q_time}, with the external test fields $F_{x/z}=\SI{1}{V/\angstrom}$.
We benchmark the prediction with a linear regression using the fit formula
\begin{align}
    f(x) = m x + c \, .
\end{align}
Hence, the monopole density would be a perfect representation if we had $R^2=1$, $m=1$, and $c=0$, where $R^2$ is the coefficient of determination. As shown in Figure \ref{fig:bench_multi_as_mono}, we are close to an ideal representation. 
Hence, we can employ the fast method. 
We solve the SCF Equations \eqref{eq:dip_scf} and \eqref{eq:quad_scf}, for each frame numerically using the iterative procedure 
\begin{align}
    \mu^{n,l+1}_i(\vec\Omega) &= \gamma \alpha^{n,\mathrm{DD}}_{ij} (\vec \Omega) \left( E_j^{n}\left[\vec\mu^{1,l}(\vec\Omega),...\vec\mu^{N_\mathrm{mol},l}(\vec\Omega),\vec Q^{1,l}(\vec\Omega),...\vec Q^{N_\mathrm{mol},l}(\vec\Omega)\right] +F_j^\mathrm{TEST}  \right) + (1-\gamma)\mu^{n,l}_i(\vec\Omega)  \\
     Q^{n,l+1}_{ij}(\vec\Omega) &= \gamma \alpha^{n,\mathrm{QD}}_{ijk} (\vec \Omega) \left( E_k^{n}\left[\vec \mu^{1,l}(\vec\Omega),...\vec\mu^{N_\mathrm{mol},l}(\vec\Omega),\vec Q^{1,l}(\vec\Omega),... \vec Q^{N_\mathrm{mol},l}(\vec\Omega)\right]+F_k^\mathrm{TEST}  \right) + (1-\gamma) Q^{n,l}_{ij}(\vec\Omega) \, .
\end{align}
Here $F_i^\mathrm{TEST}$ is the test field defined in Equations \eqref{eq:scf_mu_time_2} and \eqref{eq:scf_Q_time_2}, and
the update parameter $\gamma$ is set to $\gamma=0.75$ and $l$ denotes the iteration step. We set $F^\mathrm{TEST}=\SI{1}{V\angstrom^{-1}}$. However, as the induced multipoles are linear in the test field the actual value does not matter. 
We verify convergence by computing the maximum distance between the induced molecular electric dipole moments between two iteration steps,  divided by the average induced molecular electric dipole moment, defining the cost function
\begin{align}
\mathrm{cost}_l=\frac{\max{|{\vec{\mu}^{n,l}}-\vec{\mu}^{n,l-1}|}}{\frac{1}{N_\mathrm{mol}} \sum\limits_n^{N_\mathrm{mol}} |\vec{\mu}^{n,l}|} \, ,
\label{eq:cost}
\end{align}
where $l$ denotes the iteration step and $n$ the molecule. 
The function $\max x^n$ extracts the maximal value of $x^n$ from all molecules.
This quantity is also testing the convergence of the electric quadrupoles, as both are linear functions in the local E-field $E_i^{n}$.
\begin{figure}
\centering
\includegraphics[width=0.4\textwidth]{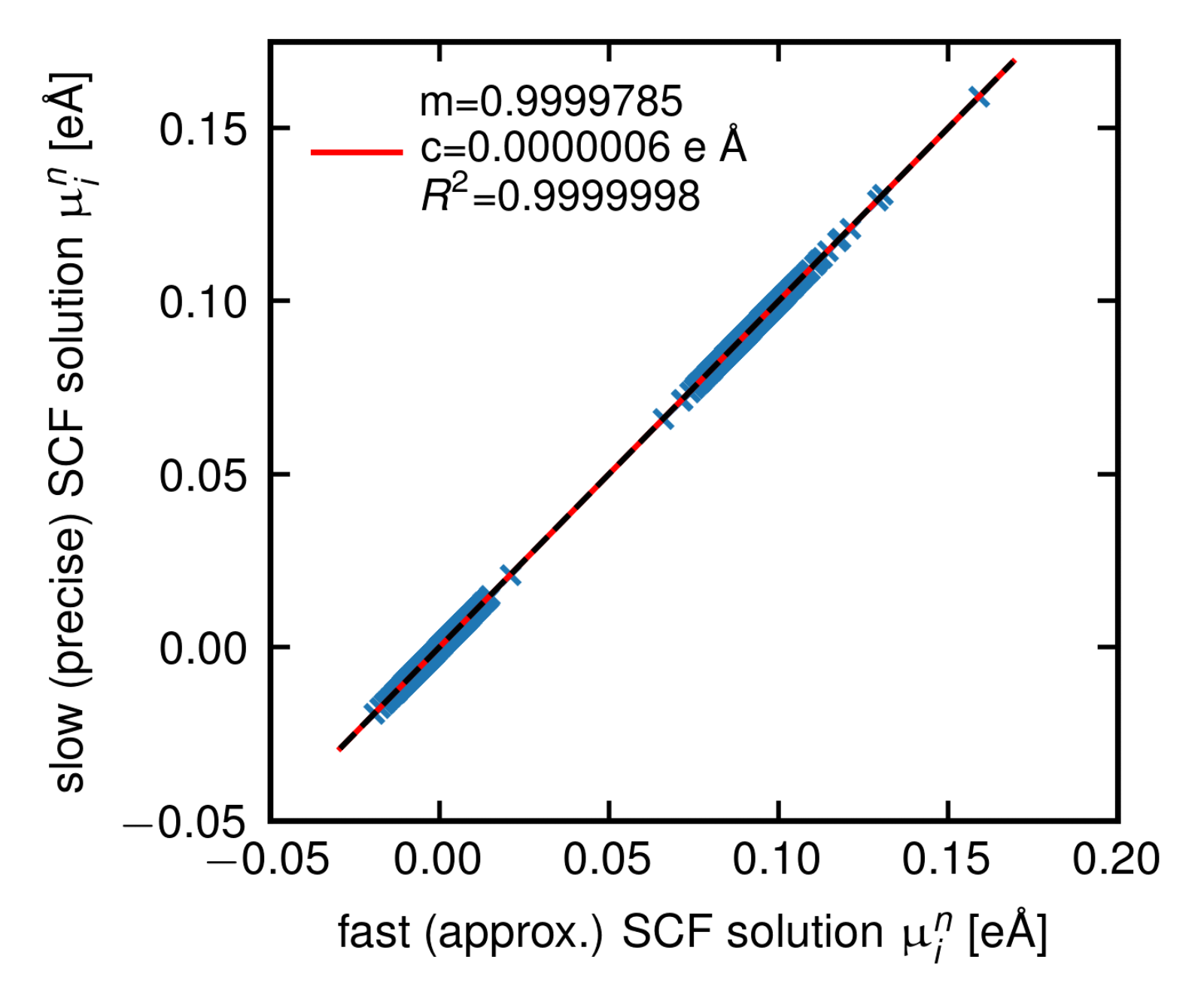}
\caption{Estimating the upper bound of the error induced by the fast method. The solution of the SCF Equations~\eqref{eq:scf_mu_time} and~\eqref{eq:scf_Q_time} for the frame with the highest final cost value ($\mathrm{cost}_{50} = 1.5 \times 10^{-4}$) is compared to the solution obtained using the slow but more precise method, which was iterated until $\mathrm{cost}_l < 10^{-6}$. As shown, the agreement is very good, demonstrating that even the frame with the highest final cost value provides an accurate estimate of the induced electric dipoles and quadrupoles. The benchmarking is performed analogously to Figure \ref{fig:bench_multi_as_mono}.} 
\label{fig:benchmark_methods}
\end{figure}
With the fast method, the cost function plateaus at values in the range of $10^{-4}$ to $10^{-5}$, depending on the molecular configuration, which is of the same order of magnitude as the error in the prediction of the local electric field presented in Figure \ref{fig:bench_multi_as_mono}. 
We iterate each frame until $\mathrm{cost}_l<10^{-4}$, or alternatively for 50 iteration steps.
The highest final cost value of $\mathrm{cost}_{50}$ observed in all frames is $\mathrm{cost}_{50}=1.5 \times 10^{-4}$. 
To estimate an upper bound on the error introduced by the fast method, we compare the electric dipoles obtained from the frame with the highest final cost value of the entire data to those predicted by the slow method, which was iterated until $\mathrm{cost}_l < 10^{-6}$. 
We compare the sets of induced dipoles $\mu^n_i$ predicted by both methods in Figure~\ref{fig:benchmark_methods}, where we benchmark them following the same procedure as in Figure~\ref{fig:bench_multi_as_mono}.
As evident, even the frame with the largest final cost using the fast method yields predictions that agree well with those from the slow and more precise approach.
The values of $m$, $c$, and $R^2$ in Figures~\ref{fig:bench_multi_as_mono} and~\ref{fig:benchmark_methods} are of the same order of magnitude, indicating that the SCF equation is well conditioned, which means that an error in the electric fields acting on the molecular centers does not lead to an amplified error in the prediction of the electric multipoles. The previous benchmark presented in Figure \ref{fig:bench_multi_as_mono} uses the same molecular configuration $\vec \Omega$.
On the other hand, the pure electric dipole contribution $\tilde{S}^{(2,\mathrm{DD})}_{ijk}(\omega^\mathrm{IR})$ does not require computation of the electric field imposed by the electric quadrupoles and the electric field imposed by the electric dipoles was calculated using the self-written electric dipole Ewald summation in Equation \eqref{eq:Ewald_dip}. 
These were iterated until the convergence criterion 
$
\frac{1}{N_\mathrm{mol}}\sum\limits_n^{N_\mathrm{mol}} \left| \mu^{n,l+1} - \mu^{n,l} \right|^2 < \SI{1e-12}{e^2 \angstrom^2}
$ was satisfied. 
Here, the external field was set to $F_i^\mathrm{TEST}= \SI{0.1}{V/\angstrom}$ and henceforth this criterion corresponds to an average agreement to approximately the eighth significant digit.
This simple convergence criterion was later replaced with the cost function \eqref{eq:cost} to make it independent of the amplitude of the applied external field.
Now we describe the parameterization of the molecular polarizabilities defined in Equation \eqref{eq:polarizabilities}.
We express the molecular polarizabilities in the molecular Eckart frame  $\bar{\alpha}^{(NM)}_{j_1 ... j_{M+N}}$ as a function of the symmetry coordinates $S^n_1,S^n_2,S^n_3$, which specify the displacements of the nuclei within the molecular frame \cite{adler-goldenFormulasTransformingInternal1985,reyTransformationInternalCoordinates1998}, i.e.
\begin{align}
    \alpha^{n, NM}_{i_1...i_{N+M}}&=\bar{\alpha}^{n, NM }_{j_1 ... j_{M+N}}(S^n_1,S^n_2,S^n_3)  D^n_{i_1 j_1}... D^n_{i_{M+N} j_{M+N}}  \, .
\end{align}
Here, we rotate the tensor from the molecular to the laboratory frame using the elements of the direction cosine matrix $D^n_{ij}$, defined in Equation \eqref{eq:direction_cosine}.
\begin{figure}
\centering
\includegraphics[width=0.5\textwidth]{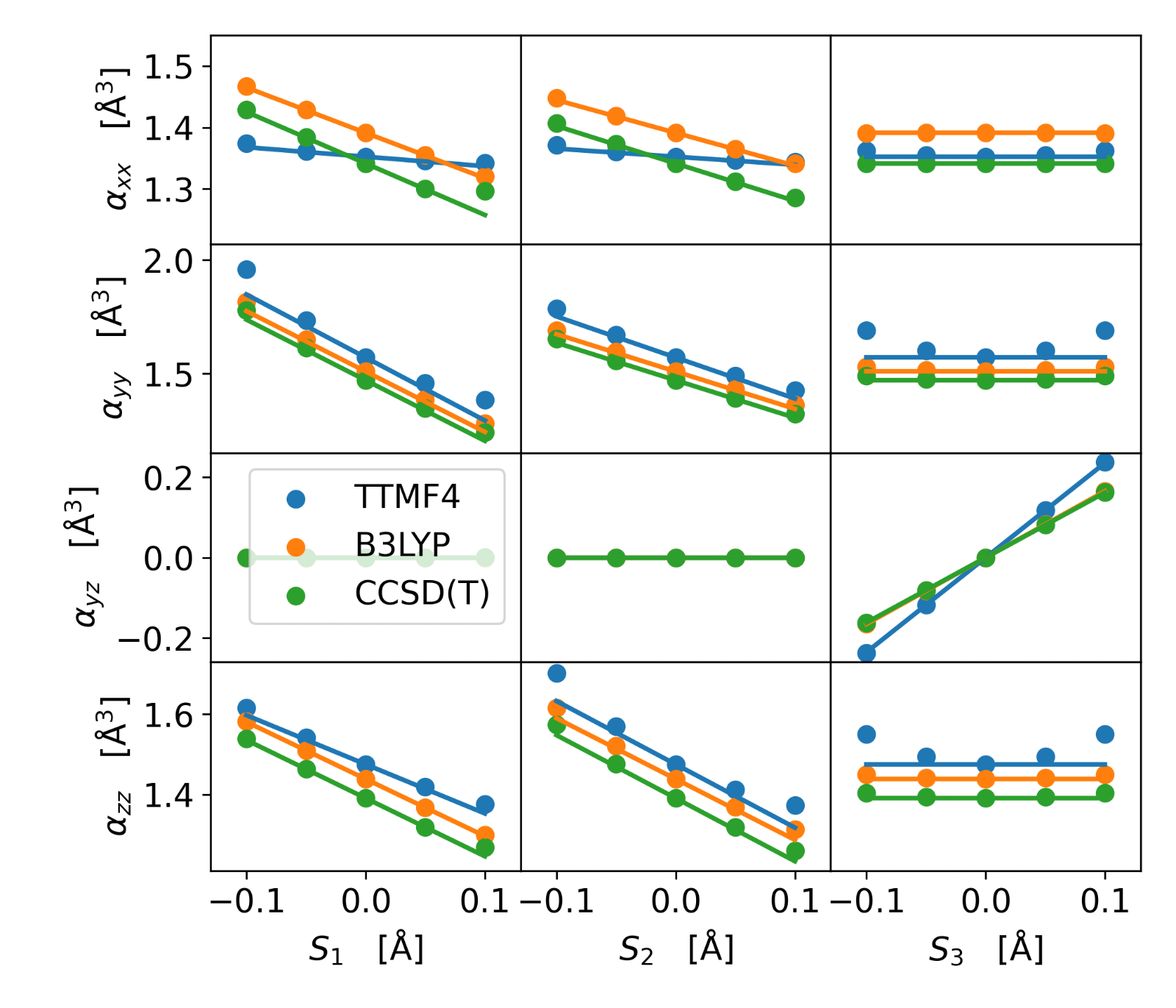}
 \caption{Polarizability tensor of the water molecule in the molecular Eckart frame as a function of the symmetry coordinates $S_1,S_2,S_3$. Three different methods of calculating the polarizability tensor are compared. The TTM4F \cite{burnhamVibrationalProtonPotential2008} model is compared to the CCSD(T) and B3LYP calculations with the aug-cc-pVTZ basis set. 
 The points correspond to the data points, and the lines to the first-order Taylor expansion of the molecular polarizability. }  
\label{fig:raman_tensor}
\end{figure} 
To capture the functional dependence of the polarizability tensor on the set of symmetry coordinates $S^n_1, S^n_2, S^n_3$, we perform a first-order Taylor expansion of the polarizability tensors
\begin{multline}    
    \bar\alpha^{n,\mathrm{NM}}_{i_1 i_2, ..i_{N+M}}(S^n_1, S^n_2, S^n_3)  \approx \bar\alpha^{n,\mathrm{NM}}_{i_1 i_2, ..i_{N+M}}(0,0,0) \\+ S^n_1 \frac{\partial}{\partial  S_1}\bar\alpha^{n,\mathrm{NM}}_{i_1 i_2, ..i_{N+M}}(S_1, 0,0)\big|_{S_1=0} + S^n_2\frac{\partial}{\partial  S_2}\bar\alpha^{n,\mathrm{NM}}_{i_1 i_2, ..i_{N+M}}(0, S_2, 0)\big|_{S_2=0} + S^n_3\frac{\partial}{\partial  S_3}\bar\alpha^{n,\mathrm{NM}}_{i_1 i_2, ..i_{N+M}}(0,0,S_3)\big|_{S_3=0} \, .
    \label{eq:TaylorPol}
\end{multline}
The numerical derivatives of the generic function $f(x)$ can be computed with the use of the central differences scheme
 \begin{align}
  \frac{f(\Delta x/2) - f(-\Delta x/2)}{ \Delta x} = 
  \frac{\partial}{\partial x}f(x)\big|_{x=0} + \frac{\Delta x^2}{24}  \frac{\partial^3}{\partial x^3}f(x)\big|_{x=0} + ...\approx  \frac{\partial}{\partial x}f(x)\big|_{x=0}  \, ,
     \label{eq:differentiate}
 \end{align}
where we plugged in the Taylor expansion of $f(x)$ around $x$ to relate it to the analytic derivative. 
We set $\Delta S_i = \SI{0.05}{\angstrom}$ for computing the derivatives in the Taylor expansion in Equation~\eqref{eq:TaylorPol} and we employ the central difference scheme.
The electric dipole and electric quadrupole polarizabilities are computed with the Gaussian 16 software \cite{frischGaussian16Rev2016}. 
The electric dipole - electric dipole polarizability $\alpha^{n,\mathrm{DD}}_{ij}$ can be directly predicted by the Gaussian~16 software.  
We compare the prediction of  $\alpha^{n,\mathrm{DD}}_{ij}$ with the modified TTMF4 model included in MB-Pol \cite{burnhamVibrationalProtonPotential2008,babinDevelopmentFirstPrinciples2013} and single-molecule quantum chemistry predictions on the level of B3LYP/aug-cc-pVTZ and CCSD(T)/aug-cc-pVTZ. 
The TTMF4 model fails to accurately capture the dependence of $\alpha^{n,\mathrm{DD}}_{xx}$ on $S_1$. 
The differences between the B3LYP/aug-cc-pVTZ and CCSD(T)/aug-cc-pVTZ levels in predicting the polarizability tensor are primarily characterized by a small and constant shift. However, the dependence of the polarizability tensor on $S_1$, $S_2$, and $S_3$ is nearly identical, as evidenced by the orange and green lines that remain almost parallel throughout Figure~\ref{fig:raman_tensor}. The dots represent the numerical values, and the straight lines represent the first-order Taylor expansion, indicating that the dependence of $\alpha^{n,\mathrm{DD}}_{ij}$ on the set of symmetry coordinates is quite linear for moderate displacements. 
The components of the electric quadrupole - electric dipole polarizability tensor defined in Equation \eqref{eq:polarizabilities}, are determined by the derivative of the electric quadrupole moment with respect to an external field
\begin{align}    \bar{\alpha}^{n,\mathrm{QD}}_{ijk} = \frac{\partial}{\partial F_k} Q^n_{ij}|_{\vec F=0} \, .\label{eq:compute_quad_pol}
\end{align}
We compute the derivative in Equation \eqref{eq:compute_quad_pol} numerically by applying a finite external field $\Delta F_i=\SI {0.514}{V\angstrom^{-1}}$. 
This amplitude is set relatively large, as the response of the electric quadrupole tensor to external fields is relatively weak, and we want to minimize errors due to the finite precision of the electric quadrupoles predicted by the Gaussian 16 software.
To test for errors due to the nonzero value of the external field, we compute the electric quadrupole contribution to the SFG spectrum $\tilde{S}^{(2,\mathrm{Q})}_{ijk}(\omega^\mathrm{IR})$ using Equations  \eqref{eq:chi2Q_fdt_corr_func} and \eqref{eq:S2Q_bulk} once with $\bar{\alpha}^{n,\mathrm{QD}}_{ijk}$ extracted using the forward differences scheme defined by
\begin{align}
    \frac{f(\Delta x)-f(0)}{\Delta x} = 
  \frac{\partial}{\partial x}f(x)\big|_{x=0} + \frac{\Delta x}{2}  \frac{\partial^2}{\partial x^2}f(x)\big|_{x=0} + \frac{\Delta x^2}{6}  \frac{\partial^3}{\partial x^3}f(x)\big|_{x=0} + ...\approx  \frac{\partial}{\partial x}f(x)\big|_{x=0}
\label{eq:forward}
\end{align}
and once using central differences defined in Equation \eqref{eq:differentiate}. 
As evident from comparing Equations \eqref{eq:differentiate} and \eqref{eq:forward} in the central differences scheme, higher-order derivatives,
which correspond to errors due to the nonzero field, are largely reduced.
As visible in Figure \ref{fig:central_vs_forward}, both methods produce identical SFG spectra, demonstrating that the nonzero value of $\Delta F_i$ does not produce any artifacts. 
In the remainder of this work, the tensor $\alpha^{n,\mathrm{QD}}_{ijk}$ extracted using central differences is used.
\begin{figure}
\centering
\includegraphics[width=1\textwidth]{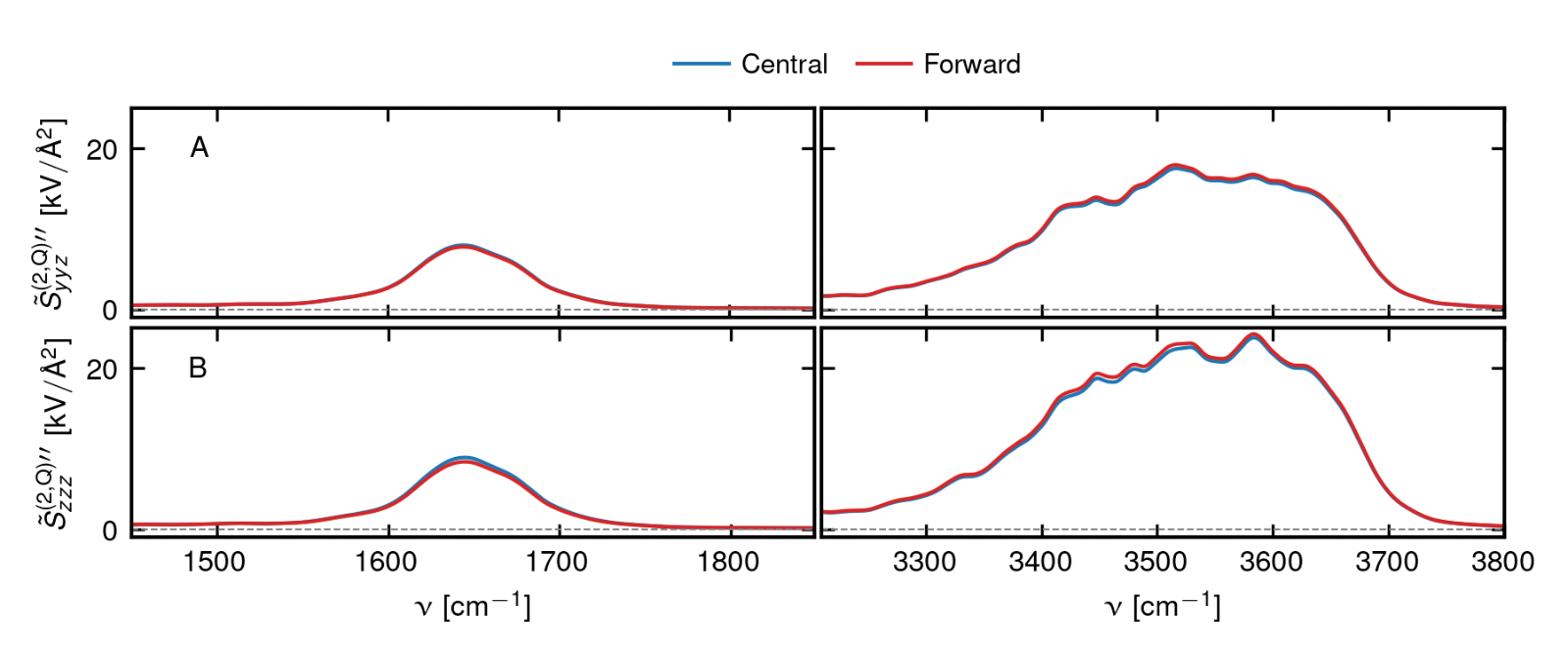}
 \caption{Comparison of the electric quadrupole contribution predicted from a simulation of bulk water using Equations \eqref{eq:chi2Q_fdt_corr_func} and \eqref{eq:S2Q_bulk} using the electric quadrupole - electric dipole polarizabilities computed via central and forward differences, defined in Equations \eqref{eq:differentiate} and \eqref{eq:forward}, respectively. 
 The yyz and zzz tensor components of the SFG spectrum $\tilde{S}^{(2,\mathrm{Q})}_{ijk}(\omega^\mathrm{IR})$ are shown in A and B, respectively. As is visible, the central and forward differences schemes produce identical SFG spectra, indicating that errors due to the nonzero field in the extraction of the electric quadrupole - electric dipole polarizability tensor are negligible.}
\label{fig:central_vs_forward}
\end{figure} 
\section{Response Functions in Periodic and Non-Periodic Systems}
\label{app:pbc}
Stern and Feller showed \cite{sternCalculationDielectricPermittivity2003} that in a system with periodic boundary conditions in all three dimensions and translational invariance in the $x,y$-plane, named periodic system, the external field $F_i$ under tinfoil boundary conditions is given by
\begin{align}
\tilde F_i (\omega)= \left( \delta_{ix} + \delta_{iy} \right) \tilde E_i (\omega)+ \varepsilon_0^{-1} \delta_{iz} \left[ \tilde D_z(\omega) - \tilde P_z(\omega) / V \right],
\label{eq:Fext_PBC}
\end{align}
where $\delta_{ix}$, $\tilde E_i (\omega)$, $\tilde D_z (\omega)$, $\tilde P_z (\omega)$ and $V$ are the Kronecker-delta, the E-field, the D-field, the total dipole moment and the volume, respectively. 
This relation holds in general beyond the electric dipole approximation \cite{bonthuisDielectricProfileInterfacial2011,gekleNanometerResolvedRadioFrequencyAbsorption2014}. 
Now we consider another system, which is not periodic in the $z$ dimension, but only in the $xy$ dimension, which corresponds to our interface system. 
The latter non-periodic system coincides with the system periodic along $z$ when an infinite amount of vacuum is added along the $z$-direction, from which follows that the field from the periodic images goes to zero ($\delta_{iz} \varepsilon_0^{-1} \tilde{P}_z(\omega)/V\rightarrow 0)$ \cite{locheCommentHydrophobicSurface2019a}.
We define the linear response function of the non-periodic system  as $\tilde{\varphi}[O(\cdot), P_i(\cdot), \omega]$. 
However, due to performance reasons, it is not feasible to simulate gigantic volumes. 
Consequently, we prefer to simulate a smaller volume and relate the response function $\tilde{\varphi}[O(\cdot), P_i(\cdot), \omega]$ to the response function of the periodic system, by
\begin{align}
\tilde{\varphi}^\mathrm{PBC}[O(\cdot), P_i(\cdot), \omega] \tilde F_i(\omega) =\tilde{\varphi}[O(\cdot), P_i(\cdot), \omega] \left[ \tilde F_i(\omega) +  \delta_{iz} \varepsilon_0^{-1} \tilde P_z(\omega) / V \right] \, .
\end{align}
We can insert the linear response relation
\begin{align}
\varepsilon_0^{-1} \tilde{P}_z (\omega)=  \tilde{S}^{(1,\mathrm{PBC})}_{zz}(\omega) \tilde{F}_z(\omega) ,
\end{align}
where 
\begin{align}
    \tilde{S}^{(1,\mathrm{PBC})}_{zz}(\omega) = \varepsilon_0^{-1} \tilde{\varphi}^\mathrm{PBC}[P_z(\cdot), P_z(\cdot), \omega] +  \varepsilon_0^{-1} A_{zz} \,
\end{align}
is the linear response of the total systems dipole density to an external field and $A_{zz}=\frac{\partial}{\partial \mathcal{F}_z} P_z\left(\vec \Omega  \right) \bigg|_{\mathcal{F}_z=0} \,$ is the effective electric dipole - electric dipole polarizability of the whole system. 
The response function of the non-periodic system is given by
\begin{align} \tilde{\varphi}[O(\cdot), P_i(\cdot), \omega]= c^{\mathrm{PBC}}_i \tilde{\varphi}^\mathrm{PBC}[O(\cdot), P_i(\cdot), \omega] \, ,
\end{align}
where the frequency-dependent factor 
\begin{align}
c^{\mathrm{PBC}}_i(\omega) = \frac{1}{1 + \delta_{iz}  \tilde{S}^{(1,\mathrm{PBC})}_{zz} (\omega) / V}
\label{eq:pbc_corr_fac}
\end{align}
 serves as a periodic boundary correction if the system of interest is non-periodic, but simulated under periodic boundary conditions. 
 This correction factor is also applied to the prediction of the local field factors $f^{n}_{ij} (t)$ defined in Equation \eqref{eq:def_local_field_factor}. 
 We can relate the local field factor \eqref{eq:def_local_field_factor} in the non-periodic system $f^{n}_{ij} (t)$ to the one which is numerically predicted in a periodic system $f^{n,\mathrm{PBC}}_{ij} (t)$ by 
 \begin{align}
     f_{ij}^n(t) = f^{n,\mathrm{PBC}}_{ij} (t)c_j^\mathrm{PBC}(\infty) \, ,
 \end{align}
 where 
 \begin{align}
     c_i^\mathrm{PBC}(\infty)=\frac{1}{1 + \delta_{iz} \varepsilon_0^{-1}  A_{zz} / V} \, .
 \end{align}
 is the off-resonant periodic boundary correction factor.
\section{Smoothing Procedure and Mean Substraction}
\begin{figure}
    \centering
    \includegraphics[scale=1]{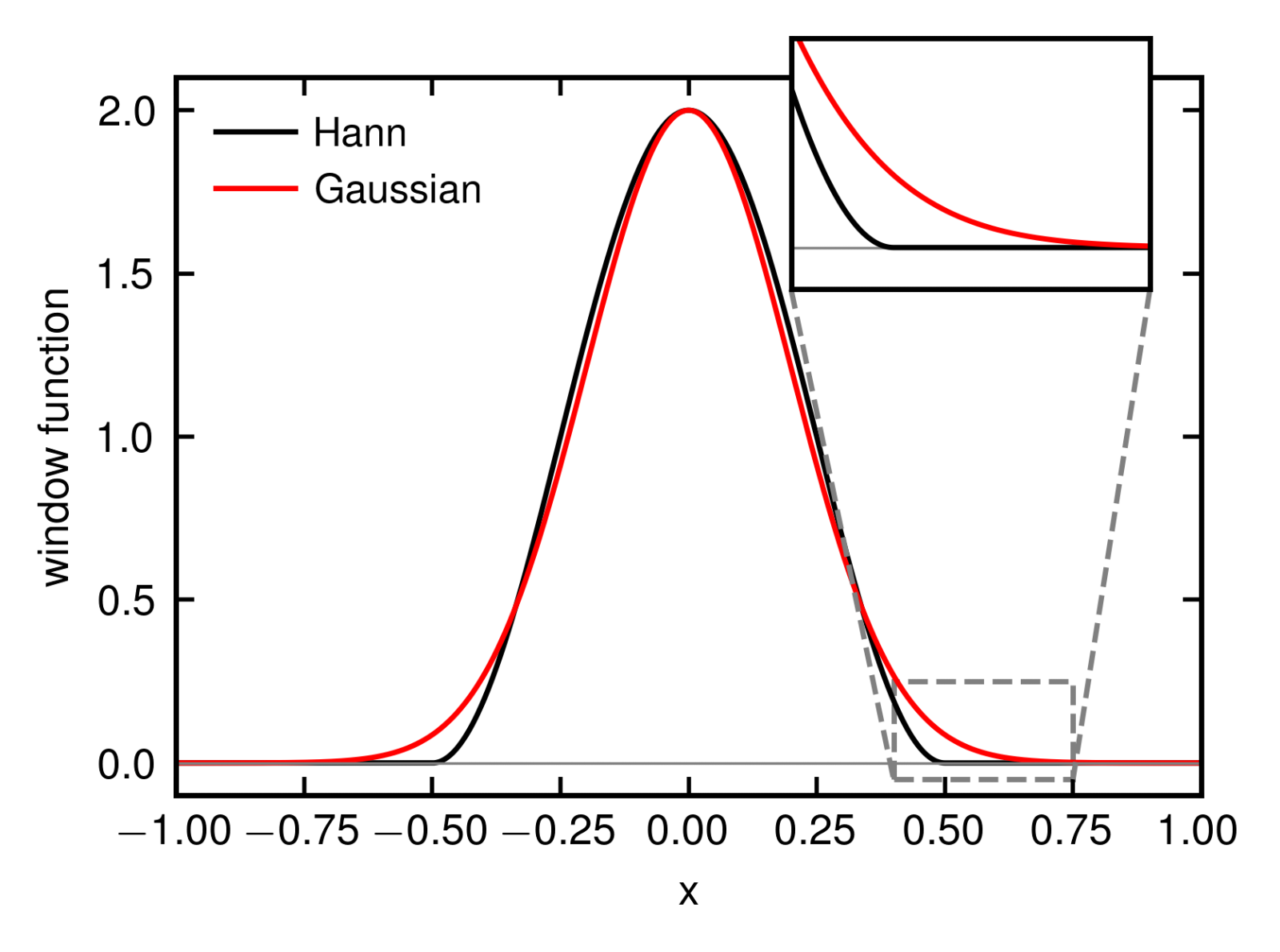}
    \caption{Comparison between the Hann and Gaussian window functions, defined in Equations \eqref{eq:Hann} and \eqref{eq:normal}, respectively. Here $\Delta x=1$ and $\sigma = \frac{\Delta x}{\sqrt{8 \pi}}\approx 0.2$ is chosen to assure that both distributions have the same peak heights.
    The Hann window is only nonzero over an interval of $\Delta x$ whereas the Gaussian window is strictly speaking everywhere nonzero. The inset shows a segment of the tail of the distributions. }    \label{fig:hann_vs_gauss}
\end{figure}
\begin{figure}
    \centering
    \includegraphics[scale=1]{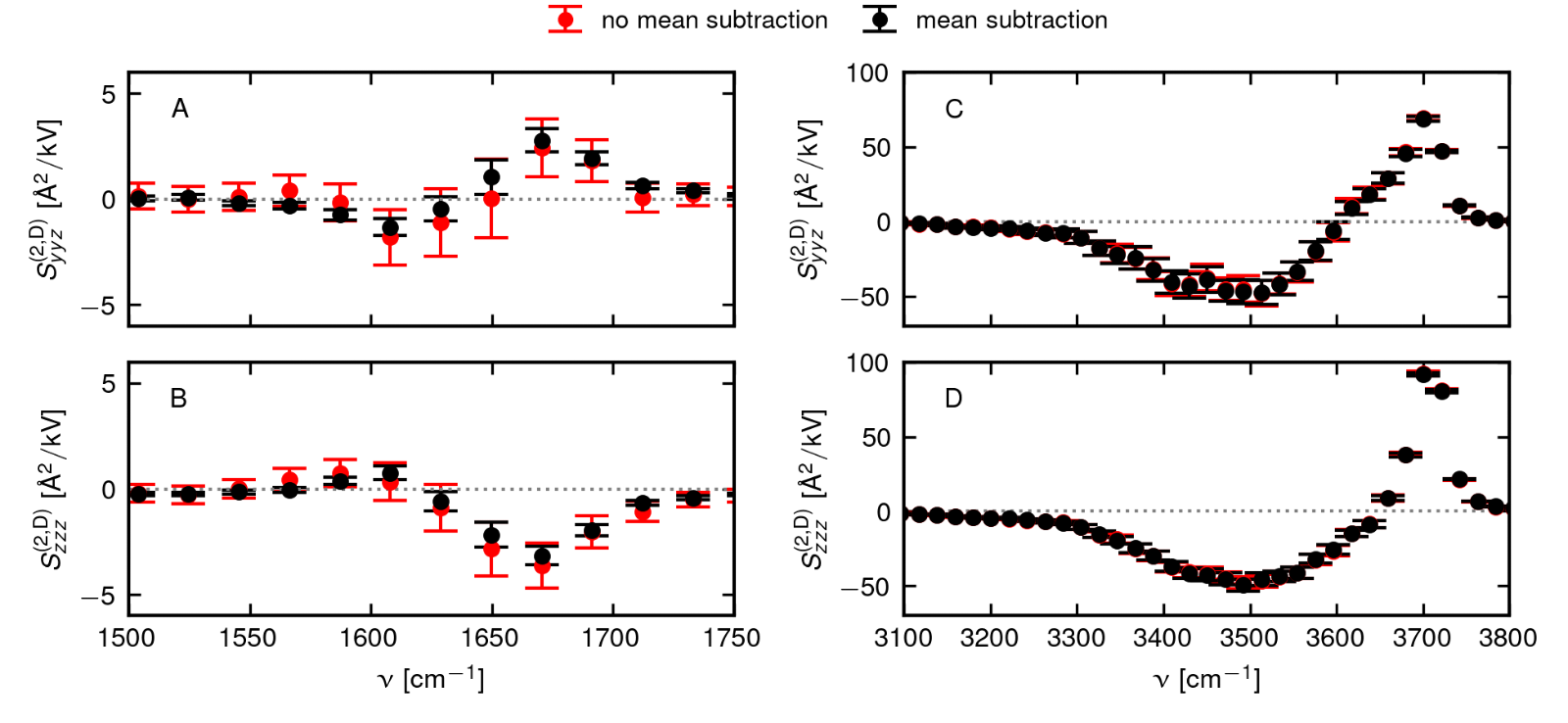}
    \caption{The electric dipole contribution to the SFG signal $\tilde{S}^{(2,\mathrm{D})}_{ijk}(\omega^\mathrm{IR})$ defined in Equation \eqref{eq:S2ijk_beta}.
   We compare the spectra calculated with 
   the left and right side of the Expression
   \eqref{eq:adij_mun_mean_subs}-\eqref{eq:aqijk_Qn_mean_sub}.
   The red dots results from the computation of the effective polarizability profile as defined in Equation \eqref{eq:adij_mun} and the black dots from the computation of the effective polarizability profile as described in Equation \eqref{eq:adij_mun_mean_subs}. The error bars correspond to a 95\% confidence interval, estimated using Equation \eqref{eq:yerr}.
   We show the bending contribution in A  \&   B and the stretching contribution in C  \&   D. 
   The subtraction of the mean leads to an increase in the signal-to-noise ratio without significant alterations of the spectral line shape.}
    \label{fig:mean_subs}
\end{figure}
Here we summarize the smoothing procedure used for the presented spectra. All spectra $\tilde{s}\left(\omega^\mathrm{IR} \right)$ are smoothed by convolution, i.e.
\begin{align}
    \tilde{s}_\mathrm{smooth} \left( \omega^\mathrm{IR} \right) = \int\limits_{-\infty}^\infty \mathrm{d} \omega  \tilde{s}\left(\omega \right) w \left( \omega - \omega^\mathrm{IR} \right) \,
\end{align}
where $w \left( \omega^\mathrm{IR} \right)$ is a window function \cite{blackmanMeasurementPowerSpectra1958}. 
Here we employ the Hann window function \cite{blackmanMeasurementPowerSpectra1958} defined by 
\begin{align}
    w^\mathrm{Hann}\left( x , \Delta x\right) = \frac{1}{\Delta x} \Pi\left( \frac{x}{\Delta x} \right)\left[ 1 +  \cos{\left(  \frac{2 \pi x}{\Delta x} \right)} \right] \, , 
    \label{eq:Hann}
\end{align}
where $\Pi\left( \frac{x}{\Delta x} \right)$ is the rectangular function defined in Equation \eqref{eq:rect_func}.
The Hann window function, for $\Delta x=1$, is in Figure \ref{fig:hann_vs_gauss} compared to a Gaussian window function 
\begin{align}
    w^\mathrm{Gauss}(x,\sigma)=  \frac{1}{\sqrt{2 \pi \sigma^2}} e^{-\frac{x^2}{2 \sigma^2}} \, ,
    \label{eq:normal}
\end{align}
where $\sigma$ is the standard deviation and chosen so that the peak heights are equal $\sigma = \frac{\Delta x}{\sqrt{8 \pi}}$. 
Both have a similar lineshape, but the Hann window function is only nonzero in the interval $-\Delta x/2<x<\Delta x/2$, whereas the Gaussian window function is nonzero everywhere.
The electric dipole contributions to the SFG spectra, presented in as Figure \ref{fig:big_multipole} A–D, consist of closely separated positive and negative peaks.
As a result, a long-tailed window function causes signal cancellation. 
Therefore, we prefer the Hann window function over the Gaussian window function in this case.
We use a window size of $\Delta \omega^\mathrm{IR} = \SI{8.0}{THz}$ for the polarization contributions to the SFG spectrum, the dielectric profile and the dielectric constant.
Hence, $\Delta\omega^\mathrm {IR}$ is smaller than the spacing between the closest peaks ($\approx\SI {11}{cm^{-1}}$) that can be found in the electric dipole contribution of the bending band.
A window size of $\Delta \omega^\mathrm{IR}=\SI{8.7}{THz}$ is chosen for the calculation of the magnetic dipole susceptibility, where the fluctuation-dissipation relations are presented in Equations \eqref{eq:chi2m_approx} and \eqref{eq:chi2m_exact}. 
All position-resolved spectra are binned using a bin size of $\SI{0.5}{\angstrom}$. Furthermore, all second-order response profiles are smoothed with a Gaussian filter with a standard deviation of $\sigma=\SI{0.25}{\angstrom}$ in space.  In position space, where there are no closely spaced double peaks, we use a Gaussian window function, as it is commonly applied to position-based quantities, such as a continuous particle density \cite{willardInstantaneousLiquidInterfaces2010} or Gaussian charges \cite{zhangDeepPotentialModel2022}.
In the calculation of the second-order response profiles $\tilde{s}^{(2,\beta)}_{ijk}(z,\omega^\mathrm{IR})$ we slice each trajectory into pieces of average length $\SI{25}{ps}$, compute the spectrum for each slice separately, and average over these spectra. 
To increase the signal-to-noise ratio, we effectively remove the contribution to the SFG spectra due to the moving mean of the molecular point polarizabilities. We achieve this by replacing the effective polarizability profiles according to
\begin{align}
    a^{n,\mathrm{D}}_{ij}(z,t) &= \frac{1}{L_x L_y} \sum\limits_n^{N_\mathrm{mol} } \alpha^{n,\mathrm{DD}}_{ik}(t) f^n_{kj}(t) \delta \left[ z-z^n(t) \right]\rightarrow \frac{1}{L_x L_y} \sum\limits_n^{N_\mathrm{mol} } \left[\alpha^{n,\mathrm{DD}}_{ik}(t) f^n_{kj}(t)  - \overline{\alpha^{n,\mathrm{DD}}_{ik}(t) f^n_{kj}(t)}\left[ z^n(t) \right] \right]\delta \left[ z-z^n(t) \right] 
    \label{eq:adij_mun_mean_subs} \\
        a^{n,\mathrm{DD}}_{ij}(z,t) &= \frac{1}{L_x L_y} \sum\limits_n^{N_\mathrm{mol} } \alpha^{n,\mathrm{DD}}_{ik}(t) f^{n,\mathrm{D}}_{kj}(t) \delta \left[ z-z^n(t) \right]\rightarrow \frac{1}{L_x L_y} \sum\limits_n^{N_\mathrm{mol} } \left[\alpha^{n,\mathrm{DD}}_{ik}(t) f^{n,\mathrm{D}}_{kj}(t)  - \overline{\alpha^{n,\mathrm{DD}}_{ik}(t) f^{n,\mathrm{D}}_{kj}(t)} \left[ z^n(t) \right] \right] \delta \left[ z-z^n(t) \right]  \label{eq:addij_mun_mean_subs} \\
        a^{n,\mathrm{Q}}_{ijk}(z,t) &= \frac{1}{L_x L_y} \sum\limits_n^{N_\mathrm{mol} } \alpha^{n,\mathrm{QD}}_{ijl}(t) f^n_{lk}(t) \delta \left[ z-z^n(t) \right]\rightarrow \frac{1}{L_x L_y} \sum\limits_n^{N_\mathrm{mol} } \left[\alpha^{n,\mathrm{QD}}_{ijl}(t) f^n_{lk}(t)  - \overline{\alpha^{n,\mathrm{QD}}_{ijl}(t) f^n_{lk}(t)}  \left[ z^n(t) \right] \right] \delta \left[ z-z^n(t) \right] \, .   \label{eq:aqijk_Qn_mean_sub}
\end{align}
Here, the overbar denotes a conditional time average, where the mean is taken over all times the molecule has a specific z-position.
This scheme does not alter the spectrum, as the center of mass does not oscillate at the frequencies of interest, as shown in Figure \ref{fig:origin_sfg_signal}, and consequently the net contribution due to the moving mean polarizability needs to be zero. 
To provide a numerical proof for this, we compare the electric dipole contribution $\tilde{S}^{(2,\mathrm{D})}_{ijk}(\omega^\mathrm{IR})$ with and without this treatment in the bending and stretching frequency region. 
This comparison is presented in Figure \ref{fig:mean_subs} A-D.
Here, we do not smooth the signal. 
Rather, we bin the spectra using a bin size of $\Delta \omega^\mathrm{IR}= \SI{3.9}{THz}$. 
We compute the binned SFG spectra for each of our 94 trajectories separately. 
We compute the 95\% confidence interval according to the well-known relation
\begin{align}
\mathrm{err}=1.96\frac{\sigma}{\sqrt{N}} \, ,
\label{eq:yerr}
\end{align}
where $\sigma$ is the standard deviation.
As becomes evident, subtracting the mean enhances the signal-to-noise ratio without statistically significant modifications of the spectral line shape. 
We conclude that the modifications in the expressions \eqref{eq:adij_mun_mean_subs}-\eqref{eq:aqijk_Qn_mean_sub} are numerically robust and do not introduce artifacts.
Of course, this is only true because the center of mass does not oscillate at the frequency $\omega^\mathrm{IR}$.
Hence, we do not subtract the mean in the test of the origin dependence in Section \ref{sec:origin_sfg}. \textcolor{changed}{As can be seen in Figure \ref{fig:mean_subs}, the signal-to-noise ratio is high, indicating that noise does not significantly affect our analysis.}
\textcolor{changed}{\section{Absolute SFG Spectra and Configuration Analysis}
\begin{figure}
 \centering 
 \includegraphics[scale=1]{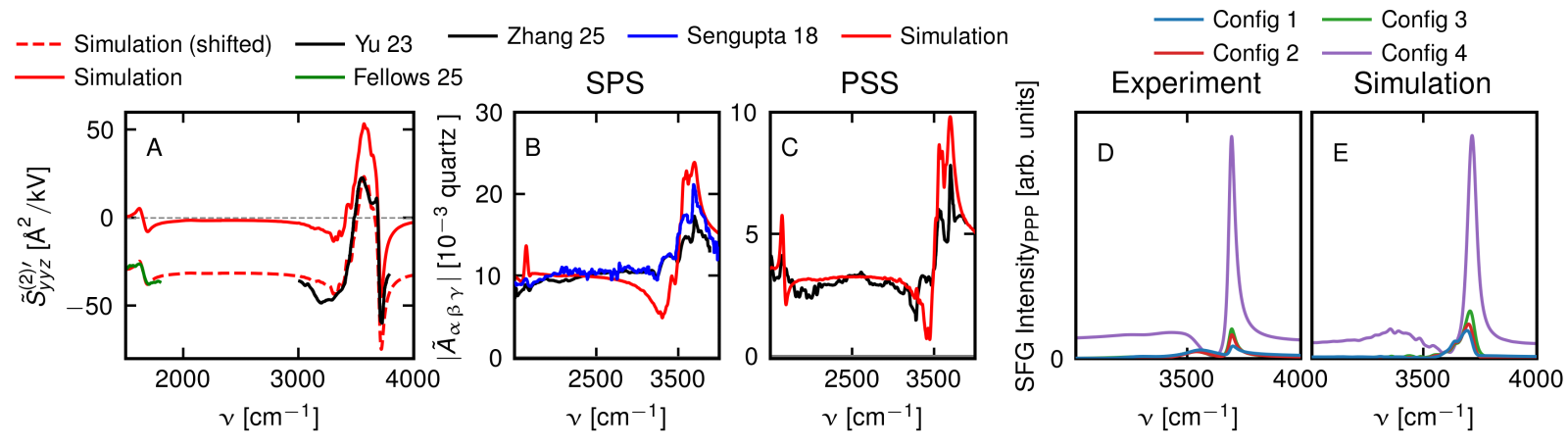}
 \caption{\textcolor{changed}{A: Graphical estimation of the baseline correction of the real part of the SFG spectrum. We use the experimental spectrum from Yu et al. \cite{yuFresnelFactorCorrection2023} as a reference and apply the baseline correction to the spectrum from Fellows et al. \cite{fellowsImportanceLayerDependentMolecular2025b} and to our simulated prediction. B \& C: Comparison of the polarization combinations SPS and PSS of the experimental absolute SFG spectra defined in Equations \eqref{eq:SFG_amplitude_SPS} and \eqref{eq:SFG_amplitude_PSS} with simulations. The experimental data are taken from Zhang \textit{et al.} \cite{zhangQuantitativeConsistencyIntensity2025a} and  Sengupta \textit{et al.} \cite{senguptaNeatWaterVapor2018a}. 
 D \& E: Configuration analysis of the experimentally detected radiation in PPP geometry.
 In D, we present an experimental configuration analysis extracted from Gan \textit{et al.} \cite{ganPolarizationExperimentalConfiguration2006}. We theoretically recreate this experimental configuration analysis in E using our theoretical SFG spectra $\tilde{S}^{(2)}_{ijk}(\omega^\mathrm{IR})$ and Equation \eqref{eq:SFG_intensity_PPP}. }}
 \label{fig:real_part_comp}
\end{figure}
Here, we list details on the comparison of absolute SFG spectra presented in Figure 1 (k) in the main text, present results for the polarization combinations SPS and PSS, and present a comparison between the experimental and the simulated configuration analysis, which is used to estimate bulk multipole contributions.
\subsection{Details on the Comparison of Absolute SFG Spectra}
We compare the absolute experimental SFG spectrum $\left|\tilde{S}^{(2)}_{yyz}(\omega^\mathrm{IR})\right|=\sqrt{\tilde{S}^{(2)\prime}_{yyz}(\omega^\mathrm{IR})^2+\tilde{S}^{(2)\prime \prime}_{yyz}(\omega^\mathrm{IR})^2}$ from various groups \cite{senguptaNeatWaterVapor2018a,yuFresnelFactorCorrection2023,fellowsImportanceLayerDependentMolecular2025b,zhangQuantitativeConsistencyIntensity2025a} with our theoretical prediction in Figure 1 (k) in the main text.
From Equation \eqref{eq:R_SFG} follows that the amplitude of the second-order radiation is proportional to the complex SFG amplitude $\tilde A_{\alpha \beta \gamma}(\omega_\mathrm{IR})$, which is given by \cite{moritaTheorySumFrequency2018a}
\begin{align}
 \tilde{A}_{\mathrm{SSP}}(\omega_\mathrm{IR}) &= \mathcal{L}^\mathrm{SFG}_y \mathcal{L}^\mathrm{VIS}_y \mathcal{L}^\mathrm{IR}_z \tilde{S}^{(2)}_{yyz}(\omega^\mathrm{IR}) \, , 
 \label{eq:SFG_amplitude_SSP} \\
 \tilde{A}_{\mathrm{PSS}}(\omega_\mathrm{IR}) &= \mathcal{L}^\mathrm{SFG}_z \mathcal{L}^\mathrm{VIS}_y \mathcal{L}^\mathrm{IR}_y \tilde{S}^{(2)}_{zyy}(\omega^\mathrm{IR}) \, , 
 \label{eq:SFG_amplitude_PSS} \\
 \tilde{A}_{\mathrm{SPS}}(\omega_\mathrm{IR}) &= \mathcal{L}^\mathrm{SFG}_y \mathcal{L}^\mathrm{VIS}_z \mathcal{L}^\mathrm{IR}_y \tilde{S}^{(2)}_{yzy}(\omega^\mathrm{IR}) \, , 
 \label{eq:SFG_amplitude_SPS} \\
 \tilde{A}_{\mathrm{PPP}}(\omega_\mathrm{IR}) &= \mathcal{L}^\mathrm{SFG}_z \mathcal{L}^\mathrm{VIS}_z \mathcal{L}^\mathrm{IR}_z \tilde{S}^{(2)}_{zzz}(\omega^\mathrm{IR}) - \mathcal{L}^\mathrm{SFG}_x \mathcal{L}^\mathrm{VIS}_x \mathcal{L}^\mathrm{IR}_z \tilde{S}^{(2)}_{yyz}(\omega^\mathrm{IR}) + \mathcal{L}^\mathrm{SFG}_z \mathcal{L}^\mathrm{VIS}_x \mathcal{L}^\mathrm{IR}_x \tilde{S}^{(2)}_{zyy}(\omega^\mathrm{IR}) - \mathcal{L}^\mathrm{SFG}_x \mathcal{L}^\mathrm{VIS}_z \mathcal{L}^\mathrm{IR}_x \tilde{S}^{(2)}_{yzy}(\omega^\mathrm{IR}) \, ,
 \label{eq:SFG_amplitude_PPP} 
\end{align}
where $\mathcal{L}^\alpha_i$ are the optical factors defined in Equations \eqref{eq:F_ext_x}-\eqref{eq:F_ext_z} and we used the rotational symmetry of the planar interface. The indices $\alpha,\beta,\gamma \in \lbrace \mathrm{S,P} \rbrace$ define the polarization of the outgoing, visible, and infrared beams, respectively.  
Here, S denotes polarization perpendicular ("senkrecht") to the plane of incidence and P polarization parallel to the plane of incidence. We use the reported experimental geometry from Sengupta \textit{et al.} \cite{senguptaNeatWaterVapor2018a} and
Zhang \textit{et al.} \cite{zhangQuantitativeConsistencyIntensity2025a} and the values $\tilde{n}_\mathrm{w}=1.33$ \cite{haynesCRCHandbookChemistry2015a} and $\tilde{n}_\mathrm{q}=1.55$ \cite{pressleyCRCHandbookLasers1971} for the refractive indices of water and quartz, respectively, in the computation of the optical factors defined in Equations \eqref{eq:F_ext_x}-\eqref{eq:F_ext_z} and the wavevectors mismatch $\Delta k_z$ defined in Equation \eqref{eq:delta_k}.
For the comparison of absolute spectra in Figure 1 (k) of the main text, we compute the absolute spectra from the phase-resolved measurements of Yu \textit{et al.} \cite{yuFresnelFactorCorrection2023} and Fellows \textit{et al.} \cite{fellowsImportanceLayerDependentMolecular2025b}.
Since Fellows \textit{et al.} \cite{fellowsImportanceLayerDependentMolecular2025b} used D$_2$O as reference, we apply a $\SI{-20}{\angstrom^2/kV}$ baseline offset to the real part. 
Similarly, we do not predict the off-resonant contribution in this work, 
which makes it necessary to add a baseline offset of $\SI{-30}{\angstrom^2/kV}$ to the theoretical prediction of $\tilde{S}^{(2)\prime}_{yyz}(\omega^\mathrm{IR})$.
These baseline offsets are graphically estimated by comparison with the phase-resolved spectrum measured by Yu \textit{et al.} \cite{yuFresnelFactorCorrection2023} and are presented in Figure \ref{fig:real_part_comp} A.
The $|\tilde{S}^{(2)}_{yyz}(\omega^\mathrm{IR})|$ spectrum reported by Zhang \textit{et al.} \cite{zhangQuantitativeConsistencyIntensity2025a} employs a different normalization convention than the other spectra presented in this work, which we account for by dividing the experimental data by $2 \tilde\varepsilon_\mathrm{eff}$, where $ \tilde\varepsilon_\mathrm{eff}=\frac{\tilde \varepsilon(\tilde \varepsilon+5)}{4\tilde \varepsilon+2}$ with $\tilde\varepsilon=1.78$ \cite{haynesCRCHandbookChemistry2015a}.
To compare with the spectrum from Sengupta \textit{et al.} \cite{senguptaNeatWaterVapor2018a}, which is presented in a.u., we normalize it such that the free-OH peak height agrees with the spectrum reported by Zhang \textit{et al.} \cite{zhangQuantitativeConsistencyIntensity2025a} and divide the spectrum by the appropriate optical factors.
While the spectrum from Zhang \textit{et al.} \cite{zhangQuantitativeConsistencyIntensity2025a} was constructed from the published fit parameters, the spectrum from Sengupta \textit{et al.} \cite{senguptaNeatWaterVapor2018a} was extracted from the published figure. 
\subsection{Results for SPS and PSS Polarizations}
We compare the SFG amplitudes $|\tilde A_{\mathrm{SPS}}(\omega_\mathrm{IR})|$ and $|\tilde A_{\mathrm{PSS}}(\omega_\mathrm{IR})|$, predicted from our multipolar theory, with experimental results from Zhang \textit{et al.} \cite{zhangQuantitativeConsistencyIntensity2025a} and Sengupta \textit{et al.} \cite{senguptaNeatWaterVapor2018a}, 
using Equations \eqref{eq:SFG_amplitude_PSS} and \eqref{eq:SFG_amplitude_SPS}. 
We avoid rescaling of experimental spectra, whenever possible.
 Therefore, as Zhang \text{et al.} \cite{zhangQuantitativeConsistencyIntensity2025a} report the absolute SFG spectra relative to the SFG amplitude of quartz, we need to divide our theoretical predictions by the geometry-dependent SFG amplitudes of quartz to allow for quantitative comparison.
 These are determined by \cite{pressleyCRCHandbookLasers1971,boydNonlinearOptics2008,thamerQuantitativeDeterminationNonlinear2019}
 \begin{align}
 \tilde A^\mathrm{q}_{\mathrm{SSP}} &= -\mathcal{L}^\mathrm{SFG,q}_y \mathcal{L}^\mathrm{VIS,q}_y \mathcal{L}^\mathrm{IR,q}_x \tilde{S}^{(2)}_{q} \\ 
 \tilde  A^\mathrm{q}_{\mathrm{PSS}} &= -\mathcal{L}^\mathrm{SFG,q}_x \mathcal{L}^\mathrm{VIS,q}_y \mathcal{L}^\mathrm{IR,q}_y \tilde{S}^{(2)}_{q} \\
 \tilde  A^\mathrm{q}_{\mathrm{SPS}} &= -\mathcal{L}^\mathrm{SFG,q}_y \mathcal{L}^\mathrm{VIS,q}_x \mathcal{L}^\mathrm{IR,q}_y \tilde{S}^{(2)}_{q} \\
 \tilde  A^\mathrm{q}_{\mathrm{PPP}} &= \mathcal{L}^\mathrm{SFG,q}_x \mathcal{L}^\mathrm{VIS,q}_x \mathcal{L}^\mathrm{IR,q}_x \tilde{S}^{(2)}_{q} \, .
 \end{align}
The SFG signal of quartz is given by
$|\tilde{S}^{(2)}_{q}|=\SI{1.2}{pm V^{-1}} (\Delta k_z)^{-1}$ \cite{boydNonlinearOptics2008}.
We compare experimental measurements of $|\tilde{A}_{\alpha \beta \gamma}(\omega^\mathrm{IR})|/|\tilde{A}^\mathrm{q}_{\alpha \beta \gamma}|$ with our multipolar predictions in Figure \ref{fig:real_part_comp} B \& C. 
 Here, the spectra reported by Sengupta \textit{et al.} \cite{senguptaNeatWaterVapor2018a} are normalized so that they have the same intensity at $\nu=\SI{2500}{cm^{-1}}$ as the spectra reported by Zhang \textit{et al.} \cite{zhangQuantitativeConsistencyIntensity2025a}. Further, the spectra by Sengupta \textit{et al.} \cite{senguptaNeatWaterVapor2018a} are divided by the optical factors of their setup and then multiplied by those of the setup used by Zhang \textit{et al.} \cite{zhangQuantitativeConsistencyIntensity2025a}, to account for the different experimental configuration. 
 We shift the predicted real parts $\tilde{S}^{(2)\prime}_{zzz}(\omega^\mathrm{IR})$, $\tilde{S}^{(2)\prime}_{yzy}(\omega^\mathrm{IR})$ and  $\tilde{S}^{(2)\prime}_{zyy}(\omega^\mathrm{IR})$ by 
 $\SI{-32}{\angstrom^2/kV}$, $\SI{-22}{\angstrom^2/kV}$ and $\SI{-7}{\angstrom^2/kV}$, respectively, to account for the non-resonant contribution that is missing in our theory. 
 These parameters reproduce the baseline around $\nu=\SI{2500}{cm^{-1}}$ of the experiments visible in Figure \ref{fig:real_part_comp} B-E.
 Although the SPS and PSS polarization combinations yield weaker SFG signals than SSP (which is proportional to $|\tilde{S}^{(2)}_{yyz}(\omega^\mathrm{IR})|$ shown in Figure 1 (k) in the main text) and are therefore more prone to errors, experiment and theory still agree quite well, as can be seen in Figure \ref{fig:real_part_comp} B \& C.
To our knowledge such a quantitative comparison of different polarizations has not been presented before.
\subsection{Configuration Analysis}
 In addition to the multipole contributions arising from the integral of multipole gradients over the interface predicted in this work, there are also bulk multipole contributions discussed in Section \ref{sec:BM}. As explained in Section \ref{sec:BM} and previously by Shiratori and Morita \cite{shiratoriTheoryQuadrupoleContributions2012}, these bulk multipole contributions do depend on the experimentally tunable wavectors. Hence, bulk multipoles can be analyzed by experimental configuration analysis, i.e., by varying the incident angles of the incoming beams and analyzing the measured SFG spectrum. 
Experimental configuration analysis was performed in 2006 by Gan \textit{et al.} \cite{ganPolarizationExperimentalConfiguration2006} in a different context. 
It is theoretically expected that bulk multipole contributions are most relevant in PPP geometry \cite{shiratoriTheoryQuadrupoleContributions2012}, for which we present the configuration analysis in Figure \ref{fig:real_part_comp} D \& E. 
 Combining Equations \eqref{eq:R_SFG} and \eqref{eq:SFG_amplitude_PPP}, the absolute value of the experimentally detectable radiation is proportional to the SFG Intensity defined by
\begin{align}
 \mathrm{SFG\,Intensity}_\mathrm{\alpha \beta \gamma} (\omega_\mathrm{IR}) = \frac{\left| \tilde{A}_{\alpha \beta \gamma}(\omega^\mathrm{IR}) \right|^2}{\cos^{2} \theta_2^\mathrm{SFG}}  \, ,
 \label{eq:SFG_intensity_PPP}
\end{align}
where $\theta_2^\mathrm{SFG}$ is the angle of refraction of the SFG beam in medium 2. 
\begin{table}
    \centering
    \caption{\textcolor{changed}{Incident angles for the visible and IR beams used in configurations 1–4 from Gan \textit{et al.} \cite{ganPolarizationExperimentalConfiguration2006}.}}
    \label{tab:angles}
    \begin{tabular}{c c c}
        \hline
        \noalign{\vskip 3pt}
        Configuration & $\theta_2^{\mathrm{VIS}}$ \, [deg] & $\theta_2^{\mathrm{IR}}$ \, [deg] \\
        \noalign{\vskip 3pt}
        \hline
        1 & 39 & 55 \\
        2 & 45 & 55 \\
        3 & 48 & 57 \\
        4 & 63 & 55 \\
        \hline
    \end{tabular}
\end{table}
We use the reported configurations 1-4 and the tabulated fit parameters from Table III of reference \cite{ganPolarizationExperimentalConfiguration2006} to compare experimental measurments and theoretical predictions of $\mathrm{SFG\,Intensity}_\mathrm{PPP}$. The incident angles in configurations 1-4 are summarized in Table \ref{tab:angles}.
Even though bulk multipoles are not considered in our theoretical framework, we can reproduce the experimental configuration analysis from Gan \textit{et al.} \cite{ganPolarizationExperimentalConfiguration2006}. This finding indicates that at OH-stretch frequencies,  interface multipole contributions are dominant compared to bulk multipole contributions. To our knowledge, there exists no published experimental configuration analysis in the bending region, which could further clarify the relevance of bulk multipoles at the air-water interface. }
\section{The Fluctuation-Dissipation Theorem}
\label{app:fd_and_kk}
\subsection{Classical Formulation}
\label{sec:classical_fdt}
In this section, we derive the relation between the classical linear response function
\begin{align}
\varphi[O(\cdot), P_i(\cdot), t ] = -\Theta(t ) \int  \mathrm{d} \vec \Omega  O( \vec \Omega )  e^{t  \lbrace H_0(\vec \Omega), \cdot \rbrace } \lbrace  P_i(\vec \Omega)  , \rho^{(0)} (\vec \Omega) \rbrace
\label{eq:response_function1}
\end{align}
and equilibrium correlation functions. A more general formulation without the classical approximation is given in the well-known publication from Kubo in 1966 \cite{kuboFluctuationdissipationTheorem1966}.
In the canonical ensemble, the equilibrium distribution is given by
\begin{align}
\rho^{(0)}(\vec \Omega) = \frac{e^{-\beta H_0 (\vec \Omega)}}{Z_{NVT}},
\label{eq:nvt_ensemble}
\end{align}
where $Z_{NVT}$ is the partition function and the inverse thermal energy is defined as $\beta = \frac{1}{k_B T}$, where $k_B$ is the Boltzmann constant.
With Equation \eqref{eq:nvt_ensemble} we evaluate the Poisson bracket in Equation \eqref{eq:response_function1} and obtain
\begin{align}
\varphi[O(\cdot), P_i(\cdot), t] = \beta \Theta(t) \int   \mathrm{d} \vec \Omega \rho^{(0)}(\vec \Omega) \dot{P}_i(\vec \Omega)  e^{-t \lbrace H_0(\vec \Omega), \cdot \rbrace } O(\vec \Omega) \, ,
\label{eq:response_function2}
\end{align}
after using Hamilton's equations and the anti-self-adjoint property of the Liouville operator. 
The integral in Equation \eqref{eq:response_function2} can be identified as the cross-correlation function, i.e.
\begin{align}
\varphi[O(\cdot), P_i(\cdot), t]  = \beta \Theta(t) \langle O(t) \dot{ P_i}(0) \rangle \, .
\label{eq:response_function3}
\end{align}
Using integration by parts one obtains the fluctuation-dissipation theorem\cite{kuboFluctuationdissipationTheorem1966}
\begin{align}
\varphi[O(\cdot), P_i(\cdot), t]  &= -\beta \Theta(t) \frac{\partial}{\partial t} \langle  O(t)  P_i(0) \rangle \\
&= -\beta \Theta(t) \dot{ C}_{O P_i} (t),
\label{eq:fdt}
\end{align}
where we introduce  $C_{O P_i}(t)$ as an abbreviation for the correlation function. 
We compute the Fourier transformation of Equation \eqref{eq:fdt} and substitute $\Theta(t)= \frac{1}{2} \left[\mathrm{sgn}(t) + 1 \right]$, which leads to
\begin{align}
\tilde{\varphi}[O(\cdot), P_i(\cdot), \omega]= -\frac{\beta}{2} \mathcal{F}\left[ \mathrm{sgn}(t) \dot C_{ O P_i}(t) \right]
- \frac{\beta}{2} \mathcal{F}\left[\dot C_{O P_i}(t) \right],
\label{eq:ft_resp}
\end{align}
Now we consider two real observables $S(t)$ and $A(t)$, with different symmetries $C_{SP_i}(t)=C_{SP_i}(-t)$ and 
$C_{A P_i}(t)=-C_{A P_i}(-t)$. In the former case we can relate the imaginary part of the linear response function $\tilde{\varphi}[S(\cdot), P_i(\cdot), \omega]''$ to the Fourier transformation of the equilibrium correlation function, i.e.
\begin{align}
\tilde{\varphi}''[S(\cdot), P_i(\cdot), \omega] &= \frac{\beta \omega}{2} \tilde{C}_{S P_i} (\omega) \\
&= \frac{\beta \omega}{2 t^\mathrm{SIG}}  \mathrm{Re}\left[ \tilde{S} (\omega)\tilde{P}_i (\omega)^* \right] \, .
\end{align}
In the latter case the real part $\tilde{\varphi}'[A(\cdot), P_i(\cdot), \omega]$ is connected to the Fourier transformation of the correlation function by
\begin{align}
\tilde{\varphi}'[A(\cdot), P_i(\cdot), \omega]&= i \frac{\beta \omega}{2} \tilde{C}_{A P_i} (\omega)\\
&= -\frac{\beta \omega}{2 t^\mathrm{SIG}} \mathrm{Im} \left[ \tilde{A}(\omega)  \tilde{P}_i(\omega)^* \right].
\end{align}
Here, we use the correlation theorem, which derivation is given in Section \ref{SI:xcorr}, and $t^\mathrm{SIG}$ is the time of measurement.
If one examines Equation \eqref{eq:ft_resp}, one sees that the real and imaginary parts of the response function are related by multiplication with the signum function in the time domain. 
In the frequency domain this corresponds to a convolution with the Fourier transformed signum function $\tilde{\mathrm{sgn}}(\omega)$, i.e. 
\begin{align}
\tilde{\varphi}'[O(\cdot), P_i(\cdot), \omega] = \int\limits_{-\infty}^\infty \mathrm{d} \omega' 
\tilde{\mathrm{sgn}}(\omega - \omega')\tilde{\varphi}''[O(\cdot), P_i(\cdot), \omega'],
\label{eq:kramers_kronig}
\end{align}
which is known as the Kramers-Kronig relation. 
However, Equation \eqref{eq:kramers_kronig} cannot be straightforwardly computed numerically. This problem can be circumvented by transforming back into the time domain and multiplying with $\mathrm{sgn(t)}$ if one wants to retrieve the real from the imaginary part or vice versa.
\textcolor{changed}{
\subsection{Approximating Quantum Response Functions}
We revisit the classical fluctuation-dissipation theorem in Section \ref{sec:classical_fdt}.
Here we compare the classical linear response function
\begin{align}
    \phi_\mathrm{C}(t)= \Theta(t) \left\langle \lbrace A(0), B(t) \rbrace \right\rangle_\mathrm{C}   
    \label{eq:classical_response_function}
\end{align}
and the quantum-mechanical equivalent \cite{kuboFluctuationdissipationTheorem1966}
\begin{align}
    \phi_\mathrm{Q}(t)= \frac{\Theta(t)}{i \hbar } \left\langle[  \hat A(0), \hat B(t) ] \right\rangle_\mathrm{Q} \, .
    \label{eq:quantum_response_function}
\end{align}
Here,  $\lbrace A, B \rbrace $ is the Poisson bracket defined in Equation \eqref{eq:poisson}, $[\hat A, \hat B]=\hat A\hat B - \hat B \hat A$ is a commutator, $A$ and $B$ are observables and the hat symbol denotes the corresponding operators. The classical thermal average is defined as 
\begin{align}
    \langle A \rangle_\mathrm{C} = \frac{1}{Z_\mathrm{C}}\int \mathrm{d} \vec \Omega e^{-\beta H_0 (\vec \Omega)} A( \vec \Omega) \, 
    \label{eq:thermal_average_C}
\end{align}
and the quantum-mechanical thermal average is defined as \cite{mahanManyParticlePhysics1990} 
\begin{align}
    \langle A \rangle_\mathrm{Q} = \frac{1}{Z_\mathrm{Q}} \sum \limits_{n=0}^\infty e^{-\beta H_n} \left\langle \Psi_n | \hat A | \Psi_n \right\rangle \, ,
    \label{eq:thermal_average_Q}
\end{align}
where $H_n$ is the energy level of the $n$-the Eigenstate with wavefunction $\Psi_n$ and $Z_\mathrm{C}$ and $Z_\mathrm{Q}$ are the classical and quantum-mechanical partition function, respectively. We also define classical and quantum-mechanical correlation functions as
\begin{align}
    C_\mathrm{C}(t) &= \langle A(0) B(t) \rangle_\mathrm{C} 
    \label{eq:correlation_function_C}  \, ,
    \\
    C_\mathrm{Q}(t) &= \langle \hat A(0) \hat B(t) \rangle_\mathrm{Q} \, .
    \label{eq:correlation_function_Q}
\end{align}
The following relations hold \cite{kuboFluctuationdissipationTheorem1966} 
\begin{align}
    \int\limits_{-\infty}^\infty \mathrm{d} t e^{-i \omega t} \left\langle \lbrace A(0), B(t) \rbrace \right\rangle_\mathrm{C}  = -i\beta \omega \int\limits_{-\infty}^\infty \mathrm{d} t \, e^{-i \omega t} C_\mathrm{C}(t) \\
    \int\limits_{-\infty}^\infty \mathrm{d} t e^{-i \omega t} \frac{1}{i \hbar } \left\langle [ \hat A(0), \hat B(t) ] \right\rangle_\mathrm{Q}  = \frac{1}{i \hbar }(1-e^{-\beta \hbar \omega}) \int\limits_{-\infty}^\infty \mathrm{d} t \, e^{-i \omega t} C_\mathrm{Q}(t)  \, .   
\end{align}
If we assume that the left-hand sides are identical, meaning that the response functions are the same, we obtain the relation between the classical and quantum Fourier-transformed correlation functions as
\begin{align}
\tilde{C}_\mathrm{Q}(\omega) \approx \frac{\beta \hbar \omega}{1 - e^{-\beta \hbar \omega}} \tilde{C}_\mathrm{C}(\omega) \, .
\label{eq:harmonic_quantum_factor}
\end{align}
Equation \eqref{eq:harmonic_quantum_factor} has led to the introduction of the harmonic quantum correction factor $Q_\mathrm{HA}=\frac{\beta \hbar \omega}{1-e^{-\beta \hbar \omega}}$ in the literature \cite{ramirezQuantumCorrectionsClassical2004a,auerVibrationalSumfrequencySpectroscopy2008a}. However, the quantity which determines the SFG response is the quantum response function $\phi_\mathrm{Q}(t)$ defined in Equation
\eqref{eq:quantum_response_function}, not the quantum correlation function  $C_\mathrm{Q}(t)$ defined in Equation \eqref{eq:correlation_function_Q}.
In the following, we present a model calculation for the classical and quantum-mechanical SFG response functions to derive the best possible estimate of  $\phi_\mathrm{Q}(t)$ from classical trajectories. \\ \\
We assume that we have a chemical bond whose length deviates by $x$ from the equilibrium length $x_0$, composed of two atoms, e.g., oxygen and hydrogen. In the Born-Oppenheimer approximation, the valence electrons wavefunction depends only parametrically on the nuclei positions. Hence, permanent and induced dipole moments are functions of the nuclei positions only. As chemical bonds are, in general, quite stiff, the dependence of dipole moment and polarizability on the bond length coordinate is well approximated by a first-order Taylor expansion 
\begin{align}
    \alpha_{ij}( x ) &\approx  \alpha_{ij}^0 + \Delta \alpha_{ij} x \\
   \mu_i( x ) &\approx  \mu_i^0 + \Delta \mu_i x \, .
\end{align}
The parameters $\Delta \alpha_{ij}$ and $\Delta \mu_i$ are known as transition polarizability and dipole moment, respectively. 
Similarly, the potential energy of a chemical bond is well-described by the leading order expansion
\begin{align}
V(x)=\frac{m}{2}\omega_0^2 x^2 - x \Delta \mu_i  E_i^\mathrm{L,IR}(t) \, ,
\end{align}
where $m$ is the mass and $\omega_0$ the eigenfrequency and $E_i^\mathrm{L,IR}(t)$ is the local electric E-field of frequency $\omega^\mathrm{IR}$.
As discussed in Section \ref{sec:lin_spons_born_opp}, the second-order dipole is determined by
\begin{align}
\mu^{(2)}_i(t) = \left[ \alpha_{ij}(t) - \alpha^0_{ij} \right] \mathcal{E}^\mathrm{L,VIS}_j e^{-i\omega^\mathrm{VIS} t}\, .
\end{align}
The IR-field driven polarizability $\alpha_{ij}(t)$ defines the resonant hyperpolarizability
\begin{align}
   \beta^\mathrm{Q}_{ijk}(t)= \frac{\Delta \alpha_{ij} \Delta \mu_k \Theta(t)}{i \hbar \varepsilon_0 } \left\langle[  \hat x(0), \hat x(t) ] \right\rangle_\mathrm{Q} \, .
   \label{eq:hyperpol_Q}
\end{align}
The Heisenberg position operator can be expressed as \cite{mahanManyParticlePhysics1990}
\begin{align}
    \hat x(t) = \sqrt{ \frac{\hbar}{2 m \omega_0}} \left( \hat{a} e^{-i \omega_0 t } + \hat{a}^* e^{i \omega_0 t } \right) \, .
\end{align}
Here $\hat{a}$ and $\hat{a}^*$ are the so-called lowering and raising operators. Using $[\hat{a},\hat{a}^*]=1$, the commutator in Equation \eqref{eq:hyperpol_Q} turns out  to be
\begin{align}
    [\hat x(0),\hat x(t)]=\frac{i \hbar}{m \omega_0} \sin{(\omega_0 t)} \, .
    \label{eq:xx_commutator}
\end{align}
The thermal averaging procedure does not modify this result, and the quantum hyperpolarizability is thus given by
\begin{align}
    \beta^\mathrm{Q}_{ijk}(t)=\frac{\Delta \alpha_{ij} \Delta \mu_k}{m \omega_0 \varepsilon_0} \Theta(t)\sin{(\omega_0 t)} \, .
    \label{eq:quantum_hyperpol}
\end{align}
An approximation for the quantum hyperpolarizability is the classical hyperpolarizability, which can be predicted by the classical fluctuation-dissipation theorem
\begin{align}
\beta^\mathrm{C}_{ijk}(t) = -\beta \Delta \alpha_{ij} \Delta \mu_k \Theta(t) \frac{\partial}{\partial t}\langle x(0)x(t) \rangle_C \, .
\end{align}
The solution for the classical observable $x(t)$ is 
\begin{align}
    x(t)= x_0 \cos{(\omega_0 t)} + \frac{p_0}{m \omega_0} \sin{(\omega_0 t)} \, ,
\end{align}
where $p(t)=m \dot{x}(t)$ is the momentum and $p_0=p(0)$ and $x_0=x(0)$ are the initial conditions. Now, we compute the thermal average in the classical canonical ensemble
\begin{align}
    \langle x(0) x(t)\rangle_\mathrm{C} &= \int \mathrm{d} x_0 \int \mathrm{d} p_0 \,  \rho^{(0)} (x_0,p_0) x_0 x(t) \\
    &= \frac{1}{\beta m \omega_0^2} \cos{(\omega_0 t)} \, .
\end{align}
Hence, the classical hyperpolarizability turns out to be equal to the quantum hyperpolarizability in Equation \eqref{eq:quantum_hyperpol}, i.e.
\begin{align}
    \beta^\mathrm{C}_{ijk}(t) = \frac{ \Delta \alpha_{ij} \Delta \mu_k}{m \omega_0 \varepsilon_0 } \Theta(t)\sin{(\omega_0 t)} = \beta^\mathrm{Q}_{ijk}(t) \, .
\end{align}
This result reflects the well-known fact that quantum and classical harmonic oscillators share the same positional response function \cite{caldeiraQuantumTunnellingDissipative1983,ramirezQuantumCorrectionsClassical2004a}. 
Harmonic quantum corrections modify the correlation functions, but these correlation functions must be used with the matching quantum or classical fluctuation–dissipation theorem \cite{kuboFluctuationdissipationTheorem1966} in order to predict the correct response functions and from that vibrational spectra.
Hence, to leading order in a Taylor expansion of the bond potential, the classical and quantum response functions used for SFG spectra prediction are identical and no quantum correction factor is present.
}
\subsection{The Correlation Theorem}
\label{SI:xcorr}
An equilibrium cross correlation function between two observables $A(t) \in \mathbb{C}$ and $B(t) \in \mathbb{C}$ is given by
\begin{align}
C_{AB}(t) = \lim\limits_{t^\mathrm{SIG}\rightarrow \infty}\frac{1}{t^\mathrm{SIG}} \int\limits_{-t^\mathrm{SIG}/2}^{t^\mathrm{SIG}/2} \mathrm d t' A(t+t') B(t'),
\label{eq:SI_cross_corr_func1}
\end{align}
where $t^\mathrm{SIG}$ is the measurement time.
We take the Fourier transform of Equation \eqref{eq:SI_cross_corr_func1}
\begin{align}
\tilde{C}_{AB}(t) = \lim\limits_{t^\mathrm{SIG}\rightarrow \infty}\frac{1}{t^\mathrm{SIG}} 
\int\limits_{-t^\mathrm{SIG} /2 }^{t^\mathrm{SIG} / 2} \mathrm d t e^{i \omega t}
\int\limits_{-t^\mathrm{SIG}/2}^{t^\mathrm{SIG}/2} \mathrm d t' A(t+t') B(t'),
\label{eq:SI_cross_corr_func3}
\end{align}
substitute $t'' = t+t'$ and obtain
\begin{align}
\tilde{C}_{AB}(\omega) = \lim\limits_{t^\mathrm{SIG}\rightarrow \infty}\frac{1}{t^\mathrm{SIG}} 
\int\limits_{-t^\mathrm{SIG} /2 }^{t^\mathrm{SIG} / 2}  \mathrm d t''  e^{i \omega t''} A(t'')
\int\limits_{-t^\mathrm{SIG}/2}^{t^\mathrm{SIG}/2} \mathrm d t'  e^{-i \omega t'}  B(t'),
\label{eq:SI_cross_corr_func4}
\end{align}
which gives us 
\begin{align}
\tilde{C}_{AB}(\omega) = \lim\limits_{t^\mathrm{SIG}\rightarrow \infty}\frac{1}{t^\mathrm{SIG}} \mathrm{FT}[\Pi(t/t^\mathrm{SIG}) A(t)] (\omega) \, \mathrm{FT}[\Pi(t/t^\mathrm{SIG}) B(t)^*](\omega)^* \, ,
\label{eq:x_corr_theorem}
\end{align}
where $\Pi(t/t^\mathrm{SIG})$ is the rectangular function defined in Equation \eqref{eq:rect_func} and $\mathrm{FT}[x(t)](\omega)=\tilde{x}(\omega)$ denotes a Fourier transformation. The discrete version of Equation \eqref{eq:x_corr_theorem} is known as the correlation theorem \cite{smithiiiMathematicsDiscreteFourier2007} and the superscript $^*$ denotes complex conjugation.
\section{Relationship Between Electric and External Fields}
\label{sec:local_and_ext} 
The second-order response profile $\tilde s^{(2)}_{ijk}\left( z,\omega^{\mathrm{VIS}}, \omega^{\mathrm{IR}} \right)$ describes the second-order electric current density, due to two-wave mixing of $z$-polarized D-fields and $x$- or $y$-polarized E-fields as described by Equations \eqref{eq:ext_field_maxwell} and \eqref{eq:s2ijk}. 
the first-order response relation between z-polarized D-fields and z-polarized E-fields is given by
\begin{align}
    \mathcal{E}^\alpha_z(z) &= \int\limits_{-\infty}^\infty \mathrm d z' \tilde \varepsilon^{-1,\mathrm{NL}}_{zz}(z,z',\omega^\alpha) \mathcal{D}^\alpha_z \, ,
    \label{eq:functional_epansion_eps}
    \end{align}
 which can be simpfield to
    \begin{align}
    \mathcal{E}^\alpha_z(z) & =\varepsilon_0^{-1} \tilde{\varepsilon}^{-1}_{zz}(z, \omega^\mathrm{\alpha})  \mathcal{D}^\alpha_z \,,
    \label{eq:def_inv_eps_zz}
\end{align}
since $z$-polarized D-fields are spatially constant.
Here, $\tilde{\varepsilon}^{-1,\mathrm{NL}}_{zz}(z,z',\omega)$ is the non-local inverse dielectric function and 
\begin{align}
\tilde{\varepsilon}^{-1}_{zz} \left( z, \omega \right) = \int \mathrm{d} z' \tilde{\varepsilon}^{-1,\mathrm{NL}}_{zz}(z,z',\omega)
\end{align} 
is the inverse dielectric profile  \cite{sternCalculationDielectricPermittivity2003, bonthuisDielectricProfileInterfacial2011, bonthuisProfileStaticPermittivity2012, bonthuisContinuumHowMolecular2013, gekleNanometerResolvedRadioFrequencyAbsorption2014, locheCommentHydrophobicSurface2019a}.
As the $x/y$ component of the E-fields are constant in space, we obtain similarly 
\begin{align}   \varepsilon_0^{-1}\mathcal{D}^\alpha_{x/y}(z) &=  \tilde{\varepsilon}_{xx/yy} \left(z, \omega^\alpha \right) \mathcal{E}^\alpha_{x/y}\, .
    \label{eq:def_eps_xx}
\end{align}
Furthermore, we define
\begin{align}    \tilde{\varepsilon}_{zz}(z, \omega  ) = \frac{1}{\tilde{\varepsilon}^{-1}_{zz}(z, \omega  ) } \, ,
    \label{eq:def_eps_zz}
\end{align}
as the dielectric profile parallel to the interface normal. 
We note that $ \tilde{\varepsilon}_{zz}(z, \omega  )$ can have poles, as $\tilde{\varepsilon}^{-1}_{zz}(z,\omega )$ can be equal to zero.
Indeed, in the static limit ($\omega = 0$), $\tilde{\varepsilon}^{-1}_{zz}(z, 0)$ crosses zero multiple times at the water–graphene interface.\cite{locheCommentHydrophobicSurface2019a}
The $zz$ component of the dielectric profile tensor is determined by the linear response of the polarization profile $\tilde{s}^{(1,\mathrm{P})}_{ij}(z, \omega)$ defined in Equation \eqref{eq:def_s1P_ij} to a $z$-polarized external field, i.e.
\begin{align}
p_z^{(1)}(z,t) = e^{-i \omega^\alpha t} \tilde{s}^{(1,\mathrm{P})}_{zz}(z, \omega^\alpha) \mathcal{D}^\alpha_z + c.c. \, .
\label{eq:first_order_pz}
\end{align}
Inserting Equations \eqref{eq:first_order_pz} and \eqref{eq:def_inv_eps_zz} into Equation \eqref{eq:def_EField} leads to the inverse dielectric profile \cite{sternCalculationDielectricPermittivity2003, bonthuisDielectricProfileInterfacial2011, bonthuisProfileStaticPermittivity2012, bonthuisContinuumHowMolecular2013, gekleNanometerResolvedRadioFrequencyAbsorption2014}
\begin{align}
\tilde \varepsilon_{zz}^{-1}(z, \omega) = 1 -\tilde{s}^{(1,\mathrm{P})}_{zz}(z, \omega) .
\label{eq:inv_eps_zz}
\end{align}
From Equation \eqref{eq:ext_field_maxwell} follows that the lateral external fields $F_{x/z}(t)$ can be identified as E-fields, leading to
\begin{align}
\varepsilon_0^{-1} p_{x/y}^{(1)}(z,t) = e^{-i \omega^\alpha t} \tilde{s}^{(1,\mathrm{P})}_{xx/yy}(z, \omega^\alpha) \mathcal{E}^\alpha_{x/y} + c.c. \, .
\label{eq:first_order_pxxyy}
\end{align}
Combining Equations \eqref{eq:first_order_pxxyy}, \eqref{eq:def_eps_xx} and \eqref{eq:def_EField} leads to
\begin{align}
\tilde \varepsilon_{xx/yy}(z,\omega) &= 1 +  \tilde s^{(1,\mathrm{P})}_{xx/yy}(z,\omega) \, .
\label{eq:lateral_EField_x}
\end{align}
Since the external field in the bulk region always equals the E-field, as stated in Equation \eqref{eq:ext_field_maxwell_bulk}, the bulk dielectric constant can be expressed analogously to Equation \eqref{eq:lateral_EField_x} as 
\begin{align}
    \tilde \varepsilon_{ij}(\omega) &= 1 + \tilde s^{(1,\mathrm{P})}_{ij}(\omega) \, ,
\label{eq:dielectric_bulk}
\end{align}
and does not depend on $z$.
In an isotropic system, the dielectric constant obeys $ \tilde \varepsilon_{ij}^\alpha= \delta_{ij}\tilde \varepsilon^\alpha$.
\bibliography{mybib}